\documentclass[a4paper,12pt,twoside,openright]{report}

\usepackage{lineno,hyperref}
\usepackage[nolist]{acronym}
\usepackage[onehalfspacing]{setspace}
\usepackage{booktabs}
\usepackage{graphicx}
\usepackage{float}
\usepackage{amsmath}
\usepackage{listings}
\usepackage{cite}
\usepackage{tikz}
\usepackage{wasysym}
\usepackage{amssymb}
\usepackage{dirtree}
\usepackage[figuresright]{rotating}

\usepackage{subcaption}
\usepackage{adjustbox}

\usetikzlibrary{matrix,calc}
\usepackage[para*]{manyfoot}
\DeclareNewFootnote[para]{A}
\let\oldfootnoteA\footnoteA
\renewcommand{\footnoteA}[1]{%
\oldfootnoteA{\makebox[.32\dimexpr\textwidth-2\footglue\relax][l]{#1}}}
\begin{document}
\thispagestyle{empty}
\begin{center}
\large{DISS.\,ETH\,NO. 25833}\\
\vfill
\Huge{\textbf{Cherenkov-Plenoscope}}\\
\large{A ground-based Approach in\\One Giga Electron Volt\\One Second Time-To-Detection\\Gamma-Ray-Astronomy}\\
\vfill
\Large
A thesis submitted to attain the degree of\\
DOCTOR OF SCIENCES of ETH ZURICH\\
(Dr. sc. ETH Zurich)
\vfill
\normalsize
presented by \\[5pt]
{\Large Sebastian Achim Mueller }\\[5pt]
M. sc. in Physics, TU Dortmund University\\
born on 5 April 1988\\[5pt]
citizen of\\
Germany\\
\vfill
accepted on the recommendation of\\
\vspace{0.5cm}
Prof. Dr. Adrian Biland\\
Prof. em. Dr. Dr. h.c. Felicitas Pauss\\
Hon.-Prof. Dr. Werner Hofmann\\
\vfill
2019
\end{center}
%
%------------------------------------------------------------------------------
\newpage
\section*{Abstract}
Telescopes -- 'far seeing' -- have since centuries revealed insights to objects at cosmic distances.
Adopted for gamma-ray-astronomy, ground based Cherenkov-telescopes image the faint Cherenkov-light of air-showers induced by cosmic gamma-rays rushing into earth's atmosphere.
In the race for the lowest possible energy-threshold for cosmic gamma-rays, these Cherenkov-telescopes have become bigger, and now reached their physical limits.
The required structural rigidity for image-quality constrains a cost-effective construction of telescopes with apertures beyond $30\,$meter in diameter.
Moreover, as the aperture increases, the narrower depth-of-field
irrecoverably blurs the images what prevents the reconstruction of the cosmic particle's properties.
To overcome these limits, we propose plenoptic-perception with light-fields.
Our proposed $71\,$meter Cherenkov-plenoscope requires much less structural rigidity and turns a narrow depth-of-field into three-dimensional reconstruction-power.
With an energy-threshold for gamma-rays of one Giga electron Volt, 20\,times lower than what is foreseen for the future planned Cherenkov-Telescope-Array (CTA), the Cherenkov-plenoscope could become the portal to enter the sub second time-scale of the highly variable gamma-ray-sky.
We present the Cherenkov-plenoscope in Part \ref{PartPlenoscope} of this thesis.\\
In Part \ref{PartPhotonStream}, we push Cherenkov-astronomy by sensing Cherenkov-light in the quantum-regime.
A key ability of Cherenkov-telescopes, and our proposed Che\-ren\-kov-plen\-os\-cope, is the detection of few Cherenkov-photons within the ever-present nightly pool of photons emitted by stars, zodiacal dust, atmospheric glow, and others.
Photo-sensors and electronics have made huge progress, and now reach a regime where the quantized nature of photons can be resolved.
We present the identification of single-photons during regular observation-conditions with the $3.6\,$meter Cherenkov-telescope named FACT on Canary island La Palma, Spain.
We implement a true single-photon-representation for air-shower-records and compare it to established representations which are usually highly entangled with the photo-sensors and electronics in use and thus usually do not have a quantized description.
Our representation contains the arrival-times of single-photons which makes it the most natural, and arguably the most interchangeable representation possible for Cherenkov-astronomy.
With the complete time-structure of the air-shower's Cherenkov-photons, our  single-photon-representation has the potential to improve the reconstruction of the cosmic particle's properties.
\newpage
\section*{Zusammenfassung}
Das Teleskop -- 'In die Ferne sehen' -- gew\"ahrt seit Jahrhunderten Einblicke auf Objekte in kosmischen Distanzen.
\"Ubernommen fuer die Astronomie der Gammastrahlen bilden bodengebundene Cherenkov-Teleskope das schwache Cherenkov-Licht ab, welches von Luftschauern ausgeht.
Luftschauer werden durch kosmische Strahlung erzeugt, wenn diese in die irdische Atmosph\"are eindringt.
Im Rennen um die kleinst m\"ogliche Energieschwelle fuer kosmische Gammastrahlen sind Cherenkov-Teleskope immer gr\"ossser geworden und stossen nun an ihre physikalischen Grenzen.
Die ben\"otigte strukturelle Steifigkeit f\"ur die Abbildungsqualit\"at verhindert einen kosteneffiziente Bau von Teleskopen mit Aperturdurchmessern von mehr als 30 Metern.
Mehr noch, mit wachsenden Aperturen verw\"ascht die immer schmalere Tiefensch\"arfe die Bilder unwiederruflich und verhindert die Rekonstruktion der kosmischen Strahlung.
Um diese Grenzen zu \"uberwinden, schlagen wir eine plenoptische Wahrnehmung und die Verwendung von Lichtfeldern vor.
Unser hier vorgeschlagenes 71\,Meter Cherenkov-Plenoskop ben\"otigt viel weniger strukturelle Steifigkeit und wandelt eine schmale Tiefensch\"arfe in drei dimensionale Rekonstruktionskraft um.
Mit einer Energieschwelle von einem Giga Elektronen Volt f\"ur Gammastrahlen, 20\,mal geringer als die voraussichtliche Energieschwelle f\"ur das geplante Cherenkov-Telescope-Array (CTA), k\"onnte unser Cherenkov-Plenoskop das Portal werden um in den h\"ochst variablen Gammastrahlenhimmel auf Zeitskalen unterhalb einer Sekunde vorzudringen.
In Teil \ref{PartPlenoscope} dieser Arbeit pr\"asentieren wir das Cherenkov-Plenoskop.
In Teil \ref{PartPhotonStream} dieser Arbeit treiben wir die Gammastrahlungsastronomie weiter voran indem wir Cherenkov-Licht im Quantenbereich erfassen.
Eine Schl\"usself\"ahigkeit von Cherenkov-Teleskopen und unserem vorgeschlagenen Cherenkov-Plenoskop ist die Detektierung von wenigen Cherenkov-Photonen inmitten des allgegenw\"artigen Sees aus n\"achtlichen Photonen ausgesandt von Sternen, Zo\-di\-ak\-staub, at\-mos\-ph\"ar\-ischem Gl\"uhen und anderem.
Photosensoren und Elektronik haben gewaltige Fortschritte gemacht und erlauben es nun, die quantisierte Natur von Photonen aufzul\"osen.
Wir pr\"asentieren die Identifikation von Einzelphotonen w\"ahrend regul\"aren Beobachtungsbedingungen auf dem 3,6\,Meter Cherenkov-Teleskop namens FACT auf der kanarischen Insel La Palma, Spanien.
Wir implementieren eine echte Einzelphotonendarstellung f\"ur Luftschaueraufnahmen und vergleichen diese mit etablierten Darstellungen, welche f\"ur gew\"ohnlich hochgradig mit den jeweilig verwendeten Photosensoren und Elektroniken verstrickt sind und darum in der Regel keine quantisierte Darstellung haben.
Unsere Darstellung enth\"alt die Ankunftszeiten von Einzelphotonen was sie zur nat\"urlichsten und best austauschbaren Darstellung f\"ur die Cherenkov-Astronomie macht.
Mit der voll\-st\"an\-digen Zeitstruktur der Cherenkov-Photonen im Luftschauer hat unsere Einzelphotondarstellung das Potential, die Re\-kon\-struk\-tion der kosmischen Strahlung zu ver\-bess\-ern.
%
%------------------------------------------------------------------------------
\tableofcontents
%------------------------------------------------------------------------------
%
\chapter*{Introduction}
Earth is not the center of the universe.
A simple conclusion drawn from observations of the night-sky that changed a complete society and marked the starting point of our modern quest for knowledge.
During this quest, the fields of physics, astronomy, and cosmology merged closer together.
Observations of the sky, beyond earth, became important for scientific progress in seemingly opponent fields which investigate the innermost structure of matter.
So let us take a closer look into the sky.
Already with our eyes we see structures like stars, the moon, the sun, and  our home galaxy named Milky-Way.
With telescopes \mbox{('far seeing')} we see more complex structures like moons of planets, nebulae, and foreign galaxies.
It is the light which provides us with insights as it travels from distant objects to our eyes.\\
However, there is light which the human-eye can not see.
New windows beyond the visible-light were opened in the past 100\,years.
For example, the window of invisible radio-light was opened and became a pillar of modern astronomy.
Now, a young, and novel window of high energetic gamma-rays is about to open to astronomy.
And we do what we can to tear open this window.
\section*{Gamma-ray-astronomy}
\label{SecGammaRayAstronomy}
Gamma-rays are single photons so energetic that they can disrupt the nuclei of an atom with bounding-energies of MeV or even GeV.
Although gamma-rays are electromagnetic radiation as visible- or radio-light, the term gamma-\textit{ray} already implies that their wave-character is hardly recognizable in most interactions.
At energies onwards from some MeV, the particle-character dominates and so we often call it gamma-ray instead of gamma-radiation.
Beside the gamma-rays produced on earth by nuclear decay, there are also gamma-rays of cosmic origin.
In fact, there are not only cosmic gamma-rays but also other high-energetic particles of cosmic origin like protons, electrons, and heavy ions.
On earth, the atmosphere shields us from cosmic particles but when we go up in altitude the increase in flux of cosmic particles becomes evident \cite{hess1912uber}.
In todays gamma-ray-astronomy, high energetic charged particles create great challenges in the observations of gamma-rays and have to be distinguished from gamma-rays.
Even more challenging, charged particles are much more abundant and outshine even the brightest sources of gamma-rays.
With this said, the three \footnote{In the regime of MeV-energies, even the polarization \cite{dean2008polarized} of a gamma-ray can be measured.} goals of gamma-ray-astronomy today are to measure
\begin{itemize}
    \item the incident-direction of a gamma-ray,
    \item the energy of a gamma-ray,
    \item and the type of particle to assert that the cosmic particle actually was a gamma-ray.
\end{itemize}
All these three measurements must be done for each individual event of an incoming cosmic gamma-ray or incoming cosmic-ray which is different from optical- and radio-astronomy.
In optical- and radio-astronomy, many cosmic photons contribute to a single measurement.
Cosmic particles can either be detected directly outside earth's shielding atmosphere, or indirectly by observing the interactions of the cosmic particles in the atmosphere.
Both of these methods are currently being investigated for gamma-ray-astronomy, and both have exclusive advantages above the other.
Similar to the spectrum\cite{olive2014Review} of energies of cosmic-rays, the spectrum of energies of gamma-rays decreases rapidly towards higher energies.
In the case of the Crab Nebula, one of the brightest known sources of cosmic gamma-rays in the sky, gamma-rays with energies of 100\,GeV are two to three orders-of-magnitude more abundant than gamma-rays with energies of 1\,TeV \cite{aleksic2015measurement}.
This steep decrease of the spectrum of the energy causes the observations of gamma-rays to be dominated by particles with energies close to the lower energy-threshold of the detectors.
\section*{Detecting gamma-rays directly}
\begin{figure}
    \begin{center}
        \includegraphics[width=1.0\textwidth]{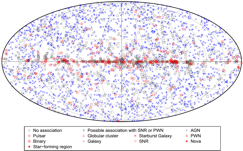}
        \caption[Gamma-ray-sky seen by the Fermi-LAT satellite]{The inert, but full-solid-angle sky of gamma-rays seen by the Fermi-LAT detector \cite{acero2015fermi3fgl}.
            AGN is short for Active-Galactic-Nucleus, PWN is short for Pulsar-Wind-Nebula, and SNR is short for Super-Nova-Remnant.
        }
        \label{FigFermi3fgl}
    \end{center}
\end{figure}

\begin{figure}
    \begin{center}
        \includegraphics[width=0.6\textwidth]{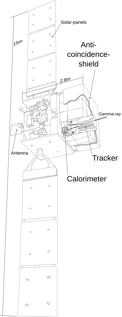}
        \caption[The Fermi-LAT satellite, schematic]{The Large-Area-Telescope (LAT) for gamma-rays named after Enrico Fermi.
        The LAT is dedicated to measure the three properties: Direction, energy, and type of cosmic particles.
        Schematics is inspired by two separate figures provided by \url{www.nasa.gov/glast}.
        }
        \label{FigFermiLatSchematic}
    \end{center}
\end{figure}

The direct detection of gamma-rays takes place in space, outside of earth's atmosphere.
Detectors in space wait for cosmic gamma-rays to interact in their mass.
Detectors in space have typically three components of which each one is dedicated to measure one of the three important properties of cosmic particles.
First, there is a tracker that reconstructs the trajectories of electrons and positrons created via pair-production $\gamma \rightarrow e^{+} + e^{-}$ to estimate the direction of the initial gamma-ray.
The trajectory of charged cosmic-rays can even be reconstructed directly by the tracker from the initial particle itself.
Second, there is a calorimeter where the secondary interaction-products can deposit all their energy in order to estimate the energy of the cosmic particle.
And third, there is an anti-coincidence-detector wrapped around the inner tracker and calorimeter.
The anti-coincidence-detector measures the type of the particle by distinguishing electrically charged cosmic-rays from neutral gamma-rays.
Figure \ref{FigFermiLatSchematic} shows all three components on the very successful Fermi-Large-Area-Telescope.\\
So detectors in space fulfill all demands of gamma-ray-astronomy.
However, detectors in space are limited in mass and volume and are unfortunately not expected to become any bigger in the foreseeable future of space-travel.
Since the mass of the detector is mandatory for the tracker and especially for the calorimeter, the effective areas for the detection of gamma-rays today are below $1\,$m$^{2}$.
Today, detectors in space are excellent to create static and wide field-of-view surveys of the inert gamma-ray-sky with years of exposure-time, see Figure \ref{FigFermi3fgl}.
Only in rare cases of an excessively high emission of gamma-rays from a temporarily flaring source, detectors in space are able to resolve time structures on the time-scale of days\footnote{%
This excludes assumptions about pulsed emission as it is commonly assumed for pulsars.
The emission-period of pulsars is typically tens of milliseconds, but a deviation from this pulsation will still only be detected after days of exposure-time.
} \cite{tavani2011discovery,abdo2011CrabFlare}.
\section*{Detecting gamma-rays indirectly}
Indirect detectors make use of the atmosphere of the earth.
When a cosmic particle or cosmic gamma-ray enters the atmosphere, it can interact with the molecules in the atmosphere.
Due to the high kinetic energy of the cosmic particle, the fragments and the newly created particles of such interactions travel further down the atmosphere almost parallel to the trajectory of the cosmic particle.
This interaction of particles continues in a cascade until the kinetic energies of the interaction-products are not sufficient anymore to further create new particles.
For energies below $\approx 80\,$MeV ionization becomes more likely than bremsstrahlung which effectively ends the cascade.
This destructive and cascaded process is called air-shower.
Depending on the initial kinetic energy of the cosmic particle, air-showers can be several kilometers long.
At energies above $10^{12}$\,eV, air-showers can reach the ground before the cascade reaches its climax.
There are two different types of air-showers.
First, there are air-showers which are dominated by electro-magnetic interactions.
Such air-showers are induced by cosmic particles of electro-magnetic type like the gamma-ray, or the electron.
Second, there are air-showers which have both hadronic and electro-magnetic interactions.
Those air-showers are induced by cosmic particles of hadronic type like the proton or the iron-nucleus.
\begin{figure}
    \begin{center}
        \includegraphics[width=1.0\textwidth]{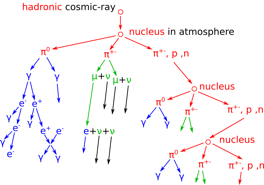}
        \caption[Interactions inside an air-shower]{The dominant interactions inside an air-shower.
            The hadronic-component is red, the electro-magnetic-component is blue, and the weak-decay-component is green.
            Figure is not to scale.
            For cosmic particles of electro-magnetic type, the air-shower starts directly inside the blue branch on the left.
        }
        \label{FigAirShowerInteractions}
    \end{center}
\end{figure}
Figure \ref{FigAirShowerInteractions} shows the two different air-shower types.
The momentum of secondary particles perpendicular to the trajectory of the cosmic particle, is larger in hadronic interactions than it is in electro-magnetic interactions.
Therefore, air-showers with hadronic components have a wider spread of secondary particles perpendicular to the trajectory of the cosmic particle.
The different geometry of narrow electro-magnetic air-showers, and wide and bushy hadronic air-showers allows us to reconstruct whether the cosmic particle was of hadronic, or of electro-magnetic type.
On ground, indirect detectors have sensors to detect the particles and radiations which are created in the air-shower and have not yet been absorbed again in the atmosphere.
Five types of interaction-products are likely to reach ground before they get absorbed in the atmosphere themselves.
First, blueish, visible Cherenkov-photons that are produced by charged particles in the air-shower that traverse the air faster than the local speed of light in this air.
Second, charged muons that originate from hadronic interactions in the air-shower are unlikely to undergo further interactions in the air.
Third, ultra-violet and visible photons produced by atoms and molecules which got ionized by the air-shower and later recombine.
Fourth, radio-waves which are produced by the separation of electric charges in the air-shower due to the magnetic-field of the earth, and the Askaryan-effect.
Fifth, neutrinos produced in e.g. the decay of pions and muons.
Only if the energy of the cosmic particle is high enough, the creation of new, short lived, particles might not have come to an end before the air-shower reaches the ground.
When the energies are high enough, any particle might be found in the so called air-shower-tail on ground.
As only secondary particles reach the sensors on ground, the measurement of the direction, the energy and the type of the cosmic particle, is not straightforward.
To measure the three properties, indirect detectors try to reconstruct the spatial geometry of the air-shower.
There are indirect detectors that only detect the charged particles which reach the ground \cite{allekotte2008surface} and are therefore referred to as air-shower-tail-detectors.
Dedicated air-shower-tail-detectors have a dense coverage on ground optimized to tell apart the type of the cosmic particle to do gamma-ray-astronomy \cite{deyoung2012hawc}.
Other indirect detectors are telescopes which record the photons produced by florescence in the air-shower horizontally from the side \cite{abraham2010fluorescence}.
Even a detector for florescence-photons which records air-showers from space while looking down onto earth is proposed \cite{ebisuzaki2014jem}.
And then there are telescopes that record the bluish Cherenkov-photons produced in the air-shower.
For gamma-rays with energies above several $10\,$GeV, Telescopes for Cherenkov-photons can measure the three features important to gamma-ray-astronomy (direction, energy, and type) and offer large collection-areas for gamma-rays of $>10^4$\,m$^2$ \cite{bernlohr2013monte}.
\section*{The Cherenkov-telescope}
Cherenkov-telescopes\footnote{Also called Imaging-Atmospheric-Cherenkov-Telescope (IACT).} measure the energy and direction of cosmic gamma-rays and other cosmic particles.
During the night, Cherenkov-telescopes record the incident-directions $c_x$, $c_y$ and the arrival-times $t$ of both night-sky-background-photons and Cherenkov-photons produced in the air-shower.
The short but intense time-structure of the flash of Cherenkov-photons on ground of $\sim 5\,$ns allows the Cherenkov-telescope to trigger and record the Cherenkov-photons within the pool of night-sky-background-photons.
By reconstructing the geometry of the air-shower from the recorded incident-directions and arrival-times of the photons, Cherenkov-telescopes estimate the three properties important to gamma-ray-astronomy: Direction, energy and type of the cosmic particle.
The more Cherenkov-photons a Cherenkov-telescope can record of an individual air-shower, the better the properties of the cosmic particle can be reconstructed.
Larger but costly apertures for Cherenkov-photons allow Cherenkov-telescopes to reconstruct air-showers induced by cosmic particles with lower energies.
Today, small Cherenkov-telescopes have $\approx 10\,$m$^2$ apertures for Cherenkov-photons and are able to detect gamma-rays with energies down to $\sim 1\,$TeV \cite{ICRC2014_fact_crab_spectrum}.
Large Cherenkov-telescopes have $\approx 200\,$m$^2$ apertures for Cherenkov-photons and are able to detect gamma-rays with energies down to $\sim 100\,$GeV \cite{tridon2010magic}.\\
So far Cherenkov-telescopes opened the high-energy $ > 100\,$GeV window of gamma-rays to astronomy \cite{kildea2007whipple} and have found $\approx 200$ sources of gamma-rays \cite{wakely2008tevcat}.
To further open the gamma-ray-window, we have to make gamma-ray-astronomy more powerful and cost effective.
\newcommand{\NameAcp}{Portal}
\newcommand{\NumPix}{8,443}
\newcommand{\NumPax}{61}
\newcommand{\NumLix}{515,023}
\newcommand{\ReflectorFocalLenght}{106.5}
\newcommand{\ReflectorDiameter}{71}
\newcommand{\NumFacets}{1,842}
%------------------------------------------------------------------------------
\part[Cherenkov-Plenoscope]
     {Cherenkov-Plenoscope\\[\bigskipamount]
      \large A ground-based Approach in\\One Giga Electron Volt\\One Second Time-To-Detection\\Gamma-Ray-Astronomy}
\label{PartPlenoscope}
%
%------------------------------------------------------------------------------
\chapter{Introduction}
\label{ChIntroduction}
Ground based Cherenkov-telescopes measure the energy and incident-direction of cosmic gamma-rays and other cosmic particles, such as protons and electrons.
Cosmic particles induce air-showers in earth's atmosphere where in turn Cherenkov-photons are emitted.
During the night, Cherenkov-telescopes record the incident-angles $c_x$, $c_y$ and the arrival-times $t$ of these Cherenkov-photons in a three-dimensional intensity-histogram \mbox{$\mathcal{I}$[$c_x$, $c_y$, $t$]} called image-sequence.
To sense the quick $\sim 1\,$ns flash of Cherenkov-photons within the pool of night-sky-background-photons, the Cherenkov-telescope records an image-sequence with $\sim 1\,$ billion images per second.
By reconstructing properties of the air-shower from the image-sequence \cite{hillas1985cerenkov}, Cherenkov-telescopes gather information about the cosmic particle to do astronomy.

The high energy-threshold of Cherenkov-telescopes is the main limit for studying astronomical emitters of gamma-rays.
For example, the study of pulsars is limited because for most of them the emission of gamma-rays is cut off at and above $\sim 10\,$GeV \cite{magic2008pulsar,abdo2009population}.
And in general the study of sources at cosmological distances like active-galactic-nuclei and gamma-ray-bursts is limited due to the absorption of high energetic gamma-rays in the extra-galactic background-light \cite{magic2008distantQuasar}.
For example, the gamma-ray-horizon \cite{magic2008distantQuasar} in the universe contains $\sim 4.7$ times more observable volume when being able to detect gamma-rays with energies as low as $100\,$GeV instead of $200\,$GeV.
Furthermore, the study of transient sources benefits from a low energy-threshold~\cite{aharonian2001}.
Currently the only way to measure gamma-rays with energies below a few tens of GeV are telescopes in space, e.g. Fermi-LAT~\cite{acero2015fermi3fgl}, which measure gamma-rays before they interact with earth's atmosphere.
The predominant limiting factor for space-telescopes is their small collection-area which unfortunately is not expected to become far bigger within the near future.
Beside their costs, space-telescopes with their wide coverage of the sky are great for static sources of gamma-rays and year long exposures~\cite{acero2015fermi3fgl}, but their $\sim 1\,$m$^2$ collection-area limits their abilities to resolve the highly variable gamma-ray-sky.

The energy-threshold of Cherenkov-telescopes itself is limited by the efficiency to detect Cherenkov-photons, as the number of the Cherenkov-photons is roughly proportional to the cosmic particle's energy~\cite{naurois2015}.
Therefore, Cherenkov-telescopes have put great effort into lowering their energy-thresholds, mainly by moving onto mountains to be closer to the air-shower, or by enlarging their aperture for Cherenkov-photons.
This way, current and upcoming Cherenkov-telescopes reach energy-thresholds as low as $\sim 20\,$GeV~\cite{magic2008pulsar,bernlohr2013monte}, while the most ambitious proposals for Cherenkov-telescopes, on the frontier of low energy-thresholds, strives to reach $5\,$GeV~\cite{aharonian2001}.
However, two physical limits reduce the maximum aperture for Cherenkov-photons on a single Cherenkov-telescope to below $\sim 30\,$m.
First, the square-cube-law~\cite{square_qube_law} makes building bigger Cherenkov-telescopes increasingly difficult due to the need for mechanical rigidity in order to keep the targeted optical geometry for the imaging-reflector and the image-sensor.
Second, the depth-of-field induced by larger apertures renders more and more parts of the recorded images blurred and thus erodes the power to reconstruct the particle type, energy and direction from the recorded Cherenkov-photons \cite{hofmann2001focus, mirzoyan1996optical}.
We discuss the origin and limitations of a narrow depth-of-field in Chapter \ref{ChDepthOfField}.
The depth-of-field-limit is a central reason why the next generation of Cherenkov-telescopes will not exceed $\sim 23\,$m aperture-diameter \cite{bernlohr2013monte}.
Furthermore there is the technological challenge of signal processing and routing which prevents us from combining individual Cherenkov-telescopes before the trigger-level into an array to lower the energy-threshold.
With todays electronics the trigger-decision in an array of Cherenkov-telescopes can not be taken on the combined aperture for Cherenkov-photons provided by all the Cherenkov-telescopes in the array, but has to be taken on each Cherenkov-telescope individually.
Although impressive efforts were made \cite{jung2005star,lopez2016topo} to take the trigger-decision based on combined information beyond the aperture of the individual Cherenkov-telescope, still the mayor part of information-reduction is made on the individual Cherenkov-telescope before the overall trigger of the array\cite{bulian1998characteristics,funk2004trigger,weinstein2007veritas,lopez2016topo}.

The potential of reaching $> 10^4\,$m$^2$ collection-areas for gamma-rays with the atmospheric Cherenkov-method motivated us to overcome these physical limits and technological challenges.
We propose to combine the atmospheric Cherenkov-method with the plenoptic-method~\cite{lippmann1908,adelson1992single,ng2005,wilburn2005high} to build the Cherenkov-plenoscope, a telescope for cosmic gamma-rays with the large collection-areas of Cherenkov-telescopes on ground and the low energy-threshold of telescopes in space.
%
%------------------------------------------------------------------------------
\chapter{Results}
\label{ChResults}
\begin{figure}
    \centering
    \includegraphics[width=1\textwidth]{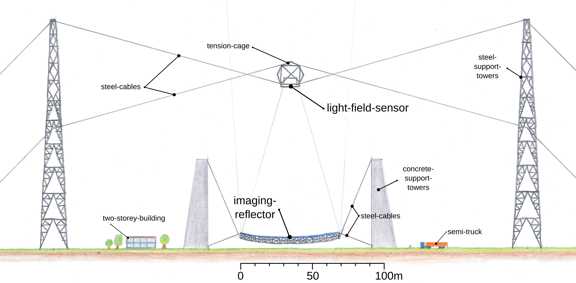}
    \caption[two-dimensional illustration of \NameAcp{}]{To scale illustration of the proposed Cherenkov-plenoscope \NameAcp{}.
    Its light-field-sensor and large imaging-reflector are suspended by cables.}
    \label{FigAcpOverview}
\end{figure}

The Cherenkov-plenoscope measures the incident-direction and energy of individual cosmic gamma-rays and cosmic-rays.
During the night, the Cherenkov-plenoscope records the air-showers induced by these cosmic particles from ground by measuring the entire (plenary) extrinsic state of the Cherenkov-photons.
It bins the photons depending on their incident-directions $(c_x, c_y)$, their support-positions on the aperture-plane $(x,y)$ and their arrival-times $t$ into a five-dimensional intensity-histogram $\mathcal{L}[c_x, c_y, x, y, t]$ called light-field-sequence.
By reconstructing the air-shower from the light-field-sequence, the Cherenkov-plenoscope gathers information about the cosmic particle to do astronomy.
To record light-fields, the Cherenkov-plenoscope uses a large imaging-reflector in combination with a light-field-sensor, see Figure \ref{FigAcpOverview}.
\section{Introducing the plenoscope's optics}
\begin{figure}
    \centering
    \includegraphics[width=0.9\textwidth]{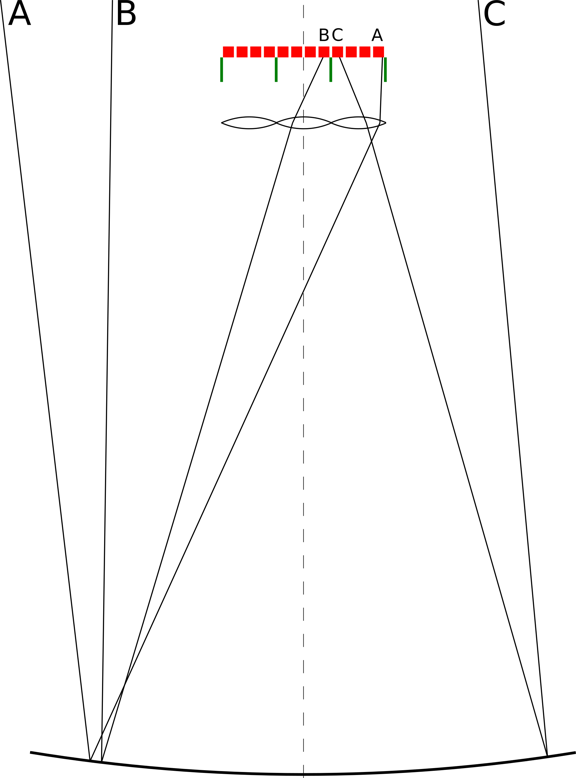}
    \caption[Optics of the plenoscope]{
    The optics of the plenoscope.
    On the bottom is the large imaging-reflector, on the top is the light-field-sensor.
    Here the light-field-sensor has $N=3$ small cameras with $M=4$ photo-sensors in each small camera.
    We demonstrate the trajectories of the photons A, B, and C.
    Photo-sensors are red, and walls in between small cameras are green.
    }
    \label{FigOpticsOverview}
\end{figure}

\begin{figure}
    \begin{minipage}{0.55\textwidth}
        \begin{figure}[H]
            \includegraphics[width=1.0\textwidth]{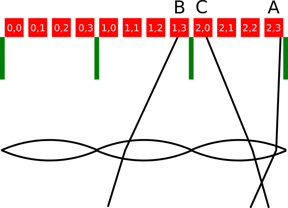}
        \end{figure}
    \end{minipage}
    \hspace{0.5cm}
    \begin{minipage}{0.4\textwidth}
        \begin{figure}[H]
            \includegraphics[width=1.0\textwidth]{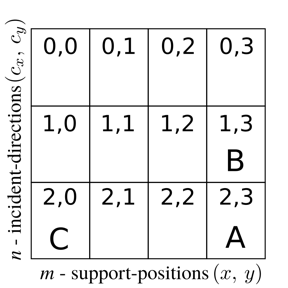}
        \end{figure}
    \end{minipage}
    \caption[Optics of the plenoscope, close-up]{
        Close-up on the optics of the light-field-sensor shown in Figure \ref{FigOpticsOverview}.
        The photo-sensors can be addressed in two dimensions $(n,m)$, see table on the right.
        The first index $n$ addresses the incident-direction.
        The second index $m$ addresses the support-position.
    }
    \label{FigOpticsOverviewCloseUp}
\end{figure}

The large imaging-reflector of the plenoscope reflects an incoming photon towards the light-field-sensor, see Figure \ref{FigOpticsOverview}.
Just as in a classic telescope, the large imaging-reflector reflects a photon towards a certain position on the sensor-plane depending on the photon's incident-direction.
On the classic image-sensor in a telescope, the photon was now absorbed by a photo-sensor and added to the recorded image.
However, the light-field-sensor of the plenoscope is a two-dimensional array of $N$ small cameras, in contrast to a classic image-sensor which is just an array of $N$ photo-sensors.
Each small camera is made out of a hexagonal lens and an image-sensor composed from $M$ photo-sensors right behind the hexagonal lens.
In the Figures \ref{FigOpticsOverview}, and \ref{FigOpticsOverviewCloseUp}, we demonstrate the observation of the three photons A, B, and C.
The photons A and B have different incident-directions and therefore are reflected onto different positions on the sensor-plane, where they enter different lenses of different small cameras.
Photon A enters the lens at $n=2$, and photon B enters the lens at $n=1$, see Figure \ref{FigOpticsOverviewCloseUp}.
However, since the support-positions of A and B are close together, they are both absorbed in the right most photo-sensor at $m=3$ on the image-sensor in their small cameras.
Photon C has a similar incident-direction as photon A and thus enters the same small camera at $n=2$.
However, since photon C has a different support-position than photon A, photon C is not absorbed by the right-most photo-sensor in its small camera, but in the photo-sensor at $m=0$.
Each photo-sensor $(n,m)$ corresponds to one specific bin in the light-field-intensity-histogram $\mathcal{L}[{{c_x}_n}, {{c_y}_n}, x_m, y_m]$, i.e. one specific light-field-cell.
Each of these light-field-cells $(n,m)$, or lixels for short, describe a bundle of photon-trajectories which can be approximated by a three-dimensional ray%
\begin{equation}
\vec{r}_{n,m}(\lambda) =
\begin{pmatrix}
x_m,\,
y_m,\,
0
\end{pmatrix}^T
+ \lambda
\begin{pmatrix}
{c_x}_n,\,
{c_y}_n,\,
\sqrt{1 - {{c_x}_n}^2 - {{c_y}_n}^2}
\end{pmatrix}^T.
\label{EqRay}
\end{equation}
Bins with the same incident-direction $n$ (${c_x}_n, {c_y}_n$) in the light-field are called a picture-cell, or a pixel for short.
And bins with the same support-position $m$ ($x_m, y_m$) in the light-field are called a principal-aperture-cell, or a paxel for short.\\
The plenoptic perception of photons has severe consequences, of which we introduce the very basics in Chapter \ref{ChDepthOfField}.
In Chapter \ref{ChOvercomingAberrations} we show how plenoptic perception can enlarge the field-of-view by overcoming the aberrations of imaging-optics \cite{hanrahan2006digital}.
Such aberrations are inevitable limits to imaging on telescopes.
In Chapter \ref{ChCompensatingMisalignmnets} we show how plenoptic perception loosens the constrains for the rigid alignment between the imaging-reflector and the sensor-plane.
Rigid alignment is a strong technological challenge on large telescopes.
In Chapter \ref{ChLightFieldGeometry} we show how we calibrate the response of the light-field-sensor in order to obtain a light-field-sequence $\mathcal{L}$ which describes photons in three-dimensional space and time.
In the Chapters \ref{ChInterpretingTheLightFieldSequence}, and \ref{ChTomography} we describe the rich and powerful ways to interpret the light-field-sequence $\mathcal{L}$ in order to reconstruct air-showers.
Finally in Chapter \ref{ChComparingOtherMethods} we compare the perception of established methods in Cherenkov-astronomy to the Cherenkov-plenoscope's perception.\\
In general, the light-field $\mathcal{L}[c_x, c_y, x, y]$ recorded by a single plenoscope is equivalent to $M$ images $\mathcal{I}_m[c_x, c_y]$ recorded by an array of $M$ telescopes located at different support-positions $(x, y)$.
Thus in the general case, a plenoscope and a dense array of telescopes can record the light-field in the same way.
However, in the particular case of quick flashes of Cherenkov-photons produced in air-showers, the Cherenkov-plenoscope has one crucial advantage over the array of Cherenkov-telescopes: Its trigger.
\section{Going around the trigger-challenge}
In an array of Cherenkov-telescopes, combining the intensity of close-by pixels $(c_x, c_y)$ from different Cherenkov-telescopes located at different positions $(x,y)$ would be crucial for the trigger-decision on the lowest possible energies.
Unfortunately this is very difficult to do.
The image-sensors of the individual Cherenkov-telescopes are separated by at least the diameter of their imaging-reflectors.
The trigger for an array of Cherenkov-telescopes needs time-delays which depend on the pointing-direction of the telescopes, and it has to reorganize all the signals from the individual telescopes from being bundled into near-by support-positions ($x,y$) to being bundled into near-by incident-directions ($c_x,c_y$).
The Cherenkov-plenoscope on the other hand has the ideal arrangement for a trigger that takes into account the full aperture of the large imaging-reflector.
In contrast to an array of Cherenkov-telescopes, the photo-sensors which represent similar incident-directions in $c_x$, and $c_y$, but belong to different support-positions in $x$, and $y$ are mechanically very close together inside the light-field-sensor ($<15\,$cm, see later Figure \ref{FigSmallCameraCluseUp}).
The plenoscope does neither need time-delays which depend on the pointing, nor does it need external, and flexible routing of signals.
In Chapter \ref{ChTrigger} we motivate the need for a trigger in Cherenkov-astronomy, and describe a possible implementation in a Cherenkov-plenoscope.
In the Sections \ref{SecSamplingTelescopeArrayTopoTrigger}, and \ref{SecSamplingTelescopeArrayDense} we show how this trigger-challenge is currently addressed on arrays of Cherenkov-telescopes.
\section{Deferring the square-cube-law}
Since the light-field-sensor records three-dimensional trajectories of photons rather than just the absorption-positions of photons, the alignment of the light-field-sensor with respect to the imaging-reflector is less constrained than the alignment of a conventional image-sensor.
As long as the actual misalignment of the light-field-sensor with respect to the imaging-reflector is known, the light-field-sensor can still record trajectories of photons.
It just samples a different region of the light-field.
And when the Cherenkov-photons of an air-shower are still within the sampled region of the light-field, a misalignment does not harm the observation-power for gamma-rays.
In Chapter \ref{ChCompensatingMisalignmnets} we discuss how the plenoscope can compensate misalignments between its light-field-sensor and its large imaging-reflector.
We further demonstrate that sharp images can still be obtained with strong misalignments.\\
The reduced demand for rigid alignment allows the Cherenkov-plenoscope to mechanically decouple the light-field-sensor from the imaging-reflector.
This way the Cherenkov-plenoscope can defer the physical limit of the square-cube-law and have larger, and more cost-effective apertures for Cherenkov-photons.
In Chapter \ref{ChCableRobotMount}, we propose a dedicated mount for the Cherenkov-plenoscope which explicitly takes advantage of these reduced demands for rigidity.
\section{Turning depth-of-field into tomography}
Again, since the Cherenkov-plenoscope records three-dimensional trajectories of photons rather than only the photons absorption-positions, the plenoscope overcomes the physical limit of the depth-of-field, as we motivate in Section \ref{SecPerceptionPlenoscope}.
However, the plenoscope not only overcomes the depth-of-field-limit but it turns the tables on it.
The narrow depth-of-field gives the plenoscope its three-dimensional reconstruction-power.
The recorded trajectories of the Cherenkov-photons can directly be used for a tomographic reconstruction of the air-shower, similar to reconstructions in light-field-microscopy\footnote{Also called focus-stack-deconvolution, or narrow-angle-tomography.} \cite{ng2006lightfieldmicroscopy}.
We discuss our first experiences with tomographic reconstructions of air-showers in Chapter \ref{ChTomography}.
Above this, all the established reconstruction-methods for air-showers can be applied as well.
For example we can project the light-field onto images focused to different object-distances \cite{ng2005}, see Section \ref{SecPostRefocusedImaging}.
On refocused images, the established \cite{hillas1985cerenkov} reconstruction of air-showers can be used with the Cherenkov-plenoscope.
We can project the light-field onto the areal intensity histogram in $x$, and $y$ to reconstruct the air-showers, as shown in \cite{chantell1998prototype, lizarazo2006data}.
And we can reconstruct the Cherenkov-light-front's three-dimensional structure in the moment when it rushes into the aperture-plane, as shown in \cite{fontaine1990aims}.
\section{Entering the \NameAcp{}, entering the 1\,GeV, 1\,s gamma-ray-sky}
To explore plenoptic perception in gamma-ray-astronomy, we propose a specific Cherenkov-plenoscope with an aperture-diameter of $71\,$m which we name \NameAcp{}, see Figure \ref{FigAcpOverview}.
With \NameAcp{} we introduce the cable-robot-mount to take advantage of the relaxed rigidity-constrains between the light-field-sensor and the imaging-reflector, see Chapters \ref{ChCompensatingMisalignmnets}, and \ref{ChCableRobotMount}.
The cable-robot-mount extensively uses computer-control to reduce the need for rigid and heavy structures.
It is inspired by the cable-suspended radio-receiver of the Arecibo-Observatory \cite{altschuler2002national}, the initial robot-crane-manipulator \cite{albus1993nist}, and the fast cable-robot-simulator \cite{miermeister2016cablerobot}.\\
\NameAcp{} uses first, a large, cable-suspended imaging-reflector with $71\,$m diameter and $106.5\,$m focal-length.
Second, \NameAcp{} uses a mechanically separated, cable-suspended light-field-sensor with $12.1\,$m diameter corresponding to $6.5^\circ$ field-of-view.
The two moving components are suspended independently of each other, such that the forces holding the light-field-sensor do not have to run through the large imaging-reflector and its mount.
\NameAcp{}'s light-field-sensor has $\NumLix{}$ light-field-cells (lixel) formed by $N=\NumPix{}$ small cameras equivalent to classical picture-cells (pixel) with $M=\NumPax{}$ principal-aperture-cells (paxel) each, see Chapter \ref{ChOpticsOfNameAcp}.
\NameAcp{} reaches zenith-distances up to $45^\circ$ without the zenith-singularity of altitude-azimuth-mounts \cite{borkowski1987near}, and has only thin cables shadowing its aperture.
During the day, \NameAcp{}'s light-field-sensor is parked on a pedestal next to the large imaging-reflector to ease service.
The independent pointing of both light-field-sensor, and large imaging-reflector without a zenith-singularity allows \NameAcp{} to point fast and hunt transient-sources.\\
\NameAcp{}'s goal is to drive the energy-threshold for gamma-rays down to $1\,$GeV to become the 'gamma-ray-timing-explorer' \cite{aharonian2001}.
To minimize losses of Cherenkov-photons \cite{aharonian2001}, and to maximize the three-dimensional reconstruction-power of air-showers, we propose to install \NameAcp{} high in altitude $\sim 5,000\,$m a.s.l\footnote{A.s.l. is short for above sea level.}.
Such exceptional dry, dark, and high sites with transparent atmospheres are already being used for astronomical instruments, such as the Llano de Chajnantor, Andes \cite{wootten2003alma}, for the southern hemisphere and Ali, Himalaya \cite{kuo2017assessments,ye2015tibet}, for the northern hemisphere.\\
In Chapter \ref{ChPictureTour}, we show pictures of the \NameAcp{} Cherenkov-plenoscope.
\section{Estimating \NameAcp{}'s sensitivity}
\label{SecPerformance}
\begin{figure}
    \centering
    \includegraphics[width=1\textwidth]{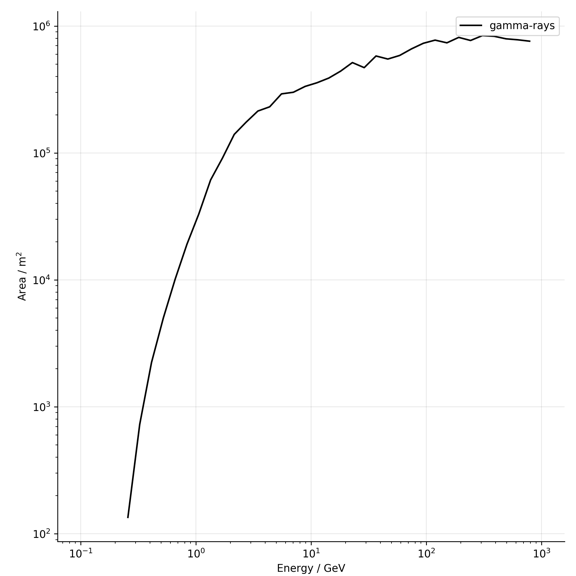}
    \caption[\NameAcp{}'s response to gamma-rays coming from a point-source]{
        Effective collection-area of \NameAcp{}'s trigger for gamma-rays coming from a point-source within the field-of-view.
        At $1\,$GeV, \NameAcp{} effectively collects gamma-rays with an area of $\approx 30,000\,$m$^2$.
    }
    \label{FigResponseGammaRays}
\end{figure}

\begin{figure}
    \centering
    \includegraphics[width=1\textwidth]{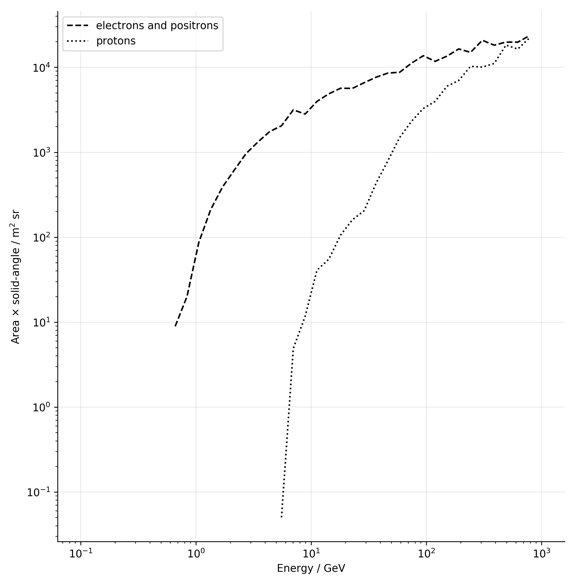}
    \caption[\NameAcp{}'s response to cosmic-rays coming from a diffuse source]{
        Effective collection-area and solid-angle of \NameAcp{}'s trigger for cosmic-rays with a diffuse distribution of incident-directions.
    }
    \label{FigResponseCosmicRays}
\end{figure}

\begin{figure}
    \centering
    \includegraphics[width=1\textwidth]{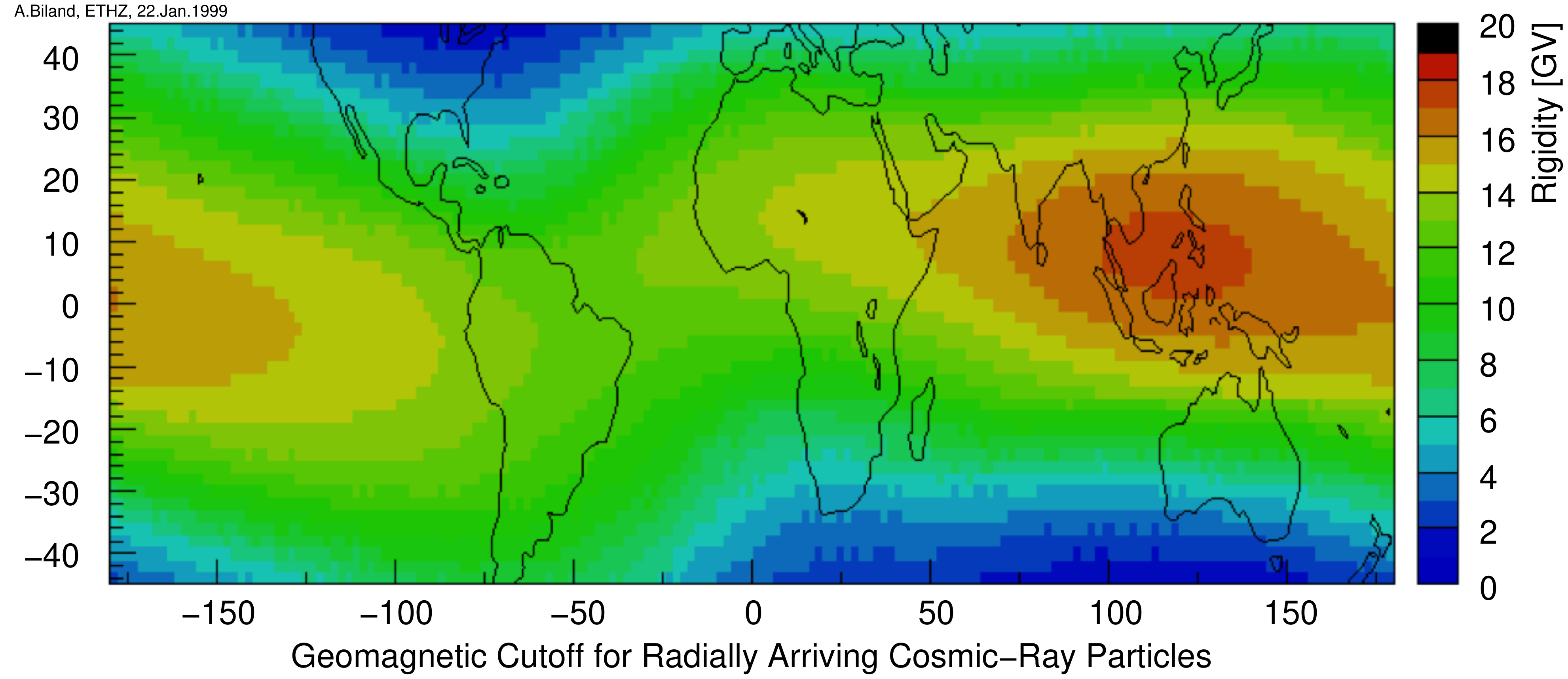}
    \caption[Geomagnetic cut-off-rigidity, global overview]{Figure provided by Adrian Biland.
        The geomagnetic cut-off-rigidity estimated based on the measured earth's magnetic-field and the simulation of cosmic-ray-trajectories.
        Different locations on earth have different cutoff-rigidities.
        In Atacama-desert in South-America, cutoff-rigidities of $\approx 10\,$GV can be reached.
    }
    \label{FigGeomagneticCutOffRigidity}
\end{figure}

\begin{figure}
    \centering
    \includegraphics[width=1\textwidth]{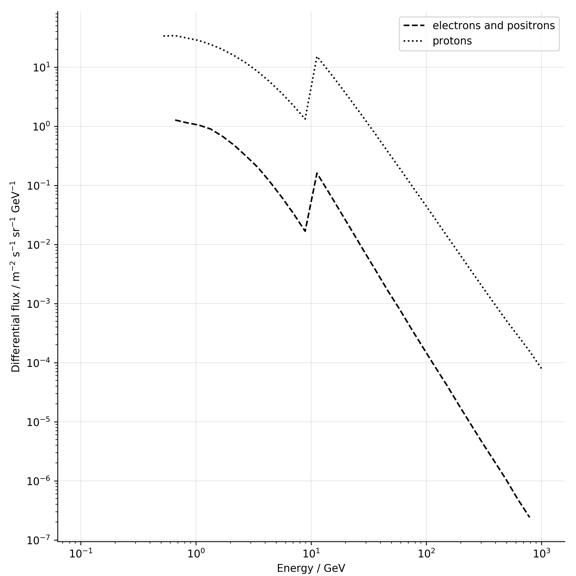}
    \caption[Flux of air-showers induced by charged particles]{The flux of air-showers initiated by charged particles in earth's atmosphere for a geomagnetic cutoff-rigidity of $10\,$GV.
        We chose a cutoff-rigidity of $10\,$GV as it is representative for a potential site in South-America, see Figure \ref{FigGeomagneticCutOffRigidity}.
        We are conservative and estimate that the flux of air-showers initiated by cosmic-rays with rigidities below $10\,$GV does not vanish completely, but is reduced to $5\%$.
        Initial flux of cosmic-rays is taken from AMS-02's precision measurements \cite{aguilar2014precision} and, \cite{aguilar2015precision}.
        For the flux of air-showers below the rigidity-cutoff we follow the ideas discussed in \cite{lipari2002fluxes}, and \cite{zuccon2003atmospheric}.
    }
    \label{FigAirShowerRateChargedCosmicRays}
\end{figure}

\begin{figure}
    \centering
    \includegraphics[width=1\textwidth]{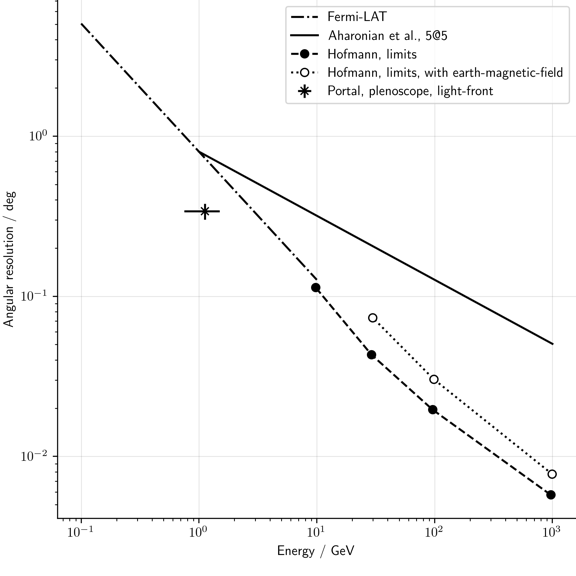}
    \caption[\NameAcp{}'s estimated angular resolution]{
        The angular resolution for the reconstructed incident-directions of gamma-rays.
        The angular resolution of Fermi-LAT \cite{fermi2010fermi}, the predicted angular resolution of the proposed 5@5 Cherenkov-telescope-array \cite{aharonian2001}, and the estimated limits for the angular resolution of ground based air-shower-observations using reconstruction-methods from the year 2006 \cite{hofmann2006performance}.
    }
    \label{FigAngularResolutions}
\end{figure}

\begin{figure}
    \centering
    \includegraphics[width=1\textwidth]{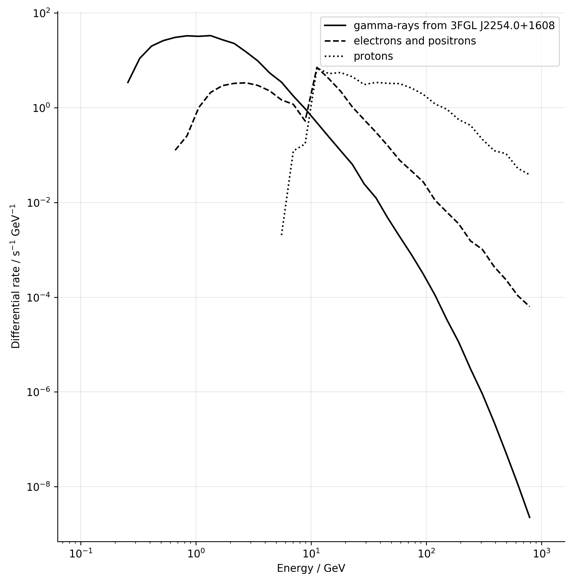}
    \caption[Expected trigger-rates on \NameAcp{}]{
        The expected trigger-rates on \NameAcp{} when observing the bright gamma-ray-source 3FGL-J2254.0+1608.
        The trigger-rates here are only for the on-region which on the sky-dome where we expect the source.
        The on-region's radius of $0.31\,^\circ$ is based on \NameAcp{}'s angular-resolution, see Figure \ref{FigAngularResolutions}.
        The expected trigger-rates are based on the fluxes of air-showers presented in Figure \ref{FigAirShowerRateChargedCosmicRays}, the flux of gamma-rays coming from 3FGL-J2254.0+1608 \cite{acero2015fermi3fgl}, and \NameAcp{}'s instrument-response-functions shown in the Figures \ref{FigResponseGammaRays}, and \ref{FigResponseCosmicRays}.
        For energies below the rigidity-cutoff, \NameAcp{} can observe the gamma-ray-sky signal dominated.
        The integrated rates in the on-region are: 94.2\,s$^{-1}$ for gamma-rays, 74.9\,s$^{-1}$ for electrons, and 458\,s$^{-1}$ for protons.
        This gives us a total trigger-rate of about $59\times10^3\,$s$^{-1}$ in the entire $6.5^{\circ}$ field-of-view.
        From the peak of the expected differential trigger-rates for gamma-rays, we find that \NameAcp{} indeed reaches an energy-threshold of $1\,$GeV.
        For more details on the trigger see Chapter \ref{ChTrigger}.
    }
    \label{FigExpectedRates}
\end{figure}

\begin{figure}[H]
    \centering
    \includegraphics[width=1\textwidth]{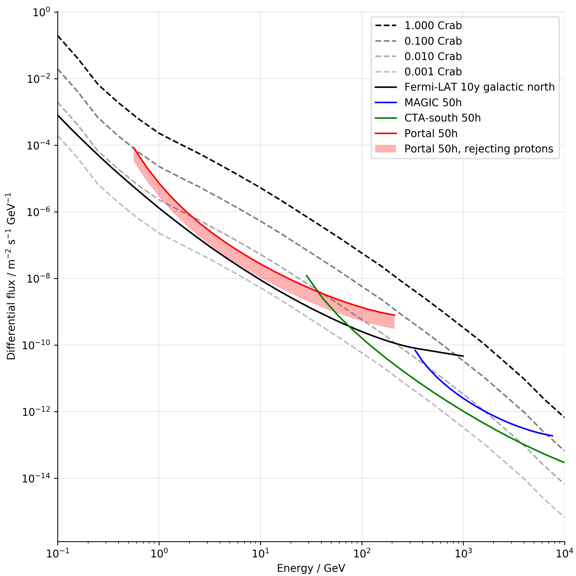}
    \caption[Sensitivity of \NameAcp{} and Fermi-LAT]{%
    The integral-spectral-exclusion-zones for the \NameAcp{} Cherenkov-plenoscope, the Fermi-LAT satellite, the MAGIC Cherenkov-telescopes, and the future planned Cherenkov-Telescope-Array (south).
    Note the different exposure-times listed in the legend.
    Every point-like gamma-ray-source with a power-law like energy-spectrum (a straight line in this log-log-scale) that touches the integral-spectral-exclusion-zones of the instruments will be detected within the listed exposure-time.
    MAGIC: \cite{ahnen2017integral}, Fermi-LAT: \cite{FermiPass8BroadbandSensitivity2016}, CTA-south: \cite{fioretti2016cherenkov}, and \cite{cta2018baseline}, Crab Nebula: \cite{aleksic2015measurement}.
    }
    \label{FigIntegralSpectralExclusionZone}
\end{figure}

We give a first estimate on \NameAcp{}'s sensitivity for cosmic gamma-rays by simulating the observations of air-showers induced by gamma-rays and charged cosmic-rays.
We run a simulation, see Chapter \ref{ChSimulation}, for the observation of individual air-showers which returns us the response of \NameAcp{}'s light-field-sensor.
Only if the Cherenkov-photons together with the night-sky-background-photons fulfill the trigger-criteria, the response of the light-field-sensor is read-out, see Chapter \ref{ChTrigger}.
We set the trigger-threshold such that the accidental trigger-rate caused by fluctuations of the night-sky-background-photons during the dark night is far ($\sim 10^{-2}$) below the expected trigger-rate for air-showers, see Figure \ref{FigTriggerRateScan}.
%
% We simulated about 1M air-showers. About 90% of the air-showers do not trigger AND have less than 20p.e. true Cherenkov-photons.
% Those approx. 900k air-showers are considered to be NSB only. The exposure-time for each air-shower is 50ns.
% This gives us an exposure-time of about 5E-8s x 9E+5 = 45E-3
%
With this trigger-threshold we estimate \NameAcp{}'s instrument-response-functions for a point-source of gamma-rays, see Figure \ref{FigResponseGammaRays}, a diffuse pool of electrons, and a diffuse pool of protons, see both in Figure \ref{FigResponseCosmicRays}.
We take the measured fluxes of cosmic electrons \cite{aguilar2014precision} and protons \cite{aguilar2015precision} and combine these with models \cite{lipari2002fluxes, zuccon2003atmospheric} regarding their efficiencies to produce air-showers in earth's atmosphere.
The earth's magnetic field effectively deflects low energetic charged particles before they can penetrate the atmosphere deep enough to interact with the air and initiate air-showers \cite{supanitsky2012earth}.
Figure \ref{FigGeomagneticCutOffRigidity} shows earth's cut-off-rigidity for cosmic-rays.
This way, we estimate not the flux of charged particles, but the flux of air-showers initiated by these charged particles in earth's atmosphere, see Figure \ref{FigAirShowerRateChargedCosmicRays}.
Knowing the true incident-directions of the simulated gamma-rays, we estimate \NameAcp{}'s angular resolution for a diffuse source, see Chapter \ref{ChAngularResolution}.
\NameAcp{}'s angular resolution reaches $\approx 0.31^{\circ}$ for an 68\% containment-radius for gamma-rays at energies between $750\,$MeV and 1,500$\,$MeV.
Such an angular resolution of \NameAcp{} would be good, but is still within the expected regime of former studies and other instruments, see Figure \ref{FigAngularResolutions}.
In this first estimate for the sensitivity we neglect that \NameAcp{}'s angular resolution will be better for higher energies.
We simulate a counting-experiment with an on-off-observation.
We define five circular regions in \NameAcp{}'s field-of-view which all have a radius equal to the estimated angular resolution.
We choose one on-region, which is centered around a hypothetical gamma-ray-source, and four additional off-regions which are centered at positions where no gamma-ray-sources are expected.
Then we count the number of events reconstructed to originate in the on- and off-regions.
With the off-regions, we estimate the expected number of background events in the on-region.
Only if the number of events in the on-region exceeds the expected fluctuations of the background by at least a factor of five standard-deviations, we claim a detection for the hypothetical source.
Figure \ref{FigExpectedRates} shows the expected trigger-rates for air-showers induced by gamma-rays and charged cosmic-rays when observing the quasar and bright gamma-ray-source 3FGL-J2254.0+1608\footnote{Also known as 3C\,454.3}.
In Figure \ref{FigExpectedRates} we find that \NameAcp{} indeed reaches its design-energy-threshold, as its differential trigger-rate for gamma-rays peaks at 1\,GeV.
With the expected trigger-rates for gamma-rays and charged cosmic-rays, we are now able to estimate \NameAcp{}'s sensitivity for the worst-case-scenario in which we do not have any separation-power to tell apart air-showers induced by gamma-rays from air-showers induced by protons.
Figure \ref{FigIntegralSpectralExclusionZone} shows the sensitivities of the \NameAcp{} Cherenkov-plenoscope, the Fermi-LAT satellite and other Cherenkov-telescopes.
Beware, as we have no energy-reconstruction for \NameAcp{} yet, we represent the sensitivity using the integral-spectral-exclusion-zone \cite{ahnen2017integral}.
This is different from the more common differential representation provided for most Cherenkov-telescopes.
In Figure \ref{FigIntegralSpectralExclusionZone} we show \NameAcp{}'s sensitivity for two scenarios.
First, a thin red curve shows \NameAcp{}'s sensitivity without any separation-power for gamma-rays and protons.
This corresponds directly to the rates shown in Figure \ref{FigExpectedRates} without further cuts.
Second, a wide red band shows \NameAcp{}'s sensitivity in the case that \NameAcp{} could separate air-showers induced by hadronic particles from air-showers induced by electromagnetic particles with similar precision as it is possible on established Cherenkov-telescopes.
The lower border of the wide red band corresponds to a rejection of $99\%$ of the hadronic air-showers.
At this point, the air-showers induced by electrons and positrons become the relevant fraction of the background.\\
Although a basic separation of electrons from gamma-rays \cite{hofmann2006performance} might be possible on \NameAcp{}, we did not investigate this option yet.\\
Note that in Figure \ref{FigIntegralSpectralExclusionZone}, the ground based instruments are listed with $50\,$h exposure-time, while the satellite Fermi-LAT is listed with its full $10\,$years exposure-time.
%
% \cite[Sec.\,3, Fig.\,4]{hofmann2006performance}
%
Since \NameAcp{}'s light-field-sequence $\mathcal{L}[c_x, c_y, x, y, t]$ contains multiple image-sequences $\mathcal{I}[c_x, c_y, t]$ of e.g. a dense array of seven large sized Cherenkov-telescopes, see Section \ref{SecExampleImagesOfAirShowers}, \NameAcp{} can always fall back to the performance of Cherenkov-telescope-arrays for energies above the geomagnetic cut-off.
In general, it can be assumed that \NameAcp{} will reach at least the sensitivity of MAGIC for energies $\gtrsim 100\,$GeV.
But in this estimate \NameAcp{} does not make use of the established analysis for air-showers on Cherenkov-telescopes which is why Figure \ref{FigIntegralSpectralExclusionZone} does not show a smooth transition of \NameAcp{}'s sensitivity into the sensitivity of MAGIC at energies $\gtrsim 100\,$GeV.
%
% Cherenkov-plenoscope and Satellites
% ------------------
%
\chapter{Outlook}
\label{ChOutlook}
\begin{figure}
    \centering
    \includegraphics[width=1\textwidth]{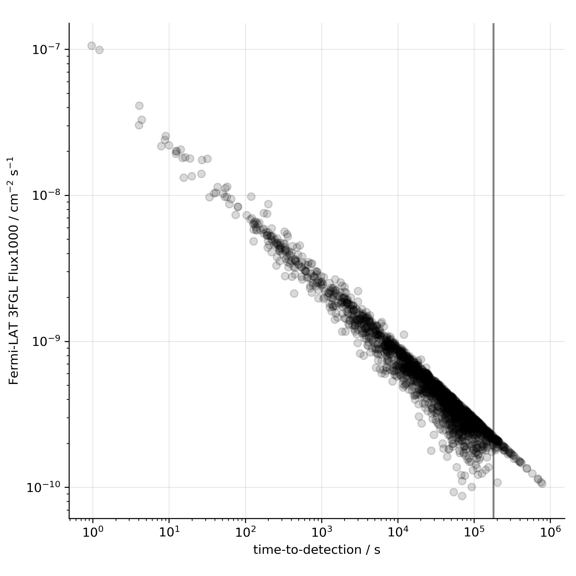}
    \caption[Flux of gamma-ray-sources vs. time-to-detections]{
        The flux of gamma-ray-sources above one GeV versus the time-to-detections with \NameAcp{}.
        These are the gamma-ray-sources listed in Fermi-LAT's 3FGL catalog \cite{acero2015fermi3fgl}, and the flux (Flux1000) listed in the catalog.
        Vertical black line marks $50\,$h.
        Almost all sources in the Fermi-3FGL-catalog can be detected by \NameAcp{} in less than $50\,$h, see dotted vertical line on the right.
        This figure corresponds to \NameAcp{}'s thin, red integral-spectral-exclusion-zone without any gamma-hadron-separation in Figure \ref{FigIntegralSpectralExclusionZone}.
        With gamma-hadron-separation, the times-to-detection reduce by a factor of $\approx 6$.
    }
    \label{FigTimesToDetection}
\end{figure}

Since satellites such as Fermi-LAT and the \NameAcp{} Cherenkov-plenoscope share the same energy-range, it would be a great opportunity for gamma-ray-astronomy to have both such complementary observatories.
All the gamma-ray-sources found in satellite-surveys in the full but inert gamma-ray-sky taken over years of exposure, are potential targets for observations with \NameAcp{}, see Figure \ref{FigTimesToDetection}.
For instance Fermi-LAT, with $\approx 20\%$ of the sky field-of-view, can observe a large number of sources with a simple monitoring strategy.
The pointing-strategy of the \NameAcp{} Cherenkov-plenoscope with only $\approx 0.1\%$ of the sky field-of-view on the other hand would best be guided to observe dedicated sources with a very large statistics of gamma-rays in a short period of time.
This is a complementary observation not possible with satellites \cite{aharonian2001}.
To point out the complementary nature of e.g. the Fermi-LAT satellite and \NameAcp{}, consider the following:
The ratio $\frac{100\%}{0.0804\%} = 1,244$, see Figure \ref{FigFieldOfViewComparison} by which Fermi-LAT's field-of-view exceeds \NameAcp{}'s field-of-view, is roughly the ratio $\frac{10\,\text{years}}{50\,\text{h}} = 1,752$ by which \NameAcp{}'s time-to-detections undercut the time-to-detections of Fermi-LAT.\\
Today, we face the challenge of the cosmic-rays origin.
We face the challenge of dark-matter-phenomena.
We face the challenge of asymmetry between matter and anti-matter.
We face the challenge of extra-galactic-background-light and extra-galactic magnetic-fields.
We face the challenge of the rapid transient-phenomena like gamma-ray-bursts, and fast-radio-bursts.
We face the challenge of electromagnetic counterparts for gravitational-waves in rapid cosmic mergers.
Designed to take on the challenges of our generation, \NameAcp{} is the most powerful gamma-ray-timing-explorer proposed yet.
As Felix Aharonian said:\\
'...\textit{the scientific reward of the implementation of ground based approach in GeV gamma-ray-astronomy will be enormous}' \cite{aharonian2005next}
%
% Dark Matter Array
% -----------------
% arXiv:astro-ph/0608407 on the Bullet cluster
% The Astrophysical Journal, 648:L109–L113, 2006 September 10
\section{Searching for dark matter}
The velocity-dispersions of stars in galaxies and the separation of baryonic matter from gravitational-lensing matter, which was observed in the colliding galaxies in the Bullet-Cluster \cite{clowe2006direct}, suggest the existence of a dark type of matter.
Together with the prediction of a lightest super-symmetric particle and the cosmological evolution of the early universe, today the scenario of the so called weakly interacting massive particle (WIMP) is frequently debated.
The non baryonic, and dark WIMP would have been created thermodynamically in the early universe and is now assumed to form gravitationally bound clumps in galaxies.
The WIMP scenario predicts gamma-ray emission from annihilation which could be visible as a halo-emission in such clumps of dark matter, and is already investigated by current instruments \cite{aharonian2006hess, bertone2010moment}.
\NameAcp{} would be ideal to not only support Fermi-LAT's quest for upper limits on such annihilation features from the WIMP but to take over and push the sensitivity-frontier since \NameAcp{}'s unmatched low energy-threshold for cosmic gamma-rays in combination with its large collection area are the key features \cite{bergstrom2011,bergstrom2013dark} to reveal the heavy sector of dark matter in an indirect search.
%
% Crab Nebula Flares
% ------------------
\section{Resolving Crab-Nebula-flares}
The discovery of powerful gamma-ray-flares above $0.1\,$GeV of the nearby super-nova-remnant SN\,1054 (Crab-Nebula) \cite{tavani2011discovery} are indicating that shorter time-to-detections will potentially reveal further insights into the production and acceleration of cosmic-rays, and the emission of gamma-rays in such extreme environments.
%
% BLACK HOLE LIGHTNING
% --------------------
% IC310 is too close to the galactic plane, but Mrk 421, Mrk 501 and
% PKS 2155-304 are honorable mentions
% PKS 2155-304 == 3FGL J2158.8-3013 ttd = 1.6s,  7.5s
% Markarjan 501 == 3FGL J1653.9+3945 -> 6.4s, 33.7
% Markarjan 421 == 3FGL J1104.4+3812 -> 0.8s, 4.0s
% /starter_kit/run/isf/time_to_detections.csv
%
\section{Investigating massive black holes}
Relativistic plasma-jets driven by super massive black holes inside active-galactic-nuclei are believed to result from the conservation of angular-momentum of in falling matter.
Although these jets extend up to distances which usually are found in between galaxies, their creation in the vicinity of the black hole remains unresolved by todays imaging-instruments.
However, fast variability in the emission of gamma-rays from these objects give hints to the particle accelerations at the base of the jets.
Only short after the first sighting of gamma-rays from Markarjan\,421 with a ground based telescope \cite{punch1992detection}, the Whipple Observatory was able to reveal flux-doubling-timescales in the $1\,$hour regime \cite{gaidos1996extremely}.
Latest observations made by the MAGIC Cherenkov-telescope with its lower energy-threshold for gamma-rays on IC\,310 were already able to reveal time-structures in the $1\,$minute regime \cite{aleksic2014blackHoleLightning}.
The flux of IC\,310 rose up to between 1 and 5 times the flux of the Crab Nebula, and MAGIC was able to provide estimates for the flux within time-bins of only $\sim 120\,$s.
The observation of flux-variabilities on such small time-scales allow insights into the structure of the bases of the jets close to the black holes which are far more precise than the structures resolved by any imaging method.
Further, flaring active-galactic-nuclei can serve as a lab to probe the energy-dependence of the speed of light as done by the H.E.S.S. Cherenkov-telescopes on PKS\,2155-304 \cite{aharonian2008limits}, and the MAGIC Cherenkov-telescopes on Markarjan\,501 \cite{albert2008probing}.
Time structures in the $1\,$minute regime were reported on PKS\,2155-304 \cite{aharonian2008limits}.
The \NameAcp{} Cherenkov-plenoscope will detect the active-galactic-nuclei Markarjan\,421 in $\approx 4.0$\,s, PKS\,2155-304 in $\approx 7.5$\,s , and Markarjan\,501 in $\approx 33.7$\,s when these are not flaring, but in their average, low states of activity which is below $0.5\,$ times the flux of the Crab Nebula \cite{acero2015fermi3fgl}.
These time-to-detections are without any gamma-hadron-separation and correspond to the solid, red line in Figure \ref{FigIntegralSpectralExclusionZone}, and the time-to-detections in Figure \ref{FigTimesToDetection}.
However, these time-to-detections go down to $0.8\,$s for Markarjan\,421, $1.6\,$s for PKS\,2155-304, and $6.4\,$s for Markarjan\,501 if gamma-hadron-separation was implemented and could be made as powerful at the low energies of the \NameAcp{} Cherenkov-plenoscope, as it has been made at $\sim 100\,$GeV on todays Cherenkov-telescopes.
%
% Mrk 421 average flux 0.446 crab units (arXiv:1310.8150, abstract)
%
% extra-galactic background light
% ------------------------------
%
\section{Sneaking below the gamma-ray-horizon -- Probing extra-galactic-background-light}
\label{SecGammaRayHorizon}
\begin{figure}
    \centering
    \includegraphics[width=0.75\textwidth]{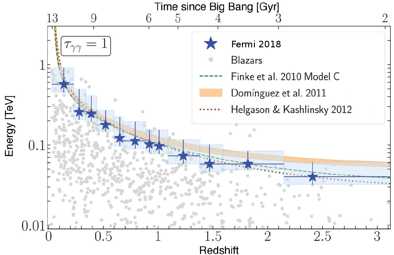}
    \caption[The gamma-ray-horizon according to Fermi-LAT]{Figure taken from \cite{fermi2018gamma}.
        The gamma-ray-horizon measured by Fermi-LAT.
        Here $\tau_{\gamma \gamma} = 1$ indicates the point after which the universe becomes opaque for gamma-rays.
        To see distant sources, the energy-threshold must be low.
    }
    \label{FigGammaRayHorizon}
\end{figure}

The red-shifted light emitted by early stars and galaxies is supposed to be the second brightest \cite{dole2006cosmic} diffuse background-radiation after the cosmic microwave-background, and yet we only know little about it.
Direct observations of this infra-red, so called extra-galactic background-light, are difficult because it is out shined by the zodiacal light and other nearby sources including the instruments themselves.
However, the attenuation of gamma-rays which interact with the extra-galactic background-light via pair production ($\gamma_\text{high energy}  +\gamma_\text{infra-red} \rightarrow e^{+} + e^{-}$) serves as an indirect measurement of the density of the extra-galactic background-light \cite{magic2008distantQuasar,aharonian2006low}.
Figure \ref{FigGammaRayHorizon} shows the attenuation of gamma-rays in the extra-galactic-background-light.
With its low energy-threshold for gamma-rays, the \NameAcp{} Cherenkov-plenoscope will not only be able to learn more about the extra-galactic background-light, but \NameAcp{} will also be able to look deeper and see more sources in the gamma-ray-sky \cite{taylor2017active} than any other existing or proposed ground based instrument.
%
% extra-galactic magnetic fields
% -----------------------------
%
\section{Probing extra-galactic magnetic-fields}
The strength and the origin of extra-galactic magnetic-fields might give valuable insights to the early formation of galaxies, but direct measurements are beyond our current reach.
However, gamma-rays can serve as an indirect probe to the strength of extra-galactic magnetic-fields \cite{neronov2010evidence}.
Strong extra-galactic magnetic-fields are expected to cause a diffuse halo-emission around distant point-sources as high energetic gamma-rays are expected to undergo pair-production with the extra-galactic-background-light to create electrons and positrons which in turn create lower energetic gamma-rays due to inverse Compton-scattering in the extra-galactic magnetic-fields.
The trajectories of the charged electrons and positrons in between this conversion are bend by the magnetic-fields, so that the extension of the halo-emission observed on earth gives an estimate on the column density of the extra-galactic magnetic field's strength.
%
% Black-hole-merger and gamma-ray-bursts
% --------------------------------------
%
\section{Resolving gamma-ray-bursts on the 10$^{-5}\,$s time-scale}
\begin{figure}
    \centering
    \includegraphics[width=0.75\textwidth]{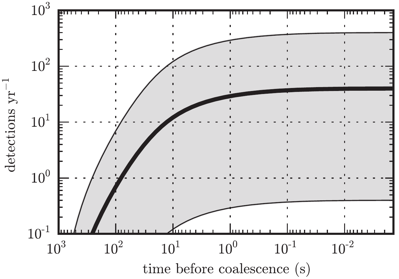}
    \caption[Detection-rate vs. time before coalescence, Advanced-LIGO]{Figure taken from \cite{cannon2012toward}.
        Advanced-LIGO's expected detection-rate versus the time before a neutron-star-neutron-star-merger.
        The thick line marks the most probable detection-rate, and the shaded areas represent its 5\% to 95\% confidence-interval.
    }
    \label{FigDetectionRateVsTimeToCoalescenceAdvancedLigo}
\end{figure}

Thanks to the tremendous efforts \cite{aasi2015advanced} put in the detection of gravitational waves, we are now able to identify the merging of e.g. two neutron-stars at cosmic distances when these heavy objects spiral into each-other.
Right from the start, electromagnetic counterparts for the short-lived gravitational-wave-transients were looked after \cite{aasi2014first}.
After the first limits \cite{savchenko2016integral}, finally the coincident detection of gravitational wave GW\,170817 and the short gamma-ray-burst GRB\,170817A \cite{abbott2017gravitational} was made.
Most likely all our current models and theories approach their limits in the extraordinary environments of cosmic mergers which makes the observation of gamma-rays emitted in e.g. neutron-star-neutron-star-mergers an outstanding probe.
Already the current generation 'Advanced-LIGO' of gravitational-wave-detectors is expected to observe about one neutron-star-neutron-star-merger per year with an alert-time of about $100\,$s before the actual merger, see Figure \ref{FigDetectionRateVsTimeToCoalescenceAdvancedLigo} \cite{cannon2012toward}.
With future gravitational-wave-detectors \cite{abbott2017exploring} there might be the opportunity of having enough early alerts, so that a guided observation for ground based instruments such as the \NameAcp{} Cherenkov-plenoscope becomes feasible.
For gamma-ray-bursts that emit gamma-rays with energies above $\sim 1$\,GeV, \NameAcp{}'s collection-area will allow to have about four orders-of-magnitude more statistics and thus time-resolution than any space-born instrument.
For example, in the long gamma-ray-burst GBR-130427A, the Fermi-LAT satellite detected \cite{ackermann2014fermi} over 500\,gamma-rays with energies above $100\,$MeV, and still $15\,$gamma-rays with energies above $10\,$GeV.
If this bright, and long gamma-ray-burst happened within \NameAcp{}'s field-of-view, \NameAcp{} would have detected it within $4.3\times10^{-5}$\,s, and \NameAcp{} would have detected gamma-rays at a rate of up to $1.7\times10^5$\,s$^{-1}$.
From \cite{ackermann2014fermi}, we conclude that the gamma-ray-flux of GBR-130427A above\footnote{Called 'Flux1000' in 3FGL \cite{acero2015fermi3fgl}, see Figure \ref{FigTimesToDetection}.} 1\,GeV is $\sim 10^{-4}$\,cm$^{-2}$\,s$^{-1}$.
Compare this flux and our estimated time-to-detection to the steady gamma-ray-sources shown in Figure \ref{FigTimesToDetection}.\\
Today, it is not known if the short gamma-ray-bursts coincident with neutron-star-neutron-star-mergers emit gamma-rays with energies above $\sim 1$\,GeV, such as it was observed in the long gamma-ray-bursts coincident with hyper-novae.
But \NameAcp{} is a good way to find out.
Assuming, without any particular model in mind, that the flux of a short gamma-ray-burst above $1$\,GeV is large enough for a satellite with $\sim 1$\,m$^2$ collection-area to detect $0.01$\,gamma-rays, \NameAcp{} will still detect a flood of $300$\,gamma-rays.
%
% which will give us a gamma-ray-light-curve with unmatched timing-resolution.
%
% Speed of light variations
% -------------------------
%
Gamma-ray-bursts also serve as test for variations of the speed-of-light \cite{abdo2009limit} where \NameAcp{}'s high timing-resolution is key to push the frontier of our models.
%
% Pulsars
% -------
%
\section{Seeing pulsars below the $10\,$GeV cut-off}
\label{SectionPulsars}
When stars run out of light elements to fuse, the outwards pushing pressure of the fusion-heat can not longer outbalance the inward pulling gravitational binding.
When gravitational pressure takes over, the electron-nuclei-plasma in the star's core condenses to neutrons while the outer shell of the star is blown away in what we observe as a super nova.
The remnant is a compact neutron-star with high magnetic field densities on its surface which rotates rapidly inside a nebula of the former outer shell.
On earth, we observe a pulsating emission of photons timed in phase with the rotation of these neutron-stars.
Therefore, we call them pulsars.
For most pulsars, the gamma-ray-emission shows a steep cutoff below $10\,$GeV \cite{aharonian2012abrupt}.
At least for the pulsar inside the Crab-nebula the energy extends, barely visible with todays instruments after $320\,$h of exposure, into the $1,000\,$GeV range \cite{ansoldi2016teraelectronvolt}.
\NameAcp{} is the first ground based instrument to measure high gamma-ray statistics in short periods of times of pulsars far below the $10\,$GeV cutoff.
In addition, \NameAcp{} at the same time can observe the high energy emission in the $1,000\,$GeV range using classic Cherenkov telescope analysis.
\NameAcp{} is ideal to extend our knowledge on the gamma-ray-emission from pulsars.
%
% Nearby pulsars and antimatter excess
% -------------------------------------
%
\section{Searching for nearby pulsars and antimatter-anisotropy}
\label{SecNearbyPulsars}
An excess of positrons in the cosmic-rays at energies above $10\,$GeV was measured by space born instruments \cite{adriani2009anomalous} and was not expected from the positron-production-efficiency of galactic propagation-models for cosmic-rays.
Beside the hype on possible explanations using dark matter, this excess might be explained with existing knowledge on nearby pulsars such as Geminga and Monogem \cite{linden2013probing}.
There might also be unknown pulsars even closer to us which we did not detect yet because their beamed emissions are missing earth.
A strong indication for the nearby-pulsar-theory would be an anisotropy in the arrival-directions of the positron flux here on earth.
Such anisotropy might be measured by current and future Cherenkov-telescopes \cite{linden2013probing} using years of exposure-time.
\NameAcp{} on the other hand can exploit the geomagnetic cutoff to have a rather pure sample of positrons \cite{supanitsky2012earth}.
The remaining background of gamma-rays could be subtracted using the static gamma-ray-sky observed by Fermi-LAT \cite{acero2015fermi3fgl}.
As the earth with its magnetic field and \NameAcp{} rotate below the sky, the flux of positrons can be probed over a wide range of galactic directions.
This combination makes \NameAcp{} an unique instrument to investigate the anisotropy of the anti-matter-positron-sky.
%
% Stellar Intensity Interferometry
% ---------------------------------
%
\section{Imaging bright stars with milli arcsecond-resolution}
\label{SecStellarIntensityInterfrometry}
Beside observing the gamma-ray-sky, the Cherenkov-plenoscope can at the same time image bright stars with angular resolutions approaching $10^{-3}$\,arcseconds.
Its plenoptic-perception and 1\,ns arrival-time-resolution for single-photons offer a unique opportunity for stellar-intensity-interferometry.
Currently, the proposed implementations of stellar-intensity-interferometers in arrays of Cherenkov-telescopes \cite{dravins2012stellar, dravins2013optical}, face large technological challenges for signal-processing and signal-transmission.
Like the trigger in Cherenkov-astronomy, a stellar-intensity-interferometer needs instant access to the photo-sensors that sample nearby incident-directions ($c_x$,\,$c_y$), but separate support-positions ($x$,\,$y$).
In telescope-arrays, such photo-sensors are housed in separate telescopes.
To correlate their signals, flexible, high bandwidth cables need to be routed over large distances.
In addition, the signals need adjustable time-delays to correct for the pointing of the telescopes.
All of this either limits the field-of-view, this is the number of photo-sensors used in each telescope, or the exposure-time.
In the \NameAcp{} Cherenkov-plenoscope on the other hand, photo-sensors that sample same incident-directions, but separate support-positions are already $\approx 15\,$cm close together inside the light-field-sensor's small cameras.
In the Cherenkov-plenoscope there is no need for adjustable time-delays, and no need for signals to leave the protective housing of the light-filed-sensor.
The unique geometry of the Cherenkov-plenoscope potentially allows to cost-efficiently install signal-correlations for stellar-intensity-interferometry in each small camera in the light-field-sensor, thus offering huge field-of-views.
Compared to Cherenkov-telescope-arrays, the \NameAcp{} Cherenkov-plenoscope can only offer a $71\,$m baseline for correlations, what is potentially enough for resolving $\approx 2\times10^{-3}$\,arcseconds.
Still, this resolution is in the regime of the European-Extremely-Large-Telescope, and the Very-Large-Telescope-
Interferometer \cite{dravins2012stellar}.\\
The Cherenkov-plenoscope offers a novel and unique trade-off for stellar-intensity-interferometry:
A limited baseline, and thus a limited angular resolution on the one hand, but much less challenging signal-transmission and signal-processing, much wider field-of-views, and unlimited exposure-times on the other hand.
%
% Chemical composition of cosmic rays
% -----------------------------------
%
\section{Probing the chemical composition of cosmic-rays}
\label{SecChemicalComposition}
To constrain the origin and the propagation of cosmic-rays, their chemical composition at energies of $\sim 10^6\,$GeV is of great interest.
At this energy, the cosmic-ray-spectrum has one of its few features, the so called knee.
Space-born detectors like AMS-01 \cite{aguilar2010relative}, and AMS-02 can measure the chemical composition of cosmic-rays precisely.
But, at energies of above $\sim 10^3\,$GeV, their small collection areas leave the chemical composition of cosmic-rays unresolved.
Ground based air-shower-tail-detectors have large collection-areas to observe cosmic-rays at energies around the knee.
They can even estimate the cosmic-ray's charge by measuring the air-shower's muon-multiplicity.
But muon-multiplicity depends on hadronic interaction-models.
Today these hadronic models need to be extrapolated far beyond the energies reached in particle-colliders.
A model-independent alternative is to observe the direct Cherenkov-light emitted by the cosmic-ray to deduce its charge \cite{kieda2001high}.
Direct Cherenkov-light-observations with Cherenkov-telescope-arrays have large collection-areas of $\sim 10^5$\,m$^2$.
But measuring the cosmic-ray's first-interaction-altitude, at which the emission of direct Cherenkov-light stops, is challenging \cite{aharonian2007first}.\\
In a first attempt \cite{engels2017master}, Axel Arbet Engels simulates an idealized Cherenkov-plenoscope with the goal to estimate the cosmic-ray's first-interaction-altitude in each individual air-shower.
He reconstructs the emission-positions of the Cherenkov-photons in three spatial dimensions from the light-field-sequence using tomography.
He uses a simple filtered-back-projection implemented by the author of this thesis (S.A.M.), see Figure \ref{FigTomographyExampleEngels} in Chapter \ref{ChTomography}.
Axel's first findings indicate a potential to resolve the first-interaction-altitude within $\approx 1,000\,$m.
This would allow a charge-resolution of $14.3\%$ for iron.
Tomographic reconstructions of air-showers with either a Cherenkov-plenoscope or an array of Cherenkov-telescopes, have the potential to reveal insights beyond the simple ellipse-models \cite{hillas1985cerenkov} often discussed in reconstructions based on imaging.
%
%------------------------------------------------------------------------------
%
%
%
%
%
%------------------------------------------------------------------------------
\chapter{Conclusion}% maximal two paragraphs
\label{ChConclusionCherenkovPlenoscope}
Before this thesis, the $1\,$GeV gamma-ray-sky with its high variability and fast transient-phenomena at and below the $\sim 1\,$second-time-scale has been far out of reach for astronomy.
But this is about to change now.\\
Our proposed $71\,$m \NameAcp{} Cherenkov-plenoscope offers $30,000\,$m$^2$ collection-area\footnoteA{Figure \ref{FigResponseGammaRays}} for cosmic gamma-rays at $1\,$GeV.
It can detect several sources in the steady, non-flaring gamma-ray-sky within seconds\footnoteA{Figure \ref{FigTimesToDetection}}.
It can, for the first time, study the gamma-ray-emission of pulsars below the crucial $10\,$GeV cutoff-energy\footnoteA{Section \ref{SectionPulsars}}.
It can look deeper into the universe, and thus choose from more potential sources, than any existing or proposed ground based instrument\footnoteA{Section \ref{SecGammaRayHorizon}, Figure \ref{FigGammaRayHorizon}}.
\NameAcp{} can sneak below the geomagnetic cutoff for charged cosmic-rays\footnoteA{Figure \ref{FigAirShowerRateChargedCosmicRays}} and study the positron-sky's anisotropy\footnoteA{Section \ref{SecNearbyPulsars}}.
Its novel cable-robot-mount has no near-zenith-singularity\footnoteA{Chapter \ref{ChCableRobotMount}} and thus can point the Cherenkov-plenoscope intrinsically faster during its hunt for transient-phenomena.
The Cherenkov-plenoscope can reconstruct the inner structures of air-showers in three spatial dimensions\footnoteA{Chapter \ref{ChTomography}}, which potentially opens a window for particle-physics.
\NameAcp{}'s field-of-view is $170$\% the solid-angle of current Cherenkov-telescopes\footnoteA{Figure \ref{FigFieldOfViewComparison}}.
And dedicated Cherenkov-plenoscopes can be build to push the current generation's field-of-view by more than one order-of-magnitude\footnoteA{Chapter \ref{ChOvercomingAberrations}, Figure \ref{FigPointSpreadVsOffAxisAngle}}.
\NameAcp{} can potentially run \NumPix{} stellar-intensity-interferometers simultaneously across its field-of-view\footnoteA{Section \ref{SecStellarIntensityInterfrometry}}, each having $\NumPax{}$ support-positions on the aperture-plane.
\NameAcp{}'s field-of-view is, for the first time in Cherenkov-astronomy, free\footnoteA{Chapter \ref{ChOvercomingAberrations}} of aberrations and distortions and thus allows a more precise reconstruction of cosmic gamma-rays than any other existing or proposed Cherenkov-telescope.
\NameAcp{} can be build now using established technology\footnoteA{Chapter \ref{ChOpticsOfNameAcp}}.
And it costs\footnoteA{Chapter \ref{ChEstimatingCost}.} only $\approx 218.5\times10^6$\,CHF, a fraction of the costs for a satellite-mission.
%
%------------------------------------------------------------------------------
%
%
%
%
%
%
%
%------------------------------------------------------------------------------
\chapter{Meeting \NameAcp{}}
\label{ChPictureTour}
\begin{figure}[H]
    \centering
    \includegraphics[width=.8\textwidth]{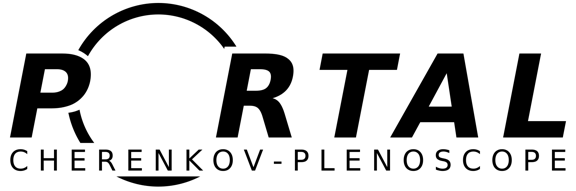}
    \caption[\NameAcp{}, logo]{The logo of the \NameAcp{} Cherenkov-plenoscope.}
    \label{FigNameAcpLogo}
\end{figure}
\begin{sidewaysfigure}
    \centering
    \includegraphics[width=1\textwidth]{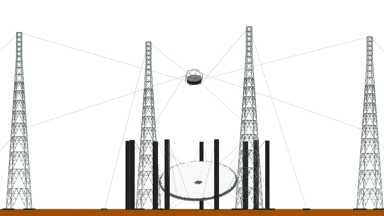}
    \caption[\NameAcp{} overview, side]{The \NameAcp{} Cherenkov-plenoscope, seen from the side. Outer masts are $162\,$m tall, and inner concrete-pillars are $64\,$m tall.}
    \label{FigNameAcpTourOverview}
\end{sidewaysfigure}
\begin{sidewaysfigure}
    \centering
    \includegraphics[width=1\textwidth]{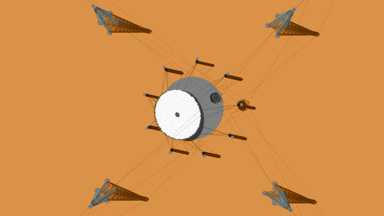}
    \caption[\NameAcp{} overview, top]{The \NameAcp{} Cherenkov-plenoscope, seen from the top. Outer mast-circle is $336\,$m in diameter, and inner concrete-pillar-circle is $128\,$m.}
    \label{FigNameAcpTourOverviewTop}
\end{sidewaysfigure}
\begin{sidewaysfigure}
    \centering
    \includegraphics[width=1\textwidth]{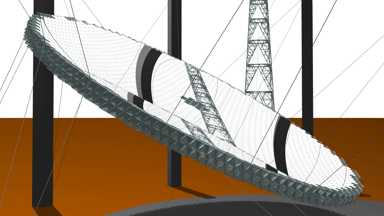}
    \caption[\NameAcp{}'s imaging-reflector]{The large, $71\,$m diameter imaging-reflector. Zenith-distance is $30^\circ$.
        Space-truss-design by civil-engineer Spyridon Daglas.
    }
    \label{FigNameAcpTourImagingReflector}
\end{sidewaysfigure}
\begin{sidewaysfigure}
    \centering
    \includegraphics[width=1\textwidth]{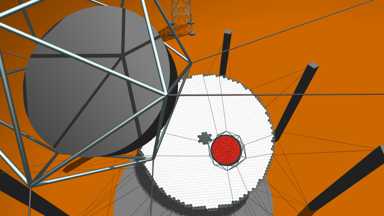}
    \caption[\NameAcp{}'s light-field-sensor and imaging-reflector, top]{Light-field-sensor and imaging-reflector seen from the top.
        The light-field-sensor looks red, because we can see the red photo-sensors through the lenses.}
    \label{FigNameAcpTourSensorReflectorTop}
\end{sidewaysfigure}
\begin{sidewaysfigure}
    \centering
    \includegraphics[width=1\textwidth]{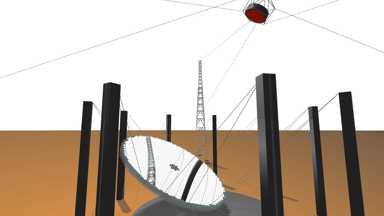}
    \caption[\NameAcp{}'s light-field-sensor and imaging-reflector, side]{Light-field-sensor and imaging-reflector seen from the side.}
    \label{FigNameAcpTourSensorReflectorSide}
\end{sidewaysfigure}
\begin{sidewaysfigure}
    \centering
    \includegraphics[width=1\textwidth]{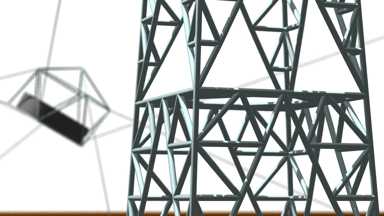}
    \caption[\NameAcp{}'s masts, close-up]{Masts and light-field-sensor in the background.
        The space-truss-design is adopted from overhead-power-lines.
        It is a fractal design composed from repeatable modules.}
    \label{FigNameAcpTourMastsCloseUp}
\end{sidewaysfigure}
\begin{sidewaysfigure}
    \centering
    \includegraphics[width=1\textwidth]{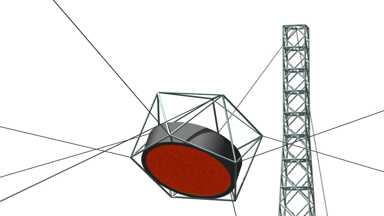}
    \caption[\NameAcp{}'s light-field-sensor]{Light-field-sensor and mast in the background.
        Compare the icosahedron-shaped cage in Figures \ref{FigCableRobotSimulatorPhotograph}, and \ref{FigCableRobotSimulatorDrawing}.
        The red light-field-sensor has $12.1\,$m diameter corresponding to $6.5^\circ$ field-of-view.
    }
    \label{FigNameAcpTourLightFieldSensor}
\end{sidewaysfigure}
\begin{sidewaysfigure}
    \centering
    \includegraphics[width=1\textwidth]{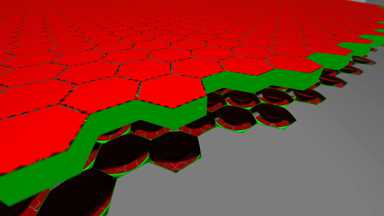}
    \caption[\NameAcp{}'s light-field-sensor, inside, close-up]{Inside the light-field-sensor. Close-up on small cameras.}
    \label{FigNameAcpTourLightFieldSensorInside}
\end{sidewaysfigure}
\begin{sidewaysfigure}
    \centering
    \includegraphics[width=1\textwidth]{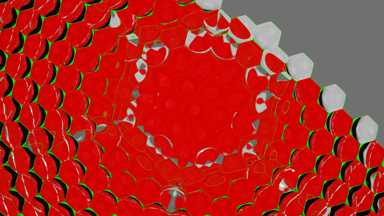}
    \caption[\NameAcp{}'s light-field-sensor, frontal]{Looking into the light-field-sensor. Looking into the densely packed array of small cameras.}
    \label{FigNameAcpTourLightFieldSensorFrontal}
\end{sidewaysfigure}
\begin{figure}[]
    \centering
    \includegraphics[width=1\textwidth]{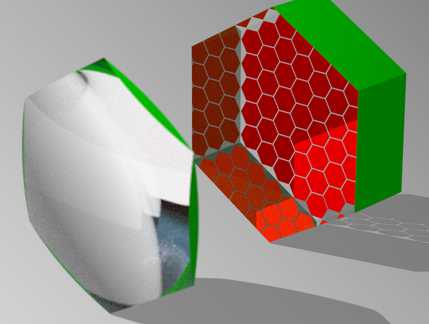}
    \caption[A small camera of \NameAcp{}'s light-field-sensor]{A small camera of \NameAcp{}'s light-field-sensor.
    A single biconvex lens with a hexagonal aperture in front of an image-sensor made out of $M=\NumPax{}$ hexagonal photo-sensors.
    The photo-sensors on the image-sensor are red.
    Surfaces touching neighboring small cameras are green.
    The red and green color-scheme allows to compare the Figures \ref{FigOpticsOverview}, and \ref{FigOpticsOverviewCloseUp}.
    For dimensions see Figure \ref{FigSmallCameraGeometry}.
    }
    \label{FigSmallCameraCluseUp}
\end{figure}
\begin{figure}
    \centering
    \includegraphics[width=1\textwidth]{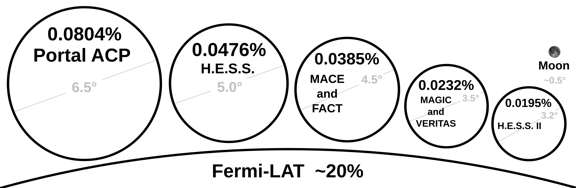}
    \caption[Field-of-view of \NameAcp{}, Fermi-LAT, and Cherenkov-telescopes]{
        The field-of-views of Cherenkov-telescopes, the \NameAcp{} Cherenkov-plenoscope, and Fermi-LAT.
        Here $100\%$ is the full $4\pi\,$sr.
        H.E.S.S.\,II \cite{cornils2005optical}, VERITAS \cite{mccann2010new}, FACT \cite{anderhub2013design}, MACE \cite{yadav2013mace}, H.E.S.S. \cite{punch2001hess,bernlohr2003optical}.
        For Fermi-LAT, the $\approx 20\%$ is the instantaneously observable field-of-view from gamma-ray-energies of $\approx 1\,$GeV to $\approx 1\,$TeV \cite{FermiPass8BroadbandSensitivity2016}.
    }
    \label{FigFieldOfViewComparison}
\end{figure}
\begin{figure}
    \centering
    \includegraphics[width=1.0\textwidth]{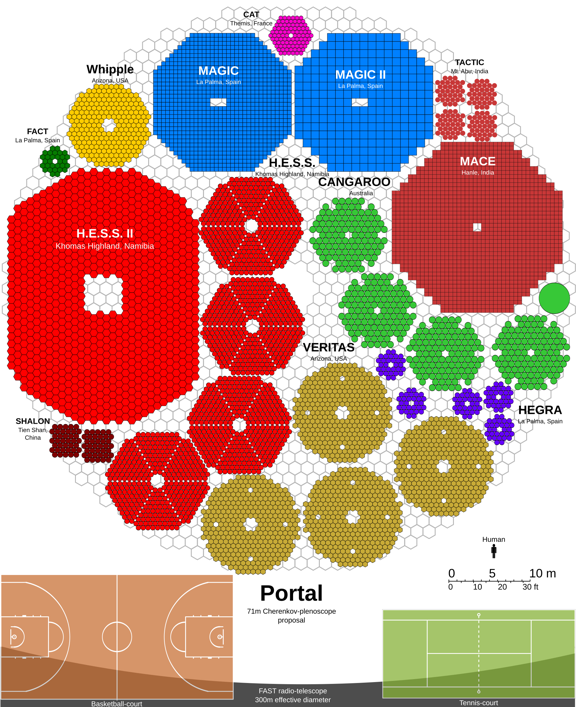}
    \caption[Apertures in atmospheric Cherenkov-astronomy]{
        The various apertures of former and current Cherenkov-telescopes in the year 2017.
        Inspired by \cite{cmglee2014comparison}.
        CANGAROO\,I \cite{roberts1998tev}, CANGAROO\,III \cite{kubo2004status}, FACT \cite{anderhub2013design}, H.E.S.S. \cite{punch2001hess,bernlohr2003optical}, H.E.S.S.\,II \cite{cornils2005optical}, MACE \cite{yadav2013mace}, SHALON \cite{sinisyna2005shalon}, TACTIC \cite{acharya2005ground}, VERITAS \cite{mccann2010new}, WHIPPLE \cite{lewis1990optical}, CAT-France \cite{barrau1998cat}.
    }
    \label{FigCherenkovAperturesComparison}
\end{figure}

To meet the \NameAcp{} Cherenkov-plenoscope, we take you on a picture-tour.
Figure \ref{FigNameAcpLogo} shows \NameAcp{}'s logo.
Since the word telescope is about far seeing, it contradicts the plenoptic method which is about close seeing in the vicinity of the aperture.
Therefore, we do not use the term 'light-field-telescope'.
Instead, we propose to call this novel class of instrument: Plenoscope.
The term is first used by Fredrik Bergholm \cite{bergholm2002plenoscope} to describe a hand-held optics with an eyepiece to explore light-fields and plenoptic-perception.
During this thesis, we created the term independently ourselves again.\\
The pictures of \NameAcp{} shown here are rendered with the same program which we use to propagate Cherenkov- and night-sky-background-photons, see Section \ref{SecMctracer}.
Here we see \NameAcp{} with its dedicated cable-robot-mount which we discuss in Chapter \ref{ChCableRobotMount}.
The author of this thesis (S.A.M.) proposes the concept of a cable-robot-mount.
And in his master-thesis \cite{daglas2015master}, civil-engineer Spyridon Daglas works out the details of the cable-robot-mount shown here.\\
Figure \ref{FigNameAcpTourOverview} shows the \NameAcp{} Cherenkov-plenoscope from the side in a distance of $\sim 1\,$km.
The four $162\,$m tall masts supporting the light-field-sensor potentially will become quite a landmark.
In Figure \ref{FigNameAcpTourOverviewTop} we see \NameAcp{} from $\sim1\,$km above.
The large $71\,$m imaging-reflector is enclosed by rectangular concrete-pillars in a circle with $128\,$m diameter.
The four outer masts are on a circle with $336\,$m diameter.
In Figure \ref{FigNameAcpTourImagingReflector} we see \NameAcp{}'s large, $71\,$m diameter imaging-reflector.
It is composed from small $2\,$m$^{2}$ mirror-facets mounted on a three layer space-truss made out of carbon-fiber-tubes.
Figure \ref{FigNameAcpTourSensorReflectorTop} shows the interplay of imaging-reflector and light-field-sensor from the top, and Figure \ref{FigNameAcpTourSensorReflectorSide} shows it from the side.
The two independent mounts supporting the two components always try to establish the desired target-geometry between the two.
Depending on the desired default-focus, see Section \ref{SecAfterthoughtTargetSensorDistance}, the light-field-sensor is $\approx 105\,$m away from the imaging-reflector.
The light-field-sensor looks red, because we can see the red photo-sensors through the lenses.
The light-field-sensor is $12.1\,$m in diameter what corresponds to $6.5^\circ$ field-of-view.
In Figure \ref{FigNameAcpTourMastsCloseUp} we see the space-truss structure of a mast, and the light-field-sensor in the background.
The space-truss-design is adopted from wide spread overhead-power-lines.
Figure \ref{FigNameAcpTourLightFieldSensor} shows the light-field-sensor inside its icosahedron-shaped cage and one of the four masts supporting it in the background.
The cage-design is adopted from the cable-robot-simulator, see Figures \ref{FigCableRobotSimulatorPhotograph}, and \ref{FigCableRobotSimulatorDrawing}.
Figure \ref{FigNameAcpTourLightFieldSensorInside} shows the densely packed small cameras inside \NameAcp{}'s light-field-sensor.
Same as in the conceptual Figures \ref{FigOpticsOverview}, and \ref{FigOpticsOverviewCloseUp}, the photo-sensors are red, and the walls separating the small cameras are green.
In Figure \ref{FigNameAcpTourLightFieldSensorFrontal} we look straight into the light-field-sensor from a close distance of $\sim 2\,$m.
Through the lenses, we see the red photo-sensors and the green walls.
A single, isolated small camera is shown in Figure \ref{FigSmallCameraCluseUp}.
The dimensions of the small camera are discussed in Chapter \ref{ChOpticsOfNameAcp}, and shown in Figure \ref{FigSmallCameraGeometry}.\\
In Figure \ref{FigFieldOfViewComparison} we compare the field-of-views of current Cherenkov-telescopes, the Fermi-LAT satellite, and the \NameAcp{} Cherenkov-plenoscope.
Using plenoptic-perception to overcome aberrations, see Chapter \ref{ChOvercomingAberrations}, \NameAcp{}'s field-of-view could be made even larger.
The only reason here to limit \NameAcp{}'s field-of-view is cost-efficiency for being a gamma-ray-timing-explorer that will focus on individual sources.
In Figure \ref{FigCherenkovAperturesComparison} we compare the aperture of \NameAcp{}'s imaging-reflector to past and present apertures in Cherenkov-telescopes.
%
%------------------------------------------------------------------------------
%
%
%
%
%
%
%
%------------------------------------------------------------------------------
\chapter{Focusing and a narrow depth-of-field}
\label{ChDepthOfField}
Every telescope with an extended aperture has the limitation of focusing and the limitation of a narrow depth-of-field.
The bigger the aperture, the narrower becomes the depth-of-field, and the bigger becomes the need for focusing.
In this Chapter we demonstrate and discuss the shortcomings of imaging on Cherenkov-telescopes.
First, we remind ourselves what imaging is all about.
Second, we discuss the theory behind focusing and the depth-of-field.
And third, we present example air-shower-images recorded with different aperture-sizes and different focuses.
\section{Defining imaging}
Imaging is about filling an intensity-histogram based on the incident-directions of incoming photons.
The resulting intensity-histogram is called image or picture and its bins are often called picture-cells, or pixels for short%
\footnote{
    Many imaging-systems, e.g. Cherenkov-telescopes are designed so that the photo-sensors in their image-sensors directly correspond to a pixel.
    However, we define a pixel to be a bin in an intensity-histogram based on the incident-directions of photons, and not to be a physical device like a photo-sensor.
}.
The most simple model for imaging is the pin-hole-camera.
\subsection*{Point like apertures}
\begin{figure}
    \centering
    \includegraphics[width=1\textwidth]{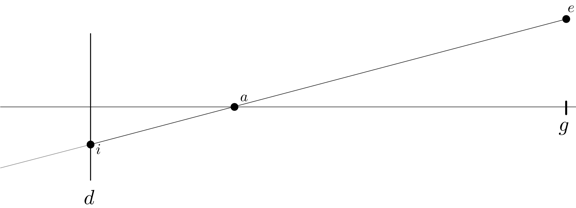}
    \caption[The pin-hole-camera]{
        The pin-hole-camera.
        Long horizontal line is the optical-axis, point $a$ is the hole in the aperture.
        Thin vertical line in distance $d$ is the sensor-plane.
        A photon emitted in object-distance $g$ passes through point $a$ on the aperture-plane and is absorbed on the sensor-plane in point $i$.
    }
    \label{FigPinHoleCamera}
\end{figure}

In a pin-hole-camera, all photons pass through a single point in the aperture-plane, see Figure \ref{FigPinHoleCamera}.
The intercept-theorem tells us where a photon is going to hit the sensor-plane when we know the photon's incident-direction.
On the pin-hole-camera, the image is sharp for all objects in the scenery.
Regardless of the object-distance $g$ of the object, the object will only illuminate a single point on the sensor-plane.
Since all objects in all object-distances are always sharp, there is no need for focusing, and there is no narrowing of the depth-of-field.
There is just one problem with the pin-hole-camera.
A point like aperture will not collect any photons.
To collect photons, we need an extended aperture.
\section{Extended apertures}
\begin{figure}{}
    \centering
    \includegraphics[width=1\textwidth]{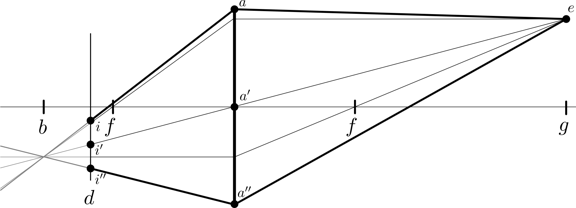}
    \caption[Extended apertures, focusing]{
        An imaging-system with focal-length $f$.
        Long horizontal line is the optical-axis, thick vertical line is the principal-aperture-plane.
        The three principal-rays of a light-source in object-distance $g$ are shown to converge in image-distance $b$.
        However, the principal-rays do not converge on the sensor-plane since it is in sensor-plane-distance $d \neq b$.
        The image of the object is not sharp as it stretches from $i$ to $i''$.
        Here the imaging-system passes the photons through the principal-aperture-plane, like a lens.
        For imaging-reflectors, the figure is mirrored along the principal-aperture-plane.
    }
    \label{FigThinLens}
\end{figure}

Real telescopes have extended apertures.
And when the aperture is extended, photons with same incident-directions, this is photons which will be assigned to the same pixel in the image, might enter the aperture at different support-positions.
Such two parallel photons can not be emitted from the same point in space (from the same object).
Thus the image will be blurred.
The extension of the aperture allows us to collect photons, but it is the reason why not all objects in an image can be sharp at the same time.
Extended apertures are described by the Thin-lens-equation
\begin{eqnarray}
\frac{1}{f} &=& \frac{1}{g} + \frac{1}{b},
\label{EqThinLens}
\end{eqnarray}
and the intercept-theorem, see Figure \ref{FigThinLens}.
The Thin-lens-equation describes in which image-distance $b$ the sensor-plane must be in order to record a sharp image of an object in object-distance $g$ when the focal-length of the imaging-system is $f$.
An image of an object is sharp when the sensor-plane is positioned such that all the photons which passed the aperture and came from the object (from a point in space), converge on the sensor-plane.
In Figure \ref{FigThinLens} we find that the image $i,i',i''$ is a scaled projection of the aperture-function $a,a',a''$.
We can describe the images of objects that are not in focus as a sharp image, recorded by a pin-hole-camera, that got convolved with the scaled aperture-function of the extended camera.
This blurring caused by the aperture-function is often discussed as Bokeh \cite{merklinger1997bokeh, ahnen2016bokeh}, where the author of this thesis (S.A.M.) took the leadership of the investigations for \cite{ahnen2016bokeh}.
In Figure \ref{FigThinLens}, the narrowness of the depth-of-field can be described as the angle between the line $\overline{ai}$ and the line $\overline{a''i''}$.
If the angle between $\overline{ai}$ and $\overline{a''i''}$ is small, the depth-of-field is wide.
Objects in a wide range of object-distances will appear sufficiently sharp in the image.
On the other hand, if the angle between $\overline{ai}$ and $\overline{a''i''}$ is large, only objects from a narrow region of object-distances (depth-of-field) will appear sufficiently sharp in the image.
So we find that focusing is about adjusting the distance between the sensor-plane and the principal-aperture-plane such that a desired object is 'sharp' in the image.
And a narrow depth-of-field is about the problem that we can not have all objects from different object-distances sharp in the image at the same time.
In the introduction of \cite{bernlohr2013monte}, we find\footnote{In \cite{bernlohr2013monte}, the authors use $d$ instead of $g$ as variable-name for the object-distance.} an estimate
%
% bernlohr2013monte, Introduction
%
% g(1 + pg/(2fD)) to g(1 - pg/(2fD))
%
% f = focal-lenght = 106.5m
% g = object-distance = 10km
% p = pixel-diameter = f*tan(0.067) = 0.1245m
%
\begin{eqnarray}
g_\pm &=& g(1 \pm pg/(2fD))
\label{EqDepthOfField}
\end{eqnarray}
for the start-object-distance $g_{-}$ and end-object-distance $g_{+}$ of the depth-of-field on a Cherenkov-telescope which is based on the thoughts of \cite{hofmann2001focus} where also the Thin-lens-equation \ref{EqThinLens} is used.
Here $p$ is the extent of a pixel projected onto the image-sensor, and $D$ is the aperture-diameter of the imaging-reflector.
A Cherenkov-telescope of the same size of \NameAcp{} would have a depth-of-field extending from $g_{-} = 9.2\,$km to $g_{+} = 10.8\,$km for an focus set to an object-distance of $g = 10.0\,$km.
It means, that only the narrow range of an air-shower between $9.2\,$km and $10.8\,$km above the principal-aperture-plane will be 'sharp', and the rest is blurred.
Here the depth-of-field is only $1.6\,$km.
Depending on the energy, air-showers can have extensions in the atmosphere which exceed the depth-of-field by about an order-of-magnitude.
\section{The Cherenkov-telescope's perception}
\label{SecPerceptionTelescope}
Imaging with Cherenkov-telescopes runs into a physical limit when we want to lower the energy-threshold for cosmic particles.
To lower the energy-threshold for cosmic particles, Cherenkov-telescopes need larger apertures, need to move further up in altitude to get closer to the air-shower, and need higher angular resolution.
But all these three measures:
\begin{itemize}
    \item larger apertures
    \item closer to the air-shower
    \item higher angular resolution
\end{itemize}
are also the key measures to narrow the depth-of-field which will blur the images.
First, in Figure \ref{FigThinLens} we see that when the aperture is enlarged, the points $a$, $a'$, and $a''$ will move further apart, and thus the image $i$, $i'$, and $i''$ will be spread out even more to blur the image.
Second, in the Depth-of-field-equation \ref{EqDepthOfField} we find that the depth-of-field, this is the difference between $g_+$ and $g_-$, becomes narrower the closer we move the Cherenkov-telescope to the air-shower, this is the smaller the object-distance $g$ becomes.
And third, from Figure \ref{FigThinLens} we conclude that when the angular resolution of the pixels is increased, a spreading of the points $i$, $i'$, and $i''$ will be more apparent and thus renders the additional angular resolution useless by blurring the image.\\
Imaging itself becomes a physical limit which prevents us from observing low energetic cosmic gamma-rays in the regime below $25\,$GeV with Cherenkov-telescopes.
\section{The Cherenkov-plenoscope's perception}
\label{SecPerceptionPlenoscope}
But what if the sensor-plane in Figure \ref{FigThinLens} not only knew that three photons arrived in the points $i$, $i'$, and $i''$.
What if it knew that these three photons traveled on the trajectories $\overline{ia}$, $\overline{i'a'}$, and $\overline{i''a''}$.
In this case we knew based on the Thin-lens-equation \ref{EqThinLens}, and the intercept-theorem that the photons approached the aperture on the trajectories $\overline{ae}$, $\overline{a'e}$, and $\overline{a''e}$.
In this case we had a strong hint that there were photons produced in the point $e$.\\
This is plenoptic perception \cite{lippmann1908}, this is what the Cherenkov-plenoscope senses.
With plenoptic perception, the Cherenkov-plenoscope turns the limitations of imaging into three-dimensional reconstruction-power.
With the Cherenkov-plenoscope, the three measures needed for lowering the energy-threshold (larger apertures, closer to the air-shower, higher angular resolution), all improve the three-dimensional reconstruction-power for air-showers.
First, when the aperture-diameter of the imaging-reflector is increased, the baselines for three-dimensional reconstructions are enlarged.
This extends the reconstructible volume of atmosphere further up in front of the aperture-plane.
Second, when the Cherenkov-plenoscope is build higher in altitude, the air-shower will be closer to the aperture where the three-dimensional reconstruction-power is largest due to the finite baseline and finite angular resolution.
Third, when the angular resolution of the pixels in increased, again the reconstructible volume of atmosphere extends further up in front of the aperture-plane.\\
Where the telescope works best with small apertures, the plenoscope works best with large apertures.
The Cherenkov-plenoscope will probably take over the performance of Cherenkov-telescopes at aperture-diameters of about $25\,$m for the reasons of limited perception due to imaging discussed in \cite{bernlohr2013monte}, and \cite{hofmann2001focus}.
\section{Example images of air-showers}
\label{SecExampleImagesOfAirShowers}
We demonstrate the need for focusing and the limitations of a narrow depth-of-field using five simulated observations of air-showers.
For each of the five simulated gamma-ray-events, we compile a collection of four different classes of figures.\\
The first class of figures shows the image of the air-shower recorded with a giant Cherenkov-telescope of the same size of \NameAcp{} with an aperture-diameter of $71\,$m.
The second class of figures shows an array of seven images from the same air-shower recorded by a dense array of seven large Cherenkov-telescopes with an aperture-diameter of $23.7\,$m each.
Figure \ref{FigPortalApertureSegmentationSevenTelescopes} shows how the apertures of the seven large Cherenkov-telescopes are positioned in the aperture of the giant Cherenkov-telescope.
The third class of figures shows eight images from the same air-shower again recorded with the giant $71\,$m Cherenkov-telescope, but this time the focus is set to eight different object-distances.
The fourth class of figures shows the distribution of the true emission-positions of the Cherenkov-photons which were detected by the instruments.
The example figures are grouped as shown in Table \ref{TabFigures}.
%
%The air-showers chosen here were induced by cosmic gamma-rays of energies above $100\,$GeV which run parallel to the optical-axis, but can be off the optical axis.
%
Such high energetic air-showers will be rare in the observations of \NameAcp{}, but serve well as a demonstration for imaging.
If not explicitly stated differently, the images of the Cherenkov-telescopes shown here are focused to an object-distance of $10\,$km.
The figures show only the intensity of photons which were classified to be Cherenkov-photons, see Chapter \ref{ChClassifyingCherenkovPhotons}.
\subsection*{Redefining imaging -- Cherenkov-plenoscope}
\begin{figure}
    \centering
    \includegraphics[width=1\textwidth]{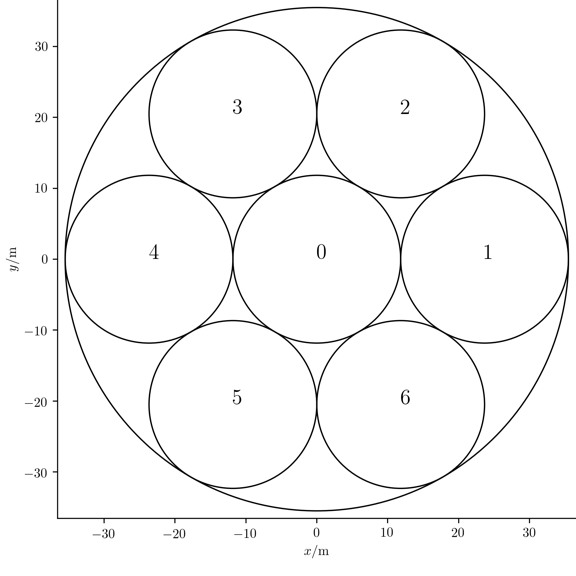}
    \caption[Segmenting \NameAcp{}'s aperture into seven large telescopes]{To demonstrate the effect of different aperture-sizes and different support-positions, we comfortably fit seven $23\,$m Large-Size-Telescopes \cite{cta2013introducing} of the upcoming Cherenkov-Telescope-Array into the aperture of \NameAcp{}.
        The numbers of the seven large Cherenkov-telescopes (0 to 6) correspond to the numbers in the lower left corners of the images in the Figures \ref{FigGammaRay006163Segmented}, \ref{FigGammaRay006106Segmented}, \ref{FigGammaRay006554Segmented}, \ref{FigGammaRay006678Segmented} and \ref{FigGammaRay006912Segmented}.
    }
    \label{FigPortalApertureSegmentationSevenTelescopes}
\end{figure}

All the figures in this chapter show the images of air-showers exactly the way a classic Cherenkov-telescope would have observed them.
However, we create these images from projections of the light-field observed by our \NameAcp{} Cherenkov-plenoscope.
As we discuss in Chapter \ref{ChInterpretingTheLightFieldSequence}, we can project the light-field of the Cherenkov-plenoscope onto images which correspond to images of Cherenkov-telescopes with different support-positions, different aperture-diameters, and different focuses.
For the demonstration of the effects of a narrow depth-of-field on images of air-shower taken by Cherenkov-telescopes, it is not relevant that the images were actually projections of a light-field recorded by a Cherenkov-plenoscope.
But we point this out here to demonstrate that the Cherenkov-plenoscope can always fall back to all the reconstruction-methods for air-showers which were developed for Cherenkov-telescopes and arrays of Cherenkov-telescopes.
Our \NameAcp{} has not only seven but \NumPax{}\,paxels to segment its aperture.
But for the purpose of this demonstration we integrate over these \NumPax{}\,paxels using the mask shown in Figure \ref{FigPortalApertureSegmentationSevenTelescopes} to obtain seven paxels corresponding to an array of seven $23.7\,$m Cherenkov-telescopes.
Compare this to the $23\,$m Large-Size-Telescope \cite{cta2013introducing} of the upcoming Cherenkov-Telescope-Array.
\begin{table}
\begin{center}
    \begin{tabular}{rrrrrrrr}
        gamma-ray&
        energy/GeV&
        $x$/m&
        $y$/m&
        \rotatebox{90}{\parbox{6cm}{Figures\\$71\,$m telescope}}&
        \rotatebox{90}{\parbox{6cm}{Figures\\$23\,$m telescope-array}}&
        \rotatebox{90}{\parbox{6cm}{Figures\\$71\,$m telescope refocused}}&
        \rotatebox{90}{\parbox{6cm}{Figures\\true emission-positions}}\\
        \toprule
        1 &
            121.8&
            -69.7&
            4.5&
            \ref{FigGammaRay006163Full} &
            \ref{FigGammaRay006163Segmented} &
            \ref{FigGammaRay006163Refocused} &
            \ref{FigGammaRay006163True}\\
        2 &
            113.6&
            14.0&
            69.7&
            \ref{FigGammaRay006106Full} &
            \ref{FigGammaRay006106Segmented} &
            \ref{FigGammaRay006106Refocused} &
            \ref{FigGammaRay006106True}\\
        3 &
            197.1&
            51.8&
            -1.0&
            \ref{FigGammaRay006554Full} &
            \ref{FigGammaRay006554Segmented} &
            \ref{FigGammaRay006554Refocused} &
            \ref{FigGammaRay006554True}\\
        4 &
            230.0&
            -10.2&
            56.4&
            \ref{FigGammaRay006678Full} &
            \ref{FigGammaRay006678Segmented} &
            \ref{FigGammaRay006678Refocused} &
            \ref{FigGammaRay006678True}\\
        5 &
            308.0&
            10.2&
            2.4&
            \ref{FigGammaRay006912Full} &
            \ref{FigGammaRay006912Segmented} &
            \ref{FigGammaRay006912Refocused} &
            \ref{FigGammaRay006912True}\\
    \end{tabular}
    \end{center}
    \caption[Properties of gamma-rays in air-shower-images]{Basic properties of the gamma-rays shown in the example images.
        All gamma-rays come from zenith, and run parallel to the optical-axis of the telescopes.
        Here $x$, and $y$ is the positions where the elongated trajectory of the cosmic gamma-ray intersects the principal-aperture-plane.
        Also the corresponding figure-numbers are shown.}
    \label{TabFigures}
\end{table}
\section{Discussion}
\label{SecDepthOfFieldDiscussion}
\begin{figure}
    \centering
    \includegraphics[width=0.7\textwidth]{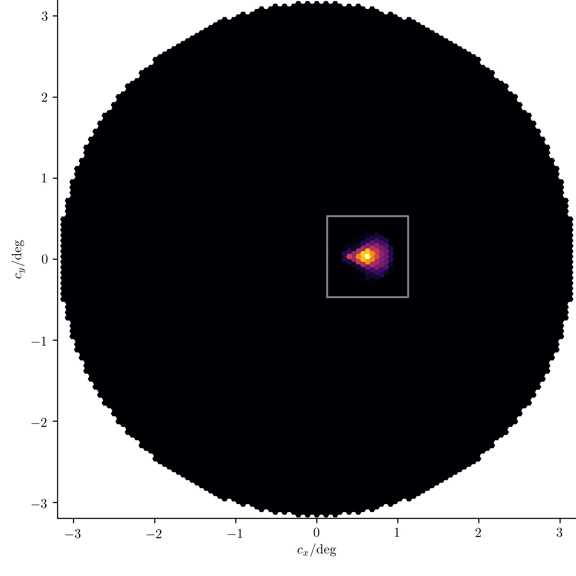}
    \caption[Example image of gamma-ray, full $71\,$m aperture]{Image of a $71\,$m Cherenkov-telescope.
    }
    \label{FigGammaRay006163Full}
\end{figure}

\begin{figure}
    \centering
    \includegraphics[trim=13 32 13 32, clip, width=.8\textwidth]{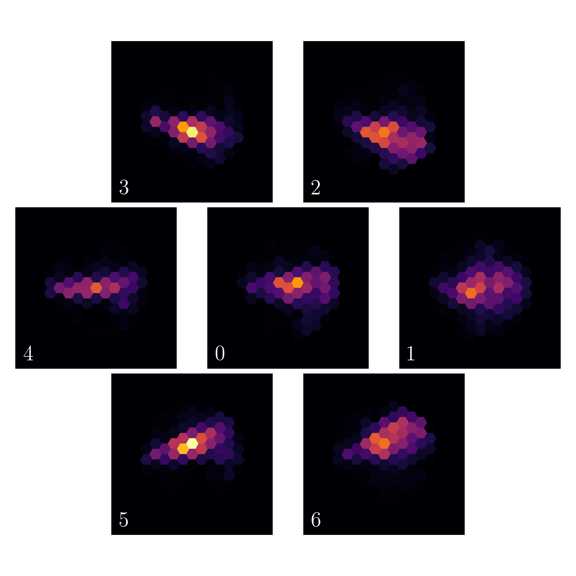}
    \caption[Example image of gamma-ray, full $71\,$m aperture]{Seven classic images from an array of Cherenkov-telescopes, see Figure \ref{FigPortalApertureSegmentationSevenTelescopes}.
    }
    \label{FigGammaRay006163Segmented}
\end{figure}

\begin{figure}
    \centering
    \includegraphics[width=1\textwidth]{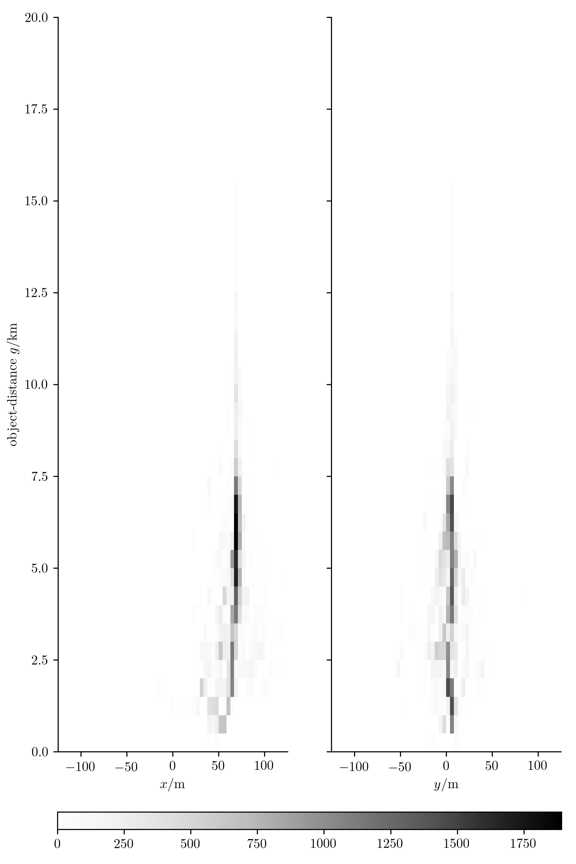}
    \caption[Example image of gamma-ray, truth]{The true emission-distribution of Cherenkov-photons which were detected.
    }
    \label{FigGammaRay006163True}
\end{figure}

\begin{figure}
    \centering
    \begin{minipage}{0.42\textwidth}
        \includegraphics[width=1\textwidth]{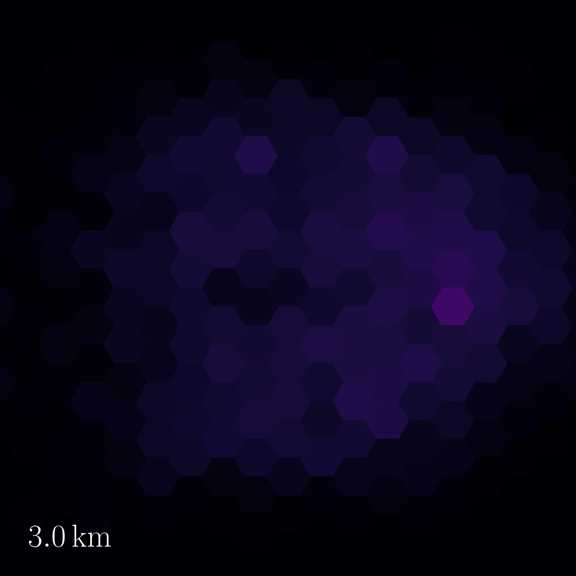}
    \end{minipage}
    \begin{minipage}{0.42\textwidth}
        \includegraphics[width=1\textwidth]{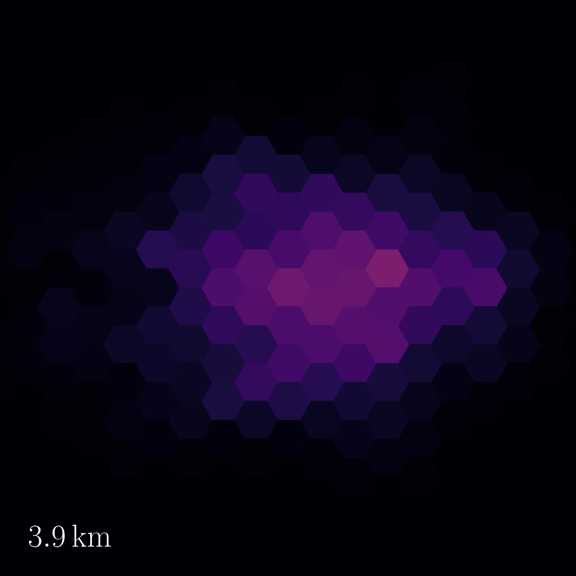}
    \end{minipage}\\
    \vspace{0.1cm}
    \begin{minipage}{0.42\textwidth}
        \includegraphics[width=1\textwidth]{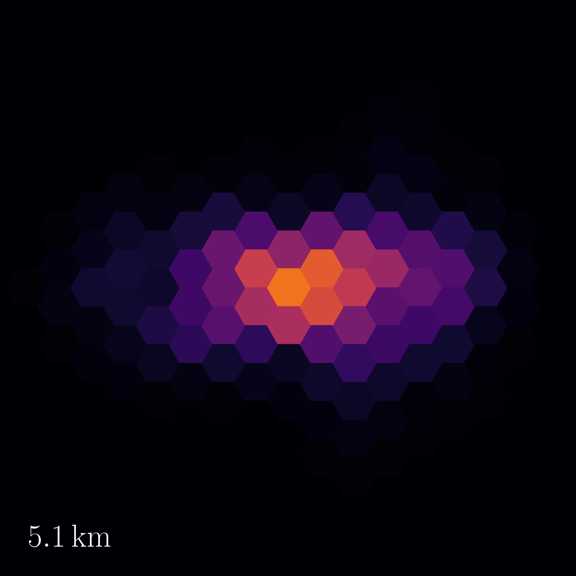}
    \end{minipage}
    \begin{minipage}{0.42\textwidth}
        \includegraphics[width=1\textwidth]{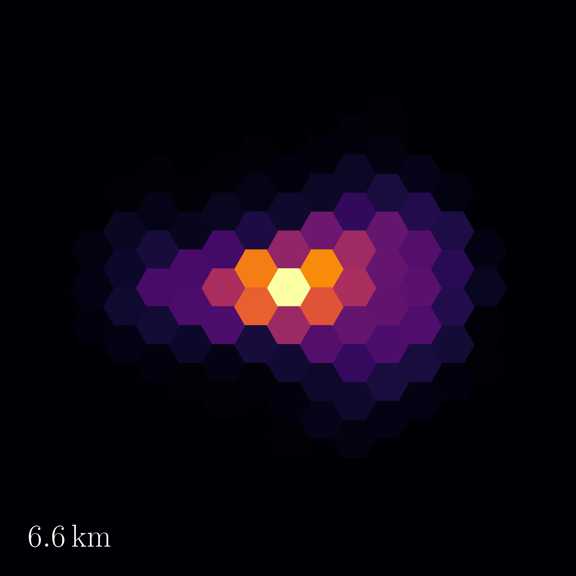}
    \end{minipage}\\
    \vspace{0.1cm}
    \begin{minipage}{0.42\textwidth}
        \includegraphics[width=1\textwidth]{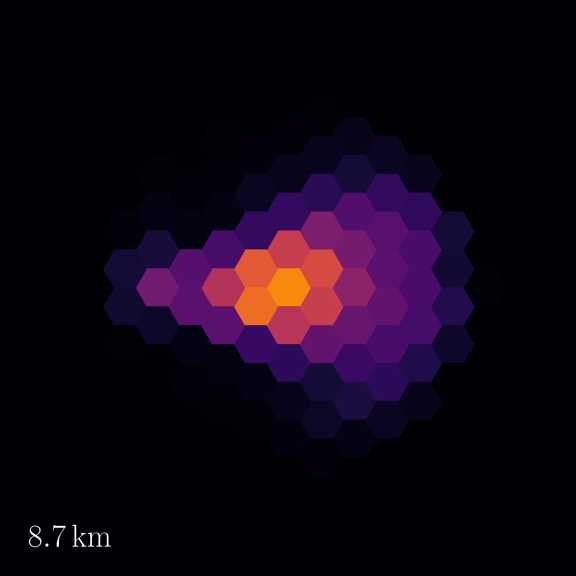}
    \end{minipage}
    \begin{minipage}{0.42\textwidth}
        \includegraphics[width=1\textwidth]{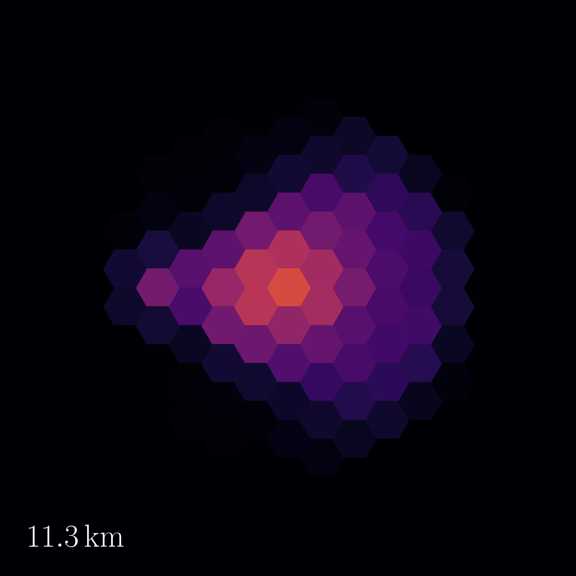}
    \end{minipage}\\
    \vspace{0.1cm}
    \begin{minipage}{0.42\textwidth}
        \includegraphics[width=1\textwidth]{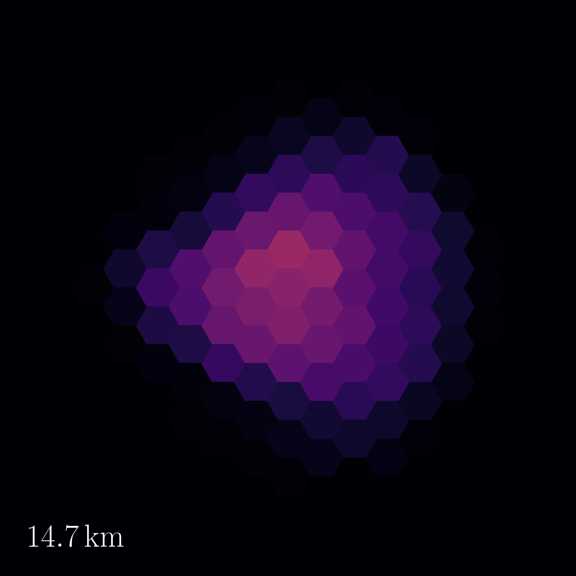}
    \end{minipage}
    \begin{minipage}{0.42\textwidth}
        \includegraphics[width=1\textwidth]{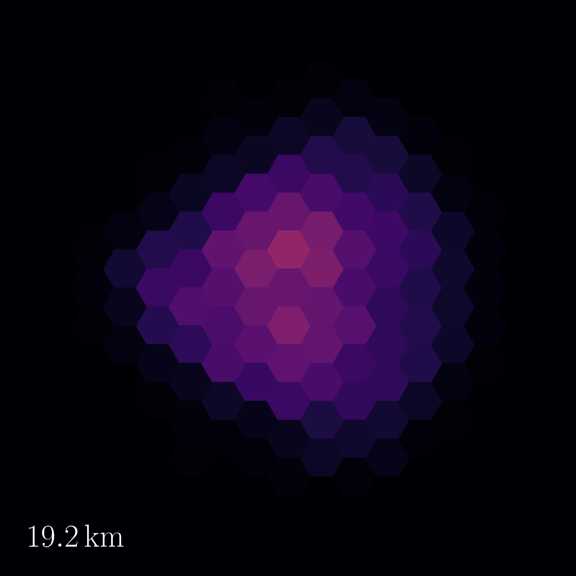}
    \end{minipage}\\
    \caption[Example image of gamma-ray, refocused]{Image taken with full $71\,$m aperture refocused to different object-distances $g$, see lower left corner of each image.
    }
    \label{FigGammaRay006163Refocused}
\end{figure}

\begin{figure}
    \centering
    \includegraphics[width=0.7\textwidth]{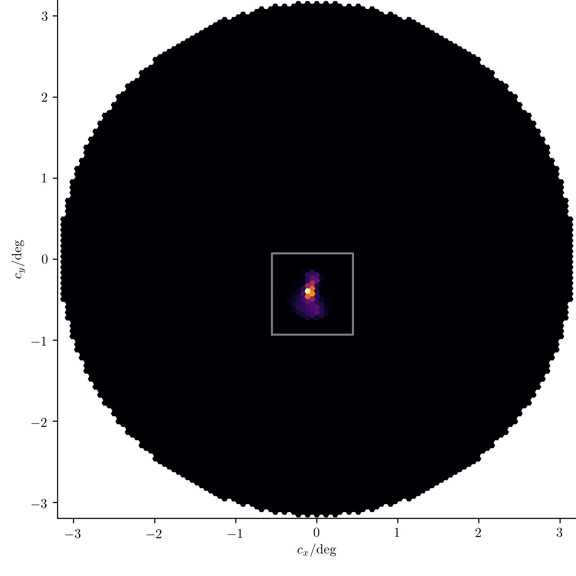}
    \caption[Example image of gamma-ray, full $71\,$m aperture]{Image of a $71\,$m Cherenkov-telescope.
    }
    \label{FigGammaRay006106Full}
\end{figure}

\begin{figure}
    \centering
    \includegraphics[trim=13 32 13 32, clip, width=.8\textwidth]{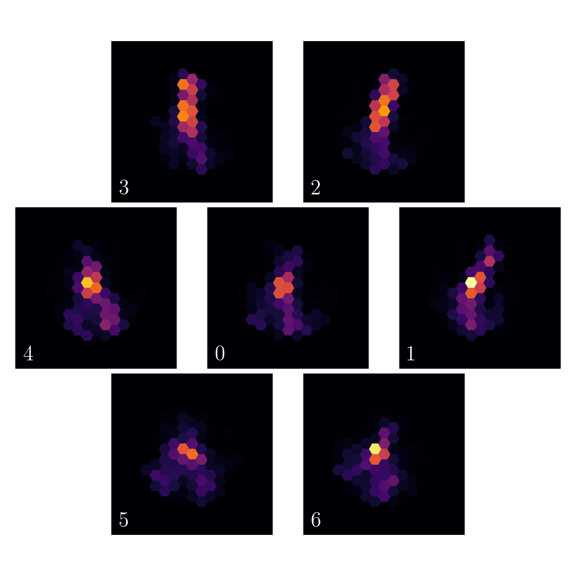}
    \caption[Example image of gamma-ray, full $71\,$m aperture]{Images of seven $23\,$m Cherenkov-telescopes.
    }
    \label{FigGammaRay006106Segmented}
\end{figure}

\begin{figure}
    \centering
    \includegraphics[width=1\textwidth]{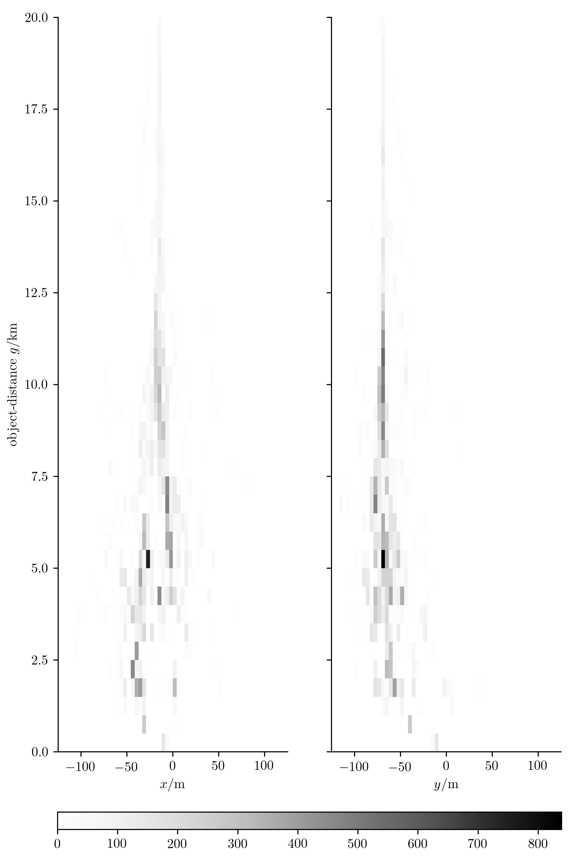}
    \caption[Example image of gamma-ray, truth]{The true emission-distribution of Cherenkov-photons which were detected.
    }
    \label{FigGammaRay006106True}
\end{figure}

\begin{figure}
    \centering
    \begin{minipage}{0.42\textwidth}
        \includegraphics[width=1\textwidth]{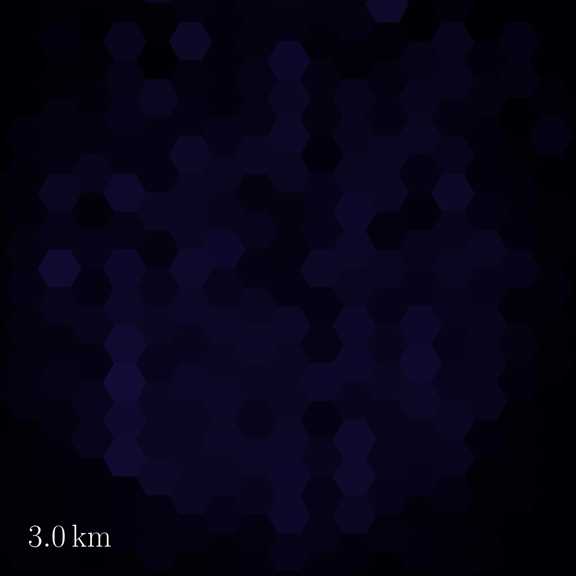}
    \end{minipage}
    \begin{minipage}{0.42\textwidth}
        \includegraphics[width=1\textwidth]{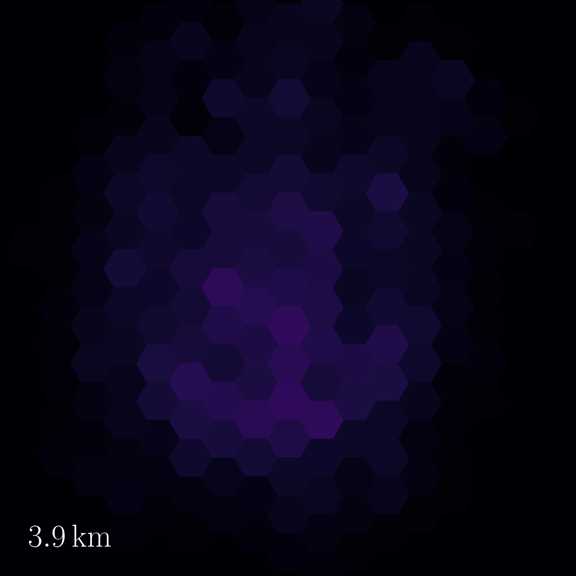}
    \end{minipage}\\
    \vspace{0.1cm}
    \begin{minipage}{0.42\textwidth}
        \includegraphics[width=1\textwidth]{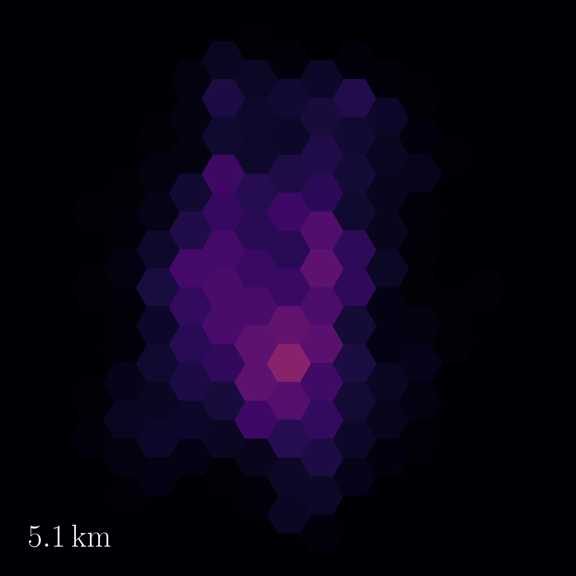}
    \end{minipage}
    \begin{minipage}{0.42\textwidth}
        \includegraphics[width=1\textwidth]{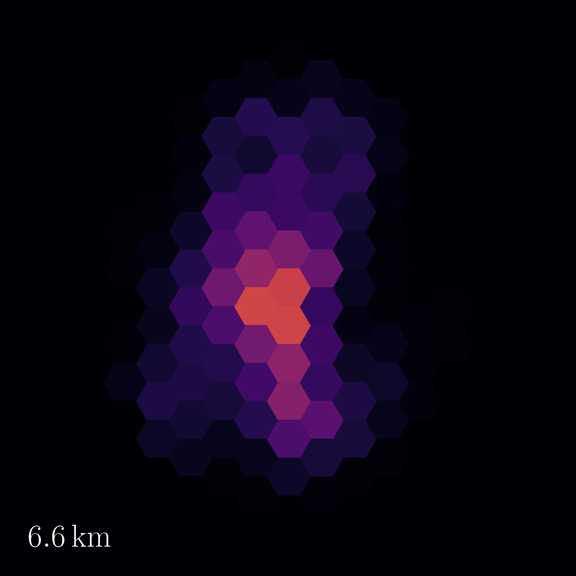}
    \end{minipage}\\
    \vspace{0.1cm}
    \begin{minipage}{0.42\textwidth}
        \includegraphics[width=1\textwidth]{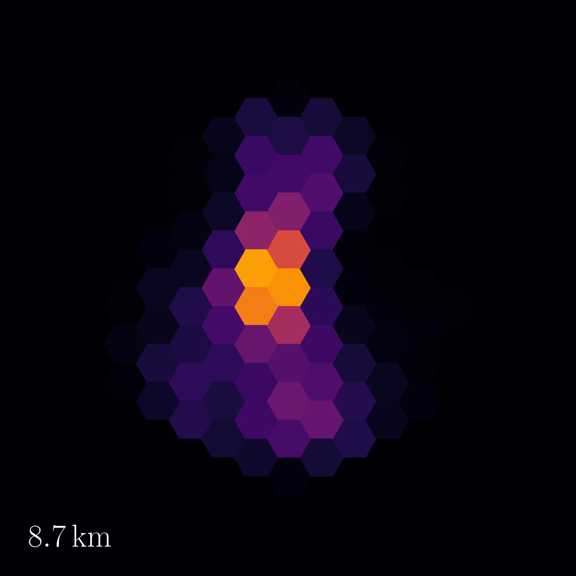}
    \end{minipage}
    \begin{minipage}{0.42\textwidth}
        \includegraphics[width=1\textwidth]{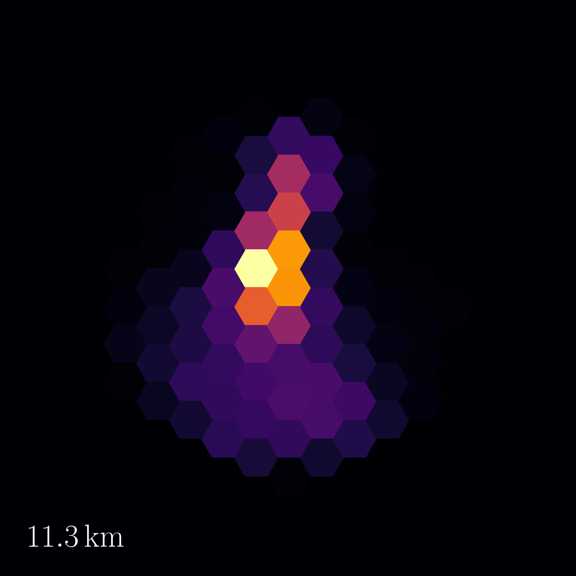}
    \end{minipage}\\
    \vspace{0.1cm}
    \begin{minipage}{0.42\textwidth}
        \includegraphics[width=1\textwidth]{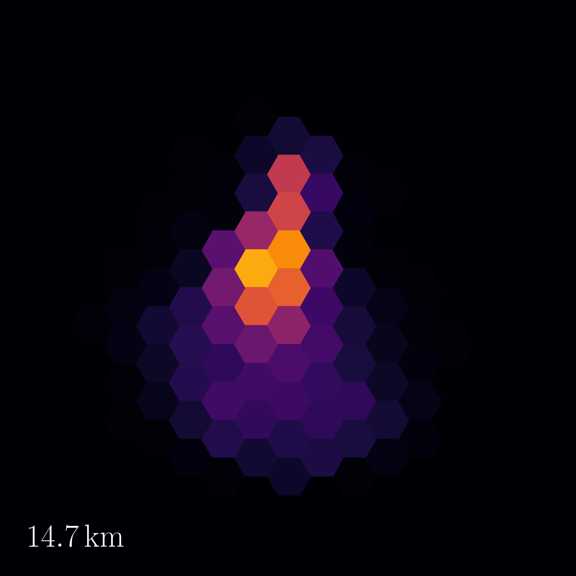}
    \end{minipage}
    \begin{minipage}{0.42\textwidth}
        \includegraphics[width=1\textwidth]{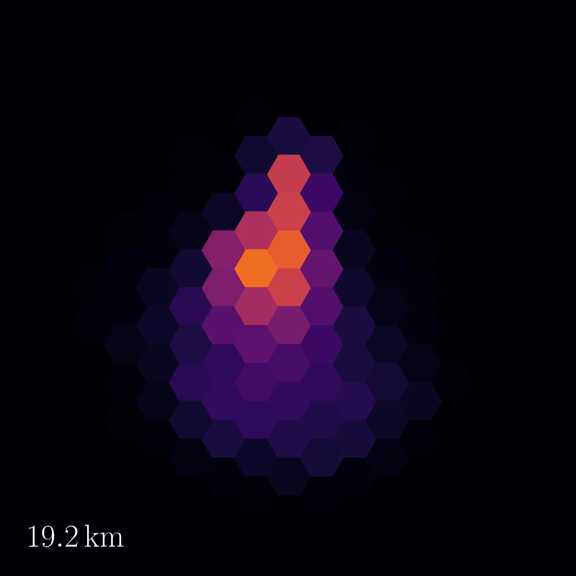}
    \end{minipage}\\
    \caption[Example image of gamma-ray, refocused]{Same event as Figure \ref{FigGammaRay006106Full}.
        Image taken with full $71\,$m aperture refocused to different object-distances $g$, see lower left corner of each image.
    }
    \label{FigGammaRay006106Refocused}
\end{figure}

\begin{figure}
    \centering
    \includegraphics[width=0.7\textwidth]{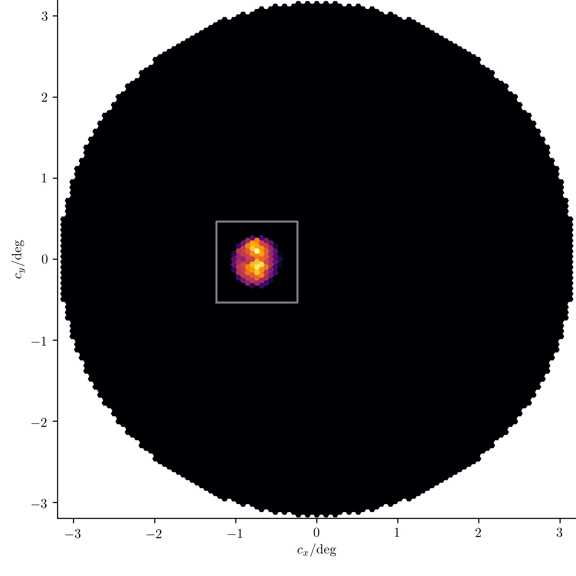}
    \caption[Example image of gamma-ray, full $71\,$m aperture]{A gamma-ray coming from zenith, the elongated trajectory of the cosmic gamma-ray intersects the aperture plane $72\,$m off the optical-axis.
        Full $71\,$m aperture.
    }
    \label{FigGammaRay006554Full}
\end{figure}

\begin{figure}
    \centering
    \includegraphics[trim=13 32 13 32, clip, width=.8\textwidth]{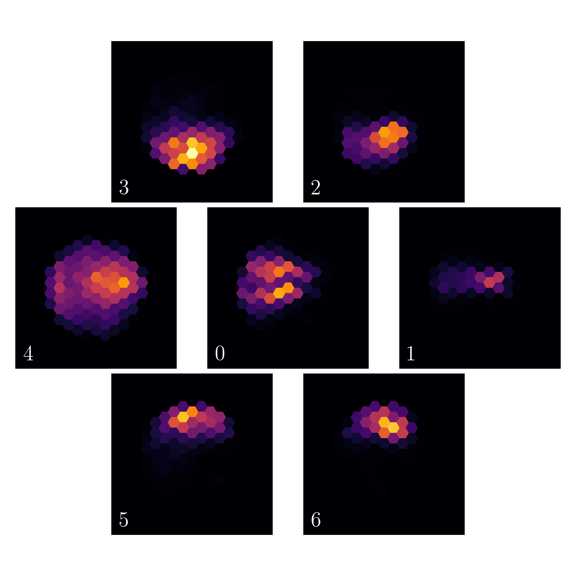}
    \caption[Example image of gamma-ray, full $71\,$m aperture]{A gamma-ray coming from zenith, the elongated trajectory of the cosmic gamma-ray intersects the aperture plane $72\,$m off the optical-axis.
        Full $71\,$m aperture.
    }
    \label{FigGammaRay006554Segmented}
\end{figure}

\begin{figure}
    \centering
    \includegraphics[width=1\textwidth]{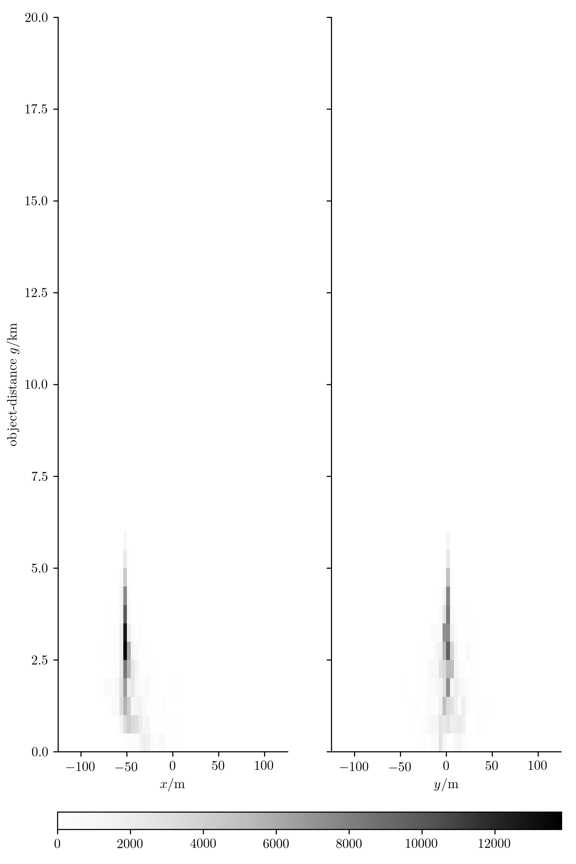}
    \caption[Example image of gamma-ray, truth]{The true emission-distribution of Cherenkov-photons which were detected.
    }
    \label{FigGammaRay006554True}
\end{figure}

\begin{figure}
    \centering
    \begin{minipage}{0.42\textwidth}
        \includegraphics[width=1\textwidth]{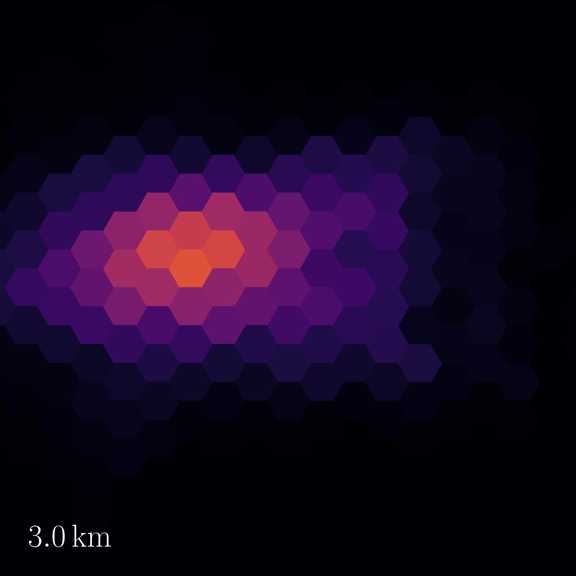}
    \end{minipage}
    \begin{minipage}{0.42\textwidth}
        \includegraphics[width=1\textwidth]{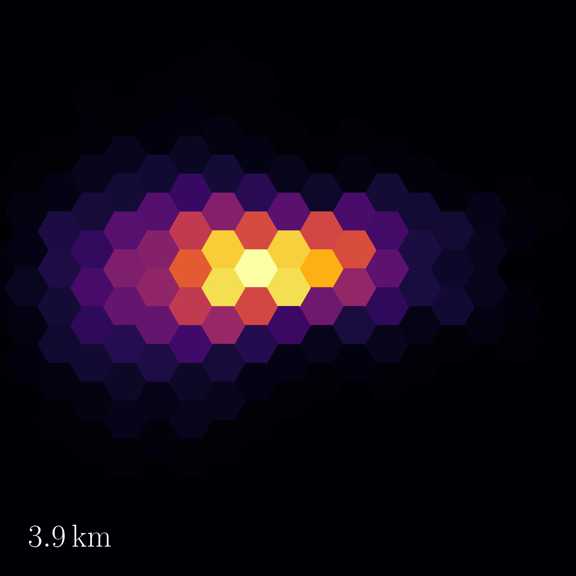}
    \end{minipage}\\
    \vspace{0.1cm}
    \begin{minipage}{0.42\textwidth}
        \includegraphics[width=1\textwidth]{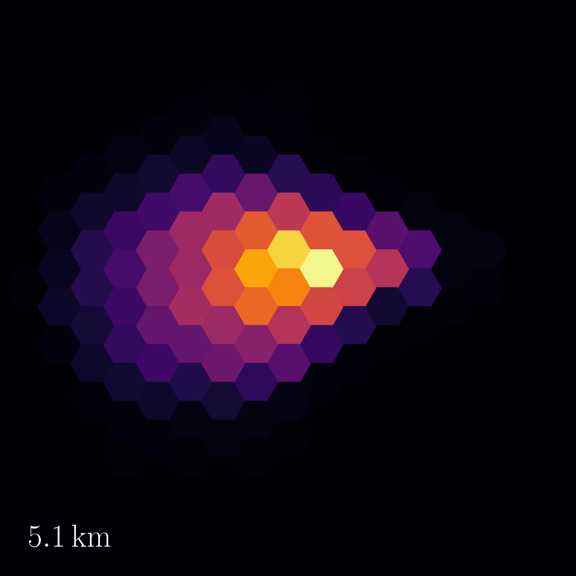}
    \end{minipage}
    \begin{minipage}{0.42\textwidth}
        \includegraphics[width=1\textwidth]{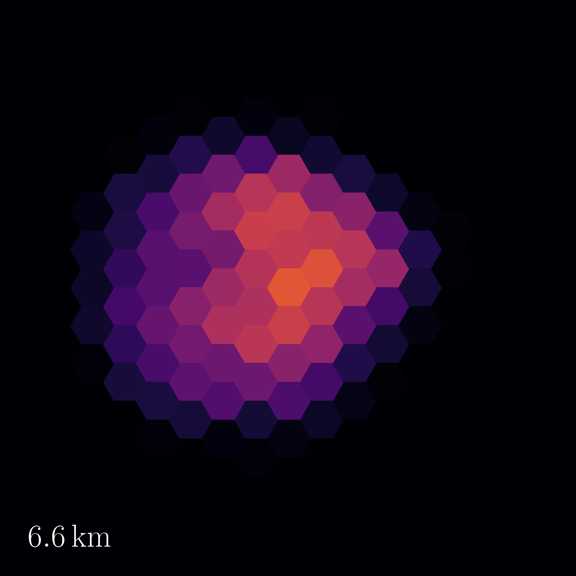}
    \end{minipage}\\
    \vspace{0.1cm}
    \begin{minipage}{0.42\textwidth}
        \includegraphics[width=1\textwidth]{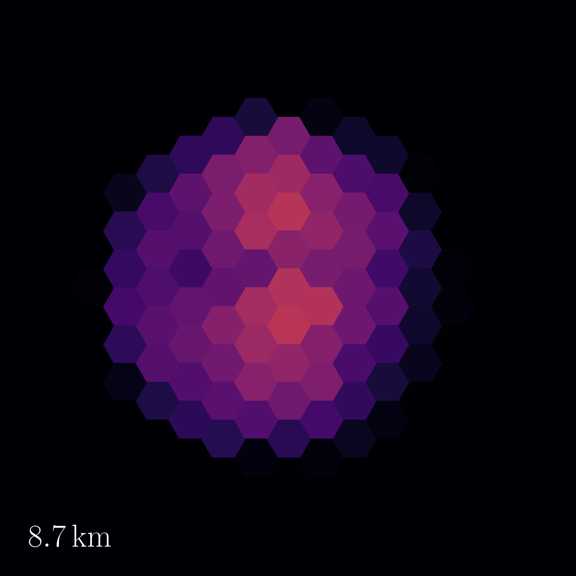}
    \end{minipage}
    \begin{minipage}{0.42\textwidth}
        \includegraphics[width=1\textwidth]{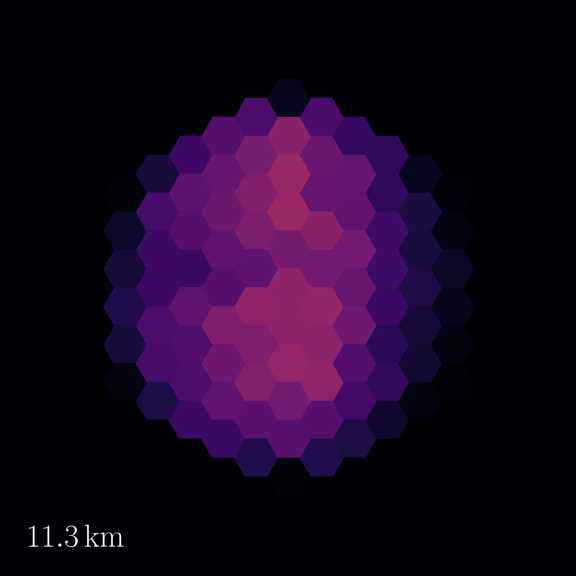}
    \end{minipage}\\
    \vspace{0.1cm}
    \begin{minipage}{0.42\textwidth}
        \includegraphics[width=1\textwidth]{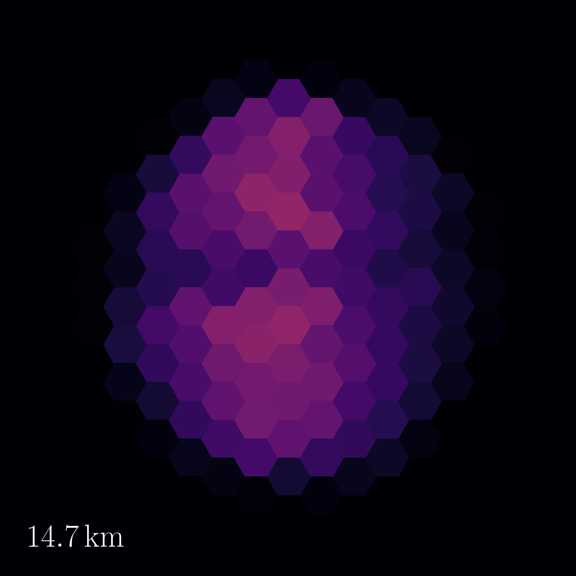}
    \end{minipage}
    \begin{minipage}{0.42\textwidth}
        \includegraphics[width=1\textwidth]{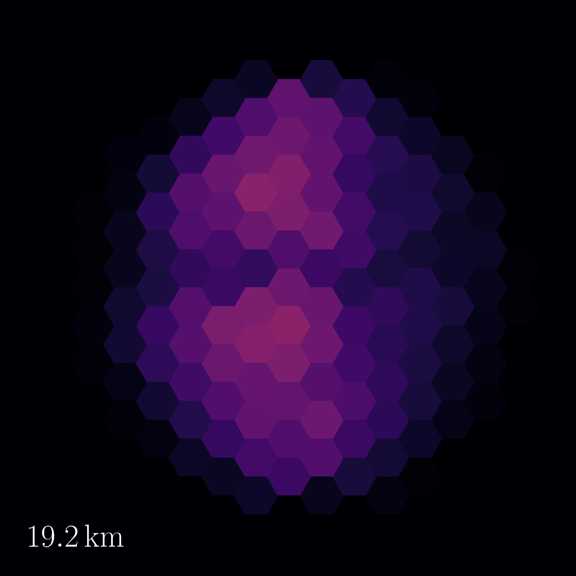}
    \end{minipage}\\
    \caption[Example image of gamma-ray, refocused]{Same event as Figure \ref{FigGammaRay006554Full}.
        Image taken with full $71\,$m aperture refocused to different object-distances $g$, see lower left corner of each image.
    }
    \label{FigGammaRay006554Refocused}
\end{figure}

\begin{figure}
    \centering
    \includegraphics[width=0.7\textwidth]{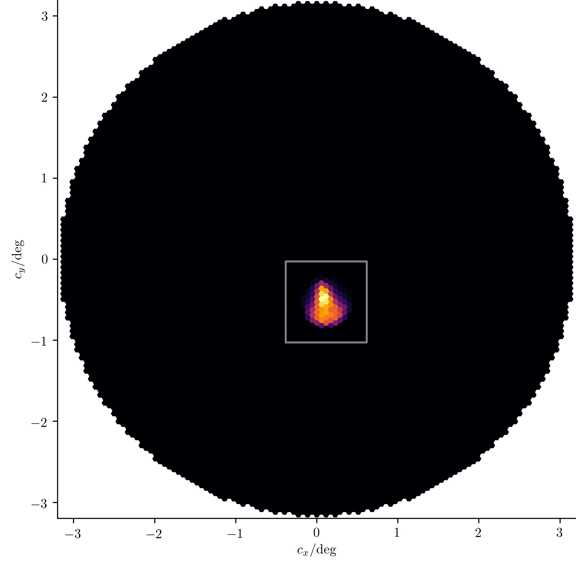}
    \caption[Example image of gamma-ray, full $71\,$m aperture]{A gamma-ray coming from zenith, the elongated trajectory of the cosmic gamma-ray intersects the aperture plane $72\,$m off the optical-axis.
        Full $71\,$m aperture.
    }
    \label{FigGammaRay006678Full}
\end{figure}

\begin{figure}
    \centering
    \includegraphics[trim=13 32 13 32, clip, width=.8\textwidth]{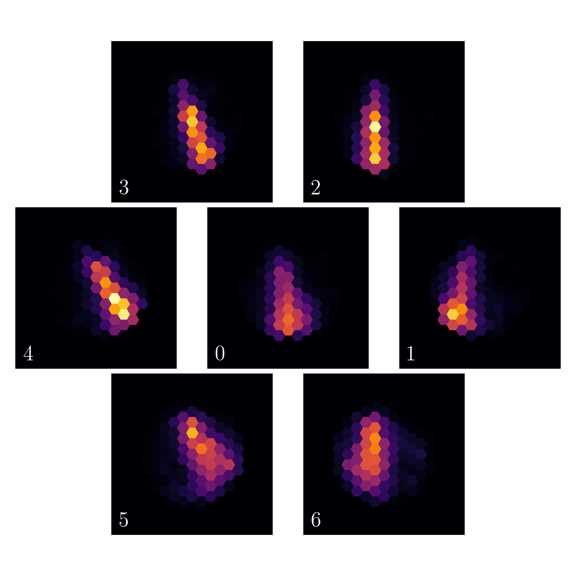}
    \caption[Example image of gamma-ray, full $71\,$m aperture]{A gamma-ray coming from zenith, the elongated trajectory of the cosmic gamma-ray intersects the aperture plane $72\,$m off the optical-axis.
        Full $71\,$m aperture.
    }
    \label{FigGammaRay006678Segmented}
\end{figure}

\begin{figure}
    \centering
    \includegraphics[width=1\textwidth]{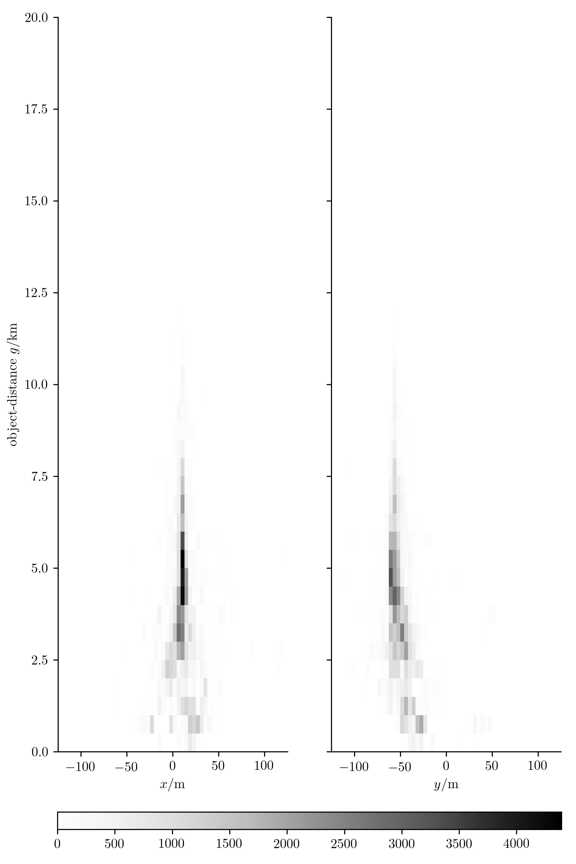}
    \caption[Example image of gamma-ray, truth]{The true emission-distribution of Cherenkov-photons which were detected.
    }
    \label{FigGammaRay006678True}
\end{figure}

\begin{figure}
    \centering
    \begin{minipage}{0.42\textwidth}
        \includegraphics[width=1\textwidth]{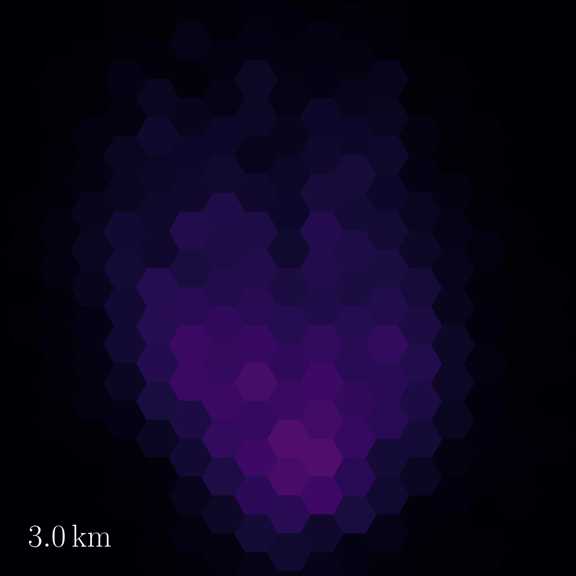}
    \end{minipage}
    \begin{minipage}{0.42\textwidth}
        \includegraphics[width=1\textwidth]{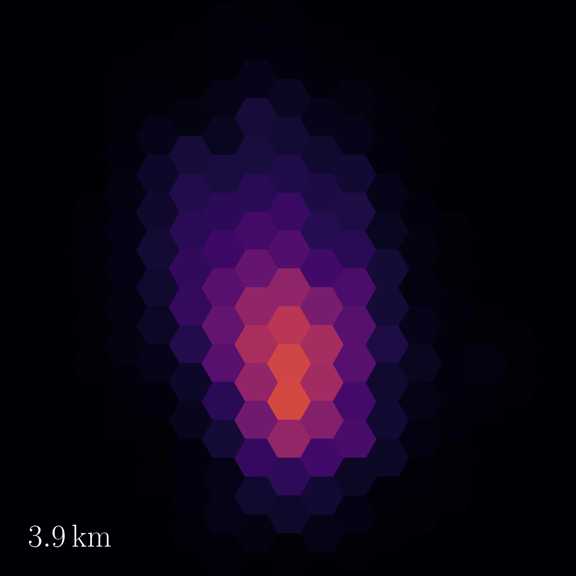}
    \end{minipage}\\
    \vspace{0.1cm}
    \begin{minipage}{0.42\textwidth}
        \includegraphics[width=1\textwidth]{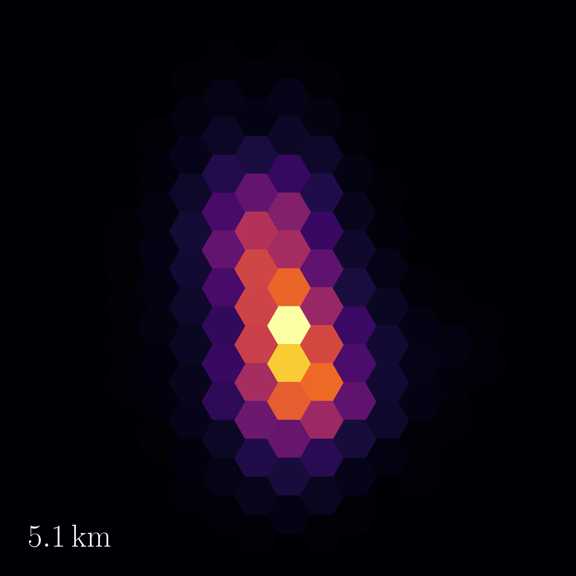}
    \end{minipage}
    \begin{minipage}{0.42\textwidth}
        \includegraphics[width=1\textwidth]{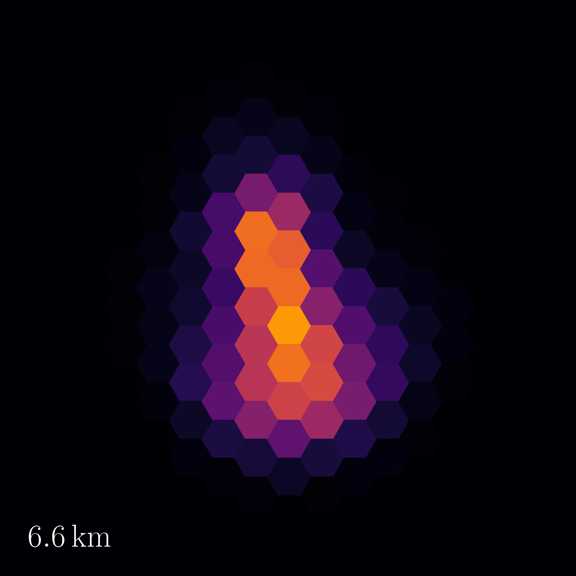}
    \end{minipage}\\
    \vspace{0.1cm}
    \begin{minipage}{0.42\textwidth}
        \includegraphics[width=1\textwidth]{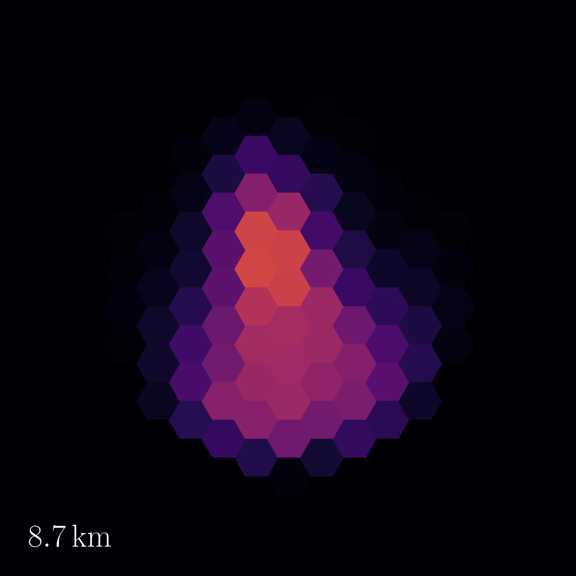}
    \end{minipage}
    \begin{minipage}{0.42\textwidth}
        \includegraphics[width=1\textwidth]{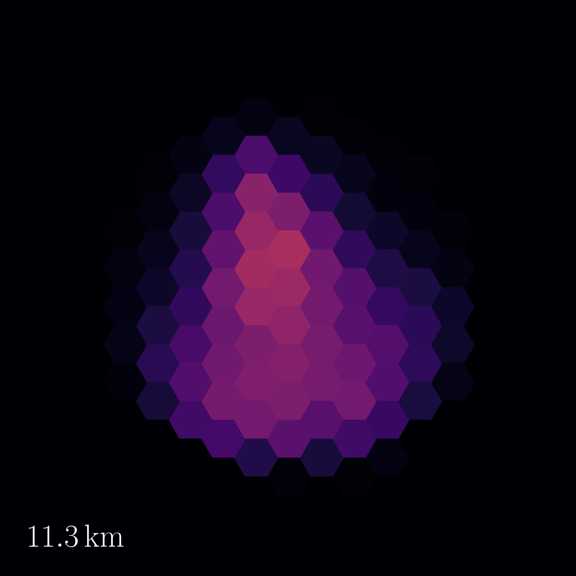}
    \end{minipage}\\
    \vspace{0.1cm}
    \begin{minipage}{0.42\textwidth}
        \includegraphics[width=1\textwidth]{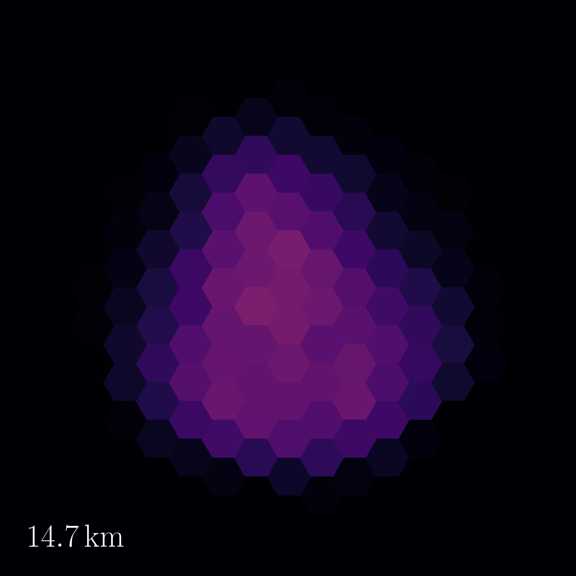}
    \end{minipage}
    \begin{minipage}{0.42\textwidth}
        \includegraphics[width=1\textwidth]{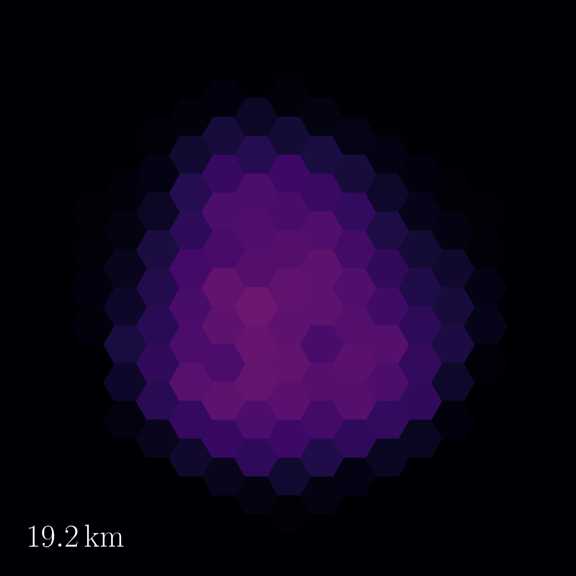}
    \end{minipage}\\
    \caption[Example image of gamma-ray, refocused]{Same event as Figure \ref{FigGammaRay006678Full}.
        Image taken with full $71\,$m aperture refocused to different object-distances $g$, see lower left corner of each image.
    }
    \label{FigGammaRay006678Refocused}
\end{figure}

\begin{figure}
    \centering
    \includegraphics[width=0.7\textwidth]{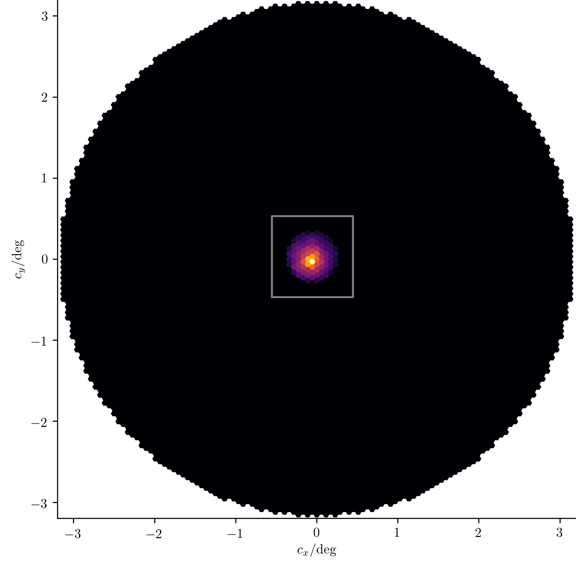}
    \caption[Example image of gamma-ray, full $71\,$m aperture]{A gamma-ray coming from zenith, the elongated trajectory of the cosmic gamma-ray intersects the aperture plane $72\,$m off the optical-axis.
        Full $71\,$m aperture.
    }
    \label{FigGammaRay006912Full}
\end{figure}

\begin{figure}
    \centering
    \includegraphics[trim=13 32 13 32, clip, width=.8\textwidth]{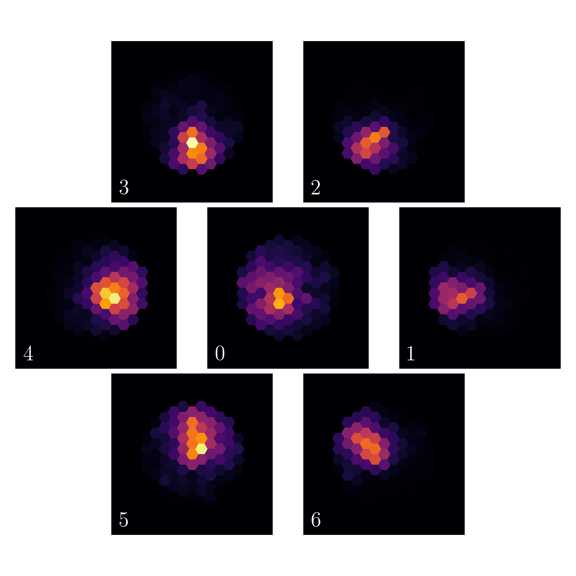}
    \caption[Example image of gamma-ray, full $71\,$m aperture]{A gamma-ray coming from zenith, the elongated trajectory of the cosmic gamma-ray intersects the aperture plane $72\,$m off the optical-axis.
        Full $71\,$m aperture.
    }
    \label{FigGammaRay006912Segmented}
\end{figure}

\begin{figure}
    \centering
    \includegraphics[width=1\textwidth]{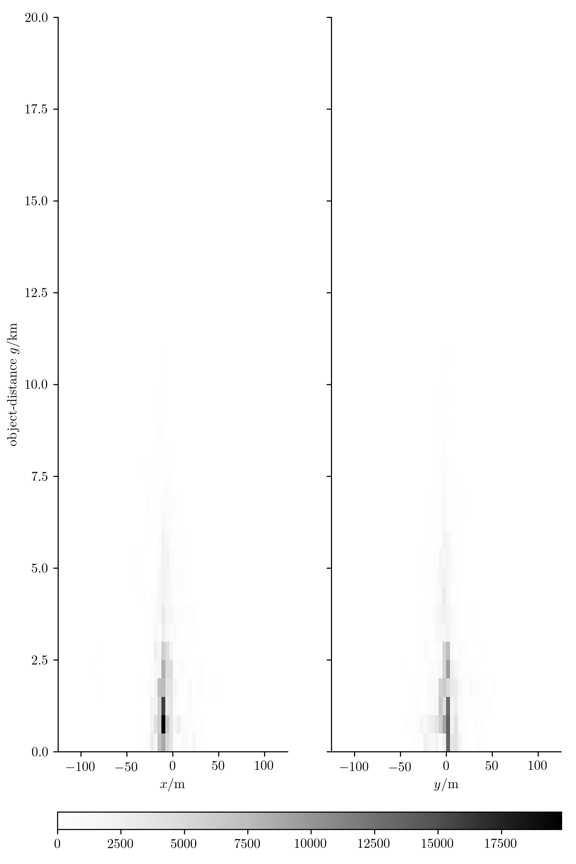}
    \caption[Example image of gamma-ray, truth]{The true emission-distribution of Cherenkov-photons which were detected.
    }
    \label{FigGammaRay006912True}
\end{figure}

\begin{figure}
    \centering
    \begin{minipage}{0.42\textwidth}
        \includegraphics[width=1\textwidth]{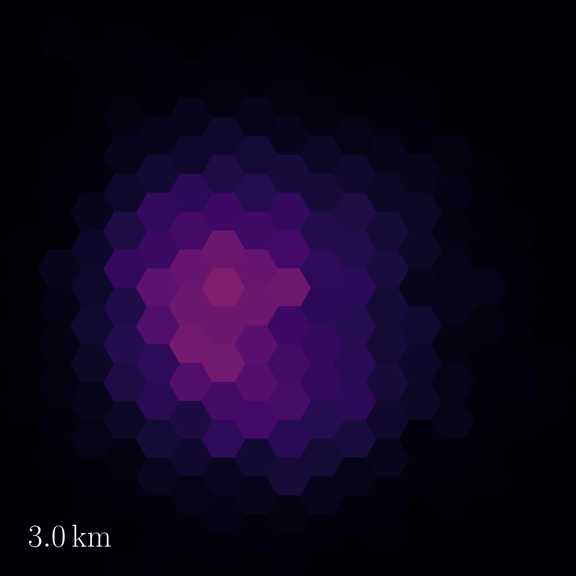}
    \end{minipage}
    \begin{minipage}{0.42\textwidth}
        \includegraphics[width=1\textwidth]{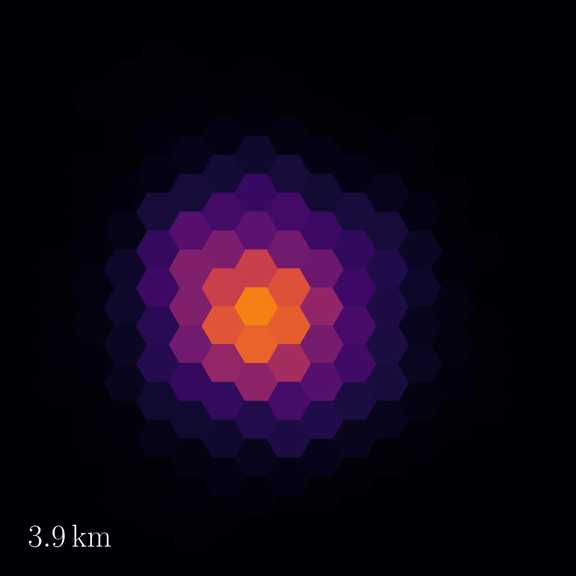}
    \end{minipage}\\
    \vspace{0.1cm}
    \begin{minipage}{0.42\textwidth}
        \includegraphics[width=1\textwidth]{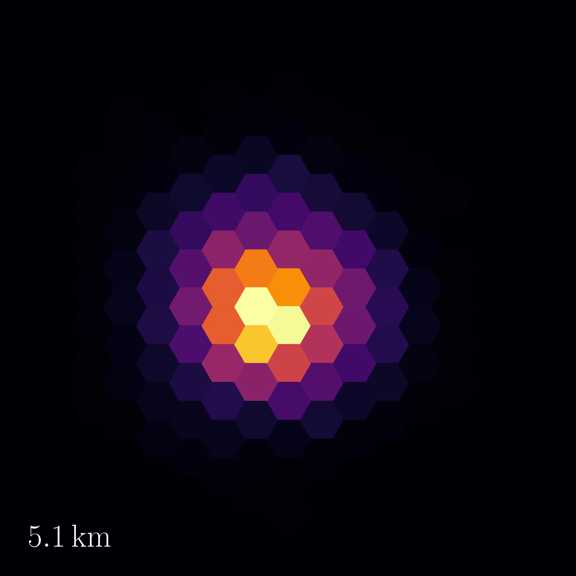}
    \end{minipage}
    \begin{minipage}{0.42\textwidth}
        \includegraphics[width=1\textwidth]{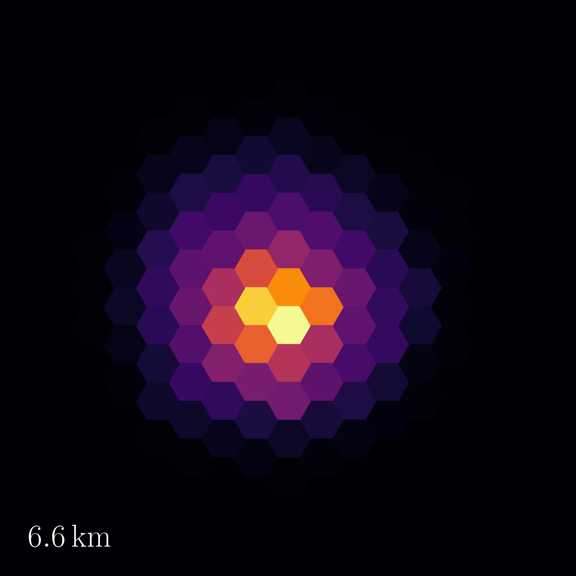}
    \end{minipage}\\
    \vspace{0.1cm}
    \begin{minipage}{0.42\textwidth}
        \includegraphics[width=1\textwidth]{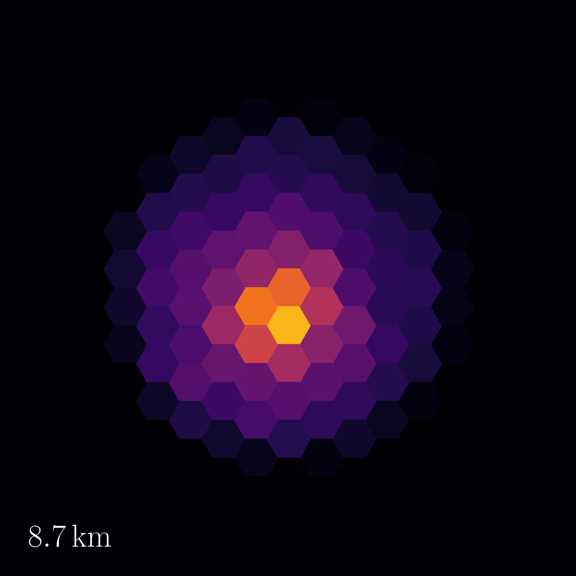}
    \end{minipage}
    \begin{minipage}{0.42\textwidth}
        \includegraphics[width=1\textwidth]{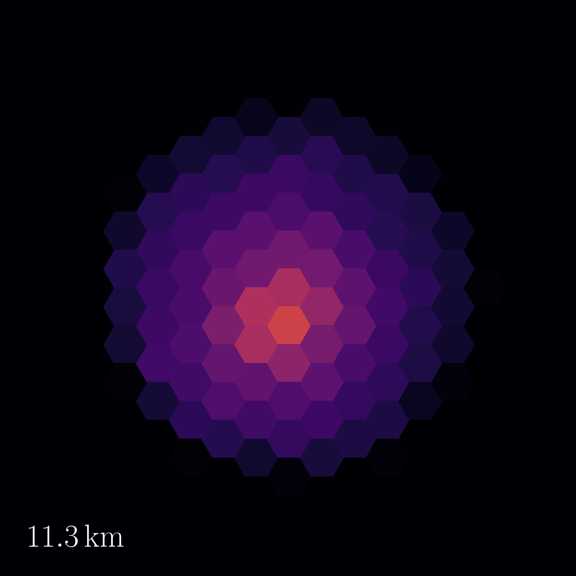}
    \end{minipage}\\
    \vspace{0.1cm}
    \begin{minipage}{0.42\textwidth}
        \includegraphics[width=1\textwidth]{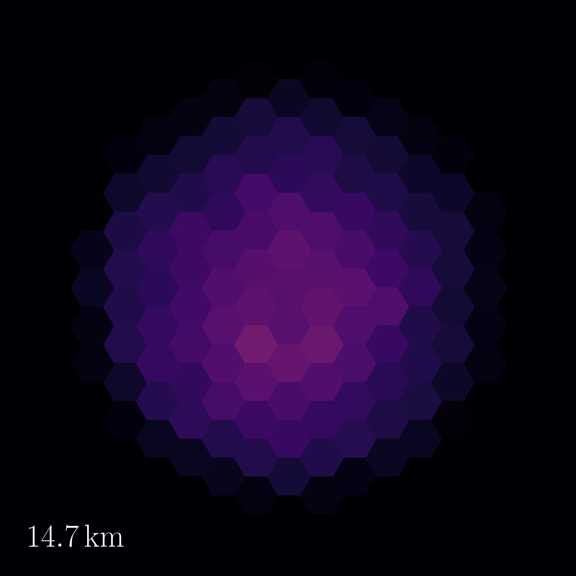}
    \end{minipage}
    \begin{minipage}{0.42\textwidth}
        \includegraphics[width=1\textwidth]{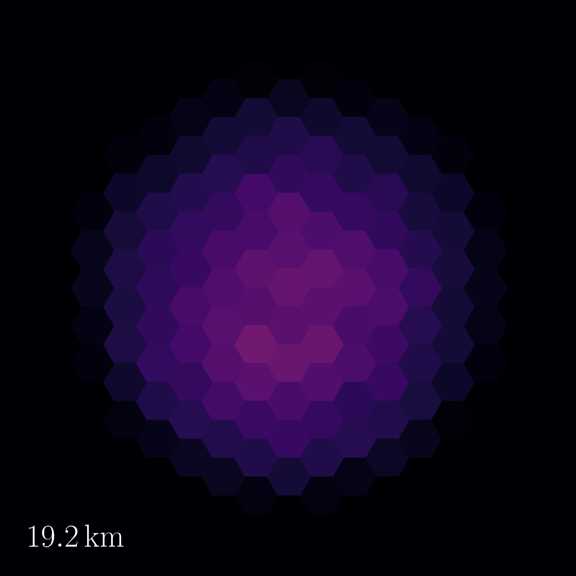}
    \end{minipage}\\
    \caption[Example image of gamma-ray, refocused]{Same event as Figure \ref{FigGammaRay006912Full}.
        Image taken with full $71\,$m aperture refocused to different object-distances $g$, see lower left corner of each image.
    }
    \label{FigGammaRay006912Refocused}
\end{figure}

In the images recorded by a giant $71\,$m telescope shown in the Figures %
\ref{FigGammaRay006163Full}, \ref{FigGammaRay006106Full}, \ref{FigGammaRay006554Full}, \ref{FigGammaRay006678Full}, and \ref{FigGammaRay006912Full}, %
we find that the air-showers do not look like ellipses anymore as it is described by Hillas' model \cite{hillas1985cerenkov}.
Depending on the distance between the cosmic particle's trajectory and the optical-axis, the air-shower-images either have a triangular, or a circular shape, but not the shape of an ellipse.
On the other hand, the air-shower-images recorded by the seven $23\,$m telescopes in the Figures %
\ref{FigGammaRay006163Segmented}, \ref{FigGammaRay006106Segmented}, \ref{FigGammaRay006554Segmented}, \ref{FigGammaRay006678Segmented}, and \ref{FigGammaRay006912Segmented},
do look much more like symmetric ellipses according to Hillas.
In the seven images recorded by the seven $23\,$m telescopes we find, as expected from stereoscopic arrays of Cherenkov-telescopes, that the main-axes of the ellipses in the individual images intersect in one point.
Since the seven $23\,$m telescopes are located at different support-positions, see Figure \ref{FigPortalApertureSegmentationSevenTelescopes}, they observe the air-showers from different perspectives which allows for stereoscopic reconstructions of the air-showers.
Now the images of the giant $71\,$m telescope are blurred and difficult to interpret because these are the sum of those seven images recorded by seven individual $23\,$m telescopes.
The image of the $71\,$m telescope, can not be separated again into the individual images of the seven $23\,$m telescopes.
All this information about the individual main-axes of the ellipses, which is so powerful to reconstruct the incident-direction of the cosmic gamma-ray, is lost in the image of the $71\,$m telescope.\\
In the Figures %
\ref{FigGammaRay006163Refocused}, \ref{FigGammaRay006106Refocused}, \ref{FigGammaRay006554Refocused}, \ref{FigGammaRay006678Refocused}, and \ref{FigGammaRay006912Refocused},
we show how drastically the air-shower-images change when we choose to focus the $71\,$m telescope to different object-distances.
To be clear, on a real Cherenkov-telescope there is no way to record such a series of images with the focus set to different object-distances.
But here we can simulate the same air-shower multiple times, while the telescope focus is varied.
Now when we look at the stack of refocused air-shower-images, we find that there are images in the middle of these stacks where the density of Cherenkov-photons is highest, while in return the density is lowest for the images at the lower and upper end of the stack.
When we look at the density of the true emission-positions of these photons in the Figures %
\ref{FigGammaRay006163True}, \ref{FigGammaRay006106True}, \ref{FigGammaRay006554True}, \ref{FigGammaRay006678True}, and \ref{FigGammaRay006912True},
we find that the object-distances for the air-shower-images in the focus-stack with the highest photon-densities, correlate with the object-distances for the highest density of true emission-positions.
Although all images in the focus-stack contain the same Cherenkov-photons, these images look different because each one focuses to just one thin slice of the air-shower along the optical-axis.
A stack of refocused images can be used for the three-dimensional, tomographic reconstruction of air-showers, see Chapter \ref{ChTomography}.
\section{Conclusion}
\label{SecDepthOfFieldConclusion}
The Cherenkov-plenoscope not only overcomes the physical limit of the narrow depth-of-field, but it turns the tables on it.
All the three measures proposed to lower the energy-threshold on Cherenkov-telescopes: Larger apertures, closer to the air-shower, and higher angular resolution; all toughen the depth-of-field-limitations on Cherenkov-telescopes, but strengthen the reconstruction-power of the Cherenkov-plenoscope.\\
%
%In our discussion, we only talk about images recorded by Cherenkov-telescopes, and we even point out that the focus-stack could not be recorded with an Cherenkov-telescope because of the short lived air-showers.
%
%However, the $71\,$m \NameAcp{} Cherenkov-plenoscope can provide all these images and even more based on a single recorded light-field-sequence of an air-shower.
%
\newpage
%------------------ 006163_000719_000016_01 ------------------------
%
%
%------------------ 006106_000711_000026_01 ------------------------
%
%
%------------------ 006554_000774_000048_01 ------------------------
%
%
%------------------ 006678_000792_000037_01 ------------------------
%
%
%------------------ 006912_000826_000015_01 ------------------------
%
%
%------------------------------------------------------------------------------
%
%
%
%
%
%
%
%------------------------------------------------------------------------------
\chapter{Introducing \NameAcp{}'s optics}
\label{ChOpticsOfNameAcp}
Although the Cherenkov-plenoscope's perception is fundamentally different from the Cherenkov-telescope's perception, its optical components are not.
There are only two different types of optical components in \NameAcp{}.
First mirror-facets, and second lenses.
Both are mass fabricated with purely spherical surfaces.
Their identical copies are used over and over again.
We list optical properties and motivate design-choices.
\section{Imaging-reflector}
\label{SecOpticsImagingReflector}
\NameAcp{}'s large imaging-reflector has $f = 106.5\,$m focal-length and $D = 71\,$m outer diameter resulting in a focal-ratio $F = f/D = 1.5$.
Because of the lenses in the light-field-sensor, we do not want to go much below this focal-ratio, see later Section \ref{SecOpticsLightFieldSensor}.
To be light-weight and cost-efficient, \NameAcp{} has a segmented imaging-reflector which is composed from many identical mirror-facets.
These mirror-facets are the first of the two optical components used in \NameAcp{}.
The Figures \ref{FigAcpOverview}, \ref{FigNameAcpTourOverview}, and \ref{FigNameAcpTourImagingReflector} show schematics and renderings of \NameAcp{}'s large imaging-reflector.
\subsection*{Mirror-facets}
The \NumFacets{} mirror-facets are also imaging-reflectors themselves with the same focal-length $f$ as the overall large imaging-reflector.
The reflective surface of each mirror-facet has the shape of a sphere with a curvature-radius of $2f$.
The perimeter of the mirror-facets is hexagonal to fill the overall aperture with only small gaps in between the edges of the facets.
The outer perimeters of all mirror-facets are fully enclosed by the $71\,$m diameter of the large imaging-reflector.
Since the center of the large imaging-reflector is shadowed by the light-field-sensor, we do not put mirror-facets within an inner diameter of $5\,$m but leave room for e.g. calibration-equipment.
Each mirror-facet provides $1.97\,$m$^2$ reflective area, compare CTA-LST-facet in Table \ref{TabMirrorFacets}.
Between the edges of neighboring mirror-facets is a gap of $2.5\,$cm for clearance.
In total, the reflective area of \NameAcp{}'s large imaging-reflector is $3,628.7\,$m$^2$.
Figure \ref{FigReflectionCtaMstDielectricMirrors} shows the mirror-facets reflectivity used in the simulations.
This is the reflectivity of a durable coating for mirror-facets developed for the upcoming Cherenkov-Telescope-Array (CTA) \cite{pareschi2013status}.
We choose a large $1.97\,$m$^2$ mirror-facet which can be mass produced today in a cost-efficient way \cite{pareschi2013status}.
Larger mirror-facets reduce the complexity of their orientational fine alignment, and are favored by the supporting space-truss-lattice, see Figure \ref{FigSpirosHexagonalMirrorFacetsSpaceTrussLattice}.
\subsection*{Principal-aperture-plane}
Without loss of generality, we describe the geometry of \NameAcp{}'s imaging-reflector with respect to the principal-aperture-plane \cite{wohler20123d,forsyth2003modern}, compare Figure \ref{FigThinLens}.
The principal-aperture-plane in $x$ and $y$, together with the optical-axis of the imaging-reflector in $z$ define the origin of the imaging-reflector's frame.
We describe the geometry of the imaging-reflector with respect to the principal-aperture-plane, because the principal-aperture-plane is also our reference to describe the light-field, see Chapter \ref{ChLightFieldGeometry}, and Figure \ref{FigRaysOnPrincipalAperturePlane} in particular.
Describing the light-field with respect to the principal-aperture-plane simplifies the representation of the light-field, as it abstracts away the complex surface-geometry of the imaging-reflector, see Chapter \ref{ChLightFieldGeometry}.
For reasons that we will discuss next, all of the mirror-facets of \NameAcp{}'s imaging-reflector are positioned above the principal-aperture-plane.
The principal-aperture-plane is not the surface of the mirror-facets, but an abstract plane.
\subsection*{Geometry}
The center-positions of the mirror-facets are embedded on a paraboloid to optimize isochronous imaging as much as possible before we have to fall back to the novel reconstruction-methods for the photon-arrival-times which are possible with the light-field-sensor, see Section \ref{SecIsochronousImaging}.
In contrast to embedding the mirror-facets center-positions on a sphere with the focal-point $\vec{f} = (0,0,f)$ in its center, as it is done in the Davis-Cotton-geometry \cite{Davies_Cotton_1957}, the paraboloid induces a wider spread in photon-incident-directions, but a narrower spread in photon-arrival-times.
For \NameAcp{}, we choose to go for the smallest possible spread in photon-arrival-times.\\
On \NameAcp{}'s imaging-reflector, the center-position of the $i$-th mirror-facet
\begin{eqnarray}
\vec{m_i} &=& \left(x_i,\,\, y_i,\,\,\frac{r_i^2}{4f} + c\right)^T
%\begin{pmatrix}
%r_i \cos \phi_i\\
%r_i \sin \phi_i\\
%r_i^2/(4f) + c
%\end{pmatrix}
\end{eqnarray}
is restricted in its $x$, and $y$-component by the hexagonal grid of the mirror-facets.
The $z$-component depends on the distance $r_i = \sqrt{x_i^2 + y_i^2}$ of the mirror-facet to the optical-axis, the focal-length $f$, and an offset $c$ which is chosen globally for all mirror-facets to fulfill
\begin{eqnarray}
f &\overset{!}{=} \frac{1}{Q} \sum_{i=0}^Q& \left\vert \vec{f} - \vec{m_i} \right\vert_2.
\end{eqnarray}
Here $Q$ is the number of mirror-facets.
The offset $c$ rises all the mirror-facets along the optical-axis up and above the principal-aperture-plane so that the average of all the distances from the mirror-facets to the focal-point equals the focal-length.
The target-orientations of the mirror-facets are such that the central spot on each mirror-facet reflects photons running parallel to the optical-axis towards the focal-point $\vec{f}$.
%
% unit_z = np.array([0.0, 0.0, 1.0])
% connection = focal_piont - facet_center
% connection /= np.linalg.norm(connection)
% rotation_axis = np.cross(unit_z, connection)
% angle_to_unit_z = np.arccos(np.dot(unit_z, connection))
% ideal_angle = angle_to_unit_z/2.0
% reflector2facet = HomTra()
% reflector2facet.set_translation(facet_center)
% reflector2facet.set_rotation_axis_and_angle(rotation_axis, ideal_angle)
% return reflector2facet
%
\begin{figure}
    \centering
    \includegraphics[width=1\textwidth]{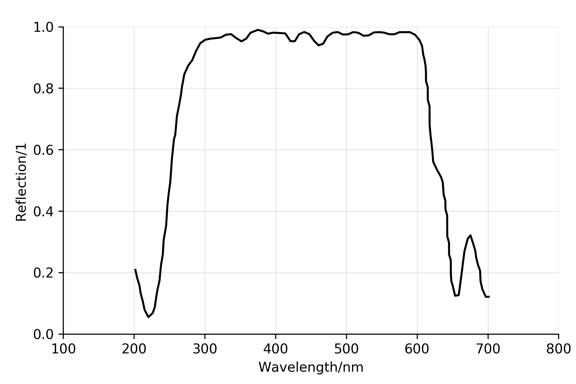}
    \caption[Reflection-coefficient of the mirror-facets, CTA-MST dielectric]{The reflection-coefficient used to simulate the mirror-facets on the imaging-reflector of \NameAcp{}.
        This is the measured reflection-coefficient of the so called 'dielectric' mirror-facets of the Medium-Size-Telescope (MST) in the Cherenkov-Telescope-Array (CTA) \cite{pareschi2013status,pareschi2013statusArxiv}.
    }
    \label{FigReflectionCtaMstDielectricMirrors}
\end{figure}
%
%------------------------------------------------------------------------------
\section{Aligning the mirror-facets on \NameAcp{}}
Just as on Cherenkov-telescopes, orientation-alignment for the \NumFacets{} mirror-facets is needed in a regime of $\approx 0.035^\circ$, which is half a pixel's field-of-view.
Since \NameAcp{} uses many more mirror-facets than existing telescopes, the alignment-method for \NameAcp{} must have an execution-time which does not scale too fast with the number of mirror-facets.
Fortunately, the alignment-methods for Cherenkov-telescopes \cite{mccann2010new, ahnen2016normalized} which are based on the characterization of solar-concentrators during the night \cite{arqueros2003novel}, have a constant execution-time with respect to the number of mirror-facets.
The author of this thesis (S.A.M.) took the leadership in the investigations of \cite{ahnen2016normalized}.
We propose to adopt such alignment-methods, and increase the robustness by using more cameras, if needed.
Currently these alignment-methods only use a single CCD-camera\footnote{%
    These cameras do not necessarily have a Charged-Coupled-Device (CCD).
    The term CCD-camera just happens to be often used for industrial cameras with sizes similar to the human eye.
    We use it here to prevent naming-collisions with the small cameras in the light-field-sensor of \NameAcp{}.
}%
in the focal-point of the imaging-reflector.
But alignment-methods can use plenoptic-perception as well by e.g. replacing the single CCD-camera with an array of CCD-cameras.
This way, a wider range of possible miss-orientations of the mirror-facets could be covered in shorter time, and possible misalignments between the large imaging-reflector and the light-field-sensor could be compensated.
\section{Light-field-sensor}
\label{SecOpticsLightFieldSensor}
\begin{figure}
    \centering
    \includegraphics[width=1\textwidth]{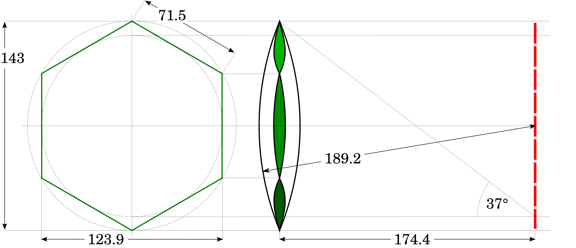}
    \caption[A small camera of \NameAcp{}'s light-field-sensor, dimensions]{
        Dimensions in millimeter.
        The geometry of the small cameras inside \NameAcp{}'s light-field-sensor.
        On the left is the view from the front, on the right is the view from the side.
        Just as in the Figures \ref{FigOpticsOverview}, \ref{FigOpticsOverviewCloseUp}, and \ref{FigSmallCameraCluseUp}, the sensitive surfaces of the photo-sensors are red and the faces of the lens which are touching neighboring lenses are green.
    }
    \label{FigSmallCameraGeometry}
\end{figure}

\begin{figure}
    \centering
    \includegraphics[width=.48\textwidth]{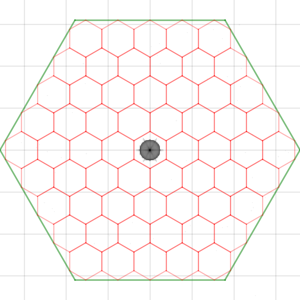}
    \hfill
    \vspace{.5cm}
    \includegraphics[width=.48\textwidth]{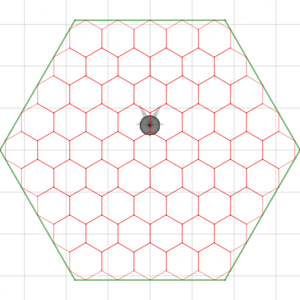}
    \includegraphics[width=.48\textwidth]{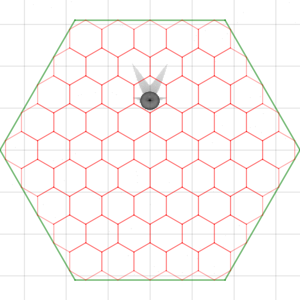}
    \hfill
    \vspace{.5cm}
    \includegraphics[width=.48\textwidth]{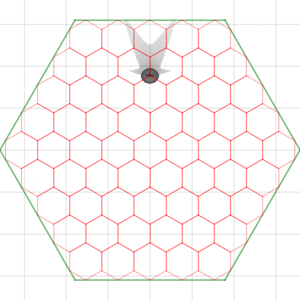}
    \includegraphics[width=.48\textwidth]{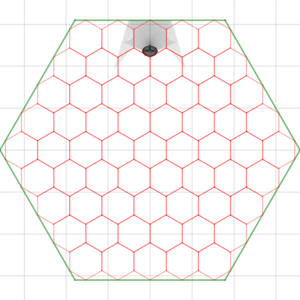}
    \hfill
   \vspace{.5cm}
    \includegraphics[width=.48\textwidth]{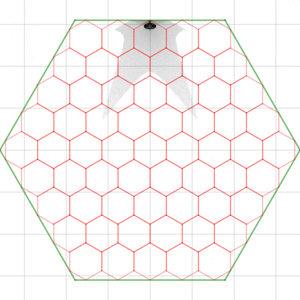}
    \includegraphics[width=.7\textwidth]{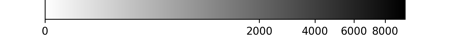}
    \caption[Point-spread-function of small camera]{
        Point-spread-functions inside the small cameras.
        From the top left to the bottom right the incident-angles of the photons are $0.0^\circ$, $3.7^\circ$, $7.4^\circ$, $11.1^\circ$, $14.7^\circ$, and $18.4^\circ$.
        Square grid has $20\,$mm spacing.
        Large hexagon is the perimeter of the small camera, compare Figure \ref{FigSmallCameraGeometry}.
        Small hexagons are the individual photo-sensors, compare Figure \ref{FigSmallCameraCluseUp}.
        Darkening is proportional to the cube-root of the intensity.
    }
    \label{FigSmallCameraPointSpreadFunction}
\end{figure}

The light-field-sensor of \NameAcp{} is a dense, two-dimensional array of identical, small cameras.
Figure \ref{FigSmallCameraGeometry} shows the dimensions of one of these small cameras.
Each small camera consists out of first a bi-convex, spherical lens made out of silica-glass, and second a two-dimensional array of photo-sensors.
This lens is the second of the two optical components used in \NameAcp{}.
In Figure \ref{FigSmallCameraGeometry} the faces of the lens which touch neighboring lenses of neighboring small cameras are shown in green.
The green faces of the lens are opaque.
The sensitive surfaces of the photo-sensors are shown in red.
The lens in the small camera is positioned such that its projection of incoming photons is sharpest in the outer region on the photo-sensors, and not in the center.
Figure \ref{FigSmallCameraPointSpreadFunction} shows the point-spread-functions inside a small camera for different incident-directions of photons coming from different positions on the large imaging-reflector.
Because the lens has a rather strong aberration in the outer region of its projection, we made the inner walls of the small camera reflective.
These inner walls have the same reflectivity as the mirror-facets of the large imaging-reflector, see Figure \ref{FigReflectionCtaMstDielectricMirrors}.
In the Figures the walls are green as they also touch neighboring small cameras just like the green surfaces of the lens.
These reflective walls actually reduce more artifacts than they create.
In Figure \ref{FigSmallCameraPointSpreadFunction} for small incident-angles up to $11.1^\circ$, we find that caustics are running out of the point-spread-function's core towards the outside.
Now with reflective inner walls, these caustics are thrown back towards the inside when the incident-angles become larger.
In any case, this correction is only minor as it effects only a small fraction of the photons.
\subsection*{Bi-convex lens with hexagonal perimeter}
\begin{figure}
    \centering
    \includegraphics[width=1\textwidth]{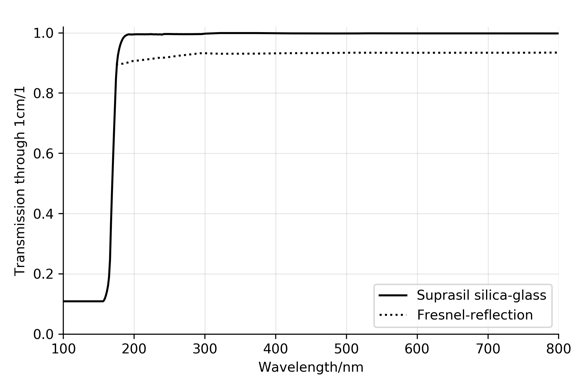}
    \caption[Transmission of silica-glass]{
        The transmission-probability of photons going through a 1\,cm thick plate of Suprasil silica-glass made by Heraeus \cite{heraeus2018quarz}.
        Reconstructed from the specifications by Heraeus based on the transmission including losses due to Fresnel-reflection, and the transmission with only the losses expected due to Fresnel-reflection.
        See also Table \ref{TabPureTransmissionSilicaGlassHeraeus}.
    }
    \label{FigTransmissionSilicaGlassHeraeus}
\end{figure}

\begin{figure}
    \centering
    \includegraphics[width=1\textwidth]{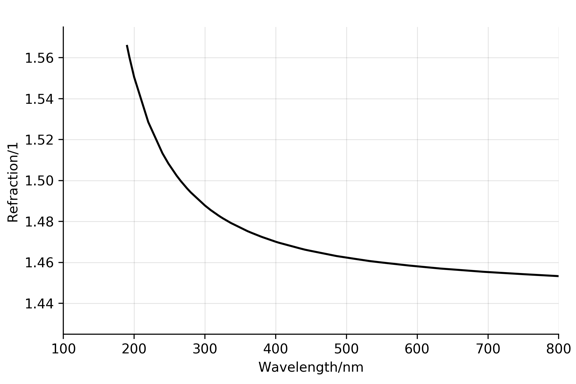}
    \caption[Refraction of silica-glass]{The refraction-index used to simulate the lenses in \NameAcp{}.
        This is the measured refraction-index of 'Suprasil' silica-glass produced by the Heraeus company \cite{heraeus2018quarz}.
    }
    \label{FigRefractionSilicaGlassHeraeus}
\end{figure}

The bi-convex lenses are made out of silica-glass.
Because of its high transmission for bluish and ultra-violet photons, we choose a silica-glass called Suprasil $311/312/313$ made Heraeus \cite{heraeus2018quarz}\footnote{
    Also other flavors of silica-glass made by other manufactures seem to have comparable properties for our needs.
    But we give credits to Heraeus here because their specifications are well documented.
}.
The Figures \ref{FigTransmissionSilicaGlassHeraeus} and the Table \ref{TabPureTransmissionSilicaGlassHeraeus} show the transparency of the silica-glass.
The lenses are so transparent that at most $\approx 1\%$ of the photons with wavelengths below $250\,$nm is absorbed inside of  them.
Of course our simulation handles the Fresnel-reflections which are much more relevant here.
Figure \ref{FigRefractionSilicaGlassHeraeus} shows the refraction-index of the lenses.
\begin{table}
\begin{center}
    \begin{tabular}{lrr}
        wavelength/nm & transmission through 1\,cm/$\%$\\
        \toprule
        193 & 98.5\\
        249 & 99.5\\
        266 & 99.9\\
        \bottomrule
    \end{tabular}
    \end{center}
    \caption[Pure transmission of silica-glass]{
        Transmission-probability of photons going through a 1\,cm thick plate of Suprasil silica-glass made by Heraeus \cite{heraeus2018quarz} excluding losses due to Fresnel-reflection.
        For wavelength above $266\,$nm and until $\approx 1,100\,$nm, the attenuation seems to be negligible for the simulations of \NameAcp{}.
        The lenses in \NameAcp{} are at most $1.5\,$cm of path-length.
    }
    \label{TabPureTransmissionSilicaGlassHeraeus}
\end{table}
\subsection*{Focal-ratio for small lenses and large imaging-reflector}
\begin{figure}
    \centering
    \includegraphics[width=1\textwidth]{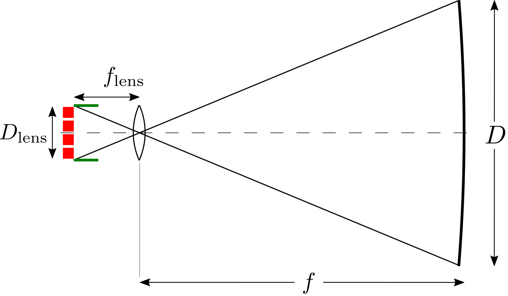}
    \caption[focal-ratio of lenses in small cameras]{The focal-ratio of the lenses should be the same as the focal-ratio of the large imaging-reflector.
        This way, the array of photo-sensors in the small camera is fully covered by the projection provided by the lens.
    }
    \label{FigLensFocalRatio}
\end{figure}

For image-quality and ease of manufacturing, we want the largest possible focal-ratios for the lenses.
But in Figure \ref{FigLensFocalRatio} we find that the largest focal-ratio for the lenses is determined by the intercept-theorem
\begin{eqnarray}
\frac{f_\text{lens}}{D_\text{lens}} &=& \frac{f}{D},
\label{EqLensFocalRatio}
\end{eqnarray}
where we assume that $f \gg f_\text{lens}$.
Since we want the small cameras to fill the sensor-plane without gaps, the array of photo-sensors in a small camera can not be larger than the lens' diameter $D_\text{lens}$.
Thus the projection provided by the lens has to fit onto an area not exceeding the lens' diameter $D_\text{lens}$.\\
However, Equation \ref{EqLensFocalRatio} is only valid for the one-dimensional case.
In the two-dimensional implementation used in \NameAcp{}'s light-field-sensor there is an additional quirk.
In Figure \ref{FigSmallCameraGeometry} we find, that the projected image of the large imaging-reflector has to fit inside the inner-diameter of the hexagonal aperture of the lens.
But the lens' optical diameter $D_\text{lens}$ is the outer diameter of the hexagonal aperture of the lens.
This is one of two reasons why on \NameAcp{} the focal-ratio for the lenses is $F_\text{lens} = 1.22$ instead of the $F = 1.5$ of the large imaging-reflector.
This is also the reason why the lenses have hexagonal apertures.
Square, or triangular apertures cause even smaller focal-ratios $F_\text{lens}$ because their inner and outer diameters deviate even more.
Figure \ref{FigSmallCameraGeometry} not only implies that an hexagonal aperture for the lenses is optimal, but also that the large imaging-reflector should have a hexagonal aperture, and not a circular one.
This way, the large imaging-reflector could be projected onto the full hexagonal array of photo-sensors inside the small camera.
In \NameAcp{}, we did not yet implement this.
\subsection*{Afterthought -- Curving the sensor-plane}
\begin{figure}
    \centering
    \includegraphics[width=1\textwidth]{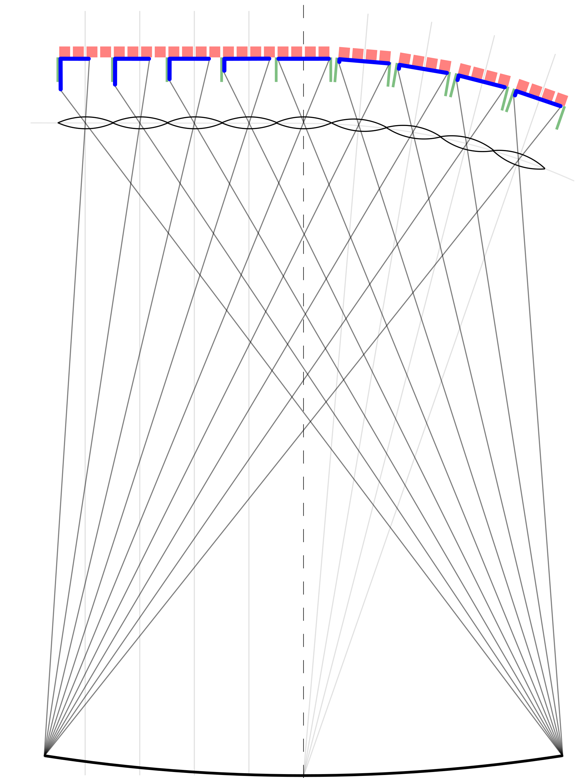}
    \caption[Curving the light-field-sensor for larger field-of-views]{We propose to curve the light-field-sensor in future studies to ease the implementation of larger field-of-views, and to allow larger focal-ratios of the lenses.
        On the left side, the plenoscope shown here has a flat light-field-sensor.
        The projections provided by the lenses are shown in dark blue.
        On the left, flat part of the light-field-sensor, these projections do not cover all photo-sensors.
        The projections move more and more to the outside the further the small camera is away from the optical-axis of the large imaging-reflector.
        On the right, curved part of the light-field-sensor, the projections still cover all the photo-sensors in the small cameras.
        Compare to the plenoscope in Figure \ref{FigOpticsOverview} which has a much smaller field-of-view.
    }
    \label{FigCurvedLightFieldSensorProposal}
\end{figure}

The \NameAcp{} Cherenkov-plenoscope simulated here uses a light-field-sensor with all its small cameras positioned in a flat plane, and with all small cameras facing parallel.
However, we see benefits for future studies when the small cameras do not face parallel, but do face towards the center of the large imaging-reflector.
This way, the projections provided by the lenses always cover the entire array of photo-sensors in each small camera.
Otherwise the projections move further off the photo-sensors the further the small cameras are off the optical-axis of the large imaging-reflector.
On \NameAcp{} this is not yet a problem because \NameAcp{}'s field-of-view is not this large, and because we naively shortened the focal-ratio of the lenses to compensate for this off-axis-effect.
This shrinks the projections provided by the lenses and is the second reason why the lenses in \NameAcp{} have smaller focal-ratios than \NameAcp{}'s large imaging-reflector.
But shortening the focal-ratios of the lenses hurts optical quality and should be avoided.
In general, and especially when going for larger field-of-views, tilting the small cameras to make them face the center of the large imaging-reflector should be a beneficial workaround.
We propose to achieve this individual tilt for the small cameras by not arranging them on a flat plane, but on a sphere.
The center of this sphere is the center of the large imaging-reflector.
And the curvature-radius of this sphere is the focal-length $f$ of the large imaging-reflector.
This way, the small cameras are all tilted to face the center of the large imaging-reflector, and the small cameras do not interferer mechanically.
The fact, that there is no flat sensor-plane anymore is fully compensated using the light-field-geometry for the same reasons we discuss in Chapter \ref{ChCompensatingMisalignmnets}.
Figure \ref{FigCurvedLightFieldSensorProposal} shows our proposed curved light-field-sensor side-by-side to a flat light-field-sensor.
\subsection*{Photo-electric-conversion}
\begin{figure}
    \centering
    \includegraphics[width=1\textwidth]{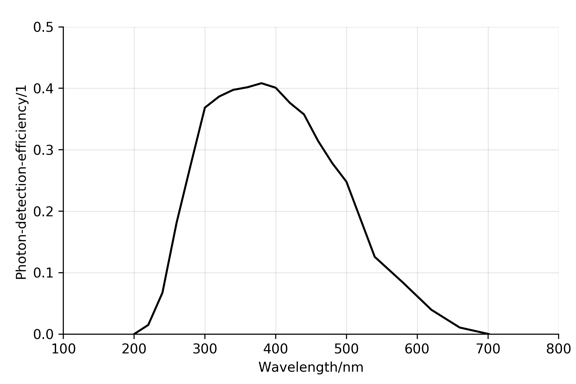}
    \caption[Photon-detection-efficiency of photo-sensors]{The photon-detection-efficiency used to simulate the photo-sensors in the light-field-sensor of \NameAcp{}.
        This is the measured photon-detection-efficiency of the photo-multiplier-tube R11920-100-05 by manufacturer Hamamatsu developed for the Large-Size-Telescope (LST) in the Cherenkov-Telescope-Array (CTA) \cite{toyama2013novel}.
    }
    \label{FigPhotonDetectionEfficiencyR11930}
\end{figure}

In \NameAcp{} we simulate photo-sensors which have the photon-detection-efficiency of the Hamamatsu R11920-100-05 photo-multiplier-tubes designed for the Large-Size-Telescope (LST) in the Cherenkov-Telescope-Array (CTA), see Figure \ref{FigPhotonDetectionEfficiencyR11930}.
This photo-sensor is optimized to detect bluish Cherenkov-photons and to reject reddish night-sky-background-photons.
Rejecting reddish night-sky-background photons eases the trigger-decision for air-showers.
For the read-out of the signals, we assume that \NameAcp{} has the read-out of the FACT\footnote{%
    This is not because we think that the read-out of the FACT Cherenkov-telescope is the best read-out possible, but because we know its shortcomings and artifacts and how to represent those in a simulation.
} Cherenkov-telescope which is discussed in Part \ref{PartPhotonStream} of this thesis.
\section{Conclusion}
\label{SecOpticsConclusion}
The \NameAcp{} Cherenkov-plenoscope only uses two types of optical components.
First, mirror-facets, and second lenses.
Both these components have purely spherical surfaces to ease their production.
Both these components are mass produced to populate the plenoscope with identical copies.
The optics of \NameAcp{} can be produced today using established methods developed for Cherenkov-telescopes.
%
%------------------------------------------------------------------------------
%
%
%
%
%
%
%
%------------------------------------------------------------------------------
\chapter{Estimating the light-field-geometry}
\label{ChLightFieldGeometry}
Here we describe in a quantitative way how the plenoscope perceives the light-field.
We do this by describing the geometry of the light-field perceived by each photo-sensor in the plenoscope.
Based on this description, we can reconstruct the three-dimensional trajectories and arrival-times of photons.
In Section \ref{SecAdressingLixel}, we introduce two addressing schemes for the photo-sensors in the light-field-sensor.
In Section \ref{SecApproximatingLixelsUsingRays}, we present our implementation of an approximative description for the geometry of the light-field.
In Section \ref{SecLightFieldGeometryOfNameAcp}, we visualize this approximative description for \NameAcp{}.
We name this approximative description the light-field-geometry of the plenoscope.\\
Further in Section \ref{SecIsochronousImaging}, we present perfect isochronous imaging with the Cherenkov-plenoscope, which overcomes a fundamental shortcoming of Cherenkov-telescopes, and is useful when projecting the light-field onto images.
\section{Addressing photo-sensors and lixels}
\label{SecAdressingLixel}
When we first introduce an addressing scheme for the photo-senors in the \NameAcp{} plenoscope in Equation \ref{EqRay} we chose a two-dimensional scheme ($n,m$) where $n$ loops over all the $N=\NumPix{}$ small cameras and $m$ loops over the $M=\NumPax{}$ photo-sensors in each of those small cameras.
In this first introduction, the small cameras can be identified as pixels, and the individual photo-sensors in the small cameras can be identified as paxels.
We did so to stress the hierarchy between the photo-sensors in the light-field-sensor presented in Figure \ref{FigOpticsOverview}.
However, the fact that such a two-dimensional addressing is possible goes back two the specific design of \NameAcp{}'s light-field-sensor.
But in general one does not have to group the photo-sensors into clusters of pixels and paxels.
For \NameAcp{} the two-dimensional addressing is only adequate as an approximation.
In general, we identify each photo-sensor with a light-field-cell (lixel) and address those in a one-dimensional scheme.
For \NameAcp{}, the number of lixels is
\begin{eqnarray}
K &=& M \times N,
\label{EqNumberLixel}
\end{eqnarray}
and the address of the lixel is composed as
\begin{eqnarray}
k &=& m + n \times M
\label{EqAddressingLixels}
\end{eqnarray}
such that in turn the addresses of the pixel and paxel are
\begin{eqnarray}
\label{EqAddressingPixel}
n &=& \text{floor}(k/M)\\
\label{EqAddressingPaxel}
m &=& k - n \times M.
\end{eqnarray}
The one-dimensional addressing is more natural when we describe advanced reconstructions such as tomography or the refocusing of images in post, while the two-dimensional addressing can be more accessible for the very basic imaging and areal sampling without corrections of the shortcomings introduced by real optical components.
The naive two-dimensional addressing of lixels can be more accessible because it allows to identify parts of the hardware as 'pixels' as it is often done in Cherenkov-telescopes.
However, prepare to break with this established mindset because otherwise we can not overcome aberrations in Chapter \ref{ChOvercomingAberrations}, compensate misalignments in Chapter \ref{ChCompensatingMisalignmnets}, or refocus images in post in Section \ref{SecPostRefocusedImaging}.
\section{Approximating lixels using rays}
\label{SecApproximatingLixelsUsingRays}
\begin{figure}
    \centering
    \includegraphics[width=1\textwidth]{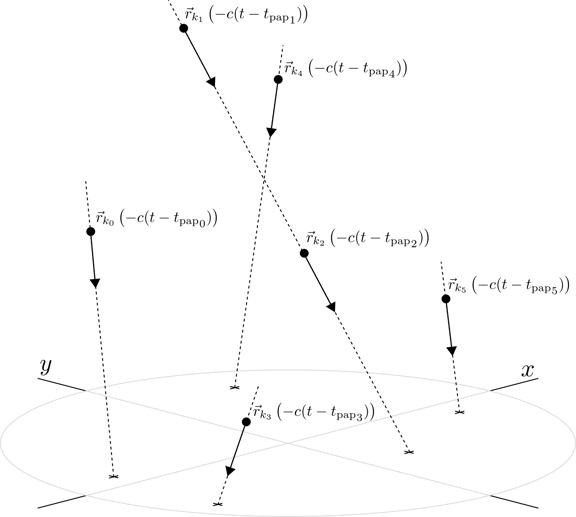}
    \caption[Photons on the principal-aperture-plane]{%
        The photons in the light-field-sequence $\mathcal{L}$.
        Black dots are photons with arrows indicating their direction of motion along their dashed trajectories.
        The calibration of the plenoscope abstracts away all the quirks of the large imaging-reflector, the lenses, the small cameras and the photo-sensors.
        What is left are single photons traveling along rays with respect to the principal-aperture-plane.
        Note that photon 1 and photon 2 occupy the same ray $k_1 = k_2$, but at different positions because of their different arrival-times ${t_\text{pap}}_1 \neq {t_\text{pap}}_2$.
        Here $c$ is the speed of light.
    }
    \label{FigRaysOnPrincipalAperturePlane}
\end{figure}

To interpret the photon-intensity recorded by a photo-sensor in the plenoscope, we have to know which part of the light-field is sampled by this photo-sensor.
We have to know the light-field-cell (lixel) to each photo-sensor.
We use the ray $\vec{r}_{k}(\lambda)$ from Equation \ref{EqRay} to approximate how photo-sensor $k$ samples the light-field.
This is an approximation because a ray only samples a singular incident-direction and a singular support-position.
But actual photo-sensors always sample an extended region of incident-directions and an extended region of support-positions.
For \NameAcp{} we estimate the rays $\vec{r}_{0}(\lambda)$ to $\vec{r}_{K-1}(\lambda)$ in a computer-simulation, where we throw many photons into the plenoscope.
We randomly draw the support-position of each photon on the principal-aperture-plane, and we randomly draw the incident-direction of each photon.
All the photons are emitted so that they would travel the same distance before they intersect the principal-aperture-plane.
When the photon is absorbed by a photo-sensor, we append the geometry of the trajectory of the photon to the according photo-sensor.
Finally we approximate the light-field-geometry of each individual photo-sensor with the 12 properties shown in Table \ref{TabCalibrationStatistics}.
Here we estimate the ray $\vec{r}_{k}(\lambda)$ using the mean support-positions $\overline{x}, \overline{y}$ and the mean incident-direction $\overline{c_x}, \overline{c_y}$ of all photons which are detected by photo-sensor $k$.
To describe the lixels in more detail, higher order statistics of the photons detected by a photo-sensor $k$ might be used such as the uncertainties listed in Table \ref{TabCalibrationStatistics}.
\begin{table}
\begin{center}
    \begin{tabular}{lrr}
        & mean & uncertainty\\
        \toprule
        Support-position in $x$ & $\overline{x}$ & $\sigma_x$\\
        Support-position in $y$ & $\overline{y}$ & $\sigma_y$\\
        Incident-direction in $x$ & $\overline{c_x}$ & $\sigma_{c_x}$\\
        Incident-direction in $y$ & $\overline{c_y}$ & $\sigma_{c_y}$\\
        arrival-time-delay (w.r.t. principal-aperture-plane) & $\overline{t}_\text{pap}$ & ${\sigma_t}_\text{pap}$\\
        Collection-efficiency & $\overline{\eta}$ & $\sigma_\eta$\\
        \bottomrule
    \end{tabular}
    \end{center}
    \caption[Approximating the geometry of a light-field-cell (lixel)]{Light-field-geometry of a single photo-sensor in the plenoscope approximated using 12 values.}
    \label{TabCalibrationStatistics}
\end{table}
Here the collection-efficiency $\eta$ is proportional to the ratio of the number of all the photons that got thrown to the number of all the photons that got absorbed in the photo-sensor.
Now we can use the resulting table
\setcounter{MaxMatrixCols}{16}
\begin{eqnarray}
    G &=&
    \begin{Bmatrix}%
        \overline{x}_0 &
        \overline{y}_0 &
        \overline{c_x}_0 &
        \overline{c_y}_0 &
        {\overline{t}_\text{pap}}_0 &
        \overline{\eta}_0 \\
        \overline{x}_1 &
        \overline{y}_1 &
        \overline{c_x}_1 &
        \overline{c_y}_1 &
        {\overline{t}_\text{pap}}_1 &
        \overline{\eta}_1 \\
        \vdots & \vdots & \vdots & \vdots & \vdots & \vdots\\
        \overline{x}_{K-1} &
        \overline{y}_{K-1} &
        \overline{c_x}_{K-1{}} &
        \overline{c_y}_{K-1} &
        {\overline{t}_\text{pap}}_{K-1} &
        \overline{\eta}_{K-1} \\
    \end{Bmatrix}
\end{eqnarray}
of all the $K$ photo-sensors to approximate the light-field-geometry of the plenoscope.
The first four columns of $G$ define the support-positions and incident-directions of the rays in Equation \ref{EqRay}.
So we can also think of
\begin{eqnarray}
    G &=&
    \begin{Bmatrix}%
        \vec{r}_0(\lambda) & {\overline{t}_\text{pap}}_{0} & \overline{\eta}_{0}\\
        \vec{r}_1(\lambda) & {\overline{t}_\text{pap}}_{1} & \overline{\eta}_{1}\\
        \vdots & \vdots & \vdots\\
        \vec{r}_{K-1}(\lambda) & {\overline{t}_\text{pap}}_{K-1} & \overline{\eta}_{K-1}\\
    \end{Bmatrix}
    \label{EqRaysOnPrincipalAperturePlane}
\end{eqnarray}
as a table of rays, arrival-time-delays, and collection-efficiencies for all the $K$ lixels in the plenoscope.
The light-field-geometry $G$ allows us to calibrate the responses of the plenoscope to abstract away from the optics, so that we only need to think about photons arriving on the flat surface of the principal-aperture-plane.
Figure \ref{FigRaysOnPrincipalAperturePlane} shows how we can think of the calibrated light-field-sequence $\mathcal{L}$.
All our further interpretation of the light-field-sequence $\mathcal{L}$ are based on the mindset presented by the Equation \ref{EqRaysOnPrincipalAperturePlane}, and Figure \ref{FigRaysOnPrincipalAperturePlane}.
\section{Isochronous imaging}
\label{SecIsochronousImaging}
\begin{figure}
    \centering
    \includegraphics[width=1\textwidth]{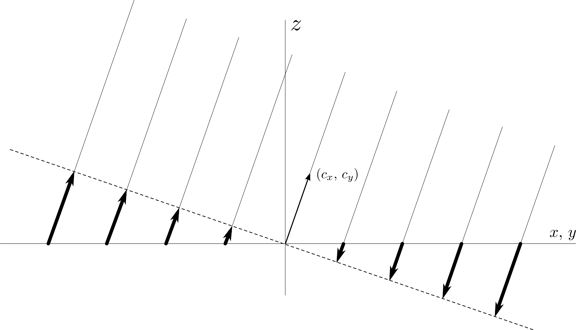}
    \caption[Time-of-flight offsets for isochronous imaging]{Time-of-flight offsets for isochronous imaging.
        Parallel rays carry photons arriving on the principal-aperture-plane $x,\,y$.
        For isochronous imaging we add the signed time-of-flight offsets (thick arrows) between the light-front (dashed-line) and the principal-aperture-plane to the arrival-times of the photons in pixel $(c_x\,,c_y)$.
    }
    \label{FigIsochronImaging}
\end{figure}

We defined the light-field-geometry $G$ with respect to the principal-aperture-plane.
However, if one intends to calculate classic image-sequences $\mathcal{I}$ from the recorded light-field-sequence $\mathcal{L}$, the arrival-time-delays ${\overline{t}_\text{pap}}$ must not be used for calibration.
One can use no arrival-time-delays and just use the raw recorded arrival-times $t_\text{raw}$ to get a good timing resolution for imaging.
This is because the coarse surface of \NameAcp{}'s large imaging-reflector follows a parabola.
However, a parabola can only do good isochronous imaging in the central part of the image, but the plenoscope can do better.
The plenoscope can synthesize the perfect isochronous image in post for the entire field-of-view.
Isochronous imaging means that incoming light-fronts, where the photons travel together in a plane perpendicular to their direction of motion, will arrive in the same pixel at the same moment.
Figure \ref{FigIsochronImaging} shows the path-length-delays which are added to the calibration used for the principal-aperture-plane.
The Figures \ref{FigLfgTImagingMeanSensorPlane} and \ref{FigLfgTImagingMeanSensorPlaneXandY} show the time-delays ${\overline{t}_\text{img}}$ which are added to the raw arrival-times $t_\text{raw}$ in order to obtain perfect isochronous images.
\section{Light-field-geometry of \NameAcp{}}
\label{SecLightFieldGeometryOfNameAcp}
\begin{figure}
    \centering
    \includegraphics[width=1\textwidth]{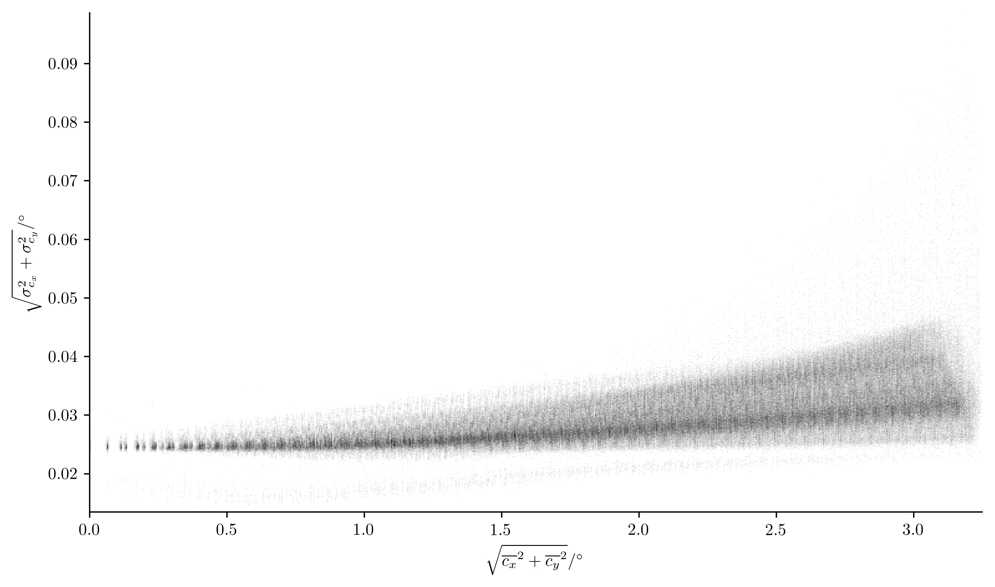}
    \caption[\NameAcp{} light-field-geometry, inverse point-spread-function]{Histogram of the mean magnitude of the incident-directions versus the spread magnitude of the incident-directions for all photo-sensors.
        The darkening in the histogram goes with the square-root of the number of lixels.
    }
    \label{FigLfgCmeanVsCstd}
\end{figure}

\begin{figure}
    \centering
    \includegraphics[width=\textwidth]{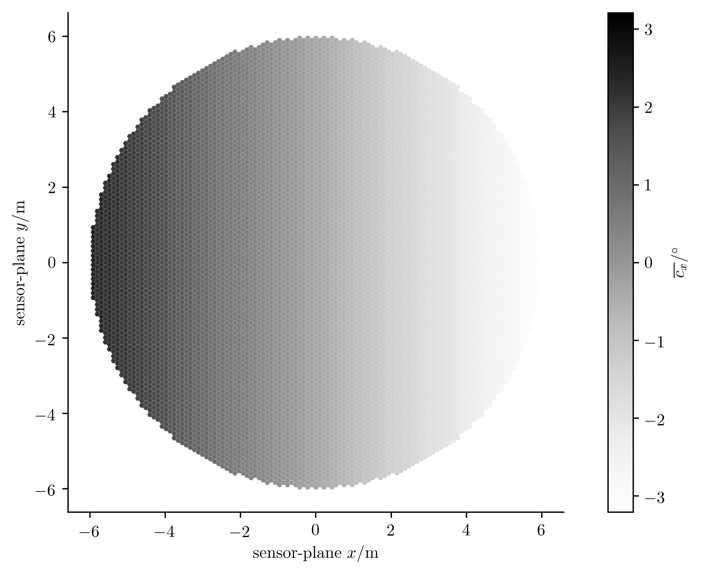}
    \includegraphics[width=\textwidth]{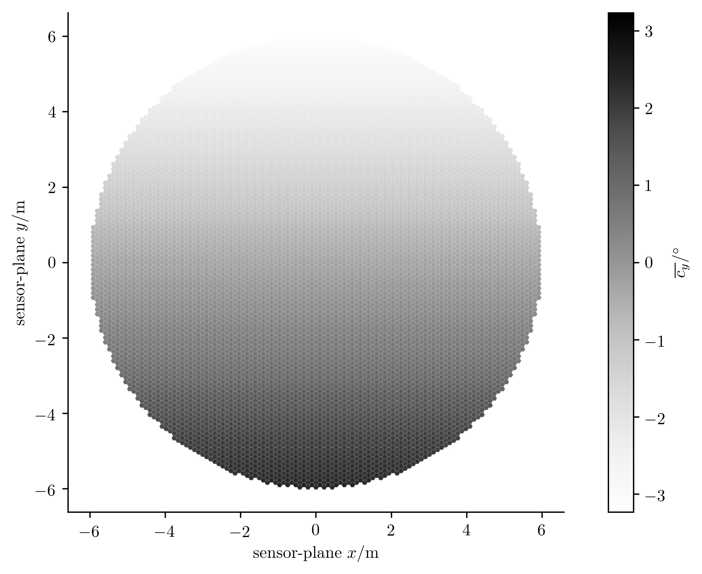}
    \caption[\NameAcp{} light-field-geometry, incident-directions on sensor-plane]{
        The photo-sensors colored according to their incident-directions $\overline{c_x}$ (top) and $\overline{c_y}$ (bottom).
    }
    \label{FigLfgCxCyMeanSensorPlane}
\end{figure}

\begin{figure}
    \centering
    \includegraphics[width=\textwidth]{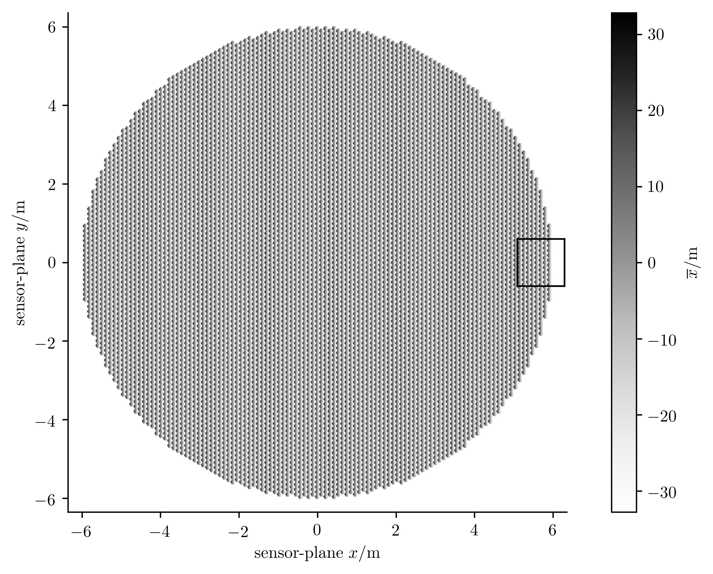}
    \includegraphics[width=\textwidth]{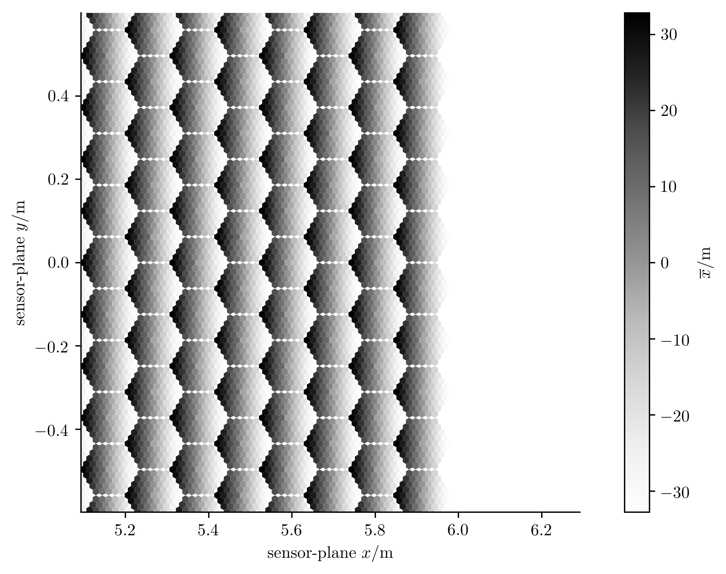}
    \caption[\NameAcp{} light-field-geometry, support-positions $\overline{x}$ on sensor-plane]{
        The photo-sensors colored according to their support-position $\overline{x}$. Full sensor-plane on top, close-up at the bottom.
    }
    \label{FigLfgXmeanSensorPlane}
\end{figure}

\begin{figure}
    \centering
    \includegraphics[width=\textwidth]{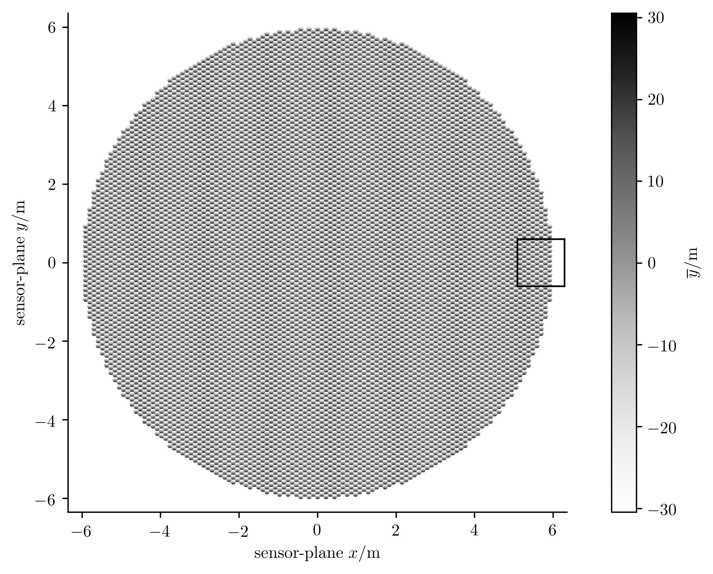}
    \includegraphics[width=\textwidth]{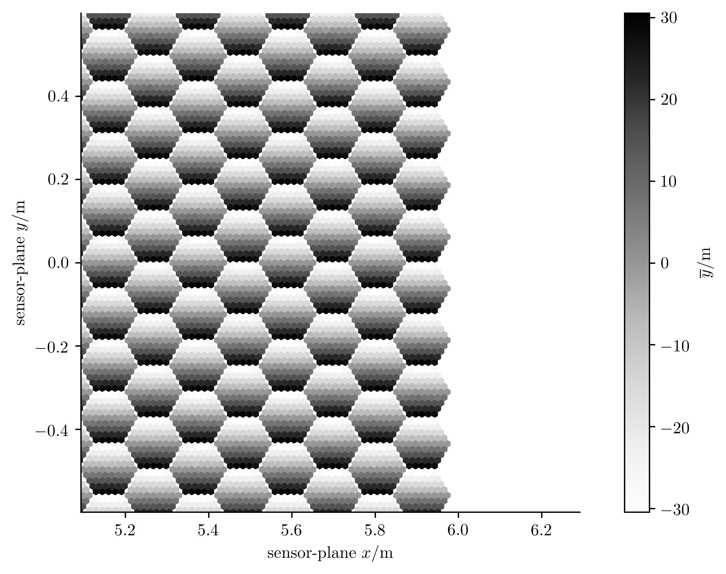}
    \caption[\NameAcp{} light-field-geometry, support-positions $\overline{y}$ on sensor-plane]{
        The photo-sensors colored according to their support-position $\overline{y}$. Full sensor-plane on top, close-up at the bottom.
    }
    \label{FigLfgYmeanSensorPlane}
\end{figure}

\begin{figure}
    \centering
    \includegraphics[width=\textwidth]{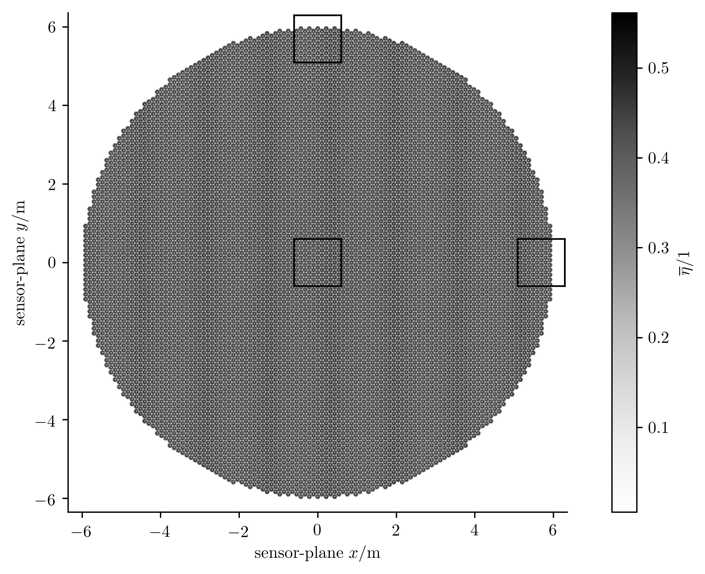}
    \includegraphics[width=\textwidth]{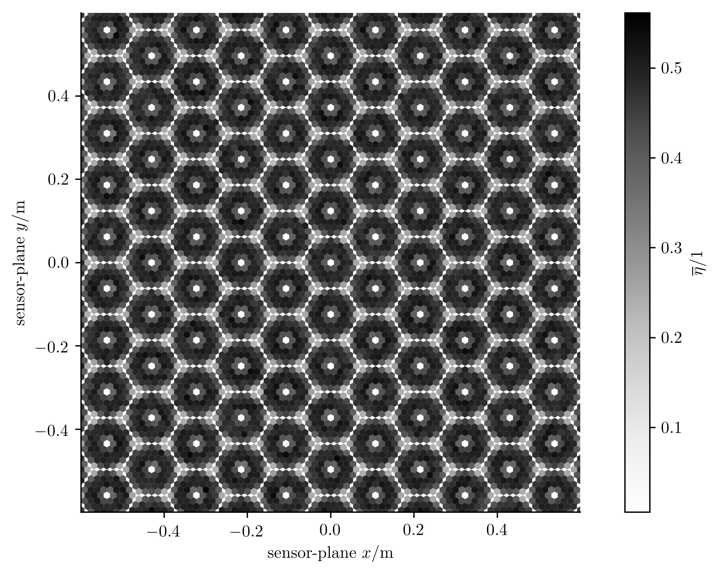}
    \caption[\NameAcp{} light-field-geometry, collection-efficiency $\overline{\eta}$ on sensor-plane]{
        The photo-sensors colored according to their collection-efficiency $\overline{\eta}$. Full sensor-plane on top, central close-up at the bottom.
    }
    \label{FigLfgEtaMeanSensorPlane}
\end{figure}

\begin{figure}
    \centering
    \includegraphics[width=\textwidth]{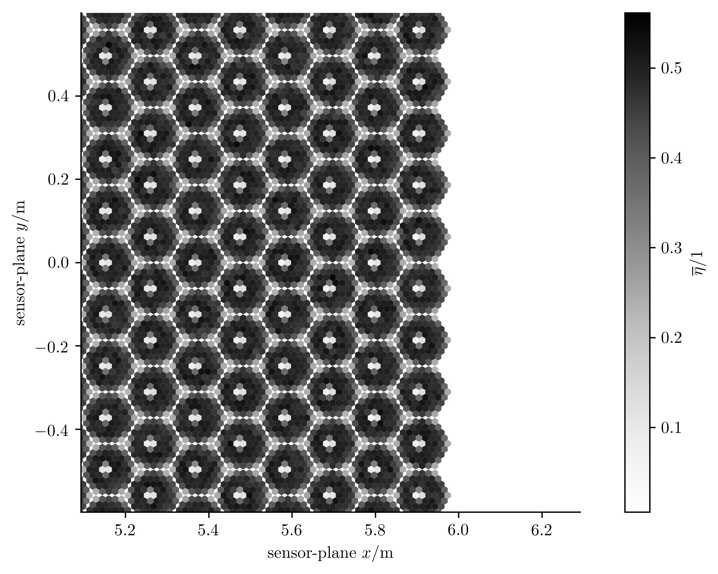}
    \includegraphics[width=\textwidth]{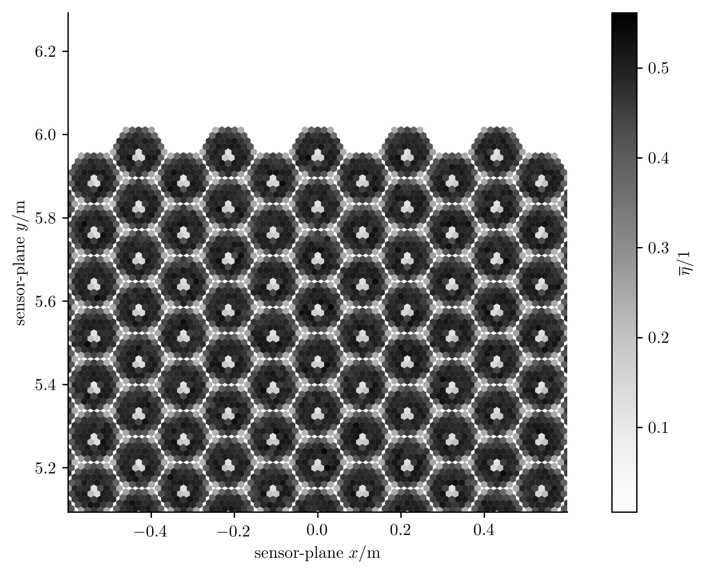}
    \caption[\NameAcp{} light-field-geometry, collection-efficiency $\overline{\eta}$ on sensor-plane, close-up]{
        Same as Figure \ref{FigLfgEtaMeanSensorPlane}.
        Close-ups on the left and upper regions of the the sensor-plane.
    }
    \label{FigLfgEtaMeanSensorPlaneXandY}
\end{figure}

\begin{figure}
    \centering
    \includegraphics[width=\textwidth]{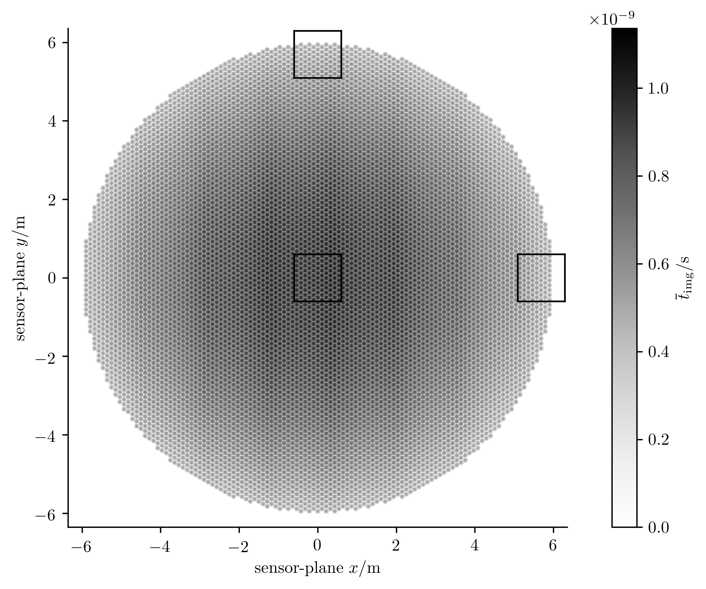}
    \includegraphics[width=\textwidth]{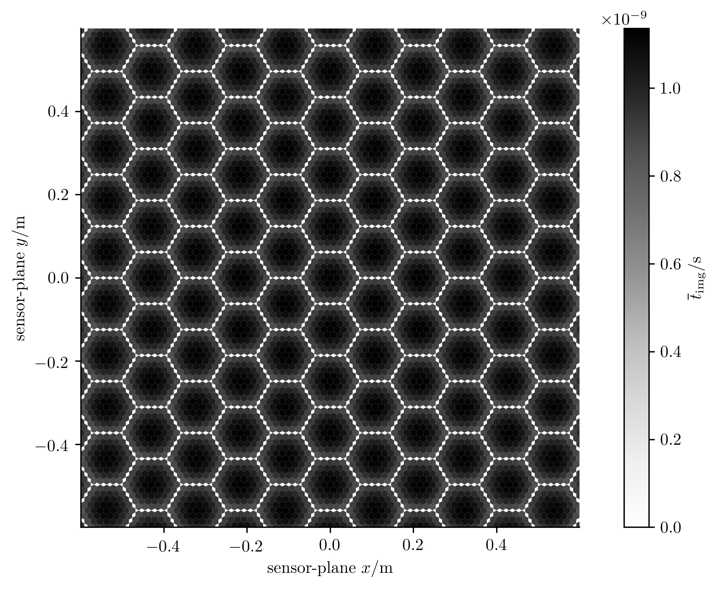}
    \caption[\NameAcp{} light-field-geometry, time-delay $\overline{t}_\text{img}$ for imaging]{
        The photo-sensors colored according to their arrival-time-delay for the image-plane $\overline{t}_\text{img}$. Full sensor-plane on top, central close-up at the bottom.
    }
    \label{FigLfgTImagingMeanSensorPlane}
\end{figure}

\begin{figure}
    \centering
    \includegraphics[width=\textwidth]{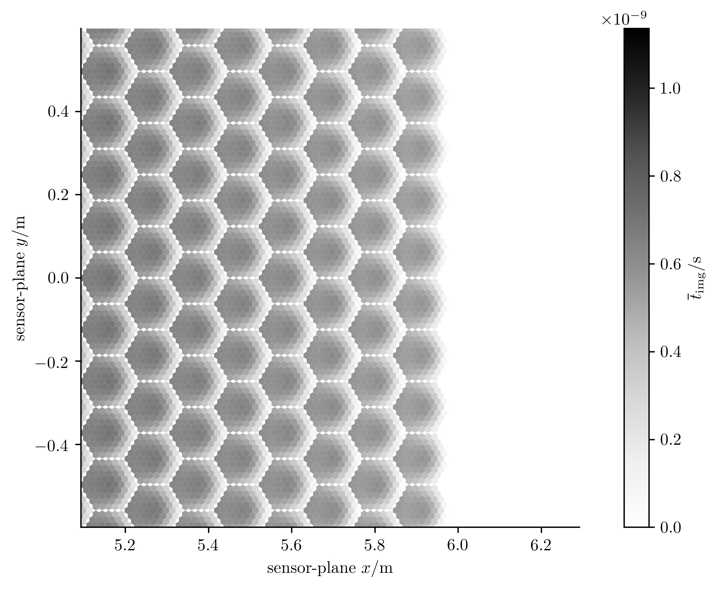}
    \includegraphics[width=\textwidth]{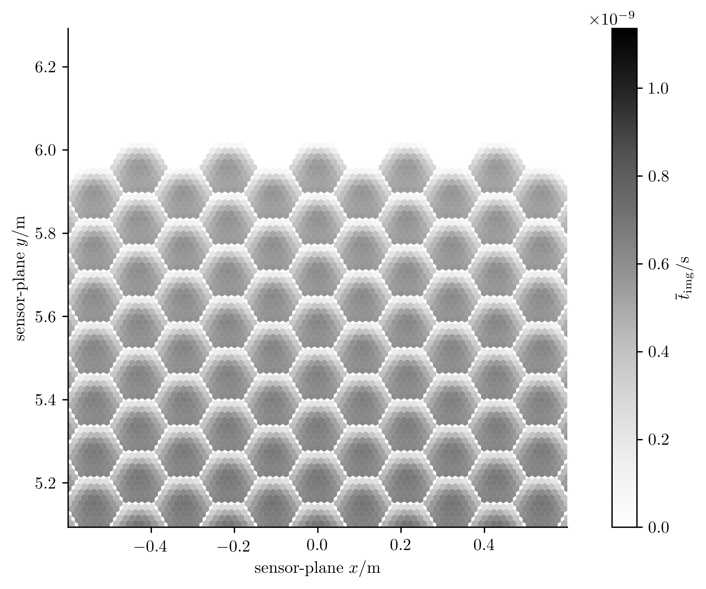}
    \caption[\NameAcp{} light-field-geometry, time-delay $\overline{t}_\text{img}$ for imaging, close-up]{
        Same as Figure \ref{FigLfgTImagingMeanSensorPlane}. Close-ups on left and upper regions of the sensor-plane.
    }
    \label{FigLfgTImagingMeanSensorPlaneXandY}
\end{figure}

\begin{figure}
    \centering
    \includegraphics[width=\textwidth]{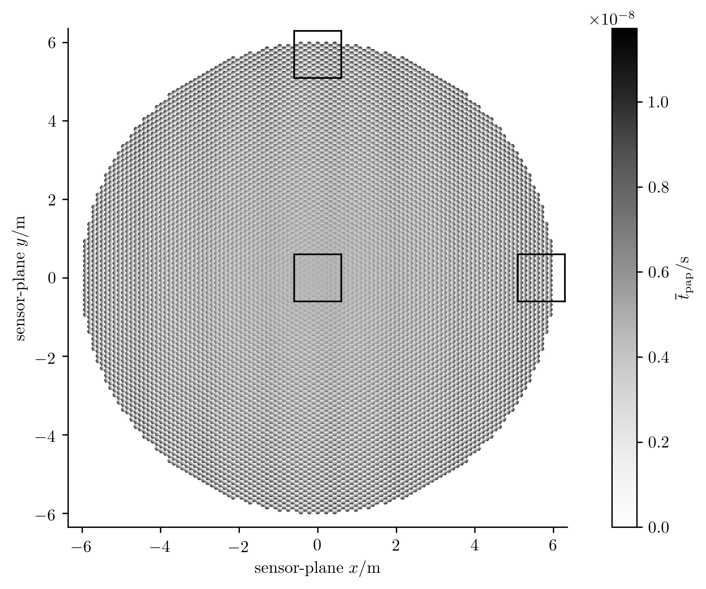}
    \includegraphics[width=\textwidth]{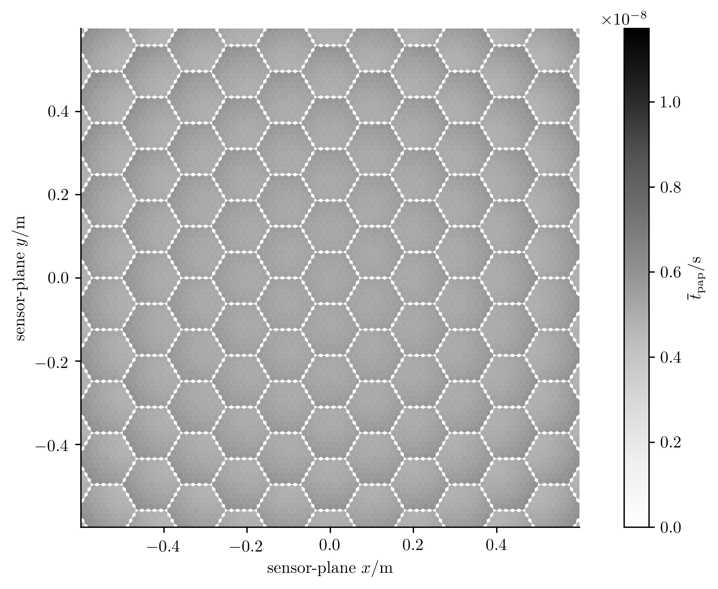}
    \caption[\NameAcp{} light-field-geometry, time-delay $\overline{t}_\text{pap}$ for aperture]{
        The photo-sensors colored according to their arrival-time-delay for the principal-aperture-plane $\overline{t}_\text{pap}$. Full sensor-plane on top, central close-up at the bottom.
    }
    \label{FigLfgTApertureMeanSensorPlane}
\end{figure}

\begin{figure}
    \centering
    \includegraphics[width=\textwidth]{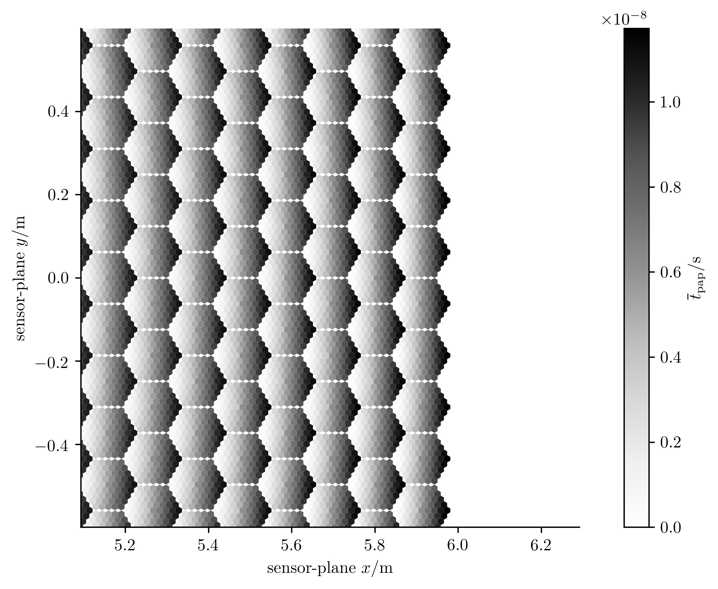}
    \includegraphics[width=\textwidth]{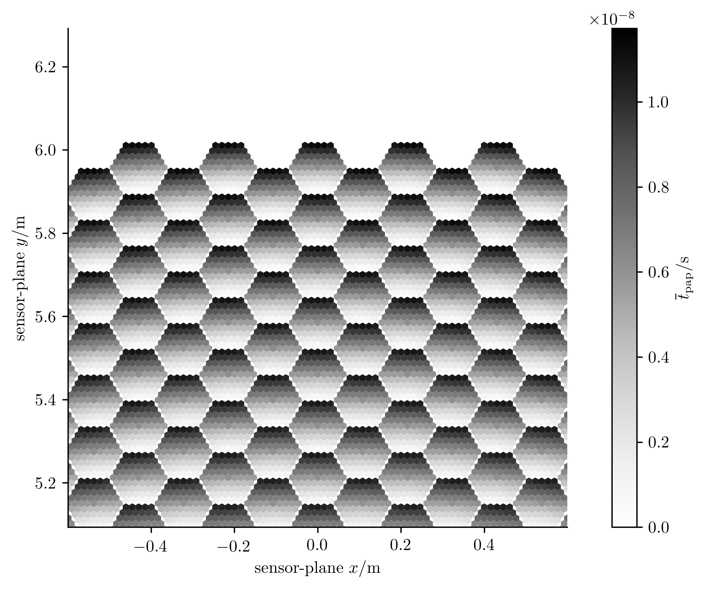}
    \caption[\NameAcp{} light-field-geometry, time-delay $\overline{t}_\text{pap}$ for aperture, close-up]{
        Same as Figure \ref{FigLfgTApertureMeanSensorPlane}. Close-ups on left and upper regions of the sensor-plane.
    }
    \label{FigLfgTApertureMeanSensorPlaneXandY}
\end{figure}

We visualize the light-field-geometry of \NameAcp{} when the large imaging-reflector and the light-field-sensor of \NameAcp{} are both in their target-geometry with respect to each other.
We propagate $10^9$ photons to estimate \NameAcp{}'s light-field-geometry.\\
In Figure \ref{FigLfgCmeanVsCstd} we find the magnitude of the spread in incident-directions versus the magnitude of the incident-directions.
This is an inverse representation of the point-spread-function discussed on  telescopes.
As expected from the large imaging-reflector, the inner region has a smaller spread in incident-directions.
The finite aperture of the small cameras in the light-field-sensor alone is expected to induce a spread in the order \footnote{Standard-deviation of a uniform distribution in a limited range: $0.067^\circ/\sqrt{12}$.} of $\approx 0.02^\circ$.
Even in the outer region of the field-of-view, the spread in incident-directions is below the spacing of the small cameras.\\
The Figures \ref{FigLfgCxCyMeanSensorPlane} to \ref{FigLfgTApertureMeanSensorPlaneXandY} show the arrangement of the photo-sensors inside the $12.1\,$m diameter light-field-sensor of \NameAcp{}.
We draw each photo-sensor as a small hexagon and color it according to its value.
We also show close-ups of the same sensor-plane in order to see the structures inside the small cameras.
Compare the arrangement of the photo-sensors with Figure \ref{FigSmallCameraCluseUp}, and \ref{FigSmallCameraPointSpreadFunction}.\\
Figure \ref{FigLfgCxCyMeanSensorPlane} shows the incident-directions in both $\overline{c_x}$, and $\overline{c_y}$ for all photo-sensors.
Same as with an image-sensor in a telescope, the positions of the photo-sensors on the sensor-plane correspond directly to the incident-directions.\\
The Figures \ref{FigLfgXmeanSensorPlane} and \ref{FigLfgYmeanSensorPlane} show the support-positions $\overline{x}$, and $\overline{y}$ for all the photo-sensors.
Unlike a telescope, the additional projection provided by the lenses in the small cameras allow the photo-sensors inside a small camera to sample the support-positions across the full $71\,$m aperture of the large imaging-reflector.
The close-ups in the Figures \ref{FigLfgXmeanSensorPlane} and \ref{FigLfgYmeanSensorPlane} show clearly the repeating pattern of the small cameras.\\
Figures \ref{FigLfgEtaMeanSensorPlane} and \ref{FigLfgEtaMeanSensorPlaneXandY} show the collection-efficiency $\overline{\eta}$ of all the photo-sensors.
As can be seen from the close-ups, the central photo-sensors inside a small camera are less efficient than the outer ones.
This is because the central photo-sensor is partly blocked due to the housing of the light-field-sensor itself.
Depending on the position of the small camera on the sensor-plane, the shadow of the housing of the light-field-sensor moves accordingly to the opposite direction.
Since the observations of air-showers approach a single photon regime, the recorded light-field-sequences are always subject to quantization noise.
This makes an application of the collection-efficiency $\overline{\eta}$ difficult or even impossible for the correction of photon-intensities.
However, we use it extensively to speed up the simulations of \NameAcp{} when we inject night-sky-background-photons into the light-field-sensor, see also Section \ref{SecSimulatingNightSkyBackgroundPhotons}.\\
The Figures \ref{FigLfgTImagingMeanSensorPlane} and \ref{FigLfgTImagingMeanSensorPlaneXandY} show the time-delays $\overline{t}_\text{img}$ of the photo-sensors which are needed for a perfect isochronous image.
The overall spread of the delays is only $\approx 1\,$ns and thus already sufficient for the observation of air-showers even if we do not correct for it.
Just as one would expect from a telescope with a parabolic imaging-reflector, the spread in the central part of the sensor-plane is lowest.\\
The Figures \ref{FigLfgTApertureMeanSensorPlane} and \ref{FigLfgTApertureMeanSensorPlaneXandY} show the time-delays $\overline{t}_\text{pap}$ of the photo-sensors which are needed for the principal-aperture-plane.\\
%
%%%%%%%%%%%%%%%%%%%%%%%%%%%%%%%%%%%%%%
%
%\begin{eqnarray}
%\mathcal{L} &:=&
%    \left\{
%    \begin{array}{l}
%        G\\
%        I_\text{stream}\\
%    \end{array}
%    \right.
%\label{EqDescribeLightFieldSequence}
%\end{eqnarray}
%%
%For each photon $j$ in $I_\text{stream}$, we know the trajectory $\vec{r}_{k=k_%j}(\lambda)$ which belongs to the lixel $k_j$.
%%
%And further we know its arrival-time
%%
%\begin{eqnarray}
%{t_\text{pap}}_j &=& {t_\text{raw}}_j - {\overline{t}_\text{pap}}_{k=k_j}
%\end{eqnarray}
%%
%on the principal-aperture-plane, and the photon's arrival-time
%%
%\begin{eqnarray}
%{t_\text{img}}_j &=& {t_\text{raw}}_j - {\overline{t}_\text{img}}_{k=k_j}
%\end{eqnarray}
%%
%in the image.
%%
%This means, that we know the position, and the direction of the $j$-th photons %at any given time $t$ when we evaluate the ray $\vec{r}_{k_j}(\lambda_j)$ with %the distance
%\begin{eqnarray}
%\lambda_j &=& {-c(t - t_\text{pap}}_j)
%\end{eqnarray}
%%
%which the photon needs to travel until it intersects with the %principal-aperture-plane\footnote{Just as in CORSIKA, our direction-vector in %the ray $\vec{r}(\lambda)$ is the incident-direction of the photon which is %the negative direction of motion of the photon.}.
%
%Here $c$ is the speed of light.
%
%Figure \ref{FigRaysOnPrincipalAperturePlane} illustrates the calibrated light-field-sequence.
%
\section{Multiple light-field-geometries on \NameAcp{}}
Since in \NameAcp{} the geometry between the large imaging-reflector and the light-field-sensor might not be fix during all time, the light-field-geometry for \NameAcp{} depends on the actual geometry between the two.
This is a trade-off that we have to do in order to build a large $71\,$m  imaging-reflector with a decoupled light-field-sensor.
During the observations with \NameAcp{}, the cable-robot-mount will always try to bring the large imaging-reflector and the light-field-sensor into their desired target geometry $\hat{T}_\text{target}$.
However, small deviations from the desired target geometry $\hat{T}_\text{target}$, which the cable-robot-mount can not correct for fast enough, might occur.
The reason for such fast deviations might be wind-gusts.
The actual geometry $\hat{T}_\text{actual}$ between the large imaging-reflector and the light-field-sensor is recorded during the observations so that dedicated light-field-geometries $G(\hat{T}_\text{actual})$ can be applied accordingly.
We describe $\hat{T}$ with the homogeneous transformation from the large imaging-reflector's frame into the light-field-sensor's frame.
\section{Conclusion}
\label{SecLightFieldGeometryConclusion}
The light-field-geometry is a powerful tool to describe and simulate any optical instrument sensing photons.
The light-field-geometry allows us to mentally decouple the implementation-details of an instrument from its actual perception.
With the light-field-geometry we can abstract away all the quirks and features of \NameAcp{}'s large imaging-reflector, and its lenses in the light-field-sensor.
All these quirks and features are taken care of in the light-field-geometry.
The light-field-geometry allows us to focus on what actually matters: Individual photons traveling along their trajectories in a three-dimensional space, see Figure \ref{FigRaysOnPrincipalAperturePlane}.\\
Having the light-field-geometry of an instrument is enough to estimate the response of that instrument for a given light-field-input.
And the other way around, when we have the response and the light-field-geometry of an instrument, we can reconstruct the light-field which the instrument observed.\\
Together with the operations formulated in Chapter \ref{ChInterpretingTheLightFieldSequence}, the light-field-geometry allows us to overcome aberrations, see Chapter \ref{ChOvercomingAberrations}, and to compensate misalignments, see Chapter \ref{ChCompensatingMisalignmnets}.
%
%------------------------------------------------------------------------------
%
%
%
%
%
%
%
%------------------------------------------------------------------------------
\chapter{Interpreting the light-field-sequence}
\label{ChInterpretingTheLightFieldSequence}
The light-field-sequence $\mathcal{L}$ of the plenoscope can be approached in various ways to reconstruct the air-shower in order to learn about the cosmic particle.
Here we demonstrate the methods we have investigated so far.
First, we show three different representations of the light-field-sequence $\mathcal{L}$.
Second, we demonstrate imaging, areal-sampling, refocused imaging, synthetic apertures, light-fronts.
Three-dimensional tomography we discuss separately in Chapter \ref{ChTomography}.
\section{Representing the light-field-sequence}
There are different ways to represent the light-field-sequence $\mathcal{L}$.
We use three different representations.
Each representation has certain advantages.
Our three representations can be converted into each other without loss.
\subsection*{Lixel-time-histogram}
First, we represent the light-field-sequence as a two-dimensional histogram
\begin{eqnarray}
\mathcal{L}[k,\,t] &=& \begin{bmatrix}%
    i_{0,\,0} & i_{0,\,1} & i_{0,\,2} & \ldots & i_{0,\,T-1}\\
    i_{1,\,0} & i_{1,\,1} & i_{1,\,2} & \ldots & i_{1,\,T-1}\\
    \vdots    & \vdots    &   \vdots  & \ddots    & \vdots\\
    i_{K-1,\,0} & i_{K-1,\,1} & i_{K-1,\,2} & \ldots & i_{K-1,\,T-1}\\
    \end{bmatrix}
\end{eqnarray}
which we call lixel-time-histogram.
The columns correspond to the $K$ lixels in the plenoscope, and the rows correspond to the $T$ time-cells.
Each entry $i_{k,\,t}$ is the intensity of photons which arrived in lixel $k$ at time-cell $t$.
On \NameAcp{} the shape of $\mathcal{L}[k,\,t]$ is $((K=\NumLix{}) \times (T=100))$.
\NameAcp{} uses $T=100$ time-cells of $500\,$ps duration each.
Usually the population of $\mathcal{L}[k,\,t]$ is very sparse.
\subsection*{Pixel-paxel-time-histogram}
For the specific design of \NameAcp{}, the two-dimensional histogram $\mathcal{L}[k,\,t]$ can also be written as a three-dimensional histogram $\mathcal{L}[n,\,m,\,t]$, where the lixels $k$ are addressed using the pixels $n$, and paxels $m$ which they approximatively correspond to, see section \ref{SecAdressingLixel}.
\begin{eqnarray}
\mathcal{L}[n,\,m\,,t] &=&
    \begin{bmatrix}%
        \begin{bmatrix}%
            i_{0,\,0,\,0} & i_{0,\,1,\,0} & \ldots & i_{0,\,M-1,\,0}\\
            i_{1,\,0,\,0} & i_{1,\,1,\,0} & \ldots & i_{1,\,M-1,\,0}\\
            \vdots    & \vdots   & \ddots    & \vdots\\
            i_{N-1,\,0,\,0} & i_{N-1,\,1,\,0} & \ldots & i_{N-1,\,M-1,\,0}\\
        \end{bmatrix},\\
        \\
        \begin{bmatrix}%
            i_{0,\,0,\,1} & i_{0,\,1,\,1} &  \ldots & i_{0,\,M-1,\,1}\\
            i_{1,\,0,\,1} & i_{1,\,1,\,1} &  \ldots & i_{1,\,M-1,\,1}\\
            \vdots    & \vdots  & \ddots    & \vdots\\
            i_{N-1,\,0,\,1} & i_{N-1,\,1,\,1} & \ldots & i_{N-1,\,M-1,\,1}\\
        \end{bmatrix},\\
        \vdots\\
        \begin{bmatrix}%
            i_{0,\,0,\,T-1} & i_{0,\,1,\,T-1} & \ldots & i_{0,\,M-1,\,T-1}\\
            i_{1,\,0,\,T-1} & i_{1,\,1,\,T-1} & \ldots & i_{1,\,M-1,\,T-1}\\
            \vdots    & \vdots    &  \ddots    & \vdots\\
            i_{N-1,\,0,\,T-1} & i_{N-1,\,1,\,T-1} & \ldots & i_{N-1,\,M-1,\,T-1}\\
        \end{bmatrix}
    \end{bmatrix}
\end{eqnarray}
When we compute the sum along two of the three axis of the three-dimensional histogram $\mathcal{L}[n,\,m\,,t]$, we can directly obtain an image, an areal-sample, or an intensity-sequence.
The Equations \ref{EqLightFieldSequenceSumPixelPaxelTime} to \ref{EqLightFieldSequenceSumPixel} illustrate the various processing options of summing up the photon-intensities along certain axes of $\mathcal{L}[n,\,m\,,t]$.
\begin{eqnarray}
\label{EqLightFieldSequenceSumPixelPaxelTime}
\sum_{n,\,m,\,t} \mathcal{L}[n,\,m,\,t] &\rightarrow& \mathcal{L},\,\,\text{Intensity, number of photons}\\
\label{EqLightFieldSequenceSumPaxelTime}
\sum_{m,\,t} \mathcal{L}[n,\,m,\,t] &\rightarrow& \mathcal{L}[n],\,\,\text{Classic image of telescope}\\
\label{EqLightFieldSequenceSumPixelTime}
\sum_{n,\,t} \mathcal{L}[n,\,m,\,t] &\rightarrow& \mathcal{L}[m],\,\,\text{Areal intensity on aperture}\\
\sum_{n,\,m} \mathcal{L}[n,\,m,\,t] &\rightarrow& \mathcal{L}[t],\,\,\text{Intensity sequence, light-curve}\\
\sum_{t} \mathcal{L}[n,\,m,\,t] &\rightarrow& \mathcal{L}[n,\,m],\,\,\text{Static light-field}\\
\label{EqLightFieldSequenceSumPaxel}
\sum_{m} \mathcal{L}[n,\,m,\,t] &\rightarrow& \mathcal{L}[n,\,t],\,\,\text{Classic image-sequence, video}\\
\label{EqLightFieldSequenceSumPixel}
\sum_{n} \mathcal{L}[n,\,m,\,t] &\rightarrow& \mathcal{L}[m,\,t],\,\,\text{Areal intensity-sequence on aperture}
\end{eqnarray}
On \NameAcp{} the shape of $\mathcal{L}[n,\,m\,,t]$ is $((N=\NumPix{}) \times (M=\NumPax{}) \times (T=100))$.
\subsection*{Stream of photons}
We can also describe the light-field-sequence $\mathcal{L}$ as a stream
\begin{eqnarray}
\mathcal{L}[j] &=& \begin{bmatrix}%
    k_{0} &,& {t_\text{raw}}_0 \\
    k_{1} &,& {t_\text{raw}}_1 \\
    \vdots & & \vdots \\
    k_{J-1} &,& {t_\text{raw}}_{J-1}\\
    \end{bmatrix}
\end{eqnarray}
of all the $J$ photons in it.
Here each row in $\mathcal{L}$ describes a single photon.
The first column addresses the lixel $k_j$ which absorbed the $j$-th photon, and the second column is the arrival-time ${t_\text{raw}}_j$ of the $j$-th photon.
The stream of photons is practical to label individual photons as it is done in the classification of Cherenkov-photons and night-sky-background-photons.
Further, the stream of photons does not assume that lixels are organized hierarchical so that the light-field $\mathcal{L}[n,m]$ becomes a rectangular histogram in $n$ and $m$.
Personally, we mostly ended up using the stream of photons to represent the light-field-sequence $\mathcal{L}$ because of its generality and efficiency due to its natural zero-suppression.
\section{Directional sampling -- Imaging}
\label{SecInterpretingDirectionalSampling}
\begin{figure}{}
    \centering
    \includegraphics[width=0.8\textwidth]{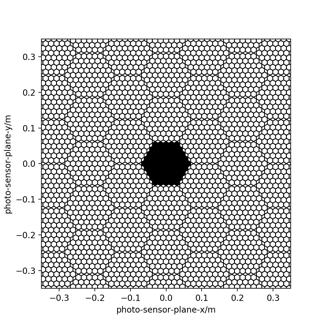}
    \caption[Directional-sampling, pixel-synthesis, sensor-plane, close-up]{
        Close-up on the sensor-plane of \NameAcp{}.
        Black photo-sensors (lixels) are summed up to synthesize the central pixel ($n = 4,221$) in the image.
        Black is a weight of $1$, and white is a weight of $0$.
    }
    \label{FigPixelLinearCombinationSensorPlane}
\end{figure}

The plenoscope can do imaging exactly like a conventional telescope.
In the special case of \NameAcp{}, we simply sum up the photon-intensities of all lixels which belong to the same pixel, see Equations \ref{EqLightFieldSequenceSumPaxelTime} and \ref{EqLightFieldSequenceSumPaxel}.
In words closer to the implementation of \NameAcp{}, we simply sum up the photon-intensities of all photo-sensors inside of each small camera, see Figure \ref{FigOpticsOverview}.
Imaging can be described as a linear-combination
\begin{eqnarray}
    \mathcal{I}[n] &=& U_\text{imaging}[n,\,k] \cdot \mathcal{L}[k]
    \label{EqImaging}
\end{eqnarray}
of the photon-intensities in the light-field.
Photon-intensities from several lixels $k$ in $\mathcal{L}[k]$ are added up in a particular pixel $n$ in the image $\mathcal{I}[n]$.
This linear-combination can be described using the imaging-matrix
\begin{eqnarray}
    U_\text{imaging}[n,\,k] &=&
    \begin{bmatrix}%
        u_{0,\,0} & u_{0,\,1} & \ldots & u_{0,\,K-1}\\
        u_{1,\,0} & u_{1,\,1} & \ldots & u_{1,\,K-1}\\
        \vdots    & \vdots    & \ddots    & \vdots\\
        u_{N-1,\,0} & u_{N-1,\,1} & \ldots & u_{N-1,\,K-1}\\
    \end{bmatrix}.
    \label{EqImagingMatrix}
\end{eqnarray}
Each matrix-element $u_{n',\,k'}$ describes how much of the photon-intensity from the lixel $k'$ in the light-field $\mathcal{L}[k]$ is added up into the pixel $n'$ in the image $\mathcal{I}[n]$.
In the special case of \NameAcp{}, we can compute the matrix-elements using Equation \ref{EqAddressingPixel}
\begin{eqnarray}
    u_{n,\,k} &=&
    \left\{
    \begin{array}{ll}
    1 & \text{if}\,\,n = \text{floor}(k/M) \\
    0 & \text{else} \\
    \end{array}
    \right.{}.
    \label{EqImagingPixelAssignmentPortal}
\end{eqnarray}
Figure \ref{FigPixelLinearCombinationSensorPlane} visualizes the matrix-elements on \NameAcp{} to synthesize the central pixel $n = 4,221$ in the image.
In general, the matrix-elements in $U_\text{imaging}[n,\,k]$ can be computed using the light-field-geometry $G$.
We can compute the matrix-elements of $U_\text{imaging}[n,\,k]$ using the angular distances between the incident-directions of the pixels $n$ in $\mathcal{I}[n]$ and a lixel $k$ in $G$.
\begin{eqnarray}
    \label{EqImagingPixelAssignmentGeneral}
    u_{n,\,k} &=&
    \left\{
    \begin{array}{ll}
    1 & \text{if}\,\,\sqrt{({c_x}_n - {c_x}_k)^2 + ({c_y}_n - {c_y}_k)^2} \leq c_\text{pixel-radius} \\
    0 & \text{else} \\
    \end{array}
    \right.{}
\end{eqnarray}
where $c_\text{pixel-radius}$ is e.g. the angular radius of a pixel in the image $\mathcal{I}[n]$.
The matrix-elements do not need to be strictly binary \mbox{(0 or 1)} as shown in the Equations \ref{EqImagingPixelAssignmentPortal}, and \ref{EqImagingPixelAssignmentGeneral}.
The weights can depend on the distances between the incident-directions, or on the integrated overlap between pixels to reduce aliasing.
The Figures \ref{FigGammaRay006163Full}, \ref{FigGammaRay006106Full}, \ref{FigGammaRay006554Full}, \ref{FigGammaRay006678Full}, and \ref{FigGammaRay006912Full} show images projected from \NameAcp{}'s light-field using Equation \ref{EqImagingPixelAssignmentPortal}.
\section{Areal-sampling}
\label{SecInterpretingArealSampling}
The plenoscope can do areal-sampling of the photon-intensities on ground similar to the modified solar-concentrators for dense areal-sampling presented in Section \ref{SecSamplingArealDense}.
An areal-sample is the photon-intensity-histogram along the support-positions $x$, and $y$ on the principal-aperture-plane.
With the special design of \NameAcp{}, we just sum the light-field-sequence $\mathcal{L}[n,\,m,\,t]$ according to the Equations \ref{EqLightFieldSequenceSumPixelTime}, or \ref{EqLightFieldSequenceSumPixel}.
In general, areal-sampling can also be formulated as a linear-combination
\begin{eqnarray}
    \mathcal{A}[m] &=& U_\text{areal}[m,\,k] \cdot \mathcal{L}[k]
    \label{EqArealSampling}
\end{eqnarray}
similar to Equation \ref{EqImaging}.
Here $\mathcal{A}[m]$ is the one-dimensional photon-intensity-histogram along the $m$ paxels on the principal-aperture-plane, and
\begin{eqnarray}
    U_\text{areal}[m,\,k] &=&
    \begin{bmatrix}%
        u_{0,\,0} & u_{0,\,1} & \ldots & u_{0,\,K-1}\\
        u_{1,\,0} & u_{1,\,1} & \ldots & u_{1,\,K-1}\\
        \vdots    & \vdots    & \ddots    & \vdots\\
        u_{M-1,\,0} & u_{M-1,\,1} & \ldots & u_{M-1,\,K-1}\\
    \end{bmatrix}
    \label{EqArealSamplinmgMatrix}
\end{eqnarray}
is the areal-sampling-matrix.
The matrix-element $u_{m',\,k'}$ expresses how much of the photon-intensity of lixel $k'$ from $\mathcal{L}[k]$, goes into paxel $m'$ in $\mathcal{A}[m]$.
In the special case of \NameAcp{}, the matrix-elements
\begin{eqnarray}
    u_{m,\,k} &=&
    \left\{
    \begin{array}{ll}
    1 & \text{if}\,\,m = k - \text{floor}(k/M) \times M \\
    0 & \text{else} \\
    \end{array}
    \right.{}
    \label{EqArealSamplingPaxelAssignmentPortal}
\end{eqnarray}
are calculated using the Equations \ref{EqAddressingPixel}, and \ref{EqAddressingPaxel}.
\begin{figure}{}
    \centering
    \includegraphics[width=0.8\textwidth]{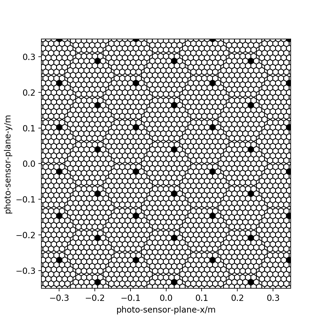}
    \caption[Areal-sampling, paxel-synthesis, sensor-plane, close-up]{
        Close-up on the sensor-plane of \NameAcp{}.
        Black photo-sensors (lixels) are summed up to synthesize a single paxel ($m = 54$) on the principal-aperture-plane.
        Black is a weight of $1$, and white is a weight of $0$.
    }
    \label{FigPaxelLinearCombinationSensorPlane}
\end{figure}
Figure \ref{FigPaxelLinearCombinationSensorPlane} shows the matrix-elements for the particular row $U_\text{areal}[m=54,\,k]$ in the special case of \NameAcp{}.
In words close to \NameAcp{}'s implementation, we sum up the intensities of each $m$-th photo-sensors in all small cameras, compare Figure \ref{FigOpticsOverviewCloseUp}.
In general, the weights
\begin{eqnarray}
    \label{EqArealSamplingPaxelAssignmentGeneral}
    u_{m,\,k} &=&
    \left\{
    \begin{array}{ll}
    1 & \text{if}\,\,\sqrt{({x}_m - {x}_k)^2 + ({y}_m - {y}_k)^2} \leq d_\text{paxel-radius} \\
    0 & \text{else} \\
    \end{array}
    \right.{}
\end{eqnarray}
can be estimated using the distance between the support-position of paxel $m$ in $\mathcal{A}$ and the support-position of lixel $k$ in the light-field-geometry $G$.
The radius of the paxels $d_\text{paxel-radius}$ can be used as a threshold here.
\section{Refocusing images in post}
\label{SecPostRefocusedImaging}
The plenoscope can project its light-field onto images which are refocused to different object-distances in post, this means after the light-field was recorded.
In Chapter \ref{ChDepthOfField} we demonstrate such refocused images on the example of the \NameAcp{} Cherenkov-plenoscope, see the Figures \ref{FigGammaRay006163Refocused}, \ref{FigGammaRay006106Refocused}, \ref{FigGammaRay006554Refocused}, \ref{FigGammaRay006678Refocused}, and \ref{FigGammaRay006912Refocused}.
Here we present a geometric description for refocusing in post which is motivated by \cite{ng2005} and Figure \ref{FigThinLens}.
First, we discuss the trajectories of the photons after they passed the principal-aperture-plane and call those trajectories image-rays.
Second, we use the image-rays and their intersections with virtual sensor-planes to formulate a projection of the light-field onto images focused to different object-distances.
\subsection*{Image-rays}
\label{SecImageRays}
\begin{figure}
    \centering
    \includegraphics[width=.5\textwidth]{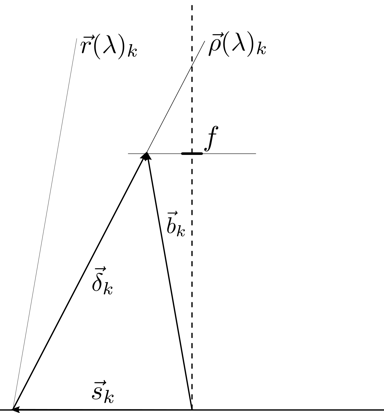}
    \caption[Image-rays]{
        A ray $\vec{r}(\lambda)_k$ is reflected on the principal-aperture-plane at the bottom of the figure.
        We call the reflected ray $\vec{\rho}(\lambda)_k$ the image-ray.
        Both rays share the same support-position $\vec{s}_k$.
    }
    \label{FigImageRay}
\end{figure}

Image-rays describe the trajectories of the photons after those have passed the principal-aperture-plane, see Figure \ref{FigImageRay}.
In Figure \ref{FigThinLens}, the image-rays are defined by their support-positions on the principal-aperture-plane $a$, $a'$, and $a''$ together with their absorption-positions on the sensor-plane $i$, $i'$, and $i''$.
Every ray $\vec{r}(\lambda)_k$ according to Equation \ref{EqRay} has a corresponding image-ray
\begin{eqnarray}
\vec{\rho}(\lambda)_k &=& \vec{s}_k + \lambda \frac{\vec{\delta}_k}{\vert \vec{\delta}_k \vert_2}.
\end{eqnarray}
Both rays $\vec{r}(\lambda)_k$, and $\vec{\rho}(\lambda)_k$ share the same support-vector
\begin{eqnarray}
\vec{s}_k &=& \left(x_k,\,\,y_k,\,\,0\right)^T
\end{eqnarray}
such that they both intersect on the principal-aperture-plane for $\lambda = 0$
\begin{eqnarray}
\vec{r}(0)_k &=& \vec{\rho}(0)_k.
\end{eqnarray}
The direction-vector $\vec{\delta}_k$ of the image-ray $\vec{\rho}(\lambda)_k$ is calculated using the Thin-lens-equation \ref{EqThinLens} and Figure \ref{FigThinLens}.
We first calculate where the image-ray would intersect with the focal-plane in focal-distance $f$ of the imaging-system
\begin{eqnarray}
\vec{b}_k &=& f \left( - \tan({c_x}_k),\,\,- \tan({c_y}_k),\,\,1 \right)^T,
\end{eqnarray}
and second we calculate the direction-vector
\begin{eqnarray}
\vec{\delta}_k &=& \vec{b}_k - \vec{s}_k
\end{eqnarray}
by subtracting the support-position $\vec{s}_k$ from the intersection point $\vec{b}_k$ in the sensor-plane.
\subsection*{Refocusing on the plenoscope}
We express the refocusing as a projection
\begin{eqnarray}
    \mathcal{I}_g[n] &=& U_\text{imaging}(g)[n,\,k] \cdot \mathcal{L}[k]
    \label{EqRefocusedImaging}
\end{eqnarray}
of the light-field onto an image, similar to regular imaging in Equation \ref{EqImaging}.
However, this time the imaging-matrix
\begin{eqnarray}
    U_\text{imaging}(g)[n,\,k] &=&
    \begin{bmatrix}%
        u_{0,\,0}(g) & u_{0,\,1}(g) & \ldots & u_{0,\,K-1}(g)\\
        u_{1,\,0}(g) & u_{1,\,1}(g) & \ldots & u_{1,\,K-1}(g)\\
        \vdots    & \vdots    & \ddots    & \vdots\\
        u_{N-1,\,0}(g) & u_{N-1,\,1}(g) & \ldots & u_{N-1,\,K-1}(g)\\
    \end{bmatrix}
    \label{EqRefocusedImagingMatrix}
\end{eqnarray}
depends on the object-distance $g$ where we want to focus on.
The matrix-elements $u_{n,\,k}(g)$ are calculated using the $k$-th image-ray $\vec{\rho}(\lambda)_k$, and the corresponding $x$, and $y$ positions of the $n$-th pixel on a virtual sensor-plane in image-distance $b$.
First, we use the Thin-lens-equation and our desired object-distance $g$ to calculate the image-distance
\begin{eqnarray}
b &=& \frac{1}{\frac{1}{f} - \frac{1}{g}}
\label{EqVirtualSensorPlaneDistance}
\end{eqnarray}
where of the virtual sensor-plane.
Second, we calculate how far
\begin{eqnarray}
\lambda_g &=& \frac{b}{\vec{z} \cdot \vec{\delta}_k}
\end{eqnarray}
we have to travel along the direction-vector $\vec{\delta}_k$ of our image-ray $\vec{\rho}(\lambda)_k$ to intersect with our virtual-sensor-plane in distance $b$.
Third, we calculate the intersection-position
\begin{eqnarray}
\vec{i}_{k,g} &=& \vec{\rho}(\lambda_g)_k
\end{eqnarray}
of the image-ray on the virtual-sensor-plane.
Fourth, we compute the position
\begin{eqnarray}
\vec{p}_{n,g} &=& \left(x_n,\,y_n,\,b\right)^T
\label{EqVirtualSensorPlaneIntersection}
\end{eqnarray}
of the $n$-th pixel in our virtual-sensor-plane.
Here $x_n$, and $y_n$ correspond to the $x$ and $y$ positions of the $n$-th pixel on the sensor-plane.
Fifth, we compute the distance
\begin{eqnarray}
o_{n,k,g} &=& \vert \vec{i}_{k,g} - \vec{p}_{n,g} \vert_2
\end{eqnarray}
between the pixel-position $\vec{p}_{n,g}$ and the intersection-position $\vec{i}_{k,g}$ in the sensor-plane.
Finally, we calculate the matrix-elements
\begin{eqnarray}
    u_{n,\,k}(g) &=&
    \left\{
    \begin{array}{ll}
    1 & \text{if}\,\,o_{n,k,g} \leq \text{pixel-radius}\\
    0 & \text{else} \\
    \end{array}
    \right.{}
    \label{EqPostRefocusedImagingPixelAssignmentGeneral}
\end{eqnarray}
by comparing the distance $o_{n,k,g}$ with e.g. the pixel-radius to decide if the $k$-th lixel is participating to the photon-intensity in the $n$-th pixel.
In Figure \ref{FigPixelLinearCombinationSensorPlane} we see a visual representation of the matrix-elements for the $n$-th pixel in \NameAcp{} when we do not refocus.
Now we can visualize the matrix-elements for the $n$-th pixel in \NameAcp{} when we refocus in post to different object-distances $g$, see Figure \ref{FigPixelRefocusedLinearCombinationSensorPlane}.
In \NameAcp{}, the default sensor-distance $d$ is equal to the focal-length $f$, such that the default object-distance where \NameAcp{} focuses on is in $g = \infty$.
\footnote{In Section \ref{SecAfterthoughtTargetSensorDistance}, we discuss possible benefits for the Chrenkov-plenoscope's trigger when setting the default object-distance not to infinity, but to the air-shower-maximum.}
Find in Figure \ref{FigPixelRefocusedLinearCombinationSensorPlane}, that the refocusing to infinity yields exactly the same matrix-elements we obtain in the case for imaging without refocusing in Figure \ref{FigPixelLinearCombinationSensorPlane}.\\
Figure \ref{FigPixelRefocusedLinearCombinationSensorPlaneColored} shows the interplay of neighboring pixels in the resulting refocused image.
We color the photo-sensors according to the seven neighboring pixels they will be added up into in the refocused image.
We find that indeed all the photo-sensors contribute to the resulting refocused images.
In the Figures \ref{FigPixelRefocusedLinearCombinationSensorPlane}, \ref{FigPixelRefocusedLinearCombinationSensorPlane2}, and \ref{FigPixelRefocusedLinearCombinationSensorPlaneColored} we find that the patterns of photo-sensors is not perfectly symmetric in all refocusing scenarios.
Due to the binary decision made in Equation \ref{EqPostRefocusedImagingPixelAssignmentGeneral}, already small statistical fluctuations in the estimation of the light-field-geometry $G$ can cause asymmetric patterns in this first implementation.
\begin{figure}{}
    \centering
    \includegraphics[width=1\textwidth]{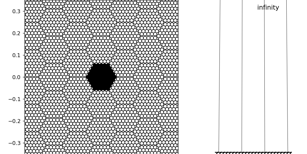}
    \vspace{0.3cm}
    \includegraphics[width=1\textwidth]{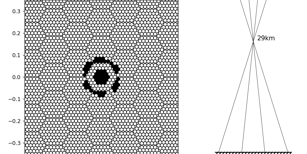}
    \vspace{0.3cm}
    \includegraphics[width=1\textwidth]{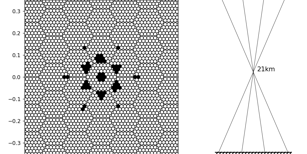}
    \caption[Refocusing, different linear-combinations of lixels, part 1\,of\,2]{Continues in Figure \ref{FigPixelRefocusedLinearCombinationSensorPlane2}.
        Part 1 of 2.
        Black photo-sensors (lixels) are summed up to synthesize a single pixel focused to different object-distances.
        The object-distance focused to is written on the right, where the rays converge.
        Here we see the central pixel $n = 4,221$ on \NameAcp{}.
    }
    \label{FigPixelRefocusedLinearCombinationSensorPlane}
\end{figure}
\begin{figure}{}
    \centering
    \includegraphics[width=1\textwidth]{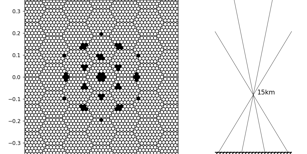}
    \vspace{0.3cm}
    \includegraphics[width=1\textwidth]{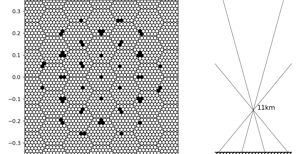}
    \vspace{0.3cm}
    \includegraphics[width=1\textwidth]{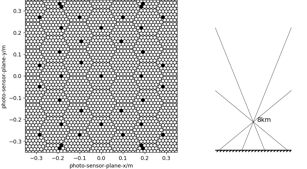}
    \caption[Refocusing, different linear-combinations of lixels, part 2\,of\,2]{
        Continuation of Figure \ref{FigPixelRefocusedLinearCombinationSensorPlane}.
        Part 2 of 2.
    }
    \label{FigPixelRefocusedLinearCombinationSensorPlane2}
\end{figure}
\begin{figure}{}
    \centering
    \includegraphics[width=1\textwidth]{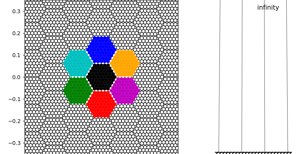}
    \vspace{0.3cm}
    \includegraphics[width=1\textwidth]{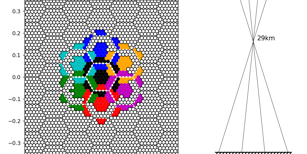}
    \vspace{0.3cm}
    \includegraphics[width=1\textwidth]{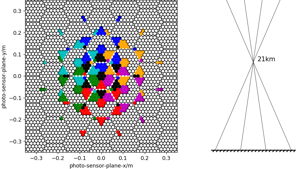}
    \caption[Refocusing, different linear-combinations of lixels]{%
        % central_seven_pixel_ids = [4221, 4124, 4222, 4220, 4125, 4317, 4318]
        % colors = ['k', 'g', 'b', 'r', 'c', 'm', 'orange']
        %
        Same as Figure \ref{FigPixelRefocusedLinearCombinationSensorPlane}, but including the neighboring pixels.
        Central, black pixel id is $4,221$,
        green is $4,124$,
        blue is $4,222$ is,
        red is $4,220$ is,
        cyan is $4,125$ is,
        magenta is $4,317$ is, and
        orange is $4,318$ is.
    }
    \label{FigPixelRefocusedLinearCombinationSensorPlaneColored}
\end{figure}
\section{Synthetic apertures}
\label{SecSyntheticApertures}
We interpret the plenoscope's light-field-sequence as a set of multiple image-sequences observed with an array of telescopes.
In Chapter \ref{ChDepthOfField} we demonstrate this on the example of \NameAcp{} in the Figures \ref{FigGammaRay006163Segmented}, \ref{FigGammaRay006106Segmented}, \ref{FigGammaRay006554Segmented}, \ref{FigGammaRay006678Segmented}, and \ref{FigGammaRay006912Segmented}.
Arrays of Cherenkov-telescopes observe gamma-rays successfully and efficiently.
Thus splitting the light-field-sequence into multiple image-sequences might be a first step to approach the light-field and the plenoscope.
We formulate the synthetic creation of telescope apertures as a projection of the light-field
\begin{eqnarray}
\label{EqSyntheticAperturesLinearCombination}
\mathcal{I}_{m}[n] &=& U^{m}[n,\,k] \cdot \mathcal{L}[k].
\end{eqnarray}
On \NameAcp{}, we can slice the light-field-sequence $\mathcal{L}[n,\,m,\,t]$ along its $M=\NumPax{}$ paxels to obtain the $M=\NumPax{}$ image-sequences $\mathcal{I}_{m}[n,\,t]$ observed by $M=\NumPax{}$ synthetic telescopes
\begin{eqnarray}
u^{m'}_{n',\,k} &=&
    \left\{
    \begin{array}{ll}
    1 & \text{if}\,\, n' = n\,\,\text{and}\,\, m' = m\\
    0 & \text{else} \\
    \end{array}
    \right.{}.
\end{eqnarray}
Here we used the Equations \ref{EqAddressingPixel}, and \ref{EqAddressingPaxel} to compute $n$, and $m$ from $k$.\\
In general we calculate the matrix-elements $u^{m}_{n,\,k}$ based on the support-positions of the rays defined in the light-field-geometry $G$.
First, we calculate the distance
\begin{eqnarray}
o_{k,\,m} &=& \sqrt{(\overline{x}_k - x_m)^2 + (\overline{y}_k - y_m)^2}
\end{eqnarray}
between the support-position of the $k$-th ray on the principal-aperture-plane at $\overline{x}_k$, $\overline{y}_k$ and the center of the $m$-th synthesized telescope-aperture $x_m$, $y_m$.
Second, we can compute the matrix-elements
\begin{eqnarray}
\label{EqSyntheticAperturesLixelAssignment}
u^{m}_{n',\,k} &=&
    \left\{
    \begin{array}{ll}
    1 & \text{if}\,\, n'=n \,\,\text{and}\,\,o_{k,\,m} \leq r_\text{m}\\
    0 & \text{else} \\
    \end{array}
    \right.{}
\end{eqnarray}
by comparing the distance $o_{k,\,m}$ with e.g. the radius $r_\text{m}$ of the $m$-th synthesized telescope-aperture.
Again we calculate $n$ from $k$ using Equation \ref{EqAddressingPixel}.
Here we describe a synthetic aperture with the shape of a disc with a radius $r_\text{m}$, but one might also use other shapes for the synthetic apertures as e.g. hexagons, squares or triangles.
In Chapter \ref{ChDepthOfField} we synthesize seven large telescopes by using the mask in Figure \ref{FigPortalApertureSegmentationSevenTelescopes} and Equation \ref{EqSyntheticAperturesLixelAssignment}.
\section{Light-front}
The plenoscope can reconstruct the light-front of the Cherenkov-photons in the three-dimensional space above its aperture, as it is shown in Figure \ref{FigRaysOnPrincipalAperturePlane}.
This might be obvious as the light-front corresponds to our very definition of the light-field-sequence in Chapter \ref{ChLightFieldGeometry} and thus does not need further interpretation, but we list it here explicitly because it opens up so many possibilities.
We use the light-front in Chapter \ref{ChAngularResolution} to reconstruct the incident-direction of very low energetic $1\,$GeV cosmic gamma-rays.
%
%------------------------------------------------------------------------------
%
%
%
%
%
%
%
%------------------------------------------------------------------------------
\chapter{Overcoming aberrations, enlarging the field-of-view}
\label{ChOvercomingAberrations}
Real imaging-optics such as lenses or imaging-reflectors suffer from aberrations which reduce the angular resolution for imaging.
The plenoscope can overcome these aberrations when they originate from the aperture of the imaging-optics being extended \cite{hanrahan2006digital}.
On a perfect imaging-optics, the absorption-position of a photon on the sensor-plane only depends on the incident-direction $c_x$, $c_y$ of the photon relative to the optical-axis of the imaging-system.
On real imaging-optics however, the photon's absorption-position on the sensor-plane also depends on the support-position $x$, $y$ where the photon passed the aperture-plane.
This results in pixels to not only collect photons with incident-directions associated with these pixels, but to also collect photons with different incident-directions.\\
Here we will demonstrate how far plenoptic perception can overcome aberrations on the example of a spherical imaging-reflector.
We will briefly inspire the drastic implications for the enlarging of the field-of-views on e.g. Cherenkov-telescopes by about one order-of-magnitude.\\
Overcoming aberrations will turn out to be already included in the light-field-geometry $G$ introduced in Chapter \ref{ChLightFieldGeometry}, and is closely related to the compensation of misalignments between the imaging-reflector and the light-field-sensor discussed in Chapter \ref{ChCompensatingMisalignmnets}.
\section{Using the light-field-geometry}
\label{SecOvercomingAberrationsIdea}
\begin{figure}{}
    \centering
    \includegraphics[width=1\textwidth]{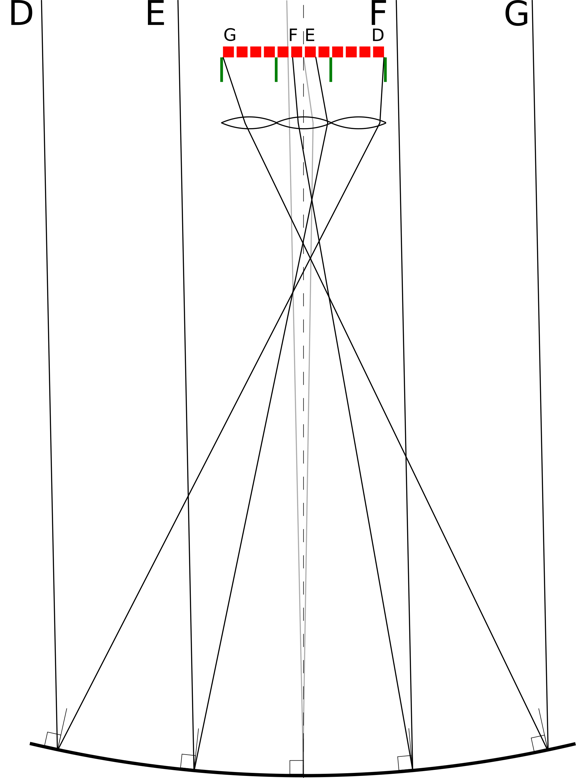}
    \caption[Aberrations and light-field-sensor]{
        Four photons D, E, F, and G with the same incident-directions on a plenoscope.
        In contrast to the imaging-reflector shown in Figure \ref{FigOpticsOverview}, here the imaging-reflector suffers from aberrations.
        To visualize the slight off-axis incident-direction, we put a fifth photon in light gray close to the optical-axis of the imaging-reflector.
        On the imaging-reflector, short lines at the intersection-positions of the photons indicate the surface-normal of the imaging-reflector.
        In Figure \ref{FigPlenoscopeWithAberrationsCloseUp} we show a close-up of the light-field-sensor.
    }
    \label{FigPlenoscopeWithAberrationsOverview}
\end{figure}

Figure \ref{FigPlenoscopeWithAberrationsOverview} shows the four parallel  photons D, E, F, and G on a plenoscope.
In Figure \ref{FigPlenoscopeWithAberrationsOverview} the imaging-reflector suffers from aberrations, in contrast to the imaging-reflector shown in Figure \ref{FigOpticsOverview} which is free of aberrations.
The aberrations of the imaging-reflector cause the photons D, and G to enter different small cameras in the light-field-sensor although the four photons D, E, F, and G have all the same incident-directions.
\begin{figure}{}
    \begin{minipage}{0.48\textwidth}
        \includegraphics[width=1\textwidth]{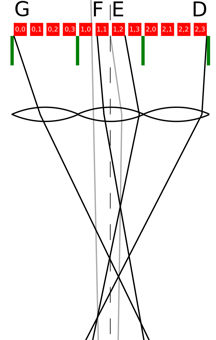}
        \caption[Aberrations and light-field-sensor, close-up]{
            Close up of the light-field-sensor shown in Figure \ref{FigPlenoscopeWithAberrationsOverview}.
        }
        \label{FigPlenoscopeWithAberrationsCloseUp}
    \end{minipage}
    \hfill
    \begin{minipage}{0.48\textwidth}
        \includegraphics[width=1\textwidth]{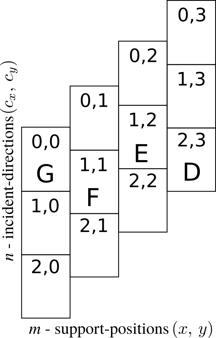}
        \caption[Aberrations and light-field-sensor, light-field-geometry]{
            The light-field-geometry of the plenoscope shown in the Figures \ref{FigPlenoscopeWithAberrationsOverview} and \ref{FigPlenoscopeWithAberrationsCloseUp} visualized in the light-field-sampling-graph, compare Figure \ref{FigLightFieldSamplingLegend} Chapter \ref{ChComparingOtherMethods}.
        }
        \label{FigPlenoscopeWithAberrationsSamplingGraph}
    \end{minipage}
\end{figure}
On a classic image-sensor, we can not resolve that the photons D, and G have the same incident-directions as the photons E, and F.
The image is blurred.
But with a light-field-sensor shown in Figure \ref{FigPlenoscopeWithAberrationsOverview} we can resolve the true incident-directions of all the photons if we know the light-field-geometry $G$ of our plenoscope.
The light-field-geometry $G$, which is introduced in Chapter \ref{ChLightFieldGeometry}, tells us for each photo-sensor in the light-field-sensor which solid-angle of incident-directions $c_x$, $c_y$, and which area of support-positions $x$, $y$ are covered by this photo-sensor.
In the close-up Figure \ref{FigPlenoscopeWithAberrationsCloseUp} we see the two-dimensional addressing of the photo-sensors in the light-field-sensor.
Analog to Figure \ref{FigOpticsOverviewCloseUp}, we see in Figure \ref{FigPlenoscopeWithAberrationsSamplingGraph} how the photo-sensors sample the light-field.
In contrast to the sampling of the plenoscope free of aberrations in Figure \ref{FigOpticsOverviewCloseUp}, here we find that the sampling of the photo-sensors is not a rectangular grid in the space of incident-directions and support-positions but a skewed grid instead.
The exact arrangement of the photo-sensors in the space of incident-directions and support-positions is what is stored in the light-field-geometry $G$.
So the sampling of the photo-sensors in Figure \ref{FigPlenoscopeWithAberrationsSamplingGraph} is a visual representation of the light-field-geometry $G$ for the plenoscope presented in Figure \ref{FigPlenoscopeWithAberrationsOverview} which has an imaging-reflector with aberrations.
In Figure \ref{FigPlenoscopeWithAberrationsSamplingGraph}, we find that the photo-sensors (0,\,0), (1,\,1), (1,\,2), and (2,\,3) sample similar incident-directions as they are arranged on a horizontal line in the space of incident-directions and support-positions.
So when we project the recorded light-field $\mathcal{L}$ onto an image using the light-field-geometry $G$, we obtain an image where the effects of the aberrations from the imaging-reflector are reduced.
The projection can be visualized in Figure \ref{FigPlenoscopeWithAberrationsSamplingGraph} when the intensity of the photo-sensors is projected onto the axis of incident-directions.
For details on projecting the light-field onto images see the Sections \ref{SecInterpretingDirectionalSampling} and \ref{SecPostRefocusedImaging}.
\section{Example -- spherical imaging-reflector}
\begin{figure}{}
    \begin{minipage}{0.48\textwidth}
        \includegraphics[width=1\textwidth, angle=90]{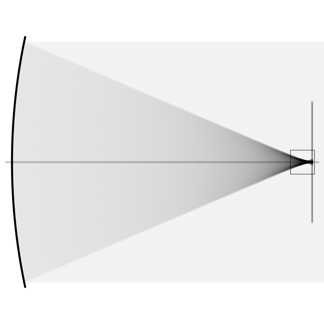}
    \end{minipage}
    \hfill
    \begin{minipage}{0.48\textwidth}
        \includegraphics[width=1\textwidth, angle=90]{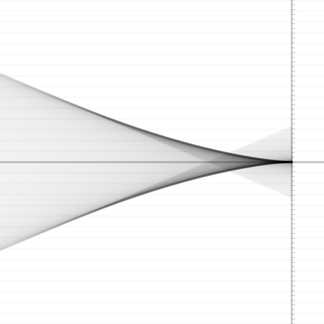}
    \end{minipage}
    \centering
    \includegraphics[width=0.8\textwidth]{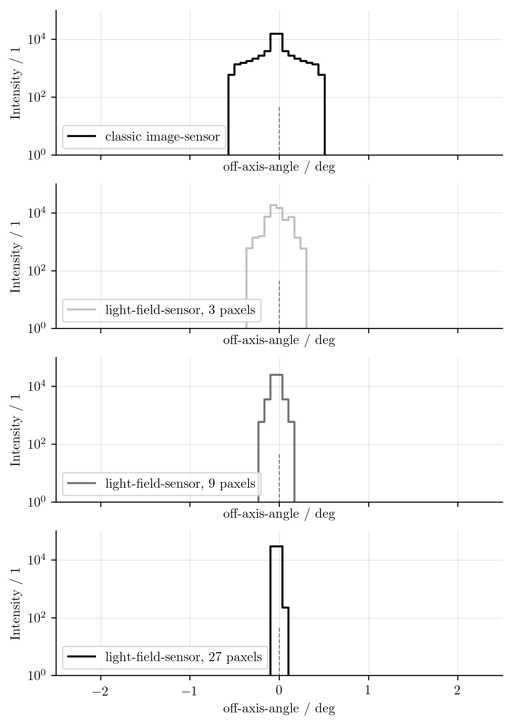}
    \caption[Overcoming aberrations, $0.0^{\circ}$]{
        $0.0^{\circ}$ incident-direction.
    }
    \label{FigSphericalAberrations0}
\end{figure}

\begin{figure}{}
    \begin{minipage}{0.48\textwidth}
        \includegraphics[width=1\textwidth, angle=90]{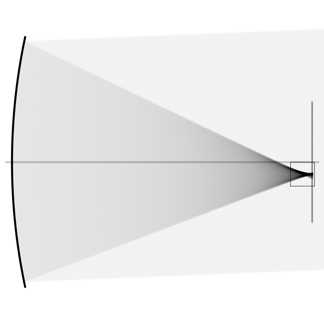}
    \end{minipage}
    \hfill
    \begin{minipage}{0.48\textwidth}
        \includegraphics[width=1\textwidth, angle=90]{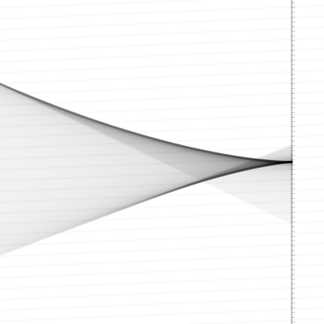}
    \end{minipage}
    \centering
    \includegraphics[width=0.8\textwidth]{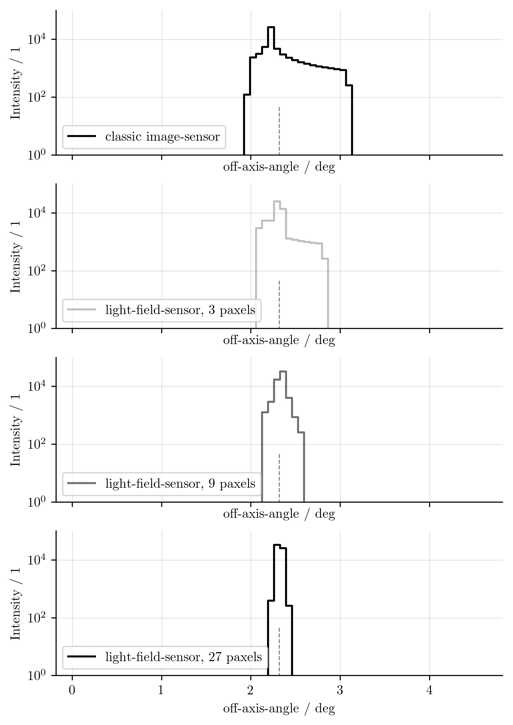}
    \caption[Overcoming aberrations, $2.3^{\circ}$]{
        $2.3^{\circ}$ incident-direction.
    }
    \label{FigSphericalAberrations3}
\end{figure}

\begin{figure}{}
    \begin{minipage}{0.48\textwidth}
        \includegraphics[width=1\textwidth, angle=90]{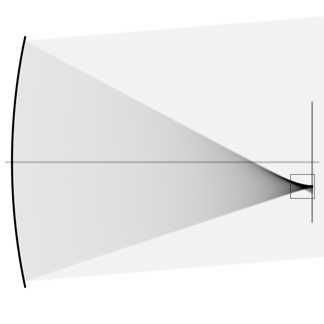}
    \end{minipage}
    \hfill
    \begin{minipage}{0.48\textwidth}
        \includegraphics[width=1\textwidth, angle=90]{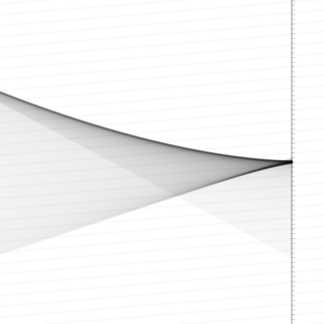}
    \end{minipage}
    \centering
    \includegraphics[width=0.8\textwidth]{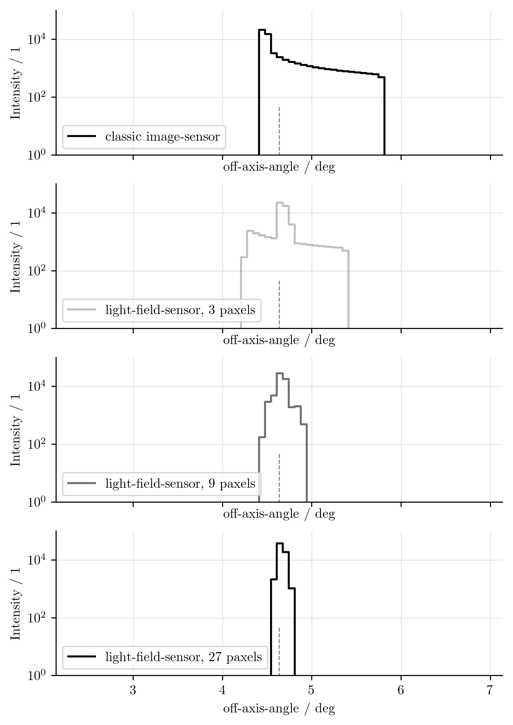}
    \caption[Overcoming aberrations, $4.6^{\circ}$]{
        $4.6^{\circ}$ incident-direction.
    }
    \label{FigSphericalAberrations6}
\end{figure}

\begin{figure}{}
    \begin{minipage}{0.48\textwidth}
        \includegraphics[width=1\textwidth, angle=90]{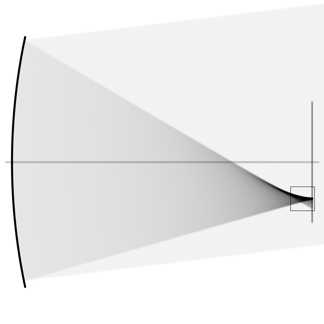}
    \end{minipage}
    \hfill
    \begin{minipage}{0.48\textwidth}
        \includegraphics[width=1\textwidth, angle=90]{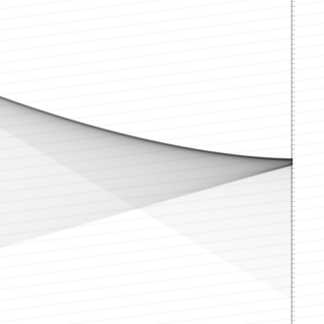}
    \end{minipage}
    \centering
    \includegraphics[width=0.8\textwidth]{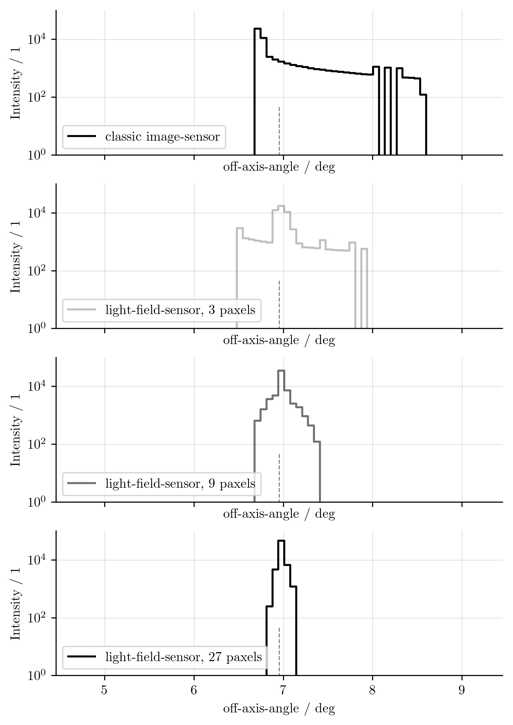}
    \caption[Overcoming aberrations, $7.0^{\circ}$]{
        $7.0^{\circ}$ incident-direction.
    }
    \label{FigSphericalAberrations9}
\end{figure}

Spherical imaging-reflectors are one of the simplest to be fabricated.
Spherical imaging-reflectors are often used to have large field-of-views \cite{fors2013telescope,cortina2016}, but like other imaging-reflectors they suffer from aberrations.
We simulate the imaging on a spherical imaging-reflector which has the same focal-length/diameter$=1.5$ ratio as \NameAcp{}'s large imaging-reflector.
One after the other, we simulate the exposures of a classic image-sensor and three different light-field-sensors.
The image-sensor and the light-field-sensor have the same pixel-sizes of $0.067^{\circ}$.
The light-field-sensors only differ in the number of paxels they use to segment the principal-aperture-plane.
We have light-field-sensors with 3, 9, and 27 paxels on the diagonal of the aperture.
This simulation is in two dimensions, and assumes that the light-field-sensors are perfect and only limited by their discrete number of paxels.\\
Before we reconstruct images from the responses of the different sensors, we estimate the light-field-geometry of each sensor, even the light-field-geometry $G_\text{spherical imaging-reflector + image-sensor}$ of the classic image-sensor.
This makes the comparison fair.
Now both the image of the classic image-sensor and the images of the light-field-sensors are corrected for distortions caused by the spherical imaging-reflector.
Distortions are when the absorption-position of a photon on the sensor-plane is not linear proportional to the incident-direction of the photon.
Distortions can be, and usually are, corrected for when reconstructing images from image-sensors.
The four Figures \ref{FigSphericalAberrations0}, \ref{FigSphericalAberrations3}, \ref{FigSphericalAberrations6}, and \ref{FigSphericalAberrations9} show the images reconstructed by both the image-sensor and the light-field-sensors for various incident-directions.
Finally, in Figure \ref{FigPointSpreadVsOffAxisAngle} we summarize our findings and show how good light-field-sensors can overcome aberrations of spherical imaging-reflectors.
\subsection*{Description for the Figures \ref{FigSphericalAberrations0}, \ref{FigSphericalAberrations3}, \ref{FigSphericalAberrations6}, and \ref{FigSphericalAberrations9}}
Each figure consists of six panels.
The two panels at the top show the spherical imaging-reflector, the optical-axis, the sensor-plane, and the trajectories of the photons.
In the left panel at the top, we find an overview of the setup, and in the right panel at the top we find a close-up on the sensor-plane.
In the left panel at the top, the close-up is indicated with a black square.
Below we have four panels showing the four one-dimensional images reconstructed by the four different sensors.
The upper most one-dimensional image is from the image-sensor, the lower ones are from the light-field-sensor with different numbers of paxels.
The four panels showing the one-dimensional images have a vertical, dashed line to indicate the true incident-direction of the photons.
\subsection*{Discussion}
\begin{figure}{}
    \centering
    \includegraphics[width=1\textwidth]{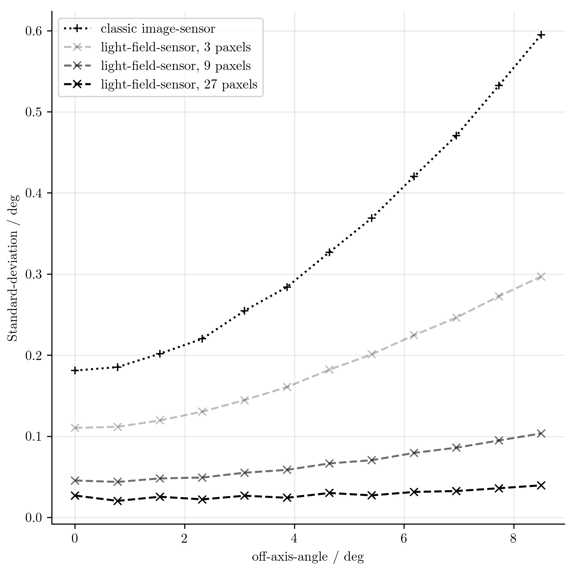}
    \caption[Overcoming aberrations, spread vs. true incident-direction]{
        The spread of the reconstructed incident-directions of photons versus the true incident-direction of the photons.
        The number of paxels here is the number of paxels on the diagonal of the aperture.
        For comparison, our \NameAcp{} Cherenkov-plenoscope has $9$ paxels across the diagonal of its large imaging-reflector.
    }
    \label{FigPointSpreadVsOffAxisAngle}
\end{figure}

From the reconstructed, one-dimensional images shown in the Figures \ref{FigSphericalAberrations0}, \ref{FigSphericalAberrations3}, \ref{FigSphericalAberrations6}, and \ref{FigSphericalAberrations9}, we find:
First, that the spread of the reconstructed incident-directions of the photons is reduced when we use light-field-sensors instead of image-sensors.
Second, that the spread of the reconstructed incident-directions of the photons becomes smaller the more paxels are used in the light-field-sensor.
And third, we find that the asymmetry of the reconstructed incident-directions of the photons is reduced on the light-field-sensors.
The binning-artifacts in the reconstructed images seen in Figure \ref{FigSphericalAberrations9} do not significantly effect our findings presented here.\\
In Figure \ref{FigPointSpreadVsOffAxisAngle}, we find that even light-field-sensors with low numbers of paxels provide a strong reduction of the effect of aberrations on imaging.
From a geometric point-of-view it is only natural, that when the number of paxels is increased further, aberrations of imaging-optics are overcome completely.
This might become more clear when we describe the light-field-geometry $G$ with other words:
In the case of the plenoscope composed from an light-field-sensor and an imaging-reflector, the light-field-geometry $G$ describes the surface-normal of the aperture, see the small surface-normals indicated in Figure \ref{FigPlenoscopeWithAberrationsOverview}.
So when we have more paxels in the light-field-sensor, we have a more dense sampling of the aperture's surface-normals, and so we can correct better for aberrations which originate from these surface-normals.
\section{Enlarging the field-of-view}
\label{SecEnlargingTheFieldOfView}
In Figure \ref{FigPointSpreadVsOffAxisAngle} we find that the spread of the reconstructed incident-directions is much reduced and becomes more constant over the range of true incident-directions when we use light-field-sensors instead of image-sensors.
Of course more detailed studies are needed, but Figure \ref{FigPointSpreadVsOffAxisAngle} suggests that a Cherenkov-plenoscope could be build with an enormous $16^{\circ}$ diagonal field-of-view.
A $201\,$deg$^2$ field-of-view would be an improvement of more than one order-of-magnitude over the field-of-views on todays Cherenkov-telescopes with about $16\,$deg$^2$, see Figure \ref{FigFieldOfViewComparison}.
And it would still be an improvement by a factor of $4$ over the planned Schwarzschild-Couder-Cherenkov-Telescopes which are dedicated for wide field-of-views \cite{cta2013introducing} and also reach angular resolutions in and below the $0.1^{\circ}$ regime.\\
The bottom-line is that Cherenkov-plenoscopes become interesting independent of their benefits for observations of low energetic gamma-rays, but because they can enlarge the field-of-view which we currently can not achieve by other means.
\section{Upgrading the Cherenkov-telescope}
\label{SecUpgradingTheCherenkovTelescope}
We suggest to equip the Cherenkov-telescope\footnote{In general. This includes all existing and future Cherenkov-telescopes.} with a light-field-sensor not to gain plenoptic perception, but to improve image-quality.
With imaging-reflectors with diameters below $\approx 23\,$m, there is not much plenoptic perception-power for air-showers to be gained anyhow, compare Chapter \ref{ChDepthOfField}.
However, the size of the field-of-view, the sharpness of the images, and especially the consistency of the images across the field-of-view would be improved dramatically.
In case one fears the costs for additional read-out-channels in a light-field-sensors, we propose a way to mutilate the light-field-sensor in order to reduce the number of read-out-channels to the same number needed for image-sensors, while still profiting from corrections for imaging provided by plenoptic perception.
We can hardwire the light-field-geometry $G$ into the light-field-sensor to sum up the photo-sensors which belong to a reconstructed pixel before these photo-sensors are fed into individual read-out-channels.
Projecting the light-field onto an image is about summing up the correct photo-sensors as we show in the Sections \ref{SecInterpretingDirectionalSampling}, and \ref{SecPostRefocusedImaging}.
To be clear: The number of individual photo-sensors will increase to the number of lixels, but the number of read-out-channels for digitization will stay the lower number of pixels.
A Cherenkov-telescope equipped with such a mutilated light-field-sensor would have the optics of a plenoscope, but the read-out of a telescope.
The resulting Cherenkov-telescope would not record light-fields, but it could record images with much larger field-of-views, and much reduced aberrations.
%
%
%Probably there would probably only be spherical optical components in such an inter-plenoptic-Cherenkov-telescope.
%
%And now, that we all have day-dreamed of the bright future for gamma-ray-astronomy, let us briefly remind ourselves of the current quest for larger field-of-views in a field where it actually matters\footnote{No offense, I personally love gamma-ray-astronomy. But you have to know your place in the food-chain.}, in optical astronomy.
%
%Yes I am looking at you Large-Synaptic-Survey-Telescope (\textit{LssT}) \cite{ivezic2008lsst}!
\section{Overcoming aberrations in \NameAcp{}}
\begin{figure}{}
    \centering
    \includegraphics[width=1\textwidth]{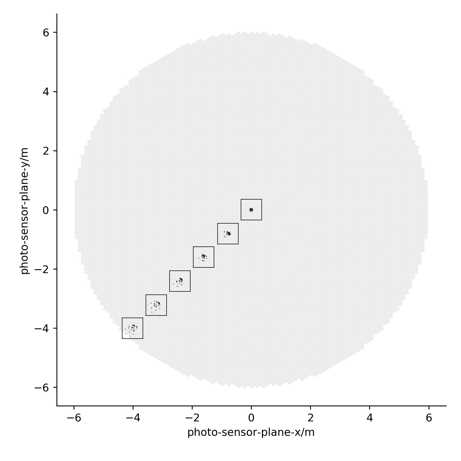}
    \caption[Aberrations on \NameAcp{}, overview]{
        Black photo-sensors are summed to synthesize the six pixels with incident-directions $0.0^\circ$, $0.6^\circ$, $1.2^\circ$, $1.8^\circ$, $2.4^\circ$, and $3.0^\circ$.
        The black boxes indicate the range of the close-ups shown in Figure \ref{FigAberrationsNameAcpCloseUp}.
    }
    \label{FigAberrationsNameAcpOverview}
\end{figure}

\begin{figure}{}
    \centering
    \includegraphics[width=.49\textwidth]{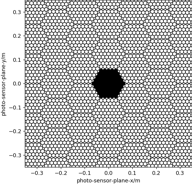}
    \includegraphics[width=.49\textwidth]{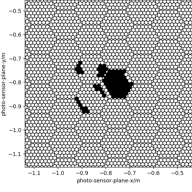}
    \includegraphics[width=.49\textwidth]{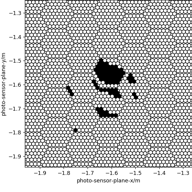}
    \includegraphics[width=.49\textwidth]{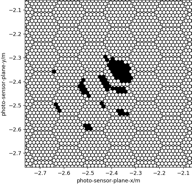}
    \includegraphics[width=.49\textwidth]{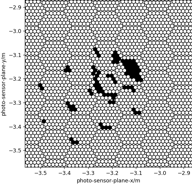}
    \includegraphics[width=.49\textwidth]{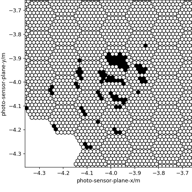}
    \caption[Aberrations on \NameAcp{}, $0.0^\circ$, $0.6^\circ$, $1.2^\circ$, $1.8^\circ$, $2.4^\circ$, and $3.0^\circ$. ]{
        Black photo-sensors are summed to synthesize pixels in the image because their corresponding lixels have incident-directions close to these pixels.
        Figure \ref{FigAberrationsNameAcpOverview} shows the summation of all the six different pixels at once.
        Upper, left shows $0.0^\circ$, %
        upper, right shows $0.6^\circ$, %
        middle, left shows $1.2^\circ$, %
        middle, right shows $1.8^\circ$, %
        lower, left shows $2.4^\circ$, and %
        lower, right shows $3.0^\circ$.
    }
    \label{FigAberrationsNameAcpCloseUp}
\end{figure}

The imaging-reflector of \NameAcp{} is not free of aberrations.
We can visualize this aberrations using the light-field-geometry of \NameAcp{}.
When we describe the projection of the light-field onto images as a matrix-multiplication, as we do in Section \ref{SecInterpretingDirectionalSampling}, we can visualize the columns of the imaging-matrix $U_\text{imaging}[n,\,k]$ just as we do in Figure \ref{FigPixelLinearCombinationSensorPlane}.
The aberrations should show up as asymmetric patterns in the visualizations of the columns of the imaging-matrix $U_\text{imaging}[n,\,k]$.
Now in the Figures \ref{FigPixelLinearCombinationSensorPlane}, \ref{FigPixelRefocusedLinearCombinationSensorPlane}, and \ref{FigPixelRefocusedLinearCombinationSensorPlane2}, we hardly see any hint for this asymmetry because the columns visualized in these figures correspond to the central pixel $n = 4,221$ in the field-of-view.
For the central pixel, the aberrations induced by the imaging-reflector of \NameAcp{} are only minor.
However, aberrations can be seen for the outer regions in the field-of-view.
Here we visualize six of these matrix-columns for six pixels ranging from the center of the field-of-view at $0.0^\circ$ to the outer perimeter of the filed-of-view at $3.0^\circ$.
We compute the imaging-matrix $U_\text{imaging}[n,\,k]$ according to Equation \ref{EqImagingPixelAssignmentGeneral}.
Table \ref{TabAberrationsNameAcpOverview} shows the incident-directions of the six different pixels.
Figure \ref{FigAberrationsNameAcpOverview} shows the photo-sensors participating to the six pixels on the photo-sensor-plane of the light-field-sensor in \NameAcp{}.
The Figures \ref{FigAberrationsNameAcpOverview}, and \ref{FigAberrationsNameAcpCloseUp} do not show the point-spread-function of the imaging-reflector, but they visualize which photo-sensors are summed to synthesize a pixel because their corresponding lixels have near-by incident-directions as the pixel.
Based on the light-field-geometry for \NameAcp{}, we can overcome a large part of the aberrations induced by the large imaging-reflector.
In Chapter \ref{ChCompensatingMisalignmnets}, reconstructed images of a phantom-source observed with the \NameAcp{} plenoscope, and a telescope are shown side-by-side.
The reduction of aberrations can be seen in the reconstructed images recorded by the plenoscope, in contrast to the images reconstructed from the image-sensor of the telescope.
\begin{table}
\begin{center}
    \begin{tabular}{rcccrr}
        $n$ & $c_x$/deg & $c_y$/deg & $\sqrt{{c_x}^2 + {c_y}^2}$/deg & $x$/m & $y$/m\\
        \toprule
4,221 & 0.00 & 0.00 & 0.0 & -0.00 & 0.00\\
3,541 & 0.42 & 0.42 & 0.6 & -0.79 & -0.81\\
2,779 & 0.85 & 0.85 & 1.2 & -1.61 & -1.60\\
2,135 & 1.27 & 1.27 & 1.8 & -2.41 & -2.41\\
1,448 & 1.70 & 1.70 & 2.4 & -3.22 & -3.22\\
902 & 2.12 & 2.12 & 3.0 & -4.01 & -4.01\\
    \end{tabular}
    \end{center}
    \caption[Aberrations on \NameAcp{}, overview]{
        The six example pixels with indexes $n$ represent various incident-directions $c_x$, and $c_y$.
        Here $x$ and $y$ are the average positions of the corresponding photo-sensors in the photo-sensor-plane, see Figure \ref{FigAberrationsNameAcpOverview}.
    }
    \label{TabAberrationsNameAcpOverview}
\end{table}
%
%------------------------------------------------------------------------------
%
%
%
%
%
%
%
%------------------------------------------------------------------------------
\chapter{Compensating misalignments between the light-field-sensor and the imaging-reflector}
\label{ChCompensatingMisalignmnets}
On telescopes, the alignment of the image-sensor with respect to the imaging-optics has a strong effect on image-quality.
Building telescope-mounts and support-structures which keep the optics aligned becomes an expensive challenge on larger telescopes.
However, if a light-field-sensor is used instead of an image-sensor, and if the actual alignment of the light-field-sensor with respect to the imaging-optics is known, the quality of the images reconstructed from the light-field is little to none effected by misalignments.\\
Compensating the misalignments of the light-field-sensor is a key-feature of the \NameAcp{} Cherenkov-plenoscope.
It allows us to mechanically separate the light-field-sensor from the large imaging-reflector in order to defer the physical limit of the square-cube-law, and to build larger apertures, see Chapter \ref{ChCableRobotMount}.
We show that the compensation of misalignments is already included in the light-field-geometry $G$ which we introduce in Chapter \ref{ChLightFieldGeometry}, and is closely related to the overcoming of aberrations discussed in Chapter \ref{ChOvercomingAberrations}.\\
First, we organize the misalignments into four components and discuss their effects on imaging.
Second, we show how the light-field-geometry $G$ can be used to compensate such misalignments.
And third, we simulate the imaging of the \NameAcp{} Cherenkov-plenoscope and a same sized Cherenkov-telescope when both observe a three-dimensional phantom-source located up in the atmosphere while the sensor-planes of both the telescope and the plenoscope are misaligned in various ways.
\section{Organizing components of misalignments}
\label{SecOrganizingComponentsOfMisalignments}
\newcommand{\RotPara}[1][0.08]{
\includegraphics[angle=90, width=#1\textwidth]{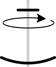}
}
\newcommand{\RotPerp}[1][0.08]{
\includegraphics[angle=90, width=#1\textwidth]{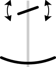}
}
\newcommand{\TransPara}[1][0.08]{
\includegraphics[angle=90, width=#1\textwidth]{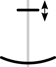}
}
\newcommand{\TransPerp}[1][0.08]{
\includegraphics[angle=90, width=#1\textwidth]{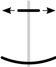}
}
We organize the misalignments between the sensor-plane and the imaging-optics into four components listed below.
We use the frame of the imaging-optics as a reference, and translate and rotate only the image-sensor.
\subsection*{$\TransPara{}$ Translations parallel to the optical-axis}
\begin{figure}{}
    \centering
    \includegraphics[width=1\textwidth]{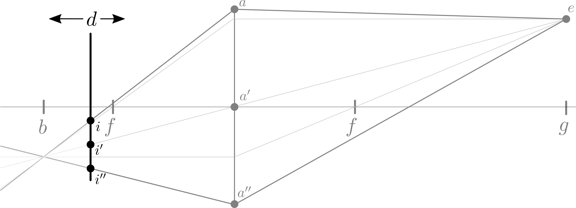}
    \caption[Misalignment, translations parallel to optical-axis]{
        Translations parallel to the optical-axis.
        The photons $\overline{ea}$, $\overline{ea'}$, and $\overline{ea''}$ only converge in a single point on the image-sensor when the sensor-plane-distance $d$ is equal to the image-distance $b$.
        But when the image-sensor is translated away from its target-position, the absorption-positions $i$, $i'$, and $i''$ do not converge and the image of $e$ is blurred.
    }
    \label{FigMisalignmentTrnslationParallel}
\end{figure}
When we translate the sensor-plane parallel to the optical-axis of the imaging-optics, see Figure \ref{FigMisalignmentTrnslationParallel}, we change the sensor-plane-distance $d$ introduced with the thin-lens in Figure \ref{FigThinLens}.
A parallel translation corresponds to focusing the image to different object-distances.
On Cherenkov-telescopes we have to choose one optimal object-distance to focus on \cite{hofmann2001focus,trichard2015enhanced} before the image of the air-shower is recorded.
On image-senors, we can not correct for parallel translations, not even when we know the actual parallel translation.
Image-sensors do not measure the trajectories $\overline{ia}$, $\overline{i'a'}$, and $\overline{i''a''}$, but only the absorption-positions $i$, $i'$, and $i''$.
However, light-field-sensors on the other hand do measure the trajectories $\overline{ia}$, $\overline{i'a'}$, and $\overline{i''a''}$ and thus can reconstruct the intersection-positions of such trajectories on a virtual sensor-plane to restore the image.
We discuss the restoring of images focused to any object-distance in Section \ref{SecPostRefocusedImaging}.\\
We estimate the acceptable tolerances for parallel translations before the effect shows up in the reconstructed images for a Cherenkov-telescope of the same size as the \NameAcp{} Cherenkov-plenoscope.
When we target to focus on an object-distance $g_\text{target} = 10.0\,$km, our target-sensor-plane-distance is $d_\text{target} = 107.64\,$m according to the Thin-lens-equation \ref{EqThinLens}.
According to the Depth-of-field-equation \ref{EqDepthOfField} the depth-of-field spans from $g_- = 9.18\,$km to $g_+ = 10.8\,$km.
The corresponding sensor-plane-distances to focus exactly on the start and end of the depth-of-field are $d_- = 107.75\,$m, and $d_+ = 107.56\,$m respectively.
%
% 1/f = 1/g + 1/b
%
% 1/b = 1/f - 1/g
%
We identify the difference between our targeted sensor-plane-distance $d_\text{target}$ and the sensor-plane-distances which would move our focus beyond our depth-of-field $d_\pm$ as the acceptable tolerance for parallel translations
\begin{eqnarray}
\Delta_\parallel^\text{trans} &=& d_\text{target} - d_\pm\\
d_\pm &=& \frac{1}{\frac{1}{f} - \frac{1}{g_\pm}}.
\label{EqToleranceTranslationParallel}
\end{eqnarray}
The sensor-plane must not be translated further than $+104\,$mm and $-88\,$mm parallel to the optical-axis before the image differs recognizably.
In units of the focal-length this tolerance is $8.5 \times 10^{-4}\,f$.
\subsection*{$\TransPerp{}$ Translations perpendicular to the optical-axis}
\begin{figure}{}
    \centering
    \includegraphics[width=1\textwidth]{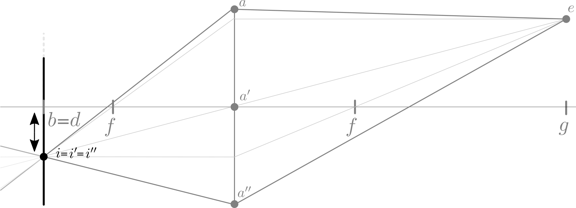}
    \caption[Misalignment, translations perpendicular to optical-axis]{
        Translations perpendicular to the optical-axis.
        The photons $\overline{ea}$, $\overline{ea'}$, and $\overline{ea''}$ will still converge in the absorption-points $i = i' = i''$.
        Although the photons enter different photo-sensors when the image-sensor is translated perpendicular to the optical-axis, they still can be assigned to the pixel corresponding to their incident-directions.
    }
    \label{FigMisalignmentTrnslationPerpendicular}
\end{figure}
When we translate the sensor-plane perpendicular to the optical-axis of the imaging-optics, see Figure \ref{FigMisalignmentTrnslationPerpendicular}, the image will simply show a different region of incident-directions.
As long as the actual perpendicular translation is known, and the field-of-view is large enough so that our region-of-interest does not drop out of it, then translations perpendicular to the optical-axis do not effect the image-quality.
On real imaging-optics unfortunately, distortions and aberrations dilute the angular resolution the larger the incident-direction of the photons becomes.
So on real telescopes, we expect that translations perpendicular to the optical-axis have an effect on the image-quality which becomes larger, the larger the perpendicular translation becomes.
However, with light-field-sensors on the other hand we can even compensate the increasing aberrations of the imaging-optics, see Chapter \ref{ChOvercomingAberrations}, so that the quality of the images reconstructed from light-field-sensors is not effected by perpendicular translations.\\
If we would not correct for perpendicular translations, a Cherenkov-telescope of the same size of \NameAcp{} could tolerate perpendicular translations up to about the extend $p$ of the projection of a pixel onto the sensor-plane
\begin{eqnarray}
\Delta_\perp^\text{trans} &=& \pm \frac{p}{2},
\label{EqToleranceTranslationPerpendicular}
\end{eqnarray}
which gives $\Delta_\perp^\text{trans} \approx \pm 63\,$mm, and is related to $\Delta_\parallel^\text{trans}$ by a factor of $f/D$.
In units of the focal-length, this tolerance is $5.8 \times 10^{-4}\,f$.
\subsection*{$\RotPara{}$ Rotations parallel to the optical-axis}
\begin{figure}{}
    \centering
    \includegraphics[width=1\textwidth]{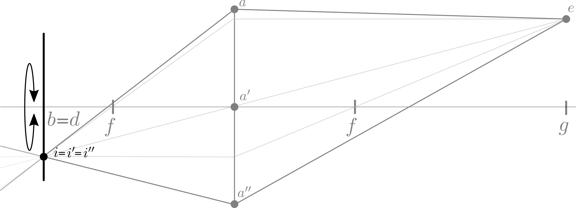}
    \caption[Misalignment, rotations parallel to the optical-axis]{
        Rotations parallel to the optical-axis.
        All photons coming from $e$ still converge in one absorption-position $i = i' = i''$.
        Just as for perpendicular translations, shown in Figure \ref{FigMisalignmentTrnslationPerpendicular}, the photons might be absorbed by a different photo-sensor now, but they still can be assigned to the pixel corresponding to their incident-directions.
    }
    \label{FigMisalignmentRotationParallel}
\end{figure}
When we rotate the image-sensor parallel to the optical-axis of the imaging-optics, see Figure \ref{FigMisalignmentRotationParallel}, we simply end up with an image rotated around its center.
We can fully compensate rotations parallel to the optical-axis in post when we know the actual parallel rotation.\\
Again we can estimate the tolerance for a parallel rotation angle
\begin{eqnarray}
\Delta_\parallel^\text{rot} &=& \frac{\Delta_\perp^\text{trans}}{f \tan(\theta_\text{FoV}/2)}
\label{EqToleranceRotationParallel}
\end{eqnarray}
when we would not correct for it in post.
We use the acceptable tolerance for perpendicular translations $\Delta_\perp^\text{trans}$ and divide it by the radius of the sensor-plane.
For the example of the Cherenkov-telescope of the same size as \NameAcp{} we get $\Delta_\parallel^\text{rot} = 0.59^\circ$.
\subsection*{$\RotPerp{}$ Rotations perpendicular to the optical-axis}
\begin{figure}{}
    \centering
    \includegraphics[width=1\textwidth]{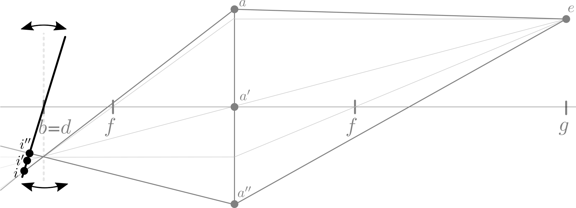}
    \caption[Misalignment, rotations perpendicular to the optical-axis]{
        Rotations perpendicular to the optical-axis.
        The absorption-positions $i$, $i'$, and $i''$ are effected in a similar way as shown in the component for parallel translations, see Figure \ref{FigMisalignmentTrnslationParallel}.
    }
    \label{FigMisalignmentRotationPerpendicular}
\end{figure}
When we rotate the sensor-plane perpendicular to the optical-axis of the imaging-optics, one region on the sensor-plane will be closer to the imaging-optics than the opposite region on the sensor-plane, see Figure \ref{FigMisalignmentRotationPerpendicular}.
The two regions effectively have different sensor-plane-distances, what corresponds to the two regions being focused to different object-distances.
So we expect similar effects on the image-quality as for translations parallel to the optical-axis, but more complicated since the image is not evenly out-of-focus, but differently for different regions.
Just as for the parallel translations, on image-sensors we can not compensate the effects of this perpendicular rotations in post, not even when we know the actual perpendicular rotation.
Again, this is because an image-sensor does not measure the trajectories $\overline{ia}$, $\overline{i'a'}$, and $\overline{i''a''}$, but only the absorption-positions $i$, $i'$, and $i''$.
For the same argument we pointed out for parallel translations, a light-field-sensor can compensate perpendicular rotations when the rotation is known.\\
We can estimate the tolerance for perpendicular rotation angles
\begin{eqnarray}
\Delta_\perp^\text{rot} &=& \frac{\Delta_\parallel^\text{trans}}{f \tan(\theta_\text{FoV}/2)}
\label{EqToleranceRotationPerpendicular}
\end{eqnarray}
based on the tolerance for parallel translations $\Delta_\parallel^\text{trans}$ and the radius of the sensor-plane.
This gives us a tolerance for perpendicular rotations of $\pm 0.83^\circ$ for a Cherenkov-telescope with the size of \NameAcp{} before the image differs recognizably.
\subsection*{Summarizing misalignment-components}
Based on the Figures \ref{FigMisalignmentTrnslationParallel}, \ref{FigMisalignmentTrnslationPerpendicular}, \ref{FigMisalignmentRotationParallel}, and \ref{FigMisalignmentRotationPerpendicular}, we summarize our findings in Table
\ref{TabMisalignmentComponents}.
%
%All misalignments which contain components of a parallel translations $\TransPara{}$ or perpendicular rotations $\RotPerp{}$, we can not compensate with image-sensors, even not when the actual misalignment is known.
%
%Misalignments which only contain components of perpendicular translations $\TransPerp{}$ and parallel rotations $\RotPara{}$, we can compensate in post when we know the actual misalignment.
%
%However, light-field-sensors can compensate any misalignment in post as long as the actual misalignment is known.\\
%
However, Table \ref{TabMisalignmentComponents} addresses perfect light-field-sensors.
As we will see next, real light-field-sensors are limited in resolution or in the acceptance-angles for incoming photons.
\begin{table}[H]
\begin{center}
    \begin{tabular}{llcc}
        misalignment- &
        acceptable tolerances
        & \multicolumn{2}{c}{can be compensated on:}\\
        component &
        without compensation &
        image-sensor &
        light-field-sensor\\
        \toprule
        $\TransPara{}$ &
        $\Delta_\parallel^\text{trans} = -88\,$mm to $+104\,$mm &
        No &
        Yes\\
        $\TransPerp{}$ &
        $\Delta_\perp^\text{trans} = \pm 63\,$mm &
        Yes, partly &
        Yes\\
        $\RotPara{}$ &
        $\Delta_\parallel^\text{rot} = \pm 0.59^\circ$ &
        Yes &
        Yes\\
        $\RotPerp{}$ &
        $\Delta_\perp^\text{rot} = \pm 0.83^\circ$ &
        No &
        Yes\\
    \end{tabular}
    \end{center}
    \caption[Misalignment-components, summary]{
        Summary of the four misalignment-components.
        The acceptable tolerances are estimated for a Cherenkov-telescope with the size of \NameAcp{}.
        The restriction 'partly' here refers to aberrations caused by real imaging-optics.
    }
    \label{TabMisalignmentComponents}
\end{table}
\section{Using the light-field-geometry}
\begin{figure}{}
    \centering
    \includegraphics[width=1\textwidth]{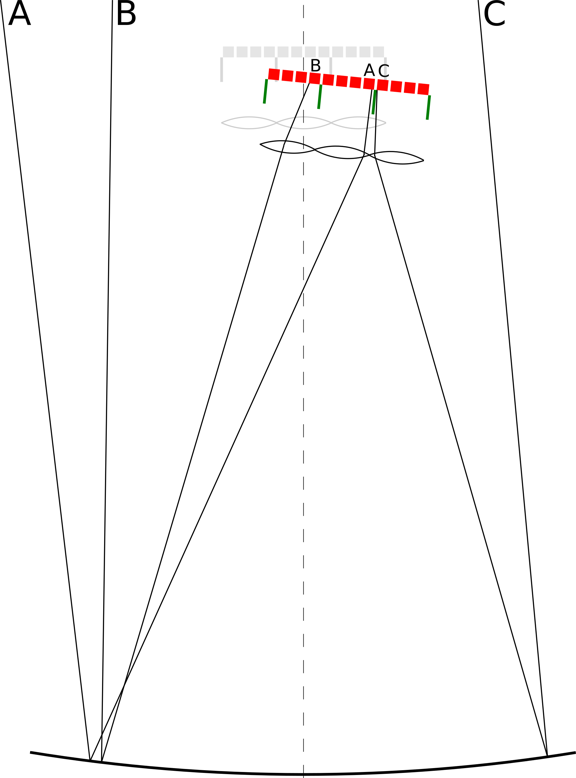}
    \caption[Misalignments between imaging-reflector and light-field-sensor]{
        The same three photons A, B, and C from Figure \ref{FigOpticsOverview}, but this time the light-field-sensor is misaligned in $\hat{T}_\text{actual}$ with respect to the imaging-reflector.
        In bright gray we indicate the target-alignment $\hat{T}_\text{target}$ of the light-field-sensor.
        Figure \ref{FigPlenoscopeMisalignmentsCloseUp} shows a close-up of the light-field-sensor.
    }
    \label{FigPlenoscopeMisalignmentsOverview}
\end{figure}

\begin{figure}{}
    \begin{minipage}{0.56\textwidth}
            \centering
            \includegraphics[width=1\textwidth]{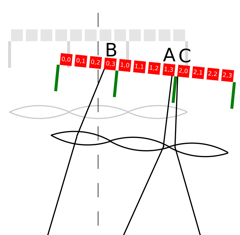}
            \caption[Misalignments between imaging-reflector and light-field-sensor, close-up]{
                Close-up of Figure \ref{FigPlenoscopeMisalignmentsOverview}.
                Figure \ref{FigPlenoscopeMisalignmentsLightFieldGeometry} shows the resulting light-field-geometry $G(\hat{T}_\text{actual})$.
            }
            \label{FigPlenoscopeMisalignmentsCloseUp}
    \end{minipage}
    \hfill
    \begin{minipage}{0.40\textwidth}
            \centering
            \includegraphics[width=1\textwidth]{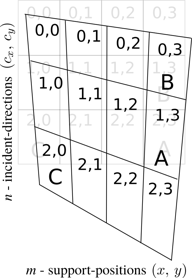}
            \caption[Misalignments between imaging-reflector and light-field-sensor, light-field-geometry]{
                A visual representation of the light-field-geometry $G(\hat{T}_\text{actual})$ corresponding to the misalignment $\hat{T}_\text{actual}$ shown in Figures \ref{FigPlenoscopeMisalignmentsOverview}, and \ref{FigPlenoscopeMisalignmentsCloseUp}.
            }
            \label{FigPlenoscopeMisalignmentsLightFieldGeometry}
    \end{minipage}
\end{figure}

Figure \ref{FigPlenoscopeMisalignmentsOverview} shows a plenoscope observing the three photons A, B, and C while its light-field-sensor is misaligned with respect to the imaging-reflector.
Despite the misalignment, the three photons are absorbed by the photo-sensors inside the light-field-sensor, see Figure \ref{FigPlenoscopeMisalignmentsCloseUp}.
When we know the actual alignment $\hat{T}_\text{actual}$ between the light-field-sensor and the imaging-reflector, we can estimate the specific light-field-geometry $G(\hat{T}_\text{actual})$ for the whole misaligned plenoscope, see Figure \ref{FigPlenoscopeMisalignmentsLightFieldGeometry}.
Although the photons A, and C have the same incident-directions, they enter different small cameras in the light-field-sensor.
But using the actual light-field-geometry $G(\hat{T}_\text{actual})$, we can still reconstruct the actual incident-directions of A, and C.
In the Figures \ref{FigPlenoscopeMisalignmentsCloseUp}, and \ref{FigPlenoscopeMisalignmentsLightFieldGeometry} we also show the target-alignment $\hat{T}_\text{target}$ of the light-field-sensor in light gray.
We find that in Figure \ref{FigPlenoscopeMisalignmentsLightFieldGeometry}, the black Photons A, B, and C which are detected by the misaligned light-field-sensor are close to their light gray counterparts which are detected by the light-field-sensor in its target-alignment.\\
From our thoughts summarized in Table \ref{TabMisalignmentComponents}, we already concluded that the superior knowledge of the light-field-sensor about the trajectories of the photons will allow it to compensate misalignments.
In Figure \ref{FigPlenoscopeMisalignmentsLightFieldGeometry} we see how this superior knowledge shows up as a distortion of the light-field-geometry $G(\hat{T}_\text{actual})$.
But from the plenoscope shown in the Figures \ref{FigPlenoscopeMisalignmentsOverview}, and \ref{FigPlenoscopeMisalignmentsCloseUp} we also see that real light-field-sensors have limited resolutions.
They have a limited number of small cameras, and a limited number of photo-sensors inside the small cameras.
In the next Section \ref{SecCompnensatingMisalignmentsPhantomSource} we take a look at the actual compensation-power for misalignments on the \NameAcp{} Cherenkov-plenoscope which has such limited resolutions as it is meant to be fabricated in a cost-effective way.
\section{Imaging a phantom-source with \NameAcp{}}
\label{SecCompnensatingMisalignmentsPhantomSource}
\begin{figure}{}
    \begin{minipage}{0.6\textwidth}
            \centering
            % trim={<left> <lower> <right> <upper>}
            \includegraphics[trim={4cm 5cm 2cm 9cm}, width=1\textwidth]{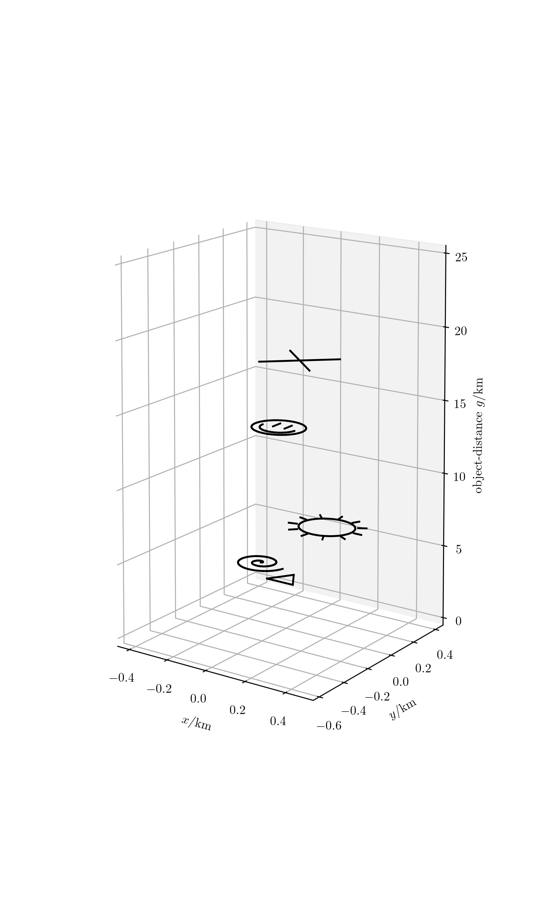}
            \caption[Phantom-source, three-dimensional]{
                The phantom-source in three dimensions consists of a triangle-wire, a spiral-wire, and a sun-symbol-wire.
                The wires emit photons.
                At the bottom is the principal-aperture-plane of the imaging-reflector.
                The symbols scale with with the object-distance they are located in.
                This way, they appear with similar sizes in the images.
                Note, the axis of the object-distance is compressed, and not to scale.
            }
            \label{FigPlenoscopeMisalignmentsPhantom3d}
    \end{minipage}
    \hfill
    \begin{minipage}{0.37\textwidth}
            \centering
            \includegraphics[width=0.9\textwidth]{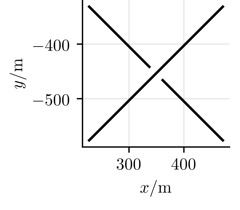}
            \includegraphics[width=0.9\textwidth]{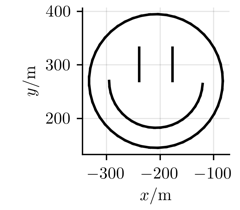}
            \includegraphics[width=0.9\textwidth]{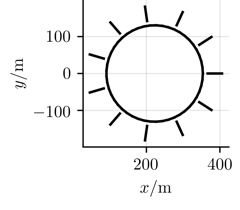}
            \includegraphics[width=0.9\textwidth]{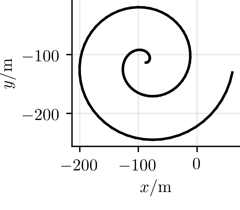}
            \includegraphics[width=0.9\textwidth]{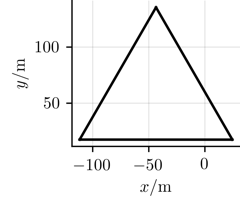}
            \caption[Parts of phantom-source]{
                The symbols are located in different object-distances above the aperture:\\
                $g_\text{triangle} = 2.5\,$km\\
                $g_\text{spiral} = 4.2\,$km\\
                $g_\text{sun-symbol} = 7.1\,$km\\
                $g_\text{smiley} = 11.9\,$km\\
                $g_\text{cross} = 20.0\,$km\\
            }
            \label{FigPlenoscopeMisalignmentsPhantomElements}
    \end{minipage}
\end{figure}

To get a qualitative feeling for the effects of misalignments on imaging, and to get a qualitative feeling for the compensation-power for misalignments of \NameAcp{}'s light-field-sensor, we reconstruct images of a phantom-source.
The phantom-source has a well defined shape in the three-dimensional space, and is described in the Figures \ref{FigPlenoscopeMisalignmentsPhantom3d}, and \ref{FigPlenoscopeMisalignmentsPhantomElements}.
The phantom-source extends from an object-distance of $2.5\,$km up to $20.0\,$km, and all its photons reach the principal-aperture-plane in a $1\,$ns time-window, similar to Cherenkov-photons produced in air-showers.
There are no night-sky-background-photons, but only the photons from the phantom-source.
In our simulation we set up the \NameAcp{} Cherenkov-plenoscope and a Cherenkov-telescope of the same size as \NameAcp{}.
The plenoscope records a light-field-sequence over a period of $50\,$ns, and the telescope records an image-sequence also over a period of $50\,$ns.
Both instruments take $2 \times 10^9$ samples per second.
However, for the static flash of photons from the phantom-source we just look at the integrated light-field and image respectively.
Both instruments have their sensor-plane-distances set to $d_\text{target} = 107.65\,$m to focus to an object-distance of $g_\text{target} = 10\,$km.
The images reconstructed from the telescope are corrected as much as possible for the given situation, e.g. they are corrected for perpendicular translations, parallel rotations, and distortions caused by the imaging-reflector.
The images reconstructed from the light-field from the plenoscope are in addition also corrected for the remaining components of misalignments such as parallel translations and perpendicular rotations.
One might argue that with a target object-distance of $g_\text{target} = 10\,$km, the Cherenkov-telescope can never produce a fully sharp image of any layer in the phantom-source, since there is no layer exactly at this object-distance, and thus a side-by-side comparison of the telescope and the plenoscope is not fair.
Yes, it is not fair, but not because we designed the phantom-source this way, but because the plenoscope is the superior instrument.
The purpose of these instruments is to record air-showers, and just as air-showers, our phantom-source spans a wide range of object-distances which simply exceeds the depth-of-field of a Cherenkov-telescope of the size of \NameAcp{}, see Chapter \ref{ChDepthOfField}.
Focuses on different object-distances of the air-shower offer different advantages for the reconstruction of different properties of the air-shower \cite{hofmann2001focus}, and thus any focus chosen on a Cherenkov-telescope of the size of \NameAcp{} ends up to be an inferior compromise.
Actually some of the misalignments will by chance even improve the imaging on the telescope.
\subsection*{What we expect to see in the images}
\begin{enumerate}
    \item We expect the images reconstructed from the light-field recorded by the plenoscope to always look the same regardless of the actual scenario of misalignment of the light-field-sensor.
    \item We expect the plenoscope to be able to project its light-field onto images focused to different object-distances to e.g. find the sharpest response for the spiral, the sun-symbol, and the smiley in the phantom-source.
    \item We expect the images from the telescope to be only one single image for each scenario of misalignment, and not multiple images taken with different focuses.
    This might be obvious, but we want to remind ourselves that like the plenoscope, the telescope has only one single chance to record the air-shower.
    The telescope has to choose one focus which it can not optimize in advance for the individual air-shower.
\end{enumerate}
\subsection*{Scenarios of misalignments}
\begin{figure}{}
    \centering \TransPara{} \hspace{.5cm} \begin{LARGE}Translation parallel $-877\,$mm\end{LARGE}\\ \vspace{0.5cm}
    \begin{tabular}{c|c}
        \begin{Large}Telescope\end{Large} & \begin{Large}Plenoscope\end{Large}\\
        image-sensor & light-field-sensor \\
        \toprule
        & \includegraphics[width=0.49\textwidth]{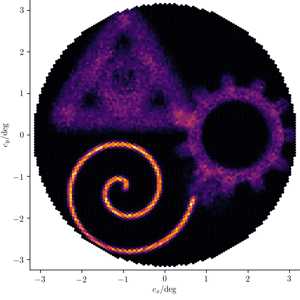}
        \\
        \includegraphics[width=0.49\textwidth]{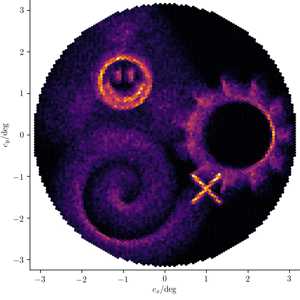}
        &
        \includegraphics[width=0.49\textwidth]{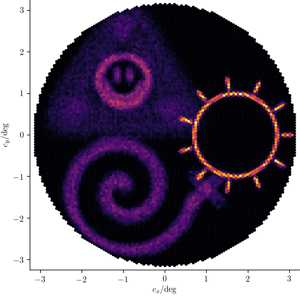}
        \\
        & \includegraphics[width=0.49\textwidth]{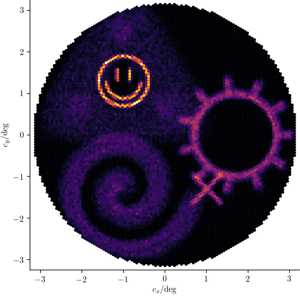}
    \end{tabular}
    \caption[Phantom source, translation parallel $-877\,$mm]{
        Sensor-plane-distance is $d = d_\text{target} -877\,$mm $ = 106.77\,$m, and corresponds to a focus on an object-distance of $42.2\,$km.
    }
    \label{FigPlenoscopeMisalignmentsTransMinus877mm}
\end{figure}

\begin{figure}{}
    \centering \TransPara{} \hspace{.5cm} \begin{LARGE}Translation parallel $-240\,$mm\end{LARGE}\\ \vspace{0.5cm}
    \begin{tabular}{c|c}
        \begin{Large}Telescope\end{Large} & \begin{Large}Plenoscope\end{Large}\\
        image-sensor & light-field-sensor \\
        \toprule
        & \includegraphics[width=0.49\textwidth]{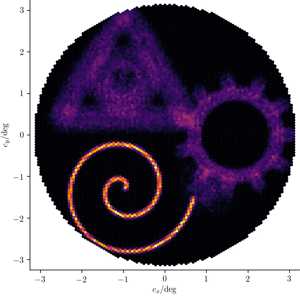}
        \\
        \includegraphics[width=0.49\textwidth]{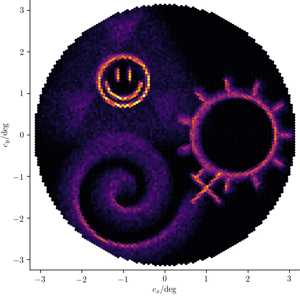}
        &
        \includegraphics[width=0.49\textwidth]{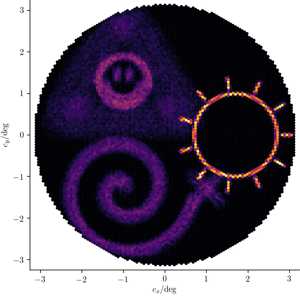}
        \\
        & \includegraphics[width=0.49\textwidth]{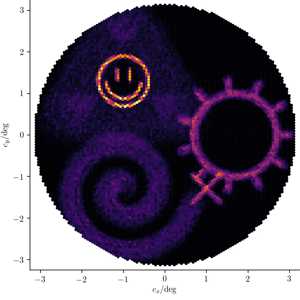}
    \end{tabular}
    \caption[Phantom source, translation parallel $-240\,$mm]{
        Sensor-plane-distance is $d = d_\text{target} -240\,$mm $ = 107.41\,$m, and corresponds to a focus on an object-distance of $12.6\,$km.
    }
    \label{FigPlenoscopeMisalignmentsTransMinus240mm}
\end{figure}

\begin{figure}{}
    \centering \TransPara{} \hspace{.5cm} \begin{LARGE}Translation parallel $+398\,$mm\end{LARGE}\\ \vspace{0.5cm}
    \begin{tabular}{c|c}
        \begin{Large}Telescope\end{Large} & \begin{Large}Plenoscope\end{Large}\\
        image-sensor & light-field-sensor \\
        \toprule
        & \includegraphics[width=0.49\textwidth]{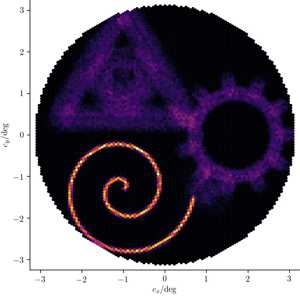}
        \\
        \includegraphics[width=0.49\textwidth]{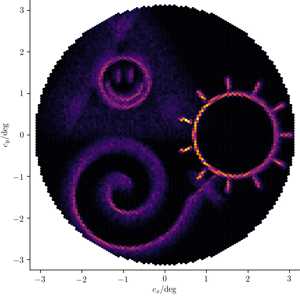}
        &
        \includegraphics[width=0.49\textwidth]{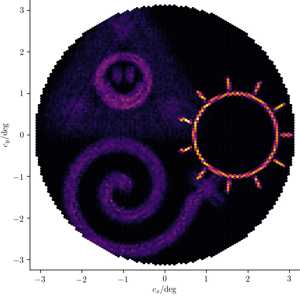}
        \\
        & \includegraphics[width=0.49\textwidth]{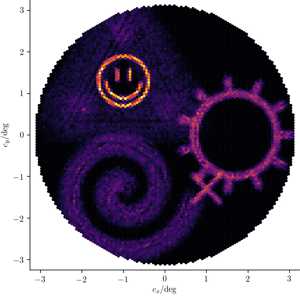}
    \end{tabular}
    \caption[Phantom source, translation parallel $+398\,$mm]{
        Sensor-plane-distance is $d = d_\text{target} + 398\,$mm $ = 108.04\,$m, and corresponds to a focus on an object-distance of $7.5\,$km.
    }
    \label{FigPlenoscopeMisalignmentsTransPlus398mm}
\end{figure}

\begin{figure}{}
    \centering \TransPara{} \hspace{.5cm} \begin{LARGE}Translation parallel $+1,035\,$mm\end{LARGE}\\ \vspace{0.5cm}
    \begin{tabular}{c|c}
        \begin{Large}Telescope\end{Large} & \begin{Large}Plenoscope\end{Large}\\
        image-sensor & light-field-sensor \\
        \toprule
        & \includegraphics[width=0.49\textwidth]{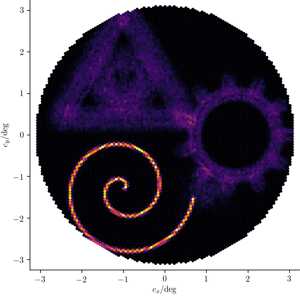}
        \\
        \includegraphics[width=0.49\textwidth]{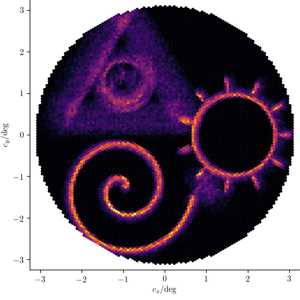}
        &
        \includegraphics[width=0.49\textwidth]{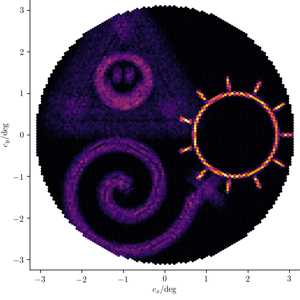}
        \\
        & \includegraphics[width=0.49\textwidth]{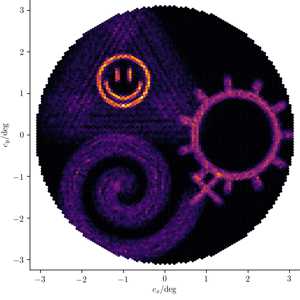}
    \end{tabular}
    \caption[Phantom source, translation parallel $+1,035\,$mm]{
        Sensor-plane-distance is $d = d_\text{target} + 1,035\,$mm $ = 108.68\,$m, and corresponds to a focus on an object-distance of $5.3\,$km.
    }
    \label{FigPlenoscopeMisalignmentsTransPlus1035mm}
\end{figure}

We set up four different scenarios of misalignments, see Table \ref{TabMisalignmentWithPhantomSource}.
First, we look at images of the phantom-source when the light-field-sensor is in its target-alignment for an object-distance of $10\,$km.
Second, we look at images taken with three different perpendicular rotations of up to $8.31^\circ$.
Third, we look at images taken with four different parallel translations from $-877\,$mm to $+1,035\,$mm.
And fourth, we look at images taken when all components of misalignment come together.
\begin{table}[]
\begin{center}
    \begin{tabular}{rrrrc}
        $\TransPara{}/$mm & $\TransPerp{}/$mm & $\RotPara{}/$deg & $\RotPerp{}/$deg & Figures\\
        \toprule
        0 & 0 & 0 & 0 & \ref{FigPlenoscopeMisalignmentsTarget}, \ref{FigTargetAlignmentRefocus1}, \ref{FigTargetAlignmentRefocus2}\\
        \hline
        0 & 0 & 0 & 2.70 & \ref{FigPlenoscopeMisalignmentsRot1}\\
        0 & 0 & 0 & 5.54 & \ref{FigPlenoscopeMisalignmentsRot2}\\
        0 & 0 & 0 & 8.31 & \ref{FigPlenoscopeMisalignmentsRot3}\\
        \hline
        -877 & 0 & 0 & 0 & \ref{FigPlenoscopeMisalignmentsTransMinus877mm}\\
        -240 & 0 & 0 & 0 & \ref{FigPlenoscopeMisalignmentsTransMinus240mm}\\
        +398 & 0 & 0 & 0 & \ref{FigPlenoscopeMisalignmentsTransPlus398mm}\\
        +1,035 & 0 & 0 & 0 & \ref{FigPlenoscopeMisalignmentsTransPlus1035mm}\\
        \hline
        +1,200 & 800 & 17.00 & 8.00 & \ref{FigPlenoscopeMisalignmentsComposition}\\
    \end{tabular}
    \end{center}
    \caption[Simulated scenarios of misalignments, overview]{
        Overview of the simulated scenarios of misalignments.
        The magnitude of the misalignments is up to one order-of-magnitude above the acceptable tolerances listed in Table \ref{TabMisalignmentComponents}.
        We concentrate on the components of misalignment which can not be compensated by image-sensors.
    }
    \label{TabMisalignmentWithPhantomSource}
\end{table}
\subsection*{Description for the Figures \ref{FigPlenoscopeMisalignmentsTarget} to \ref{FigPlenoscopeMisalignmentsComposition}}
The Figures \ref{FigPlenoscopeMisalignmentsTarget} to \ref{FigPlenoscopeMisalignmentsComposition} show images of the phantom-source.
The top of the figure shows the type and magnitude of the misalignment, compare with Table \ref{TabMisalignmentWithPhantomSource}.
The figures are split into two columns.
The left column named 'Telescope', shows the single image recorded by the Cherenkov-telescope.
The right column named 'Plenoscope' shows three images projected from the light-field recorded by the \NameAcp{} Cherenkov-plenoscope.
The three reconstructed images are focused such that the spiral, the sun-symbol and the smiley are sharp in the image.
The choice of three images here is arbitrary and could always be extended to a more detailed series of refocused images like the series shown in the Figures \ref{FigTargetAlignmentRefocus1}, and \ref{FigTargetAlignmentRefocus2}.
Always keep in mind that the images $\mathcal{I}[c_x, c_y]$ we look at are only incomplete projections of the light-field $\mathcal{L}[c_x, c_y, x, y]$, and that the only reason for doing so is to allow our limited visual perception to get an impression.
\subsection*{Discussing the reconstructed images}
\begin{figure}{}
    \centering \begin{LARGE}Target-alignment\end{LARGE}\\ \vspace{0.5cm}
    \begin{tabular}{c|c}
        \begin{Large}Telescope\end{Large} & \begin{Large}Plenoscope\end{Large}\\
        image-sensor & light-field-sensor \\
        \toprule
        & \includegraphics[width=0.49\textwidth]{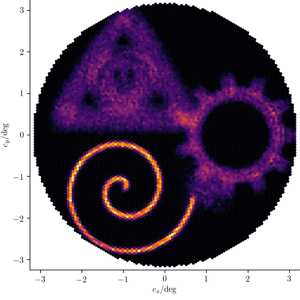}
        \\
        \includegraphics[width=0.49\textwidth]{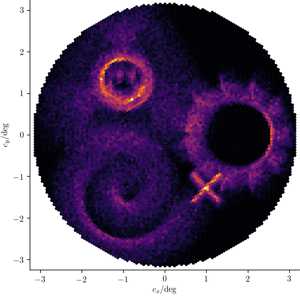}
        &
        \includegraphics[width=0.49\textwidth]{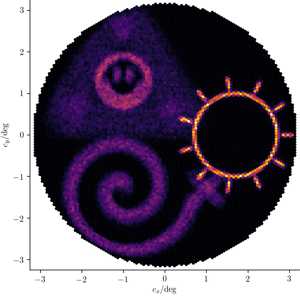}
        \\
        & \includegraphics[width=0.49\textwidth]{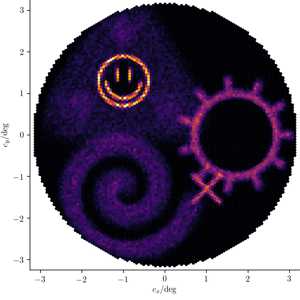}
    \end{tabular}
    \caption[Phantom source, target-alignment]{
        See a detailed series of refocused images from the plenoscope in the Figures \ref{FigTargetAlignmentRefocus1}, and \ref{FigTargetAlignmentRefocus2}.
    }
    \label{FigPlenoscopeMisalignmentsTarget}
\end{figure}

\begin{figure}{}
    \centering \RotPerp{} \hspace{.5cm} \begin{LARGE}Rotation perpendicular $y$-axis $2.77^\circ$\end{LARGE}\\ \vspace{0.5cm}
    \begin{tabular}{c|c}
        \begin{Large}Telescope\end{Large} & \begin{Large}Plenoscope\end{Large}\\
        image-sensor & light-field-sensor \\
        \toprule
        & \includegraphics[width=0.49\textwidth]{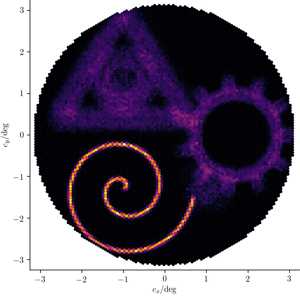}
        \\
        \includegraphics[width=0.49\textwidth]{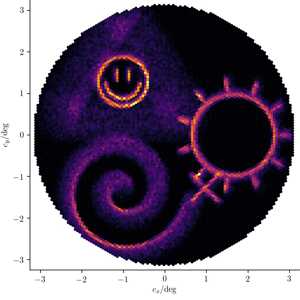}
        &
        \includegraphics[width=0.49\textwidth]{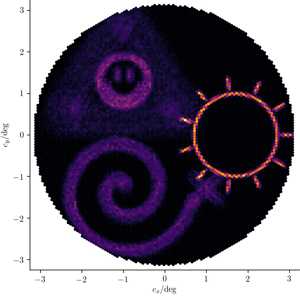}
        \\
        & \includegraphics[width=0.49\textwidth]{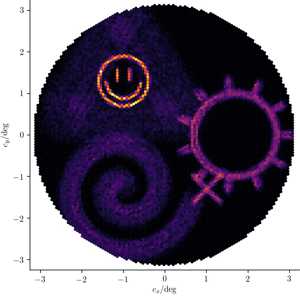}
    \end{tabular}
    \caption[Phantom source, rotation $y$-axis $2.77^\circ$]{
        By chance, the rotation helps the imaging of the smiley and the sun-symbol on the telescope.
    }
    \label{FigPlenoscopeMisalignmentsRot1}
\end{figure}

\begin{figure}{}
    \centering \RotPerp{} \hspace{.5cm} \begin{LARGE}Rotation perpendicular $y$-axis $5.54^\circ$\end{LARGE}\\ \vspace{0.5cm}
    \begin{tabular}{c|c}
        \begin{Large}Telescope\end{Large} & \begin{Large}Plenoscope\end{Large}\\
        image-sensor & light-field-sensor \\
        \toprule
        & \includegraphics[width=0.49\textwidth]{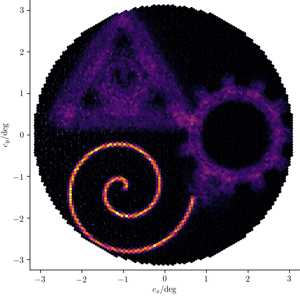}
        \\
        \includegraphics[width=0.49\textwidth]{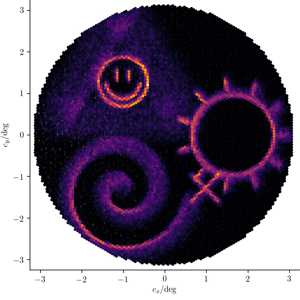}
        &
        \includegraphics[width=0.49\textwidth]{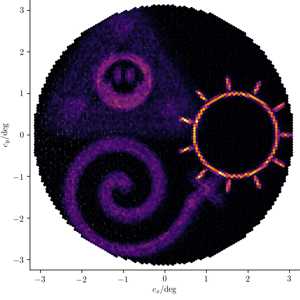}
        \\
        & \includegraphics[width=0.49\textwidth]{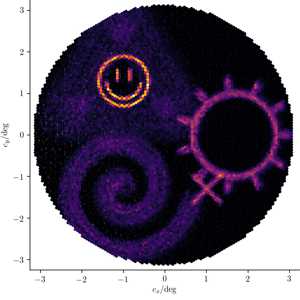}
    \end{tabular}
    \caption[Phantom source, rotation $y$-axis $5.54^\circ$]{
        Again, the rotation helps with the imaging of the smiley and the sun-symbol on the telescope, see Figure \ref{FigPlenoscopeMisalignmentsRot1}.
    }
    \label{FigPlenoscopeMisalignmentsRot2}
\end{figure}

\begin{figure}{}
    \centering \RotPerp{} \hspace{.5cm} \begin{LARGE}Rotation perpendicular $y$-axis $8.31^\circ$\end{LARGE}\\ \vspace{0.5cm}
    \begin{tabular}{c|c}
        \begin{Large}Telescope\end{Large} & \begin{Large}Plenoscope\end{Large}\\
        image-sensor & light-field-sensor \\
        \toprule
        & \includegraphics[width=0.49\textwidth]{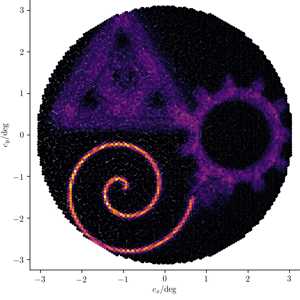}
        \\
        \includegraphics[width=0.49\textwidth]{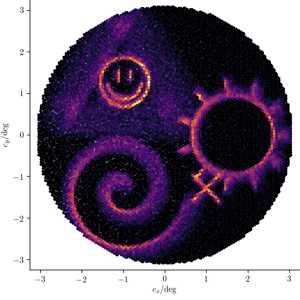}
        &
        \includegraphics[width=0.49\textwidth]{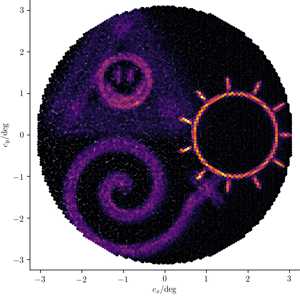}
        \\
        & \includegraphics[width=0.49\textwidth]{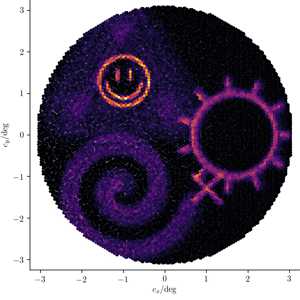}
    \end{tabular}
    \caption[Phantom source, rotation $y$-axis $8.31^\circ$]{
        Note the tiny white gaps between the pixels on the left half of the images.
        The regular grid of pixels is distorted due to the strong rotation.
    }
    \label{FigPlenoscopeMisalignmentsRot3}
\end{figure}

\begin{figure}{}
    \centering \TransPara[0.07]{} \begin{Huge}+\end{Huge} \TransPerp[0.07]{} \begin{Huge}+\end{Huge} \RotPara[0.07]{} \begin{Huge}+\end{Huge} \RotPerp[0.07]{} \hspace{.5cm} \begin{LARGE}Composition\end{LARGE}\\ \vspace{0.5cm}
    \begin{tabular}{c|c}
        \begin{Large}Telescope\end{Large} & \begin{Large}Plenoscope\end{Large}\\
        image-sensor & light-field-sensor \\
        \toprule
        & \includegraphics[width=0.49\textwidth]{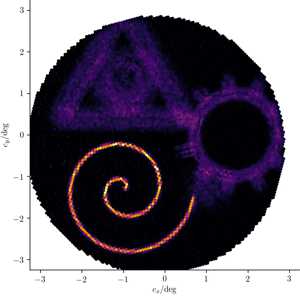}
        \\
        \includegraphics[width=0.49\textwidth]{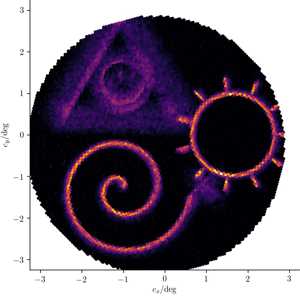}
        &
        \includegraphics[width=0.49\textwidth]{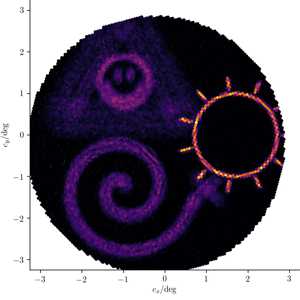}
        \\
        & \includegraphics[width=0.49\textwidth]{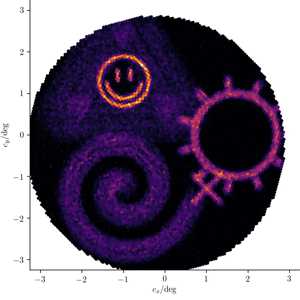}
    \end{tabular}
    \caption[Phantom source, composition of multiple misalignments]{
        The misalignment moves and rotates the image in the space of incident-directions $c_x$, and $c_y$.
        For example, the sun-symbol is not any longer fully included in the image.
    }
    \label{FigPlenoscopeMisalignmentsComposition}
\end{figure}

When we look at the images in the right plenoscope-column of the Figures \ref{FigPlenoscopeMisalignmentsTarget}, to \ref{FigPlenoscopeMisalignmentsComposition}, we find that they all look remarkably similar despite the different scenarios of strong misalignments.
Even misalignments which exceed the acceptable tolerances for image-sensors by one order-of-magnitude are apparently not an issue for the light-field-sensor in \NameAcp{}.
The spiral, the sun-symbol, and the smiley can in all scenarios be focused on sharply.
Neither the sun-symbol nor the spiral show signs of asymmetries or aberrations, in contrast to the images of the telescope in the left columns.
This is because the images reconstructed from the light-field are naturally also corrected for aberrations, see Chapter \ref{ChOvercomingAberrations}.
In fact, the images look so alike that it is hard to belief they were reconstructed from light-fields recorded in scenarios of different misalignments.
But when we take a close look on the scenario of strong perpendicular rotations of $8.31^\circ$ along the $y$-axis in Figure \ref{FigPlenoscopeMisalignmentsRot3}, we find that the hexagonal pixels in the images have small gaps in between them for negative incident-directions in $c_x$, and are compressed for positive incident-directions in $c_x$.
This irregular grid of the pixels is caused by the rotation of the sensor-plane.
Also when we look at the composition of several misalignments in Figure \ref{FigPlenoscopeMisalignmentsComposition}, we see that the reconstructed images are from a light-field recorded during this specific misalignment as we see that the sun-symbol is not fully included anymore in the field-of-view.
Also the grid of pixels is rotated in Figure \ref{FigPlenoscopeMisalignmentsComposition} to compensate the parallel rotation.
In all reconstructed images from the light-field we can see that it is possible to nail the focus on one particular part of the phantom-source.
In all the refocused images focused on the smiley we can clearly see its mouth and eyes.
In all the refocused images focused on the sun-symbol we can clearly see the flares of the sun-symbol and even see the gap between the flares and the ring.
And in all the refocused images focused on the spiral, the spiral is evenly glowing, the parts of the spiral close to the center of the image are as sharp as the parts of the spiral in the outer realm of the image.
There are no signs of aberrations.\\
Not so for the images of the telescope.
All images of the telescope show asymmetries in sharpness across the field-of-view.
Already in the target-alignment shown in Figure \ref{FigPlenoscopeMisalignmentsTarget}, the mouth and the eyes of the smiley are washed out and the spiral is almost gone.
The flares of the sun-symbol are hardly recognizable.
During the perpendicular rotation along the $y$-axis in the Figures \ref{FigPlenoscopeMisalignmentsRot1}, to \ref{FigPlenoscopeMisalignmentsRot3}, the sensor-plane happens to reach sensor-plane-distances which favor the imaging of the smiley and the sun-symbol.\\
The bottom-line here is that although the images from the telescope are not completely useless during misalignments, the images reconstructed from the light-field of the plenoscope simply nail the focus in every scenario.
However, the serious argument for the plenoscope is not that its images are sharper, but the fact that we can reconstruct the object-distances of the individual parts of the phantom-source.
Equation \ref{EqVirtualSensorPlaneDistance} states, that for any sensor-plane-distance of a virtual sensor-plane there is a corresponding object-distance.
The plenoscope can reconstruct the phantom-source in the third dimension, a dimension which is beyond the perception of the telescope.
We discuss the full potential for three-dimensional reconstructions in Chapter \ref{ChTomography} about tomography.
\begin{figure}{}
    \begin{minipage}{0.31\textwidth}
            \centering
            \includegraphics[width=1\textwidth]{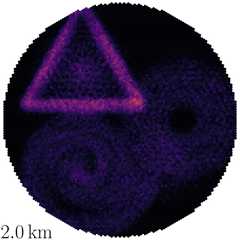}
    \end{minipage}
    \hfill
    \begin{minipage}{0.31\textwidth}
            \centering
            \includegraphics[width=1\textwidth]{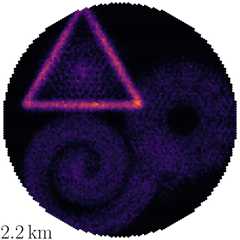}
    \end{minipage}
    \hfill
    \begin{minipage}{0.31\textwidth}
            \centering
            \includegraphics[width=1\textwidth]{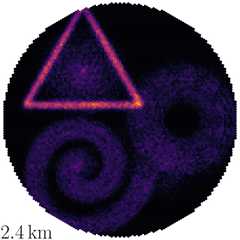}
    \end{minipage}
    \begin{minipage}{0.31\textwidth}
            \centering
            \includegraphics[width=1\textwidth]{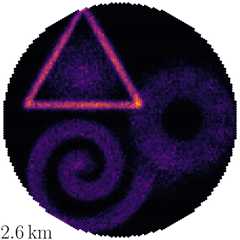}
    \end{minipage}
    \hfill
    \begin{minipage}{0.31\textwidth}
            \centering
            \includegraphics[width=1\textwidth]{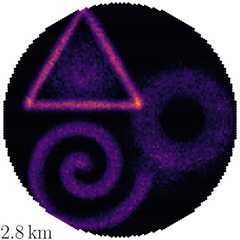}
    \end{minipage}
    \hfill
    \begin{minipage}{0.31\textwidth}
            \centering
            \includegraphics[width=1\textwidth]{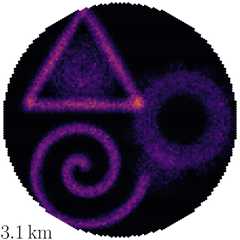}
    \end{minipage}
    \begin{minipage}{0.31\textwidth}
            \centering
            \includegraphics[width=1\textwidth]{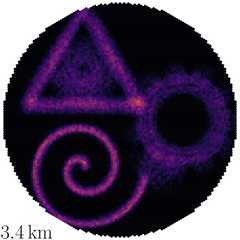}
    \end{minipage}
    \hfill
    \begin{minipage}{0.31\textwidth}
            \centering
            \includegraphics[width=1\textwidth]{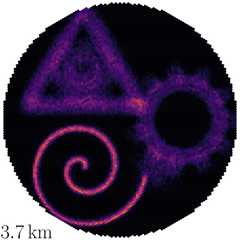}
    \end{minipage}
    \hfill
    \begin{minipage}{0.31\textwidth}
            \centering
            \includegraphics[width=1\textwidth]{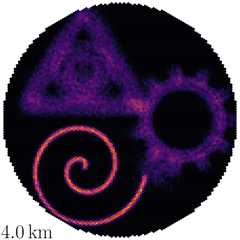}
    \end{minipage}
    \begin{minipage}{0.31\textwidth}
            \centering
            \includegraphics[width=1\textwidth]{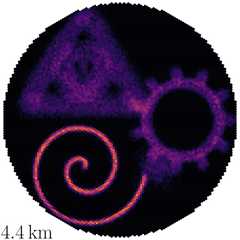}
    \end{minipage}
    \hfill
    \begin{minipage}{0.31\textwidth}
            \centering
            \includegraphics[width=1\textwidth]{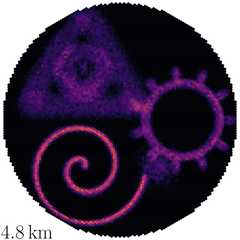}
    \end{minipage}
    \hfill
    \begin{minipage}{0.31\textwidth}
            \centering
            \includegraphics[width=1\textwidth]{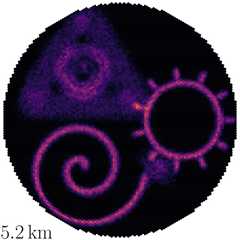}
    \end{minipage}
    \begin{minipage}{0.31\textwidth}
            \centering
            \includegraphics[width=1\textwidth]{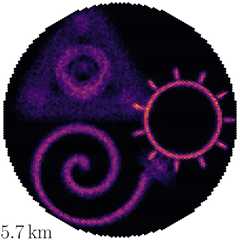}
    \end{minipage}
    \hfill
    \begin{minipage}{0.31\textwidth}
            \centering
            \includegraphics[width=1\textwidth]{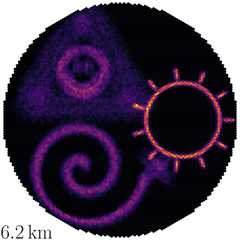}
    \end{minipage}
    \hfill
    \begin{minipage}{0.31\textwidth}
            \centering
            \includegraphics[width=1\textwidth]{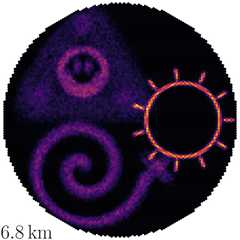}
    \end{minipage}
    \caption[Phantom source, target-alignment, refocused images part 1\,of\,2]{
        A series of images showing the phantom-source.
        The images are projections of the light-field recorded by the \NameAcp{} Cherenkov-plenoscope.
        Each image is focused to a different object-distance $g$, shown in the lower, left corner of each image.
        The series continues in Figure \ref{FigTargetAlignmentRefocus2}.
    }
    \label{FigTargetAlignmentRefocus1}
\end{figure}
\begin{figure}{}
    \begin{minipage}{0.31\textwidth}
            \centering
            \includegraphics[width=1\textwidth]{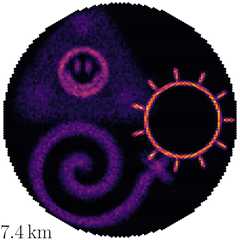}
    \end{minipage}
    \hfill
    \begin{minipage}{0.31\textwidth}
            \centering
            \includegraphics[width=1\textwidth]{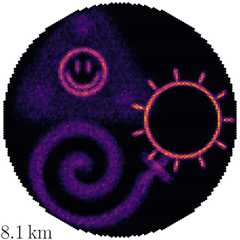}
    \end{minipage}
    \hfill
    \begin{minipage}{0.31\textwidth}
            \centering
            \includegraphics[width=1\textwidth]{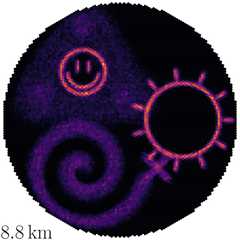}
    \end{minipage}
    \begin{minipage}{0.31\textwidth}
            \centering
            \includegraphics[width=1\textwidth]{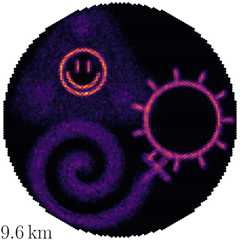}
    \end{minipage}
    \hfill
    \begin{minipage}{0.31\textwidth}
            \centering
            \includegraphics[width=1\textwidth]{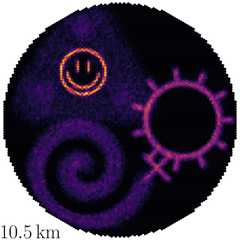}
    \end{minipage}
    \hfill
    \begin{minipage}{0.31\textwidth}
            \centering
            \includegraphics[width=1\textwidth]{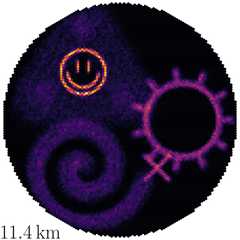}
    \end{minipage}
    \begin{minipage}{0.31\textwidth}
            \centering
            \includegraphics[width=1\textwidth]{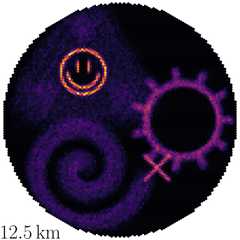}
    \end{minipage}
    \hfill
    \begin{minipage}{0.31\textwidth}
            \centering
            \includegraphics[width=1\textwidth]{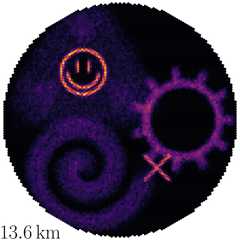}
    \end{minipage}
    \hfill
    \begin{minipage}{0.31\textwidth}
            \centering
            \includegraphics[width=1\textwidth]{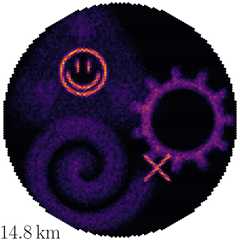}
    \end{minipage}
    \begin{minipage}{0.31\textwidth}
            \centering
            \includegraphics[width=1\textwidth]{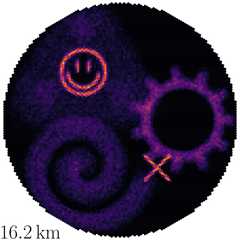}
    \end{minipage}
    \hfill
    \begin{minipage}{0.31\textwidth}
            \centering
            \includegraphics[width=1\textwidth]{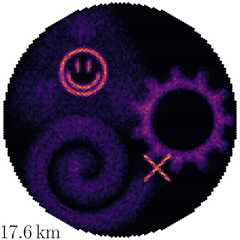}
    \end{minipage}
    \hfill
    \begin{minipage}{0.31\textwidth}
            \centering
            \includegraphics[width=1\textwidth]{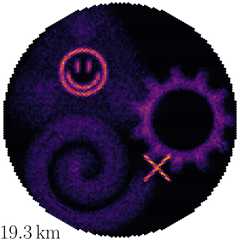}
    \end{minipage}
    \begin{minipage}{0.31\textwidth}
            \centering
            \includegraphics[width=1\textwidth]{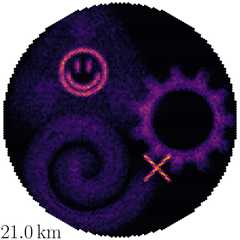}
    \end{minipage}
    \hfill
    \begin{minipage}{0.31\textwidth}
            \centering
            \includegraphics[width=1\textwidth]{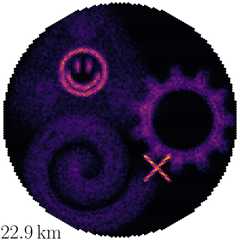}
    \end{minipage}
    \hfill
    \begin{minipage}{0.31\textwidth}
            \centering
            \includegraphics[width=1\textwidth]{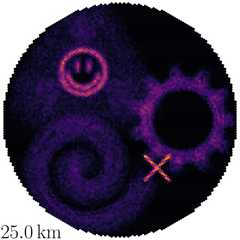}
    \end{minipage}
    \caption[Phantom source, target-alignment, refocused images part 2\,of\,2]{
        Continuation of Figure \ref{FigTargetAlignmentRefocus1}.
    }
    \label{FigTargetAlignmentRefocus2}
\end{figure}
%
%------------------------------------------------------------------------------
%
%
%
%
%
%
%
%------------------------------------------------------------------------------
\chapter{Triggering the Cherenkov-plenoscope}
\label{ChTrigger}
The continuous stream of sensor-responses in a Cherenkov-telescope or plenoscope is too large to be analyzed in detail.
Limited processing-resources force us to reject large parts of the observed sensor-responses, and only analyze a small fraction of it.
The logic which decides what parts to reject, and what parts to analyze is called the trigger.
%1\mathrm{e}{-10}
At an expected trigger-rate of $\approx59\times10^3\,$s$^{-1}$, and an exposure time of 50\,ns for each event, the trigger on \NameAcp{} rejects 99.705\,\% of the continuously observed light-field-sequence, and copies only 0.295\,\% of it to a more permanent-storage for further reconstructions of air-showers, compare Figure \ref{FigExpectedRates}.
A smart, yet simple trigger is crucial for the cost-effective detection of low energetic cosmic gamma-rays.
%
%As introduced in Section \ref{ChResults}, the Cherenkov-plenoscope does astronomy by gathering information on the cosmic particles by reconstructing their air-showers from the recorded light-field-sequences.
%
With current technology we can read-out all the photo-sensors in real-time, but we can not store the resulting light-field-sequence for further reconstructions.
We can only buffer the light-field-sequence for a short duration of $\approx 100\,$ns up to a few $1\,$us.
This is possible because all the read-out-channels can write to their own, fast but small memory-buffer in parallel.
On Cherenkov-telescopes, there exist digital, and analog implementations of this memory-buffer.
Digital implementations can use flash-analog-to-digital-converters \cite{weinstein2007veritas,bulian1998characteristics} followed by a digital memory-buffer.
Analog implementations can use e.g. a sequence of sample-and-hold-circuits\footnote{Also called domino-ring-sampler \cite{ritt2008design}, or switched-capacitor-array, or analog-ring-sampler.} \cite{funk2004trigger,sitarek2013analysis}, or even a delay-line \cite{kildea2007whipple}.
All these implementations permanently overwrite the oldest part of the recorded image-sequence with the latest one coming from the photo-sensors.
Just as in the Cherenkov-telescope, the trigger in the Cherenkov-plenoscope has to decide whether the current light-field-sequence inside the memory-buffer contains Cherenkov-photons from an air-shower worthwhile to analyze.\\
%
% total trigger-rate on NameAcp{} is 59e3 s^{-1}.
%
In this Chapter:
First, we discuss how established triggers in Cherenkov-telescopes make their decision based on the density of photons in image-sequences.
Second, we propose to adopt triggers acting on image-sequences for the Cherenkov-plenoscope, but we propose to implement multiple of such image-sequences refocused to different object-distances.
Third, we motivate the trigger-threshold chosen for the simulations of \NameAcp{}, and show the performance of the simulated trigger.
\section{Triggering Cherenkov-telescopes}
Two methods are used to make the trigger in a Cherenkov-telescope more sensitive to air-showers, and less sensitive to fluctuations of the night-sky-background-photons.
First, triggers look for a high density of photons in time.
The compact $\approx 10\,$ns arrival of Cherenkov-photons on ground clearly separates these from the steady arrival of night-sky-background-photons.
Second, triggers look for spatial patterns in the image-sequences.
Cherenkov-photons are likely to populate nearby pixels in the image-sequences because their emission-positions in the air-shower populate a rather compact volume.
Both methods are often used in parallel or in stages to make the final trigger-decision.
However, since the atmospheric Cherenkov-method approaches the regime of single-photons, there is an additional aspect relevant for the trigger-decision.
The observable fluctuations in the arrival of night-sky-background-photons can be large because the pixels sample only small solid-angles, and integrate only over short periods of $\approx 5\,$ns.
This limitation is known as the Law of Large Numbers \cite{poisson1837recherches}.
Because of this, the threshold for the trigger has to be larger then the anticipated threshold deduced from the average flux of the night-sky-background-photons.
Otherwise the trigger will store too many image-sequences only containing fluctuations of the night-sky-background.
The larger the number of photons used for the trigger-decision becomes, the smaller becomes the relative statistical fluctuation of this number.
%
%Thus the closer the trigger-threshold can be to the anticipated average flux of night-sky-background-photons.
%
%And thus the higher becomes the precision\footnote{true-positives/(true-positives + false-positives)} of the trigger to store air-showers instead of random fluctuations.
%
Therefore, some Cherenkov-telescopes sum up multiple read-out-channels before comparing the result to a threshold \cite{delagnes2006sam,vogler2011trigger,garcia2014status}.
This is called a sum-trigger.
Other Cherenkov-telescopes do compare the density of photons on individual pixels, but then demand compact regions of triggered pixels in a second stage \cite{bulian1998characteristics,weinstein2007veritas,kildea2007whipple} to counteract the fluctuation of the night-sky-background-photons.
With these methods, air-showers induced by lower energetic particles are more likely to be recorded while fluctuations of the night-sky-background-photons are more likely to be rejected.
\begin{figure}
    \centering
    \includegraphics[width=1\textwidth]{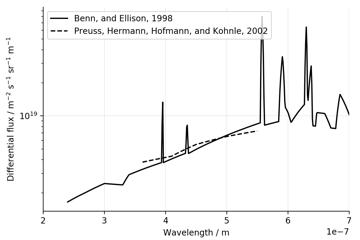}
    \caption[Flux of night-sky-background-photons, La Palma]{The differential flux of night-sky-background-photons.
        The solid line is used in the simulations of \NameAcp{}.
        We assume that the flux of night-sky-background-photons at the observatory on Roque de los Muchachos on Canary island La Palma, Spain is representative for most sites where \NameAcp{} might be build.
        Numerical values of solid line are taken from \cite{gaug2013night}, which is based on \cite{Benn1998503}.
        For comparison, the dashed line is taken from \cite{preuss2002study}.
    }
    \label{FigNightSkyBackgroundLaPalma}
\end{figure}
\section{A light-field-trigger for \NameAcp{}}
\begin{figure}
    \centering
    \includegraphics[width=1\textwidth]{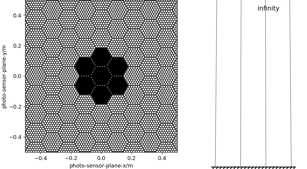}
    \caption[Central trigger-pixel on \NameAcp{}]{
        A possible trigger-pixel for the center of the field-of-view.
        Focus is set to infinity.
        A total of 427 black photo-sensors are summed up to form a trigger-pixel.
        Just like Figure \ref{FigPixelLinearCombinationSensorPlane}, this is created using Equation \ref{EqPostRefocusedImagingPixelAssignmentGeneral} but with a wider pixel-radius of $1.1 \times 0.067^{\circ}$.
    }
    \label{FigTriggerChannel4221}
\end{figure}

\begin{figure}
    \centering
    \includegraphics[width=1\textwidth]{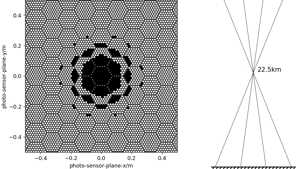}
    \includegraphics[width=1\textwidth]{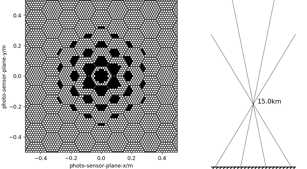}
    \includegraphics[width=1\textwidth]{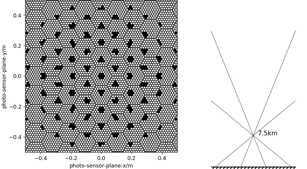}
    \caption[Three layers of trigger-pixels in \NameAcp{}]{
        Signals from black photo-sensors (lixels) will be summed up to form the central trigger-pixel in the field-of-view of \NameAcp{}.
    }
    \label{FigTriggerChannelsNameAcp}
\end{figure}

For this study of the \NameAcp{} Cherenkov-plenoscope, we adopt aspects of both trigger-implementations on Cherenkov-telescopes and create a trigger with two stages.
In the first stage, we adopt the summation of the established sum-trigger to reduce the effect of statistical fluctuations as early as possible.
We sum up $\approx 427$ lixels each into large, and overlapping trigger-pixels.
These trigger-pixels form a trigger-image-sequence.
The intensity in these trigger-pixels is then integrated continuously over a period of 5\,ns where the integral is compared with a predetermined trigger-threshold.\\
In a second stage, we adopt the search for coincident patterns and thus demand that at least two neighboring trigger-pixels must exceed their predetermined threshold in order to trigger the read-out.\\
For this, the light-field-sequence needs to be projected onto an trigger-image-sequence in real-time.
This projection can be written as a matrix-multiplication, see Equation \ref{EqRefocusedImaging} in Sections \ref{SecInterpretingDirectionalSampling}, and \ref{SecPostRefocusedImaging}.
Each trigger-pixel in the resulting image-sequence is computed from the sum of corresponding lixels (read-out-channels) in the light-field-sequence.
This means, that we can hardwire a real-time-projection of the light-field-sequence into the housing of the light-field-sensor itself using e.g. operational amplifiers to sum up the analog output-potentials of the corresponding photo-sensors.
This is for the same reasons as discussed in Section \ref{SecUpgradingTheCherenkovTelescope}.
From the implementation point-of-view, the sum-trigger for the Cherenkov-plenoscope is the same as the sum-trigger for the Cherenkov-telescope.
Both are summing up certain read-out-channels into trigger-pixels.
In the simplest case, this summation into a trigger-pixel can look like e.g. the patterns shown in the Figure \ref{FigTriggerChannel4221}.\\
A single trigger-pixel thus contains on average $7 \times \NumPax{} = 427$ read-out-channels (lixels).
As we discuss in Chapter \ref{ChOvercomingAberrations}, the (trigger-)image-sequence has much reduced aberrations and distortions.
In contrast to Cherenkov-telescopes, the \NameAcp{} Cherenkov-plenoscope profits from its plenoptic-perception here.
To take even more advantage of the plenoptic-perception, we actually implement three layers of summations, creating three (trigger-)image-sequences from one light-field-sequence in parallel.
The individual (trigger-)image-sequences are focused to object-distances of $7.5\,$km, $15\,$km, and $22.5\,$km each.
Figure \ref{FigTriggerChannelsNameAcp} shows the three different summations implemented for the central trigger-pixel ($n=4,221$) in the field-of-view.
We create trigger-pixels which include all lixels within a radius of $110\% \times 0.067^\circ$ in incident-directions.
\NameAcp{}'s trigger is able to differentiate between air-showers which reach their maximum Cherenkov-photon-production on different altitudes.
It partly overcomes the depth-of-field-limit on the trigger-level.
Table \ref{TabTriggerSummary} summarizes the basic specifications of \NameAcp{}'s trigger.
\begin{table}
\begin{center}
    \begin{tabular}{lr}
        \toprule
        Duration of sliding integration & 5\,ns\\
        Num. of refocused image-sequences & 3\\
        Object-distances of refocused image-sequences & 7.5, 15.0, and 22.5\,km\\
        Num. of trigger-pixels per trigger-image-sequence & \NumPix{}\\
        Num. of lixels summed-up into one trigger-pixel & 427\\
        Avg. nsb-photons in time-integrated trigger-pixel & 53.5\,photons\\
        Threshold for time-integrated trigger-pixel & 103\,photons\\
        Min. num. of neighboring trigger-pixels above threshold & 2\\
        Expected trigger-rate due to air-showers & $59\times10^3\,$s$^{-1}$\\
        Expected trigger-rate due to nsb-fluctuations & $\mathcal{O}(22\,$s$^{-1})$\\
        Avg. rate of nsb-photons in lixel & $25\times10^6\,$s$^{-1}$\\
        \bottomrule
    \end{tabular}
    \end{center}
    \caption[Trigger in \NameAcp{}, basic specifications]{
        The basic specifications of \NameAcp{}'s trigger.
        Here nsb is short for night-sky-background.
        We use the night-sky-background shown in Figure \ref{FigNightSkyBackgroundLaPalma}.
    }
    \label{TabTriggerSummary}
\end{table}
\section{\NameAcp{}'s trigger-threshold}
\begin{figure}
    \centering
    \includegraphics[width=1\textwidth]{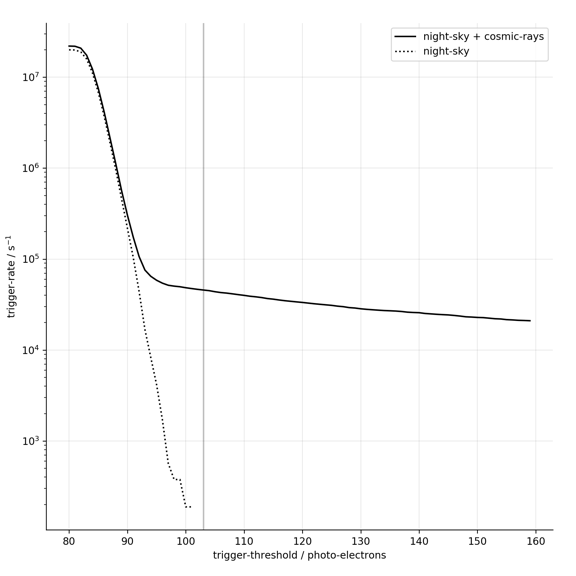}
    \caption[Trigger-rate-scan, rate vs. threshold]{
        The trigger-rate versus the trigger-threshold for the \NameAcp{} Cherenkov-plenoscope.
        This is often called rate-scan.
        Based on our simulations.
        The solid black line is the sum of triggers caused by fluctuations of the night-sky-background-photons according the flux shown in Figure \ref{FigNightSkyBackgroundLaPalma}, and charged cosmic-rays (protons, electrons, and positrons) according to the fluxes shown in Figure \ref{FigAirShowerRateChargedCosmicRays}.
        The dotted line is only the triggers caused by the night-sky-background.
        For low trigger-thresholds below $\approx 85$ photon-equivalents, our simulation runs out of statistics which is why the trigger-rate seems to saturate at $2\times10^7\,$s$^{-1}$.
        The vertical gray line marks the trigger-threshold of $103\,$photo-equivalents which we use for \NameAcp{}'s performance-estimate in Section \ref{SecPerformance}.
        The instrument-response-functions in Figures , \ref{FigResponseGammaRays} and \ref{FigResponseCosmicRays} correspond to a trigger-threshold of $103\,$photo-equivalents.
    }
    \label{FigTriggerRateScan}
\end{figure}

As the rate of air-showers with energies from $1\,$GeV to $10\,$GeV is very high, the expected trigger-rates of $\approx 59\times10^3\,$s$^{-1}$ on \NameAcp{} will be higher then the trigger-rates of existing Cherenkov-telescopes, see Figures \ref{FigAirShowerRateChargedCosmicRays}, and \ref{FigExpectedRates}.
A trigger-rate of $59\times10^3\,$s$^{-1}$ on \NameAcp{} gets close to the limits of todays implementations on Cherenkov-telescopes \cite{delagnes2006sam} which are designed to reach $100\times10^3\,$s$^{-1}$.
As a consequence, we apply a rather high trigger-threshold so that in addition to the trigger-rate induced by air-showers, the trigger-rate induced by fluctuations in the night-sky-background is low.
We set the trigger-threshold such that there are no accidental triggers caused by the night-sky-background-photons in all of the $\approx 45\,$ms of simulated observations in which the trigger is exposed.
From this we conclude that the accidental trigger-rate will not drastically exceed $22\,$s$^{-1}$.
During the dark night, there are $53.5$\,photons in a trigger-pixel integrated over $5\,$ns.
In the simulations of \NameAcp{} we use the night-sky-background-flux shown in Figure \ref{FigNightSkyBackgroundLaPalma}.
The predetermined threshold for an integrated trigger-pixel is $103$\,photons, which is $\approx 6.7$\,standard-deviations above the fluctuations of the night-sky-background.
Figure \ref{FigTriggerRateScan} shows \NameAcp{}'s trigger-rate depending on the trigger-threshold.
\section{Performance of \NameAcp{}'s trigger}
\begin{figure}
    \centering
    \includegraphics[width=1\textwidth]{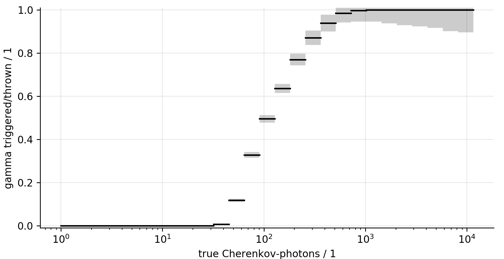}
    \caption[Trigger vs. number of true Cherenkov-photons, gamma-rays]{
        The trigger-probability for gamma-rays from a point-source versus the number of true and detected Cherenkov-photons.
    }
    \label{FigNumberTrueCherenkovPhotonsTriggerGamma}
\end{figure}

\begin{figure}
    \centering
    \includegraphics[width=1\textwidth]{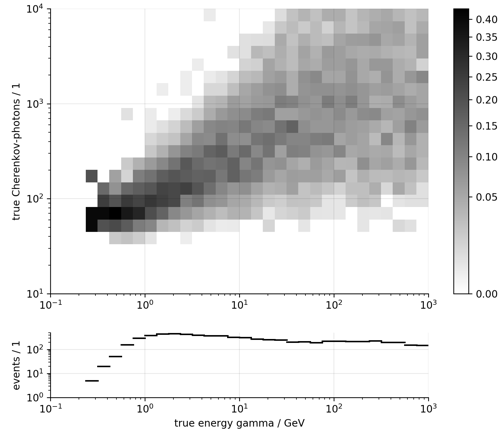}
    \caption[Number of true Cherenkov-photons vs. true energy, gamma-rays]{
        Number of true and detected Cherenkov-photons versus the true energy of gamma-rays coming from a point-source.
        The histogram in the upper panel has the sum along its columns normalized to one.
        The lower panel shows the number of events populating the columns in the upper histogram.
    }
    \label{FigNumberTrueCherenkovPhotonsEnergyGamma}
\end{figure}

\begin{figure}
    \centering
    \includegraphics[width=1\textwidth]{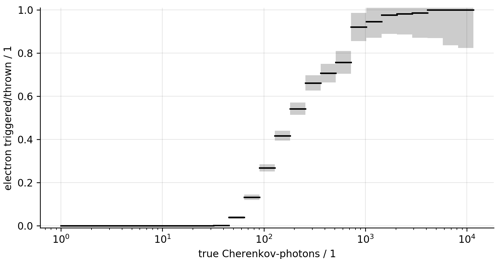}
    \caption[Trigger vs. number of true Cherenkov-photons, electrons]{
        The trigger-probability for electrons from a diffuse source versus the number of true and detected Cherenkov-photons.
    }
    \label{FigNumberTrueCherenkovPhotonsTriggerElectron}
\end{figure}

\begin{figure}
    \centering
    \includegraphics[width=1\textwidth]{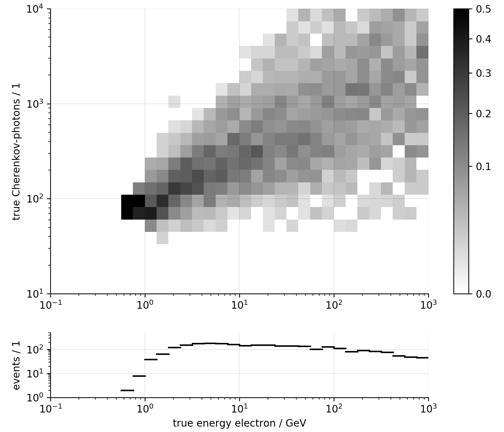}
    \caption[Number of true Cherenkov-photons vs. true energy, electrons]{
        Number of true and detected Cherenkov-photons versus the true energy of electrons coming from a diffuse source.
        The histogram in the upper panel has the sum along its columns normalized to one.
        The lower panel shows the number of events populating the columns in the upper histogram.
    }
    \label{FigNumberTrueCherenkovPhotonsEnergyElectron}
\end{figure}

\begin{figure}
    \centering
    \includegraphics[width=1\textwidth]{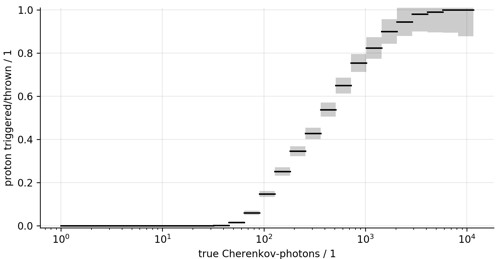}
    \caption[Trigger vs. number of true Cherenkov-photons, protons]{
        The trigger-probability for protons from a diffuse source versus the number of true and detected Cherenkov-photons.
    }
    \label{FigNumberTrueCherenkovPhotonsTriggerProton}
\end{figure}

\begin{figure}
    \centering
    \includegraphics[width=1\textwidth]{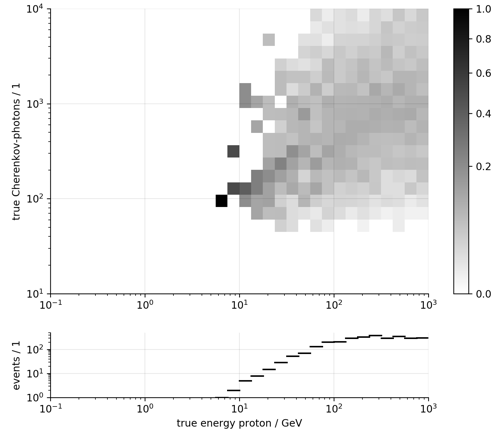}
    \caption[Number of true Cherenkov-photons vs. true energy, protons]{
        Number of true and detected Cherenkov-photons versus the true energy of protons coming from a diffuse source.
        The histogram in the upper panel has the sum along its columns normalized to one.
        The lower panel shows the number of events populating the columns in the upper histogram.
    }
    \label{FigNumberTrueCherenkovPhotonsEnergyProton}
\end{figure}

In the Figures \ref{FigNumberTrueCherenkovPhotonsTriggerGamma}, \ref{FigNumberTrueCherenkovPhotonsTriggerElectron}, and \ref{FigNumberTrueCherenkovPhotonsTriggerProton}, we show \NameAcp{}'s trigger-probability depending on the number of true and detected Cherenkov-photons.
Although both Figures \ref{FigNumberTrueCherenkovPhotonsTriggerGamma} and \ref{FigNumberTrueCherenkovPhotonsTriggerElectron} show the trigger-probability for electromagnetic air-showers, the gamma-rays in Figure \ref{FigNumberTrueCherenkovPhotonsTriggerGamma} come from a point-source, while the electrons in Figure \ref{FigNumberTrueCherenkovPhotonsTriggerElectron} come from a diffuse source.
In Figure \ref{FigNumberTrueCherenkovPhotonsTriggerGamma} we find that for gamma-rays coming from a point-source within \NameAcp{}'s field-of-view, the air-showers which have $\approx 100$ of their Cherenkov-photons detected also have a $50\%$ chance to trigger.
We also find that the probability to trigger vanishes for air-showers with only 20 or less Cherenkov-photons.
When comparing the trigger-probabilities for diffuse electrons in Figures \ref{FigNumberTrueCherenkovPhotonsTriggerElectron} with the trigger-probability for diffuse protons in Figure \ref{FigNumberTrueCherenkovPhotonsTriggerProton}, we find that protons need about twice as much Cherenkov-photons to be detected in order to trigger \NameAcp{}.
This is probably due to the wider spread of secondary particles in hadronic air-showers.
This wider spread also widens the spread of the Cherenkov-photons, and this leads to a lower density of Cherenkov-photons in the trigger-pixels.
Because of this, the correlation of the number of true and detected Cherenkov-photons with the true energy of the cosmic particle is more clear for gamma-rays in Figure \ref{FigNumberTrueCherenkovPhotonsEnergyGamma}, and electrons in Figure \ref{FigNumberTrueCherenkovPhotonsEnergyElectron}, then for protons in Figure \ref{FigNumberTrueCherenkovPhotonsEnergyProton}.
\section{Conclusion}
\label{SecTriggerConclusion}
Our first trigger for the \NameAcp{} Cherenkov-plenoscope can be implemented with established technology, and allows \NameAcp{} to reach an energy-threshold for gamma-rays of $1\,$GeV.
The sum-trigger for \NameAcp{} effectively reduces accidental triggers caused by fluctuations in the night-sky-background.
\NameAcp{}'s trigger makes its decision based on image-sequences just like the trigger in a Cherenkov-telescope, but \NameAcp{}'s trigger profits from reduced aberrations and distortions provided by the plenoptic-perception.
By summing up the responses of individual read-out-channels (lixels), it is possible to implement the projection of the light-field-sequence onto image-sequences in real-time.
The light-field-sequence can even be projected onto multiple refocused image-sequences in real-time.
Choosing between multiple layers of refocused image-sequences might allow to correct for misalignments between the light-field-sensor and the large imaging-reflector in real-time on the trigger-level.
We choose a high trigger-threshold to eliminate accidental triggers.
The lower trigger-rate due to the high trigger-threshold will ease the trigger's implementation and also the further processing of the light-field-sequences.
\section{On \NameAcp{}'s target-object-distance}
\label{SecAfterthoughtTargetSensorDistance}
The patterns shown in Figure \ref{FigTriggerChannelsNameAcp} can be improved.
In Figure \ref{FigTriggerChannelsNameAcp}, the target-geometry for \NameAcp{} is such that the sensor-plane of the light-field-sensor is exactly in a distance of one focal-length to the principal-aperture-plane of the large imaging-reflector.
Using Equation \ref{EqImagingMatrix} without the refocusing used in Equation \ref{EqRefocusedImagingMatrix}, this corresponds to a target-object-distance on infinity.
However, to optimize the trigger for closer object-distances, the light-field-sensor's target-position can be adjusted to the corresponding sensor-plane-distance in the first place.
This way, the summation-patterns shown in Figure \ref{FigTriggerChannelsNameAcp} would be much more dense in the sensor-plane what probably will ease their implementation.
In contrast, focusing to infinity would yield a much more spread summation-pattern than the one shown in Figure \ref{FigTriggerChannel4221}.
%
%------------------------------------------------------------------------------
%
%
%
%
%
%
%
%------------------------------------------------------------------------------
\chapter{Classifying Cherenkov-photons}
\label{ChClassifyingCherenkovPhotons}
About $99.9\%$ of the photons recorded by \NameAcp{}, are night-sky-background-photons which do not carry information about the origins of the cosmic particles.
It is important to tell apart Cherenkov-photons from the pool of night-sky-background-photons.
This can only be done based on the structure and density of the photons in the recorded light-field-sequence.
In this first investigation, we fall back to established methods used on Cherenkov-telescopes where photons are classified based on their density in image-sequences.
This is in the three-dimensional space of incident-directions $c_x$, $c_y$, and arrival-times $t$.
So we project \NameAcp{}'s light-field-sequence $\mathcal{L}[c_x, c_y, x, y, t]$ onto an image-sequence $\mathcal{I}[c_x, c_y, t]$ and then apply the exact same methods shown in Chapter \ref{ChaDensityBasedClustering} of Part \ref{PartPhotonStream} in this thesis.
The projection of the light-field is done using Equations \ref{EqRefocusedImaging} and \ref{EqPostRefocusedImagingPixelAssignmentGeneral}.
The only difference with the \NameAcp{} Cherenkov-plenoscope is, that we repeat the classification multiple times for a single light-field-sequence by each time refocusing the image-sequence to a different object-distance.
In the end, we classify all the dense photons found in all refocused image-sequences to be the Cherenkov-photons using a logical AND-operation.
Probably it will be worth to investigate novel methods in the future which take the light-field into account more naturally, but for this first investigation we are pleased with the classification-performance.
%
%------------------------------------------------------------------------------
%
%
%
%
%
%
%
%------------------------------------------------------------------------------
\chapter{Estimating the angular resolution for gamma-rays at 1\,GeV}
\label{ChAngularResolution}
A high angular resolution for cosmic gamma-rays is key for gamma-ray-astronomy.
In this Chapter, we show how we estimate the angular resolution of \NameAcp{} which we present in Figure \ref{FigAngularResolutions}.
In this early estimate of the angular resolution, we focus on gamma-ray-energies close to the energy-threshold of \NameAcp{} at $1\,$GeV, since those events will be most abundant in the observations.
For higher energies, we argue that \NameAcp{} will at least reach the angular resolutions reached by arrays of Cherenkov-telescopes, because the light-field-sequences recorded by \NameAcp{} can always be interpreted as multiple image-sequences recorded by an array of Cherenkov-telescopes.
Also in this early estimate, we choose to estimate the incident-direction of cosmic gamma-rays not using images, as it is usually done on Cherenkov-telescopes, but using the orientation of the Cherenkov-light-front in the recorded light-field-sequence, compare Figure \ref{FigRaysOnPrincipalAperturePlane}.
Our intent here is not to present the best method, but to present a method which makes most use of the Cherenkov-plenoscope's novel possibilities.
\section{Cherenkov-light-front}
\label{SecCherenkovPhotonFront}
\begin{figure}[]
    \centering
    %\vspace{-1cm}
    \includegraphics[width=1\textwidth]{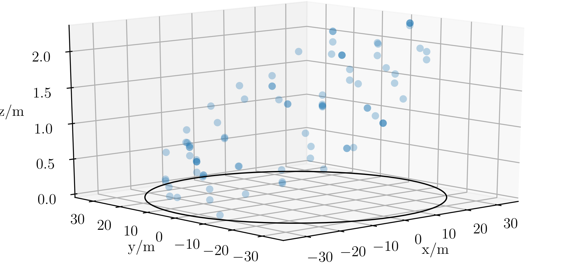}
    %
    %\vspace{-1cm}
    \includegraphics[width=1\textwidth]{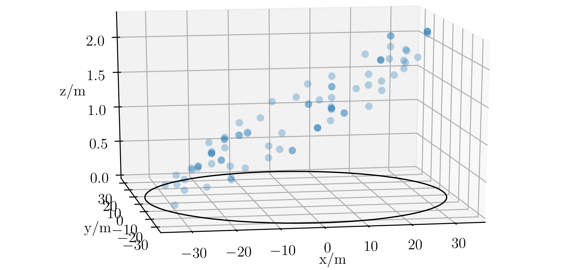}
    %
    %\vspace{-1cm}
    \includegraphics[width=1\textwidth]{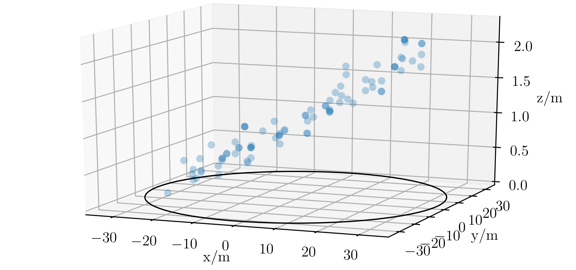}
    \caption[Cherenkov-light-front, \NameAcp{}, simulation] {
        The Cherenkov-light-front reconstructed by the \NameAcp{} Cherenkov-plenoscope for a gamma-ray with energy 997\,MeV,
        incident-direction $c_x = 1.87^\circ$, $c_y = -0.28^\circ$, and core-position $x = 34\,$m, and $x = -95\,$m.
        The same event in all three images, only the perspective is slightly rotated each time.
        Each blue point is a photon recorded by \NameAcp{} and classified to be a Cherenkov-photon from within a pool of $\approx 0.5\times10^6$ night-sky-background-photons.
    }
    \label{FigCherenkovPhotonFront169000024}
\end{figure}

When the read-out of \NameAcp{} is triggered, see Chapter \ref{ChTrigger}, \NameAcp{} records a light-field-sequence.
In this light-field-sequence, we classify Cherenkov-photons and night-sky-background-photons, see Chapter \ref{ChClassifyingCherenkovPhotons}.
Next we reconstruct the three-dimensional movement of the Cherenkov-photons above the principal-aperture-plane, see Figure \ref{FigRaysOnPrincipalAperturePlane}.
The position of a photon $j$ on its trajectory, or more precisely the position of a photon on the ray $\vec{r_k}(\lambda = - c(t - {t_\text{pap}}_{j}))$ describing the $k$-th lixel which contains the $j$-th photon, depends on the global time $t$.
The rays $\vec{r_k}(\lambda)$ are defined by the light-field-geometry $G$, see Chapter \ref{ChLightFieldGeometry}.
We choose $t$ such that the point-cloud of Cherenkov-photons is centered around the principal-aperture-plane.
Figure \ref{FigCherenkovPhotonFront169000024} shows the Cherenkov-light-front created by a gamma-ray and reconstructed by the \NameAcp{} Cherenkov-plenoscope.
\section{Limitations due to timing-resolution}
\begin{figure}[]
    \centering
    \includegraphics[width=1\textwidth]{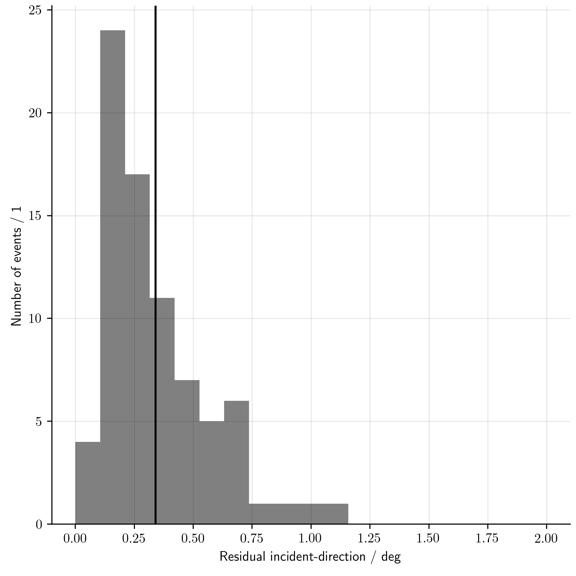}
    \caption[Angular resolution for gamma-rays at $1\,$GeV] {
        The residual incident-directions of cosmic particles inducing electro-magnetic air-showers.
        The true energies of the cosmic particles here is restricted from $750\,$MeV to $1,500\,$MeV.
        The true incident-directions of the particles are diffusely distributed inside and beyond the field-of-view of \NameAcp{}.
        A total of $68\%$ of the reconstructed events are left of the vertical black line at $0.31^\circ$.
        This vertical black line marks the $1\sigma$-containment-radius which is shown in Figure \ref{FigAngularResolutions}.
    }
    \label{FigAngularResolutionCherenkovPhotonFront}
\end{figure}

We simulate \NameAcp{}'s timing-resolution (standard-deviation) for the arrival-time of single photons to be $\Delta_t \approx 1\,$ns according to our findings for the timing-resolution of todays Cherenkov-telescopes, see Chapter \ref{ChPerformanceOfExtraction} in Part \ref{PartPhotonStream} of this thesis.
This finite timing-resolution sets a limit to the angular resolution for the reconstruction of a Cherenkov-light-front's surface-normal.
The paxels with the furthest distance in the \NameAcp{} Cherenkov-plenoscope have a baseline of $L \approx 65\,$m.
Within the duration of the timing-resolution, a photon travels a distance of $\Delta l = c \Delta_t \approx 0.3\,$m along its trajectory.
Here $c$ is the speed-of-light.
Thus the angular resolution will be limited to about $\arctan(\Delta l/L) = 0.26^\circ$.
Although the statistics of many Cherenkov-photons in the light-front help to improve the angular-resolution, we find in Figure \ref{FigAngularResolutionCherenkovPhotonFront} an angular resolution of $0.31^\circ$ which is already close to the limitations caused by a finite timing-resolution.
%
%------------------------------------------------------------------------------
%
%
%
%
%
%
%
%------------------------------------------------------------------------------
\chapter{Reconstructing air-showers using tomography}
\label{ChTomography}
To reconstruct the type, energy, and direction of a cosmic particle from the light-field-sequence we have multiple options.
We might perform a feature-generation similar to the generation of Hillas-features for images \cite{hillas1985cerenkov} but in higher dimensions for the light-field.
We also might have a look-up-table\footnote{Also called 'template-matching', or 'log-likelihood-fitting'.} \cite{le1998new} where a recorded light-field is compared to a collection of simulated light-fields for which the cosmic particle's properties are known.
Both these methods work fine with imaging on Cherenkov-telescopes and are probably a good starting-point.\\
However, with the observation-power of the Cherenkov-plenoscope it might be possible to do particle-physics within the air-showers themselves as individual jets emerge.
For particle-physics, we propose to use tomography to reconstruct the air-shower itself in three spatial dimensions.
From the light-field-sequence, we already know the three-dimensional trajectories and the arrival-times of the Cherenkov-photons.
But we still do not know the photon's production-positions.
We try to estimate the Cherenkov-photon's production-positions using an assumption.
We assume that a volume-cell in the atmosphere in which many trajectories of Cherenkov-photons come close to each other, has a high probability to also contain the production-positions of these Cherenkov-photons.
The method to estimate this probability from a light-field is known as tomography\footnote{Also called 'focus-stack-deconvolution', or 'light-field-microscopy', or 'narrow-angle-tomography'.} \cite{ng2006lightfieldmicroscopy}.\\
Here we present our first steps into tomography with the Cherenkov-plenoscope.
\section{Motivation}
\label{SecTomographyMotivation}
The reconstruction of three-dimensional densities from either a stack of refocused images \cite{mcnally1999three}, or from a light-field itself \cite{ng2006lightfieldmicroscopy} is a well established method in microscopy.
In microscopy, the density of light-emitting structures of biological cells is studied.
Biological cells are so small that their structures appear to be almost transparent when imaged with a microscope.
Often the cells are not shine through with an external light-source, but are made to shine themselves by fluorescence.
In a light-field-mindset, fluorescence-emission in biological-cells is very similar to Cherenkov-emission in air-showers.
In both scenarios the vast majority of the observed medium is transparent with only small regions emitting photons.
In both scenarios, we are interested in the production-positions of the photons.
In both scenarios, large apertures induce narrow depth-of-fields.
Here large does not mean the absolute size, but the size relative to the objects to be observed.
Microscopy motivates us to investigate tomography for the Cherenkov-plenoscope.
However, there are also differences between fluorescence-emission and Cherenkov-emission.
Most of all, unlike the isotropic emission of fluorescence-photons, Cherenkov-photons are only emitted in a narrow cone.
Also in Cherenkov-astronomy, we might profit from the measurement of the photons arrival-times.
\section{Basic Method}
\label{SecTomographyBasic}
\begin{figure}
    \centering
    \includegraphics[width=0.5\textwidth]{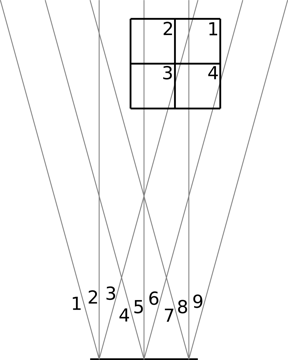}
    \caption[Basic tomography with rays and volume-cells]{Basic tomography with rays and volume-cells.
        Equation \ref{EqSystemMatrix} shows the corresponding system-matrix.
    }
    \label{FigSystemMatrix}
\end{figure}

Tomography starts with a light-field-sequence $\mathcal{L}[c_x, c_y, x, y, t]$ recorded by a plenoscope, or an array of telescopes\footnote{In tomography this is also called a line-integral, as it represents the intensity integrated along the light-field-cells which can be approximated by rays.}, and a three-dimensional structure inside a three-dimensional volume which has emitted the photons recorded in $\mathcal{L}$.
Now to reconstruct the three-dimensional structure which has emitted the photons, we have to relate the light-field-geometry $G$ of our instrument with the volume it is observing.
We divide the three-dimensional volume into densely packed volume-cells, and we approximate the light-field-cells in $G$ using rays, see Equations \ref{EqRay}, and \ref{EqRaysOnPrincipalAperturePlane}.
Figure \ref{FigSystemMatrix} shows a simplified light-field-geometry $G$  approximated with $9$ rays facing a volume divided into $4$ volume cells.
We relate the rays with the volume-cells by calculating their overlap, and represent it in a matrix
\newcommand{\gr}[1]{\textcolor[rgb]{0.7, 0.7, 0.7}{#1}}
\begin{eqnarray}
    S[v,\,k] &=& \begin{bmatrix}%
    % 1      2      3      4      5      6      7      8      9 rays  % voxels
    \gr{0}&\gr{0}&  1.1 &\gr{0}&\gr{0}&\gr{0}&\gr{0}& 1.0  &\gr{0}\\  % 0
    \gr{0}&\gr{0}&\gr{0}&\gr{0}&1.0   &\gr{0}&\gr{0}&\gr{0}&\gr{0}\\  % 1
    \gr{0}&\gr{0}& 0.6  &\gr{0}&1.0   &\gr{0}&\gr{0}&\gr{0}&\gr{0}\\  % 2
    \gr{0}&\gr{0}& 0.4  &\gr{0}&\gr{0}& 0.5  &\gr{0}& 1.0  &\gr{0}\\  % 3
    \end{bmatrix},
    \label{EqSystemMatrix}
\end{eqnarray}
which is often called system-matrix in tomography.
Here $S[v,\,k]$ has $V=4$ rows for the volume-cells, and $K=9$ columns for the rays.
In our representation $S[v,\,k]$ contains the Cartesian-distance a ray travels through a given volume-cell, with the edge of the volume-cell having length $1$.
From here on, we can follow different methods to reconstruct the volume-cell's intensities $D[v]$ using the system-matrix $S[v,\,k]$, and the recoded light-field $\mathcal{L}[k]$.
Here $D[v]$ is a one-dimensional vector containing the reconstructed photon-emission-intensity in each of the $V$ volume-cells, and $\mathcal{L}[k]$ is the a one-dimensional vector containing the measured intensities in each of the $K$ light-field-cells.
For the reconstruction of $D[v]$, we first explored filtered-back-projection, and then went on to iterative methods such as the (simultaneous)-algebraic-reconstruction-technique \cite{andersen1984simultaneous}.
Figure \ref{FigTomographyExampleEngels} shows reconstructions using our first implementation of filtered-back-projection.\\
For Cherenkov-astronomy it is important to note that, with the exception of recent advances in positron-emission-tomography \cite{conti2009state}, established tomography in microscopy, biology, and medicine does not take into account the photon's time-of-flight.
Equation \ref{EqSystemMatrix} does not contain the distance between a volume-cell and the support-position of a ray on the aperture-plane.\\
Also, Equation \ref{EqSystemMatrix} is only approximative as the light-field-cells are only approximated using single rays.
In the future this can be improved by approximating a light-field-cell with multiple rays contained in it.
\section{\NameAcp{}'s reconstruction-power}
\begin{figure}
    \centering
    \includegraphics[width=1\textwidth]{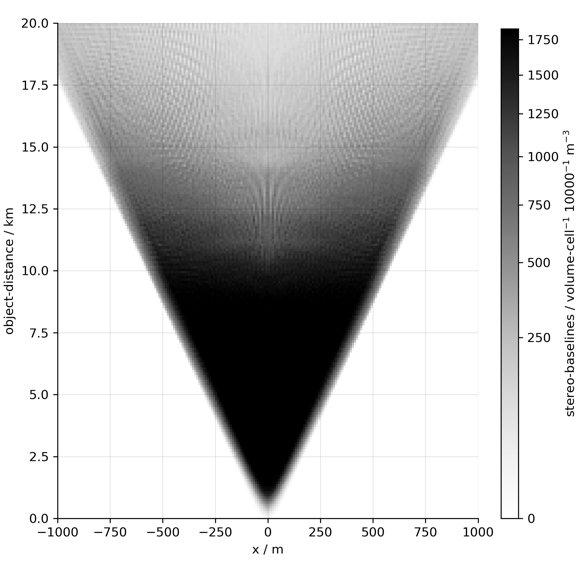}
    \caption[\NameAcp{}'s 3D reconstruction-power, stereo-baselines]{\NameAcp{}'s three-dimensional reconstruction power.
        The number of stereo-baselines in the volume-cells above \NameAcp{}'s aperture-plane.
        \NameAcp{}'s aperture-plane goes along the bottom of the figure at object-distance = 0\,km.
        Here we see only the few volume-cells in the $x$-object-distance-slice for $y = 0$.
        Each volume-cell is $10\,$m in $x$, $10\,$m in $y$, and $100\,$m in object-distance.
        The darkening here is proportional to the square-root of the number of stereo-baselines $N_\text{Baselines}$, which is proportional to the number of unique support-positions in a volume-cell $N_\text{Supports}$, see Equation \ref{EqStereoBaseline}.
        }
    \label{FigThreeDReconstructionPower}
\end{figure}

Figure \ref{FigThreeDReconstructionPower} shows \NameAcp{}'s three-dimensional reconstruction-power relative to the atmospheric volume.
The atmospheric cube with a volume of $2 \times 2 \times 20$\,km$^3$ above \NameAcp{} is divided into $200\times200\times200$ volume-cells.
Each volume-cell is $10 \times 10 \times 100$\,m$^3$.
The figure shows the number of stereo-baselines in each volume-cell.
The number of stereo-baselines in a volume-cell is
\begin{eqnarray}
    N_\text{Baselines} &=& \frac{1}{2}(N_\text{Supports}^2 - N_\text{Supports})
    \label{EqStereoBaseline}
\end{eqnarray}
where $N_\text{Supports}$ is the number of unique support-positions that emit rays into this volume-cell.
For \NameAcp{} the maximum number of stereo-baselines in a volume-cell is 1,830 when all of \NameAcp{}'s \NumPax{} support-positions (paxels) emit at least one ray into this volume-cell.
In Figure \ref{FigThreeDReconstructionPower} we find that \NameAcp{}'s reconstruction-power vanishes towards larger object-distances.
At the given binning of the atmospheric-volume, \NameAcp{}'s reconstruction-power is optimal up to object-distances of $\approx 10\,$km.
When pointing to zenith, this corresponds to an altitude of $15\,$km a.s.l. when building \NameAcp{} on $5,000\,$m a.s.l.
We also find that there is no reconstruction-power in the first few $100\,$m above \NameAcp{}'s aperture as the bundles of rays originating in each support-position have not yet overlapped.
Naturally we find that there is only reconstruction-power within \NameAcp{}'s view-cone.
This very asymmetric population of rays inside volume-cells increases the chance for unresolvable artifacts, and is often discussed as 'narrow-angle-tomography'.
From Figure \ref{FigThreeDReconstructionPower} we can already see possible limitations to \NameAcp{}'s three-dimensional reconstruction-power.
Air-showers which emit Cherenkov-photons in volume-cells with low numbers of stereo-baselines will be more difficult to be reconstructed.
Further, even if an air-shower emits Cherenkov-photons within a volume-cell, this does not yet mean that those Cherenkov-photons will reach \NameAcp{}'s aperture as the emission of Cherenkov-photons is not isotropic.
\section{Reconstructing air-showers in the refocused-image-space}
\begin{figure}{}
    \centering
    \includegraphics[width=1\textwidth]{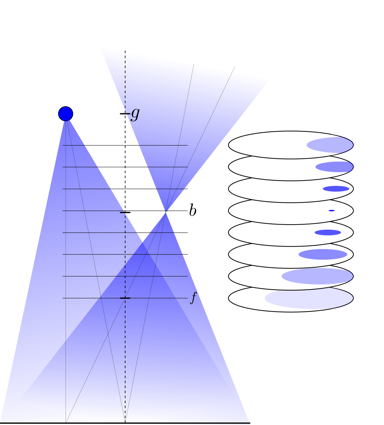}
    \caption[Tomography, refocus-stack]{
        A light-source in object-distance $g$ is observed by the plenoscope with focal-length $f$.
        The dashed line is the optical-axis and the horizontal line at the bottom is the principal-aperture-plane.
        When we refocus a stack of images from the light-field (right part of figure), the image of the light-source will be sharpest in the image for image-sensor-distance $b$ which corresponds to a focus on object-distance $g$.
        Other images for image-sensor-distances $\neq b$ will show the blurring (Bokeh) induced by the narrow depth-of-field caused by the large imaging-reflector.
        In this figure, the lowest image in the refocused stack is for a image-sensor-distance of $f$, which corresponds to a focus on $\infty$.
    }
    \label{FigRefocusStack}
\end{figure}

In Figure \ref{FigThreeDReconstructionPower} we find that the volume-cells in the Cartesian-space above the atmosphere are populated very asymmetrically.
Some have many, others have no rays passing through them.
This asymmetry can be overcome when the tomography is done in the three dimensional space of a refocused stack of images \cite{mcnally1999three}.
Instead of the Cartesian-rays defined in Equation \ref{EqRay}, we use the image-rays introduced in Section \ref{SecImageRays}.
Instead of subdividing the Cartesian-volume above the Cherenkov-plenoscope, we subdivide the three dimensional space of refocused images in incident-directions $c_x$, and $c_y$, as well as image-distances $b$,
compare Equations \ref{EqVirtualSensorPlaneDistance}, and \ref{EqVirtualSensorPlaneIntersection}.
The advantage of this refocused-image-space is, that the density of rays in the cells is almost the same for all the cells.
The plenoscope's vanishing reconstruction-power for larger object-distances is taken into account here naturally.
In our first tests with iterative reconstructions, this reduces many artifacts when estimating the emission-density of Cherenkov-photons $D[v]$.
Again, the overlap of image-rays and the pixels in the refocused-image-space can be represented in a system-matrix as shown in Equation \ref{EqSystemMatrix}, and is often called three-dimensional-point-spread-function.
Figure \ref{FigRefocusStack} illustrates the basic idea of the refocused-image-space.
\begin{figure}
    \centering
    \includegraphics[width=1\textwidth]{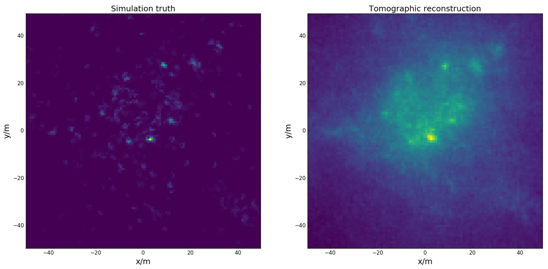}
    \includegraphics[width=1\textwidth]{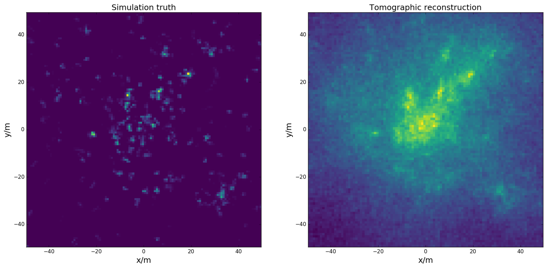}
    \caption[Tomographic reconstruction, filtered-back-projection]{Figures taken from \cite{engels2017master}.
        A tomographic reconstruction of an air-shower induced by a hadronic cosmic-ray.
        Upper figures show a $x$-$y$-slice of volume-cells at object-distance $= 10.9$\,km, and lower figures show a $x$-$y$-slice of volume-cells at object-distance $= 12.4$\,km.
        Compare the bright structures in both the simulation-truth on the left, and in the reconstruction on the right.
        The filtered-back-projection used in this figures is developed and implemented by the author of this thesis (S.A.M.).
        In his master-thesis, Axel Arbet Engels investigates the reconstruction of direct Cherenkov-photons using tomography on a perfect Cherenkov-plenoscope with an enormous $150\,$m aperture-diameter.
        The goal is to push the limits of chemical composition-measurements using direct Cherenkov-photons emitted by the cosmic nuclei itself.
    }
    \label{FigTomographyExampleEngels}
\end{figure}
\section{Conclusion}
\label{SecTomographyConclusion}
We see great potential for tomographic reconstructions of air-showers using the Cherenkov-plenoscope, or a powerful array of Cherenkov-telescopes such as the Cherenkov-Telescope-Array (CTA).
The big similarity with tomography in microscopy allows the atmospheric-Cherenkov-method to adopt many well established techniques.
For the $150\,$m diameter Cherenkov-plenoscope investigated by Axel Arbet Engels, the reconstruction-power for air-showers using the a simple filtered-back-projection is extraordinary, see Figure \ref{FigTomographyExampleEngels}.
Tomography is the generalization of the stereo-technique and will potentially enable us to observe the inner structure of air-showers.
%
%------------------------------------------------------------------------------
%
%
%
%
%
%
%
%------------------------------------------------------------------------------
\chapter{Pointing the Cherenkov-plenoscope using the cable-robot-mount}
\label{ChCableRobotMount}
We propose the cable-robot-mount to point the Cherenkov-plenoscope to different gamma-ray-sources in the sky.
Although it might be possible to mount the 71\,m diameter \NameAcp{} on a conventional altitude-azimuth-mount, we decide to investigate the cable-robot-mount as it potentially offers a more cost-effective solution.
The concept of the cable-robot-mount is proposed by the author of this thesis (S.A.M.), while the details are worked out in close collaboration with the department for civil, environmental and geomatic engineering at ETH-Zurich.
\begin{enumerate}
\item We motivate a dedicated cable-robot-mount to point the Cherenkov-plenoscope.
\item We list technical specifications, identify states of operation, and investigate environmental properties of possible sites.
\item We briefly describe how civil-engineer Spyridon Daglas together with the author of this thesis (S.A.M.) simulate the cable-robot-mount.
\item We briefly summarize the findings of Spyridon Daglas on the performance and feasibility of cable-robot-mounts ranging from $30\,$m to $100\,$m in diameter.
\end{enumerate}
\section{Motivation -- A mount for the plenoscope}
\begin{figure}
    \centering
    \includegraphics[width=1\textwidth]{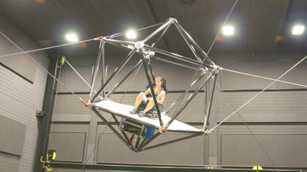}
    \caption[Cable-robot-simulator, photograph]{
    Figure taken from \cite{miermeister2016cablerobot}.
    The cable-robot-simulator by Max-Planck-Institute for Biological Cybernetics in cooperation with the Fraunhofer-Institute for Manufacturing, Engineering, and Automation.
    This impressive motion simulator can accelerate payloads of up to $500\,$kg with up to $14\,$ms$^{-2}$.
    The design-goal of the cable-robot-simulator is not to slowly move the large components of a Cherenkov-plenoscope, but to be a motion simulator for up to two passengers with an 'extraordinary' \cite{miermeister2016cablerobot} power to weight ratio for maximum acceleration.
    See also Figure \ref{FigCableRobotSimulatorDrawing}.
    For the Cherenkov-plenoscope, we adopt the icosahedron-shape of the moving platform to mount the light-field-sensor.
    Compare Figure \ref{FigNameAcpTourLightFieldSensor}.
    }
    \label{FigCableRobotSimulatorPhotograph}
\end{figure}

\begin{figure}
    \centering
    \includegraphics[width=1\textwidth]{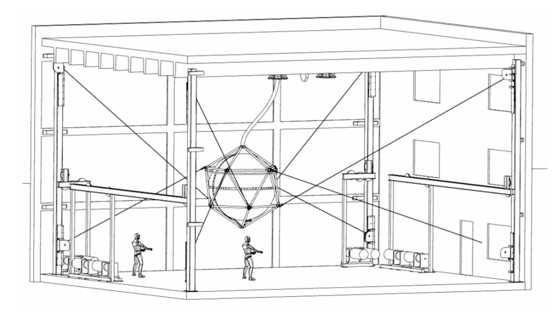}
    \caption[The cable-robot-simulator, drawing]{
    Figure taken from \cite{miermeister2016cablerobot}.
    A prototype of the cable-robot-mount is implemented in a so called 'crossed over' configuration and has eight powerful actuated winches in each corner of a large room.
    See also Figure \ref{FigCableRobotSimulatorPhotograph}.
    Compare this configuration with the cable-robot-mount for the light-field-sensor shown in Figure \ref{FigAcpOverview}.
    }
    \label{FigCableRobotSimulatorDrawing}
\end{figure}

\begin{figure}
    \centering
    \includegraphics[width=1\textwidth]{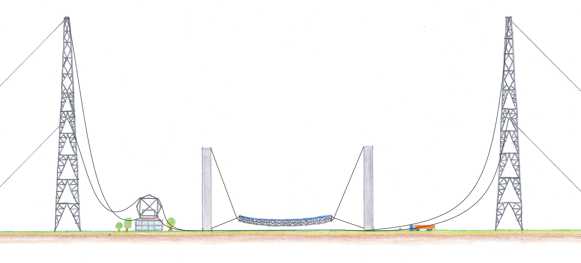}
    \caption[Cable-robot-mount, parking, drawing]{A possible parking position for the \NameAcp{} Cherenkov-plenoscope and its cable-robot-mount.
    }
    \label{FigFirstDrawingCableRobotMountParking}
\end{figure}

\begin{figure}
    \centering
    \includegraphics[width=1\textwidth]{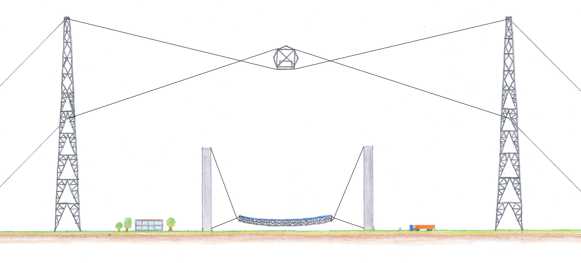}
    \includegraphics[width=1\textwidth]{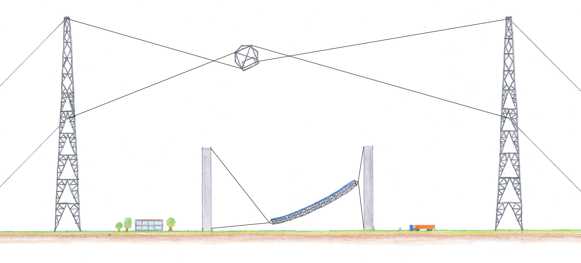}
    \includegraphics[width=1\textwidth]{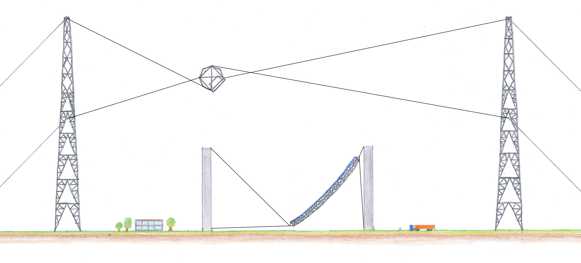}
    \caption[Cable-robot-mount, pointing, drawing]{\NameAcp{} and its cable-robot-mount pointing to different zenith-distances of $0.0^\circ$, $22.5^\circ$, and $45.0^\circ$.
    }
    \label{FigFirstDrawingCableRobotMount}
\end{figure}

\begin{figure}[H]
    \centering
    \includegraphics[width=0.8\textwidth]{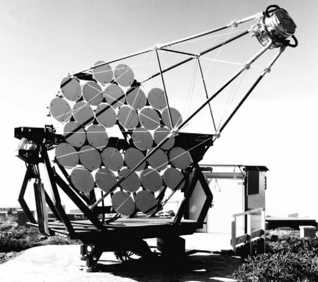}
    \caption[HEGRA, single Cherenkov-telescope]{
        Figure taken from \cite{puhlhofer2003technical}.
        %{https://www.mpi-hd.mpg.de/hfm/CT/CT.html}
        %
        One of the four Cherenkov-telescopes in the HEGRA telescope-array.
        Each telescope has $3.4\,$m effective aperture-diameter.
    }
    \label{FigHegra}
\end{figure}

\begin{figure}[H]
    \includegraphics[width=1.\textwidth]{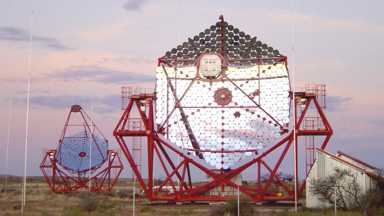}
    \caption[H.E.S.S. Cherenkov-telescope-array]{
        Figure taken from \cite{hess2018images}.
        Two of the four telescopes in H.E.S.S., Namibia.
        The telescopes have $12\,$m effective aperture-diameter and are positioned on a square with $120\,$m long edges.
    }
    \label{FigHess}
\end{figure}

\begin{figure}[H]
    \centering
    \includegraphics[width=1.\textwidth]{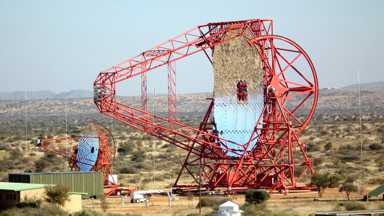}
    \caption[H.E.S.S.\,II]{
        Figure taken from \cite{hess2018images}.
        The large H.E.S.S.\,II Cherenkov-telescope with $28\,$m effective aperture-diameter.
    }
    \label{FigHessIi}
\end{figure}

\begin{figure}[H]
    \centering
    \includegraphics[width=0.65\textwidth]{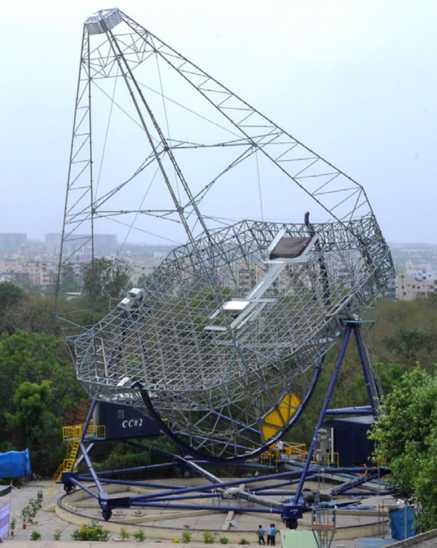}
    \caption[MACE Cherenkov-telescope]{
        Figure taken from \cite{borwankar2016simulation}.
        The MACE telescope has $21.3\,$m effective aperture-diameter.
        Here it is at its production site in Hyderabad before it is moved to Hanle at 4,270\,m a.s.l..
    }
    \label{FigMace}
\end{figure}

\begin{figure}[H]
    \centering
    \includegraphics[width=1.\textwidth]{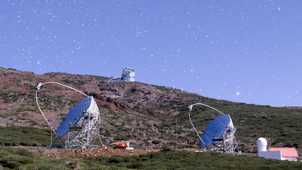}
    \caption[MAGIC Cherenkov-telescopes]{
        The MAGIC Cherenkov-telescopes.
        Photograph by Thomas Kraehenbuehl.
    }
    \label{FigMagic}
\end{figure}

To reach an energy-threshold for gamma-rays of $1\,$GeV, the \NameAcp{} Cherenkov-plenoscope needs a large imaging-reflector of $71\,$m in diameter.
Established Cherenkov-telescopes only have mounts for imaging-reflectors up to $28\,$m in diameter, see Section \ref{SecSamplingLargeTelescope}.
Although the conventional altitude-azimuth-mount of the Cherenkov-telescope might be an applicable mount for the \NameAcp{} Cherenkov-plenoscope, we see the potential for a more cost-effective mount.
On the Cherenkov-plenoscope, the demand for a rigid alignment between the imaging-reflector and the light-field-sensor is much more relaxed in contrast to the demand for a rigid alignment between the imaging-reflector and the image-sensor on a Cherenkov-telescope, see Chapter \ref{ChCompensatingMisalignmnets}.
As long as the actual alignment is known, the Cherenkov-plenoscope can compensate misalignments between its imaging-reflector and its light-field-sensor.
While on the Cherenkov-telescope rigid support-structures between the imaging-reflector and the image-sensor are mandatory, on the Cherenkov-plenoscope this rigid support-structures can be omitted.
Rigid support-structures shadow the aperture, and add up weight to the moving parts.
But the main problem that we identify with rigid support-structures is, that all the forces needed to hold the image-sensor are guided through the support-structure of the imaging-reflector, compare Figures \ref{FigWhipplePhotograph}, \ref{FigHegra}, \ref{FigHess}, \ref{FigHessIi}, \ref{FigMace}, and \ref{FigMagic}.
Since the forces holding the image-sensor change with the pointing-direction of the Cherenkov-telescope, also the shape of the imaging-reflector changes.
And such deformations of the imaging-reflector must be minimized for high optical quality.
On small Cherenkov-telescopes such deformations might be too small to effect optical quality, but on larger telescopes it becomes increasingly difficult to find cost-efficient solutions in order to keep the deformations below an acceptable limit, or to actively correct for them \cite{biland2007active}.
The physical limit of the square-cube-law \cite{square_qube_law} makes the forces needed to hold a structure grow quicker than the areas which have to tolerate these forces.
As a consequence, we can not simply scale up an existing mount.\\
Another drawback of the altitude-azimuth-mount is that it can not move fast near the zenith what conflicts the hunt for transient phenomena in the highly variable gamma-ray-sky.
The inverse kinematics of the altitude-azimuth-mount has a near-zenith-singularity \cite{borkowski1987near} which causes unacceptably large accelerations and forces when moving close to the zenith during a repositioning.\\
Motivated by the Arecibo radio-telescope \cite{altschuler2002national}, the NIST cable-robot-manipulator \cite{albus1993nist}, and the cable-robot-simulator \cite{miermeister2016cablerobot}, see Figures \ref{FigCableRobotSimulatorPhotograph}, and \ref{FigCableRobotSimulatorDrawing}, we decided to investigate a different design for the mount of \NameAcp{}.
We call it the cable-robot-mount.
Our cable-robot-mount for \NameAcp{} is separated into two independent mounts.
One mount is only actuating the large imaging-reflector, and the second mount is only actuating the light-field-sensor.
The two independent mounts always try to establish the desired target-alignment between the light-field-sensor and the large imaging-reflector, while we compensate possible misalignments due to e.g. wind-gusts later in software, see Chapter \ref{ChCompensatingMisalignmnets}.
The independent actuation of the two components reduces the forces running through the support-structure of the large imaging-reflector and thus allows us to build larger imaging-reflectors.
The cable-robot-mount supports the imaging-reflector, and the light-field-sensor with many cables in parallel which are actuated by winches in order to minimize the mass of the moving structures.
The cable-robot-mount extensively uses robotics, which is the study of controlling the actuated winches using a computer that is aware of the parallel, elastic, and inverse kinematics of the entire mount.
The cable-robot-mount is meant to trade large and costly support-structures, as well as large and expensive mechanical joints in change for computer-control and parallel actuation.
We suspect, that in contrast to the costs for support-structures and mechanical joints, the costs for computer-control and parallel actuation scale less rapidly with the size of the mount.
The cable-robot-mount only has one moving support-structure which is the dish holding the mirror-facets of the imaging-reflector.
The remaining support-structures in the cable-robot-mount are fixed towers  where the cables and winches are attached to.
Such large towers are common in e.g. overhead-power-lines what reduces the costs for their engineering and production.
Figures \ref{FigFirstDrawingCableRobotMountParking}, and \ref{FigFirstDrawingCableRobotMount} show conceptual drawings of \NameAcp{} and its cable-robot-mount.
In summary, we expect the cable-robot-mount to reduce the forces running through the support-structure of the imaging-reflector which shall allow the construction of larger imaging-reflectors.
We expect the cable-robot-mount to reposition the Cherenkov-plenoscope faster during its hunt for transient phenomena because its inverse kinematic is free of singularities.
And further, we expect the cable-robot-mount to be very cost-effective because all its fixed towers and actuated winches are mass produced components which are commonly used in e.g. overhead-power-lines, wind-turbines, and harbor-cranes.
\section{Technical specifications}
\label{SecCableRobotMountTechnicalSpecifications}
\begin{figure}
    \centering
    \includegraphics[width=1\textwidth]{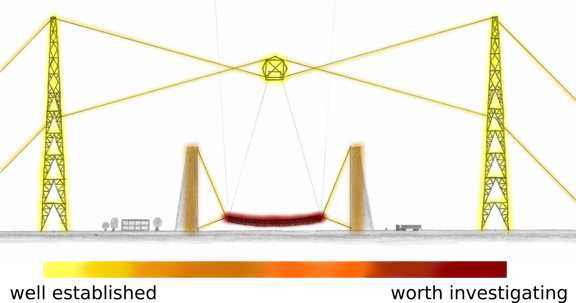}
    \caption[Cable-robot-mount, complexity]{%
        This is based on the author's interpretation of the potential for civil-engineering after discussions and studies with the experts.
        The tall masts are considered to be standard components used in overhead-land-lines.
        The cage for the light-field-sensor is also considered to be fairly standard.
        The steel-cables are considered higher standard, but still well within range of commercial products for e.g. harbor-cranes.
        The concrete-pillars are considered to be simple from a load-point-of-view, but are worth to investigate cost-effective, and wind-shielding designs.
        Clearly, the civil-engineers consider the large imaging-reflector's support-structure holding the mirror-facets to be most worthy to investigate in order to find cost-effective designs.
    }
    \label{FigCableRobotMountComplexity}
\end{figure}

To find a cost-effective design and to communicate our needs, we formulate the technical specifications for the Cherenkov-Plenoscope's mount.
With these specifications in mind, Figure \ref{FigCableRobotMountComplexity} illustrates what components of the cable-robot-mount have potential to be investigated.
Because a light-weight design for the imaging-reflector is crucial, we investigate the dimensions and masses of existing mirror-facets in Table \ref{TabMirrorFacets}.
We want to reach zenith-distances of up to $45^\circ$.
\subsection*{States of operation}
We identify three states of operation for the \NameAcp{} Cherenkov-plenoscope.
\begin{enumerate}
\item Parking.
During the day, the large imaging-reflector and the light-field-sensor are both lowered to their own pedestals for parking, see Figure \ref{FigFirstDrawingCableRobotMountParking}.
When the large imaging-reflector rests on its pedestal pointing to zenith, the intense\footnote{Solar-constant 1.3\,kW\,m$^{-2}$ $\times$ aperture-area of imaging-reflector $3,684\,$m$^2$ $\div$ area of projected image of the Sun $0.68\,$m$^2$} 7.4\,MW\,m$^{-2}$ image of the Sun can not be projected onto objects on ground.
\item Fast pointing.
The Cherenkov-plenoscope points fast from one source in the sky to another source.
Both the large imaging-reflector and the light-field-sensor are moved independent of each other.
Since there are no near-zenith-singularities, the cable-robot-mount always rotates the Cherenkov-plenoscope along one single axis only.
For fast pointing, we desire angular velocities of $90^\circ$min$^{-1}$.
In this state, the Cherenkov-plenoscope does not have to reach its target-geometry, as it will not record air-showers during fast pointing.
\item Slow tracking.
Only here the Cherenkov-plenoscope tries to reach its target-geometry.
Only during slow tracking, the Cherenkov-plenoscope records air-showers.
To counteract the rotation of the earth, an angular velocity of $0.25^\circ$min$^{-1}$ is mandatory.
To continuously spread the signal-, and background-regions across different regions inside the field-of-view \cite{finnegan2011orbit}\footnote{E.g. so called orbit-mode implemented in the VERITAS Cherenkov-telescopes.}, we desire an angular velocity of $0.75^\circ$min$^{-1}$.
Together, we conclude that low angular velocities during slow tracking can be approximated using static load-scenarios.
\end{enumerate}
\subsection*{Site and environment -- Atacama-desert, Chile}
\begin{figure}
    \centering
    \includegraphics[width=1\textwidth]{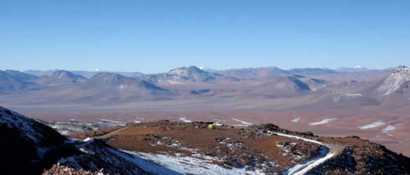}
    \caption[Panorama, Atacama-desert, Chile]{Figure taken from \cite{radford2008submillimeter}.
        Atacama-desert in Chile near the summit of Cerro Chajnantor on $5,612\,$m a.s.l..
    }
    \label{FigPanoramaAtacamaDesertChile}
\end{figure}

\begin{figure}
    \centering
    \includegraphics[width=1\textwidth]{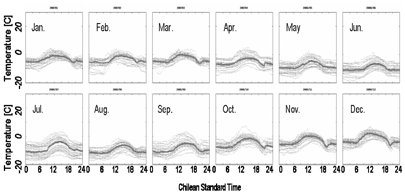}
    \caption[Temperatures, Atacama-desert, Chile]{Figure taken from \cite{miyata2008site}.
        Temperatures over the years 2006, and 2007 at the summit of Cerro Chajnantor on $5,612\,$m a.s.l., Chile.
        Temperatures for the nearby Pampa la Bola plateau on $\approx 5,000\,$m a.s.l. are reported to be $\approx 5^\circ$C higher on average.
    }
    \label{FigTemperaturesAtacamaDesertChile}
\end{figure}

\begin{figure}
    \centering
    \includegraphics[width=1\textwidth]{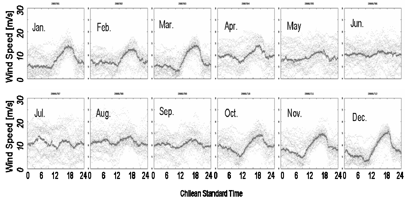}
    \caption[Temperatures, Atacama-desert, Chile]{Figure taken from \cite{miyata2008site}.
        Wind-speeds over the years 2006, and 2007 at the summit of Cerro  Chajnantor on $5,612\,$m a.s.l., Chile.
        Wind-speeds for the nearby Pampa la Bola plateau on $\approx 5,000\,$m a.s.l. are reported to be $\approx 5\,$ms$^{-1}$ lower on average.
    }
    \label{FigWindSpeedAtacamaDesertChile}
\end{figure}

We choose to adopt the environmental properties of Atacama-desert in Chile on $\approx 5,000\,$m a.s.l..
The environmental properties of this site are well documented by e.g. the European-Southern-Observatory's ALMA-telescope, and the Cherenkov-Telescope-Array.
Further, the Atacama-desert fulfills our demands for exceptional dark, and clear skies, and it is high in altitude.
Although the geomagnetic cutoff-rigidity of $\approx 10\,$GV is not the highest, compare Figure \ref{FigGeomagneticCutOffRigidity}, it is representative for our estimate of \NameAcp{}'s performance.
Figure \ref{FigPanoramaAtacamaDesertChile} shows the landscape of Atacama-desert.
And the Figures \ref{FigTemperaturesAtacamaDesertChile}, and \ref{FigWindSpeedAtacamaDesertChile}, show typical temperatures and wind-speeds for this site.
Temperature, and changes in temperature effect the choice of materials.
Wind speeds are used to formulate a worst-case load-scenario during observations.
Occasional wind-gusts are used to formulate a survival-load-scenario.
We want observations to be possible up to wind-speeds of $15$\,ms$^{-1}$ (54\,kmh$^{-1}$), and we want the mount to survive wind-speeds of up to $55\,$ms$^{-1}$ (200\,kmh$^{-1}$).
\begin{table}
    \begin{center}
        \begin{tabular}{lccrrrrr}
            type-designation & shape & area & spacing & mass & areal-density & areal-density \\
             & & & & & & with actuators\\
             & & m$^2$ & m & kg & kg\,m$^{-2}$ & kg\,m$^{-2}$ \\
            \toprule
            MAGIC 1st gen. & square  & 0.25 & 0.50 &  4.0 & 16.0 & 21.0\\
            MAGIC 2nd gen. & square  & 0.96 & 0.96 & 13.4 & 14.0 & 19.2\\
            VERITAS        & hexagon & 0.32 & 0.61 & 10.4 & 32.4 & 48.0\\
            %Veritas volume 3680cm^3, 2.82g/cm^3
            FACT/HEGRA     & hexagon & 0.32 & 0.61 &  5.5 & 17.6 & 33.7\\
            CTA-MST, INAF  & hexagon & 1.25 & 1.20 & 25.0 & 20.0 & 24.0\\
            CTA-LST, INAF  & hexagon & 1.97 & 1.50 & 45.0 & 22.8 & 25.3\\
        \end{tabular}
        \caption[Technical specifications of mirror-facets]{Mirror-facets on Cherenkov-telescopes.
            Here the spacing is the flat-to-flat-, or inner-diameter of the facets.
            We assume that the weight of a two-axis-actuator for the orientational fine-alignment is $\approx 5\,$kg/facet, regardless of the area of the facet.
            The MAGIC, and FACT facets we measured ourselves.
            The geometry of VERITAS is taken from \cite{perkins2007mirror}, while we assume that the density of the glass is $\approx 2,800\,$kg\,m$^{-3}$.
            Cherenkov-Telescope-Array (CTA) specifications are taken from \cite{pareschi2013status}.
            Mass of the CTA-LST facet is taken from \cite{teshima2011design}.
            Here LST is short for Large-Size-Telescope, MST is short for Medium-Size-Telescope, and INAF is short for Instituto-Nazionale-di-AstroFisica.
        }
        \label{TabMirrorFacets}
    \end{center}
\end{table}
\section{Designing a support-structure for the imaging-reflector}
\begin{figure}
    \centering
    \includegraphics[width=1\textwidth]{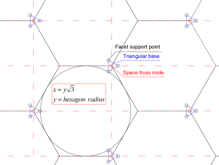}
    \caption[Hexagonal perimeter of mirror-facets and space-truss-lattice]{Figure taken from \cite{daglas2015master}.
        The hexagonal perimeter of the mirror-facets favors a rectangular type of space-truss-lattice.
        Spyridon organizes the individual space-truss-layers in a so called square-on-offset-square pattern.
    }
    \label{FigSpirosHexagonalMirrorFacetsSpaceTrussLattice}
\end{figure}

As indicated in Figure \ref{FigCableRobotMountComplexity}, we focus on investigating the imaging-reflector's support structure.
Spyridon proposes to use a space-truss to support the mirror-facets.
He proposes to use a rectangular space-truss-lattice to ease the mounting of the hexagonal mirror-facets as it is shown in Figure \ref{FigSpirosHexagonalMirrorFacetsSpaceTrussLattice}.
In this configuration, the corners of the mirror-facets line up with the nodes of the space-truss.
He disfavors a ribbed-dome space-truss, as it makes the mounting of the mirror-facets more complicated, and deforms in potentially worse shapes harming the optical quality.
After initial simulations, he identified that stiffening the outer realm of the imaging-reflector could prevent tensions from running into the inner parts.
A tension-ring in the outer realm will potentially lower the overall weight and deformations.
Such a tension-ring-design can be implemented by e.g. using stiffer bars in between the outer nodes of the space-truss.\\
Therefore, our focus is on a space-truss with an outer tension-ring.
But for example another proposal by Prof. Mario Fontana suggests to use large inflatable air-bags between the bars of the space-truss-lattice to increase the stiffness of the overall imaging-reflector.
\section{Simulating the cable-robot-mount}
\label{SecCableRobotMountComputerSimulation}
\begin{figure}
    \centering
    \includegraphics[width=1\textwidth]{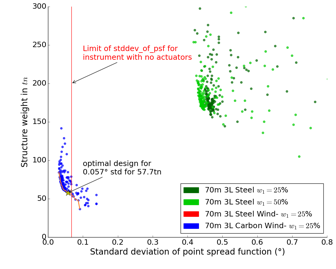}
    \caption[Cable-robot-mount, parameter-study, pareto-front]{Figure taken from \cite{daglas2015master}.
        The estimated point-spread-function of the imaging-reflector during the $45^\circ$-zenith-distance-load-scenario versus the mass of the support-structure.
        Each dot corresponds to a specific set of parameters for the imaging-reflector.
        Here all imaging-reflectors are $71\,$m in diameter and $106.5$\,m in focal-length.
        The '3L' here means $3$ layers of nodes in the space-truss, as can be seen in Figure \ref{FigNameAcpTourImagingReflector}.
        Imaging-reflectors left of the vertical red line are considered to be sufficient to observe without actuating their mirror-facets to correct for deformations.
        The yellow star marks the lightest imaging-reflector which does not need to actuate its mirror-facets yet.
    }
    \label{FigSpirosParameterStudy}
\end{figure}

We implement a simulation to investigate the deformations of \NameAcp{}'s large imaging-reflector under various load-scenarios, as well as the impact of these deformations on the optical quality.
The simulation is implemented in equal parts by Spyridon Daglas, and the author of this thesis (S.A.M.).
We quantify the optical quality using the standard-deviation of the imaging-reflector's point-spread-function for incoming photons running parallel to the optical-axis.\\
The imaging-reflector is simulated in a fully parametrized way.
This allows us to investigate the optics when we change parameters such as the diameter, the focal-length, or the type of curvature.
But it also allows us to investigate the mechanics when we change parameters such as mirror-facet-size, space-truss-designs, and material-properties.\\
We also define load-scenarios.
We simulate pointing the cable-robot-mount to various zenith-distances.
We simulate constant wind, wind-gusts, and even earth-quakes.\\
For each tuple of load-scenario, and imaging-reflector:
\begin{enumerate}
\item We simulate the deformation of the imaging-reflector under gravity when pointing to zenith.
This is the default-load-scenario.
\item We estimate the optimal orientational fine-alignment for each mirror-facet under the default-load-scenario.
This is the default-alignment.
\item We point the imaging-reflector to its target zenith-distance, and simulate its deformation under this load-scenario.
\item We estimate the point-spread-function of the imaging-reflector while it is under the actual load-scenario, and while its mirror-facets have the default-alignment.
\end{enumerate}
With this procedure, and the point-spread-function as a feedback, Spyridon explores the parameter-space for different imaging-reflectors using a particle-swarm-optimization \cite{daglas2015master}.
He uses the estimated point-spread-function as a feedback for the next iteration of the imaging-reflector's parameters.
Figure \ref{FigSpirosParameterStudy} shows the parameter-search done by Spyridon for imaging-reflectors with a space-truss and tension-ring design.
The figure shows a search for an imaging-reflector with $71\,$m diameter of reflective surface.
However, the tension-ring-design made the space-truss-lattice more like $80\,$m in diameter, with an outer region of $\approx 5\,$m not covered by mirror-facets.
This solution was easy for us to implement in our first investigation, but is not recommended for the actual implementation.
For the time being, we consider it to be an additional safety-margin.
\section{Conclusion}
\label{SecCableRobotMountConclusion}
The cable-robot-mount for the $71\,$m \NameAcp{} Cherenkov-plenoscope can be build today.
Spyridon estimates that somewhere between aperture-diameters of $50$\,m to $70\,$m is the point where the cable-robot-mount takes over the altitude-azimuth-mount in cost-effectiveness.
Spyridon further estimates that a $100\,$m diameter Cherenkov-plenoscope is potentially in reach with the cable-robot-mount.\\
From Spyridon's point-of-view, the support-structure for the imaging-reflector is most worthy to investigate further.
The remaining components of the cable-robot-mount can be adopted from existing solutions in industry.\\
Spyridon finds that the space-truss-lattice is unnecessary dense when coupling it to 'small' $2\,$m$^2$ mirror-facets.
He proposes to use larger mirror-facets, or to have space-truss-layers of different densities.
A dense layer to mount the mirror-facets, and more coarse layers below.\\
Spyridon further finds confirmation for his tension-ring-design.
The outer realm of the space-truss shall be made stiffer than the light-weight inner part.\\
Spyridon also finds that the deformations in the imaging-reflector scale rapidly for the zenith-distance above $30^\circ$.
He estimates that a mount pointing to maximum zenith-distances of $30^\circ$ can be implemented with much simpler solutions.\\
For the materials, Spyridon finds that, except for the imaging-reflector, concrete and steel are the best choices.
Only for the support-structure of the imaging-reflector, carbon-fiber-reinforced-polymers are the better choice because of their stiffness.\\
To be clear, we only explore the cable-robot-mount for static load-scenarios.
Although we consider all states of operation slow enough to be static, further discussions with experts \cite{miermeister2016cablerobot} in robotics are needed.
%
%------------------------------------------------------------------------------
%
%
%
%
%
%
%
%------------------------------------------------------------------------------
\chapter{Estimating \NameAcp{}'s cost}
\label{ChEstimatingCost}
Naturally, the $71\,$m \NameAcp{} Cherenkov-plenoscope has its costs.
We estimate that \NameAcp{} will cost about $218.5\times10^6$\,CHF.\\
We split the cost-estimate into three parts:
First, the costs for the cable-robot-mount, which is mainly worked out by civil-engineer Spyridon Daglas \cite{daglas2015master}.
Second, the costs for the optics and electronics.
And third, the costs for organization and execution.
\section{Cable-robot-mount}
Spyridon estimates that the material needed for \NameAcp{}'s cable-robot-mount will cost about $34.0\times10^6$\,CHF\footnote{In his master-thesis \cite{daglas2015master}, Spyridon reports the total costs for the 71\,m Cherenkov-plenoscope to be $170\times10^6$CHF, where he assigns $34\times10^6$CHF (20\%) to the costs of the structure's materials.
In his master-thesis, Spyridon reports only $170\times10^6$CHF based on early cost-estimates provided by us for the electronics and optics.}.
This costs include the cable-robot-mount for the large imaging-reflector, its concrete-pillars, the actuated winches, the cables and the space-frame where the mirror-facets (not included) will be mounted on.
It also includes the cable-robot-mount for the light-field-sensor, its four high steel-frame-towers, the actuated winches, the cables, and the light-field-sensor's housing where the electronics (not included) will be mounted.
%
%For the raw materials, Spyridon uses the costs listed in Table \ref{TabCostsMountMaterial}.
%
%\begin{center}
%    \begin{table}
%        \begin{tabular}{lllr}
%            component & unit-costs & demand & total/$10^6$\,CHF\\
%            %
%            \toprule
%            %
%            Steel-tubes &
%            $1.8$\,CHF\,kg$^{-1}$ & & \\
%            % full dish is 57,700kg, of which 3650m^2 are mirror-facets
%            Carbon-fiber-tubes &
%            $9.0$\,CHF\,kg$^{-1}$ &
%            57,700\,kg &
%            $0.52$\\
%            %
%            Steel-cables &
%            $150$\,CHF\,m$^{-1}$ & & \\
%            %
%            Concrete &
%            $0.3$\,CHF\,kg$^{-1}$ & & \\
%            %
%            \bottomrule
%        \end{tabular}
%        \caption[Unit-costs for materials in the cable-robot-mount]{
%           Unit-costs for materials in the cable-robot-mount according to \%cite{daglas2015master}.
%           %
%           Steel-cables: Bridon-Bekaert, The Ropes group.
%           %
%           The carbon-fiber-tubes in the support-structure for the large %imaging-reflector are 57,700\,kg carrying an additional 73,000\,kg %of mirror-facets.
%        }
%        \label{TabCostsMountMaterial}
%    \end{table}
%\end{center}
%
\section{Optics and electronics}
We estimate that \NameAcp{}'s light-field-sensor will cost about $99.6\times10^6$\,CHF.
\NameAcp{}'s mirror-facets and actuators will cost about $13.0\times10^6$\,CHF.
Table \ref{TabCostsOpticsAndElectronics} lists our estimated costs for the  individual components.
Photo-sensors: Private communication with sales-engineer Marius Metzger from Hamamatsu-photonics.\\
Read-out-electronics: The CTA Schwarzschild-Couder-telescope reports a cost of 80\,CHF\,channel$^{-1}$ \cite{williams2015cherenkov}.
The developers of the Domino-Ring-Sampler-4 read-out-electronics even claim a cost of $10 - 15$\,CHF\,channel$^{-1}$ \cite{ritt2011design}.
However, this does not include the integration-costs necessary for Cherenkov-astronomy such as amplifiers and analog-to-digital-converters which is why we better assume 80\,CHF\,channel$^{-1}$ instead.
Lenses: Fortunately, UV-transparent silica-glass-optics are wide spread for lasers in industry.
We found offers for lenses with similar dimensions as shown in Figure \ref{FigSmallCameraGeometry} for about 50\,CHF.\\
Mirror-facets and actuators: Private communication with Adrian Biland expert for optics in MAGIC, and also involved in CTA.
%
% R11920-100-05 has 30mm diameter -> 706mm^2 -> 1414
% Spyros says: 170e6 in total. Sensor is 64e6, Mirrors were 20e6
%
% \cite{garczarczyk2015medium}
%
%To set this in perspective: Before this thesis, the only known way to have $30,000\,$m$^2$ collection-area for a gamma-ray-burst, was to have $30,000$ satellites like Fermi-LAT in orbit which costs about $30^13$ USD.
%
%
\begin{center}
    \begin{table}
        \begin{tabular}{lllr}
            component & unit-costs & demand & total/$10^6$\,CHF\\
            \toprule
            Photo-sensors &
            $5\times10^5$\,CHF\,m$^{-2}$ &
            115\,m$^2$ &
            $57.5$\\
            Read-out-electronics &
            80\,CHF\,channel$^{-1}$ &
            \NumLix{}\,channels &
            $41.2$\\
            Lenses &
            100\,CHF\,lens$^{-1}$ &
            \NumPix{}\,lenses &
            $0.9$\\
            Mirror-facets &
            $3\times10^3$\,CHF\,m$^{-2}$ &
            3,684\,m$^2$ &
            $11.1$\\
            Mirror-facet-actuators &
            $1\times10^3$\,CHF\,facet$^{-1}$ &
            1,842\,facets &
            $1.9$\\
            \bottomrule
            Sum & & & 112.6\\
        \end{tabular}
        \caption[Unit-costs for optics and electronics]{
            Unit-costs for optics and electronics.
        }
        \label{TabCostsOpticsAndElectronics}
    \end{table}
\end{center}
\section{Organization and execution}
\label{SecCostsEngineering}
As a guidance for the total costs of such a large project, we follow the cost-estimates for the OverWhelmingly-Large-telescope (OWL) proposed by the European-Southern-Observatory (ESO) \cite{aglea2004owl}, \cite{brunetto2004progress}.
After all, we want to estimate \NameAcp{}'s costs when erected in the harsh environment of Atacama-desert in Chile.
Table \ref{TabCostsEsoEstimate} shows how we distribute the costs.
\begin{table}
    \centering
    \begin{tabular}{lrr}
         & fraction/\% & total/$10^6$\,CHF\\
        \toprule
        Cable-robot-mount & 16 & 34.0\\
        Optics and electronics & 51 & 112.6\\
        Central control-system & 5 & 10.9\\
        Project-engineering & 5 & 10.9\\
        Project-management & 13 & 28.3\\
        Site-infrastructure & 10 & 21.8\\
        \bottomrule
        Sum & 100 & 218.5\\
    \end{tabular}
    \caption[Cost-distribution for \NameAcp{}, adopted from ESO, OWL]{
        \NameAcp{}'s costs-distribution.
        We contribute the estimate for the costs of the cable-robot-mount, the optics, and the electronics.
        The remaining matters of expenses are added according to ESO's cost estimates for OWL \cite{aglea2004owl}, \cite{brunetto2004progress}.
        Here our sum of the 'cable-robot-mount', and the 'Optics and electronics' give 67\% and corresponds to ESO's sum for 'opto-mechanical subsystems' (42\%), 'instrumentation' (5\%), and 'telescope structure' (20\%) which gives also 67\%.
    }
    \label{TabCostsEsoEstimate}
\end{table}
%
%------------------------------------------------------------------------------
%
%
%
%
%
%
%
%------------------------------------------------------------------------------
\chapter{Simulating \NameAcp{}'s observations}
\label{ChSimulation}
\newcommand{\McTracer}{\texttt{merlict}}
\newcommand{\Corsika}{\texttt{CORSIKA}}
\newcommand{\Coconut}{\texttt{coconut}}
\newcommand{\EventIo}{\texttt{eventio}}
We investigate the Cherenkov-plenoscope with a computer-simulation of the observations on the event to event level.
An event starts with the entrance of a cosmic particle into earth's atmosphere, and ends with a response of the Cherenkov-plenoscope.
We simulate the observation of gamma-rays, electrons or protons.
For the majority of the events, the response of the Cherenkov-plenoscope is to not even trigger.
But for those few events which pass the trigger, the response of the Cherenkov-plenoscope is further investigated.
For example we reconstruct the incident-direction of the cosmic particle.
In our simulations, events are independent of each other.
We use established \cite{bernlohr2013monte} tools such as \Corsika{} \cite{heck1998corsika} to simulate extensive air-showers created by cosmic particles in the earth's atmosphere and the IACT/ATMO-package \cite{bernlohr2008AtmoIact} to simulate the emission and interaction of Cherenkov-photons in the atmosphere.
When the Cherenkov-photons reach the Cherenkov-plenoscope on ground, our own simulation takes over.
Our photon-propagator \McTracer{} simulates the interactions of Cherenkov-photons with the scenery of the Cherenkov-plenoscope.
Together with the Cherenkov-photons, we inject night-sky-background-photons, see Section \ref{SecSimulatingNightSkyBackgroundPhotons}.
We simulate the photo-sensors, and the analog-to-digital-conversion.
For the analog-to-digital-conversion we simulate the shortcomings of the read-out-electronics in the First Geiger-mode Avalanche photo-diode Cherenkov Telescope (FACT) which we are familiar with, see Part \ref{PartPhotonStream} of this thesis.
Depending on the size of the Cherenkov-plenoscope, a single event involves from $10^0$ to $10^6$ Cherenkov-Photons, several $10^6$ night-sky-background-photons, $10^4$ to $10^6$ pieces of geometric primitives describing the Cherenkov-plenoscope, and $10^4$ to $10^6$ read-out-channels for the light-field-sensor.
Table \ref{TabSoftware} lists the different programs used to simulate \NameAcp{}.
\begin{center}
\begin{singlespace}
    \begin{table}[H]
        \begin{tabular}{p{8.5cm} p{4cm}}
            purpose & name \\
            \toprule
            Operating System & Linux, Ubuntu\,Server\,16.04\\
            \hline
            gamma and cosmic ray injection\newline Air-shower propagation & \Corsika{} 7.56\newline (UrQMD, QGSJETII) \\
            Cherenkov photon emission and atmospheric interaction & IACT/AMTO package\\
            \hline
            Photon propagation in Cherenkov plenoscope scenery & \McTracer{} \\
            light-field calibration & \McTracer{} \\
            Night sky background light injection  & \McTracer{} \\
            \hline
            Photo electric conversion  & \McTracer{} \\
            Electronic readout & \McTracer{} \\
            light-field trigger & \McTracer{} \\
            Event building and storage & \McTracer{} \\
            \hline
            Event reading and visualization & plenopy \\
            Event digital refocusing & plenopy \\
            3D Air-shower tomography & plenopy \\
            \bottomrule
        \end{tabular}
        \caption[Programs to simulate \NameAcp{}]{
            The programs to simulate \NameAcp{} Cherenkov-plenoscope.
            \McTracer{} and plenopy are both written by the author of this thesis (S.A.M.).
        }
        \label{TabSoftware}
    \end{table}
\end{singlespace}
\end{center}
\section{Simulating air-showers -- \Corsika{}}
We use the COsmic Ray SImulations for KAscade (\Corsika{}) to simulate extensive air-showers.
\Corsika{} is developed at Karlsruhe Institute of Technology (KIT).
Listing \ref{FigCorsikaBuildOptions} identifies our specific flavor of \Corsika{} used to simulate \NameAcp{}.
\begin{figure}
\begin{lstlisting}[language=C,basicstyle=\small\ttfamily]
#define HAVE_BERNLOHR 1
#define HAVE_DLFCN_H 1
#define HAVE_INTTYPES_H 1
#define HAVE_MEMORY_H 1
#define HAVE_STDINT_H 1
#define HAVE_STDLIB_H 1
#define HAVE_STRINGS_H 1
#define HAVE_STRING_H 1
#define HAVE_SYS_STAT_H 1
#define HAVE_SYS_TYPES_H 1
#define HAVE_UNISTD_H 1
#define LT_OBJDIR ".libs/"
#define PACKAGE "corsika"
#define PACKAGE_BUGREPORT ""
#define PACKAGE_NAME "corsika"
#define PACKAGE_STRING "corsika 75600"
#define PACKAGE_TARNAME "corsika"
#define PACKAGE_URL ""
#define PACKAGE_VERSION "75600"
#define STDC_HEADERS 1
#define VERSION "75600"
#define __ATMEXT__ 1
#define __BYTERECL__ 1
#define __CACHE_ATMEXT__ /**/
#define __CACHE_CEFFIC__ /**/
#define __CACHE_CERENKOV__ /**/
#define __CACHE_IACT__ /**/
#define __CACHE_KEEPSOURCE__ /**/
#define __CACHE_NOCOMPILE__ /**/
#define __CACHE_QGSJETII__ /**/
#define __CACHE_URQMD__ /**/
#define __CACHE_VOLUMEDET__ /**/
#define __CEFFIC__ 1
#define __CERENKOV__ 1
#define __GFORTRAN__ 1
#define __IACT__ 1
#define __NOCOMPILE__ 1
#define __OFFIC__ 1
#define __QGSII__ 1
#define __QGSJET__ 1
#define __SAVEDCORS__ 1
#define __TIMERC__ 1
#define __UNIX__ 1
#define __URQMD__ 1
#define __VOLUMEDET__ 1
\end{lstlisting}
\caption[\Corsika{} build-options, \textit{config.h}]{The \Corsika{} build-options used for the simulations of \NameAcp{}.
    This are the lines which are not commented out in \textit{config.h} created by \Corsika{}'s own build-environment \Coconut{}.
}
\label{FigCorsikaBuildOptions}
\end{figure}
For thread-safety during parallel deployment, we wrap \Corsika{} with a program that gives each \Corsika{}-instance its own 'run'-directory with its own local buffer-files.
Our wrapper also allows \Corsika{} to be called system-wide, and grants write-access to the \EventIo{}-output-files.
\section{Propagating photons -- \McTracer{}}
\label{SecMctracer}
We simulate the interaction of photons with the Chrenkov-plenoscope using a computer-simulation based on ray-tracing called \McTracer{}.
The \McTracer{} was written by the author of this thesis (S.A.M.) and can efficiently propagate photons in complex sceneries.
The \McTracer{} is not specifically written for the Cherenkov-plenoscope, but for arbitrary sceneries, and arbitrary optical instruments.
To propagate photons efficiently in complex sceneries with millions of surfaces, the \McTracer{} is a so called 'non-sequential' ray-tracer which represents its scenery using a bounding-volume-hierarchy to avoid unnecessary intersection-tests.
Currently, \McTracer{} can simulate specular reflections, absorptions in volumetric objects, refraction, as well as Fresnel-reflections and Fresnel-transmissions.
The \McTracer{} can read in sceneries represented using triangle-meshes in the Stereo-lithography (STL) format\cite{roscoe1988stereolithography} which is commonly used in Computer-Aided-Design (CAD).
Also the \McTracer{} can read in photons in various formats, including the \EventIo{}-format used in the IACT/ATMO-package \cite{bernlohr2008AtmoIact} for \Corsika{}.
To render a scenery fast for live previews, the \McTracer{} uses a second propagator for simple, reverse ray-tracing of so called 'camera-rays'.
Camera-rays do not behave like photons, but are often sufficient to render images of a scenery for a visual inspection of the geometry before the actual scientific propagation of photons.
For example, the Figures \ref{FigNameAcpTourOverview} to \ref{FigNameAcpTourLightFieldSensorFrontal} are rendered images created by the \McTracer{}.
The \McTracer{} has a moderate test-coverage of $\approx 450$ unit-tests and integration-tests.
These tests range from simple checks of the internal book-keeping and linear algebra to complex tests where e.g. photons are guided through lenses and certain point-spread-functions are expected.
\section{Simulating night-sky-background}
\label{SecSimulatingNightSkyBackgroundPhotons}
For \NameAcp{} we simulate the night-sky-background-photons of a dark night at a clear site adequate for optical astronomy.
We choose the Observatory on Roque de los Muchachos on Canary island La Palma, Spain to be representative for the flux of night-sky-background-photons.
Figure \ref{FigNightSkyBackgroundLaPalma} shows the flux of night-sky-background-photons used in the simulations of \NameAcp{}.
To reduce computations, we do not propagate each night-sky-background photon through the scenery of the \NameAcp{} Cherenkov-plenoscope every time we simulate the observation of an air-shower.
Instead we inject night-sky-background-photons directly into the photo-sensors with the appropriate rate and wavelength-distribution.
Not every photo-sensor in the light-field-sensor in \NameAcp{} has the same collection-efficiency $\overline{\eta}$, compare Figures \ref{FigLfgEtaMeanSensorPlane}, and \ref{FigLfgEtaMeanSensorPlaneXandY}.
We take this into account by using $\overline{\eta}$ from the light-field-geometry $G$ to inject the correct rate of collected photons into the individual photo-sensors.
This is possible because we know the area and the solid-angle which is exposed in the estimation of the light-field-geometry $G$, and because the collection-efficiency $\overline{\eta}$ is defined with respect to this exposure.
On Cherenkov-telescopes it is not common to take collection-efficiencies of individual photo-sensors into account, because fortunately on telescopes the variations in collection-efficiency between different photo-sensors is only minor.
\section{Ensuring large enough scatter-radii and scatter-angles}
\label{SecScatterRange}
To estimate \NameAcp{}'s instrument-response-function, all possible events of cosmic particles creating air-showers have to be simulated.
For each particles-type, the energy, the scatter-angle, and the scatter-radius have to be drawn randomly.
The scatter-angle here is the angle between the cosmic particle's trajectory and the optical-axis of \NameAcp{}.
The scatter-radius here is the distance between the center of \NameAcp{}'s principal-aperture-plane and the scatter-position of the cosmic particle.
While the scatter-position is where the elongated trajectory of the cosmic particle intersects the principal-aperture-plane.\\
However, since computing-resources are limited, we can not simulate every possible event.
We have to compromise and avoid the simulation of events which are both rare to occur in nature, and unlikely to trigger the read-out of \NameAcp{}.
In this section:
First, we show the compromises used in the current simulations of \NameAcp{}.
And second, we discuss what might be improved in future simulations.
\subsection*{Compromises in \NameAcp{}'s simulations}
\begin{figure}
    \centering
    \includegraphics[width=1\textwidth]{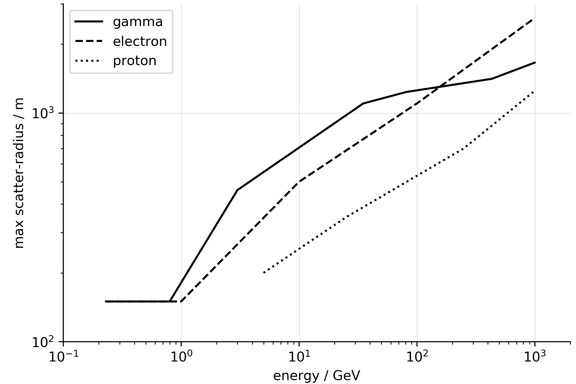}
    \caption[Maximum scatter-radius vs. energy]{The maximum scatter-radius of a particle versus its energy.
        Gamma-rays with parallel incident-directions coming from a point-source and electrons with scattered incident-directions coming from a diffuse pool.
        Protons also have scattered incident-directions coming from a diffuse pool.
        Since the detection-efficiency for small scatter-radii in the order of \NameAcp{}'s aperture-radius of $35.5\,$m is not yet explored, we better be safe and do not reduce the maximum scatter-radius below $150\,$m.
        The resulting instrument-response-functions for \NameAcp{} are shown in the Figures \ref{FigResponseGammaRays}, and \ref{FigResponseCosmicRays}.
    }
    \label{FigScatterRadiiVsEnergy}
\end{figure}

\begin{figure}
    \centering
    \includegraphics[width=1\textwidth]{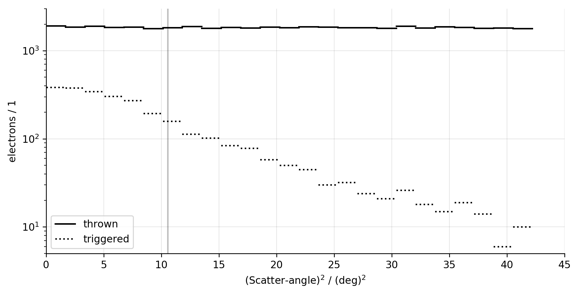}
    \includegraphics[width=1\textwidth]{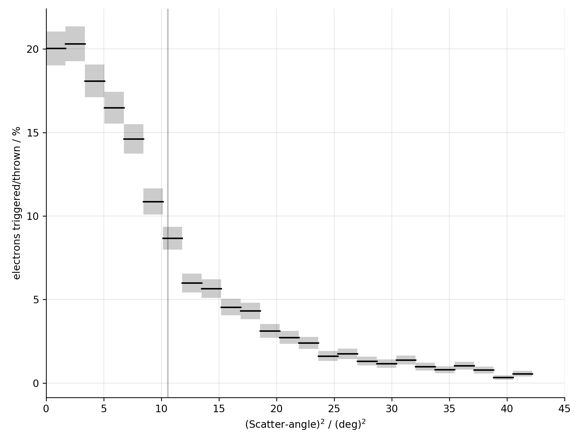}
    \caption[Scatter-angles of diffuse electrons and protons]{Absolute (upper), and relative (lower) distributions of squared
        scatter-angles of diffuse electrons, and positrons.
        Gray, vertical line indicates radius of \NameAcp{}'s field-of-view $(6.5^\circ/2)^2$.
        Integrated over all energies, and all scatter-radii.
    }
    \label{FigScatterAngleElectrons}
\end{figure}

\begin{figure}
    \centering
    \includegraphics[width=1\textwidth]{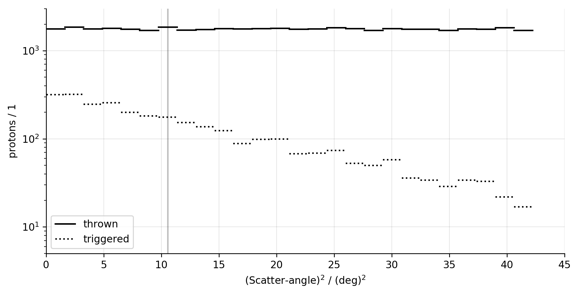}
    \includegraphics[width=1\textwidth]{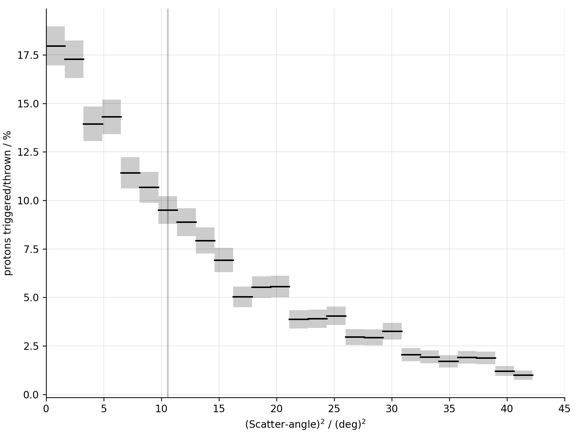}
    \caption[Scatter-angles of diffuse protons]{Absolute (upper), and relative (lower) distributions of squared
        scatter-angles for diffuse protons.
        Gray, vertical line indicates radius of \NameAcp{}'s field-of-view $(6.5^\circ/2)^2$.
        Integrated over all energies, and all scatter-radii.
    }
    \label{FigScatterAngleProtons}
\end{figure}

\begin{figure}
    \centering
    \includegraphics[width=1\textwidth]{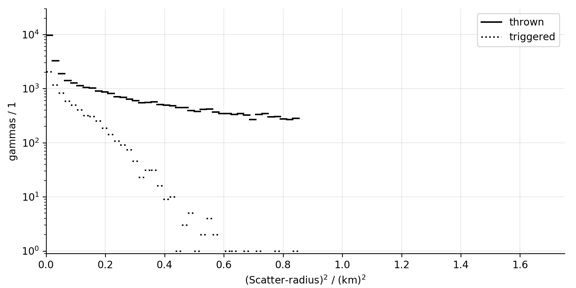}
    \includegraphics[width=1\textwidth]{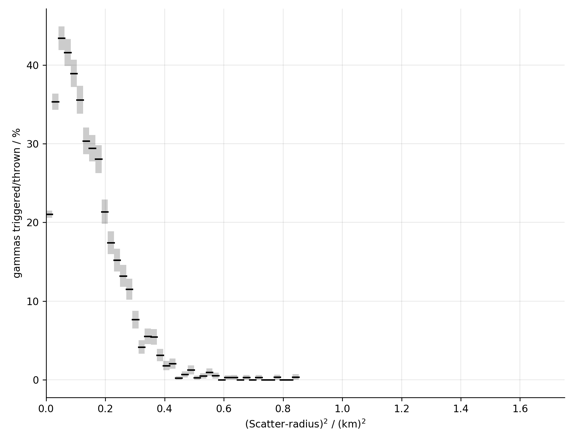}
    \caption[Scatter-radii of gamma-rays]{Absolute (upper), and relative (lower) distributions of squared
        scatter-radii of gamma-rays coming from a point-source in the center of the field-of-view.
        Integrated over all energies.
    }
    \label{FigScatterRadiusGammaRays}
\end{figure}

\begin{figure}
    \centering
    \includegraphics[width=1\textwidth]{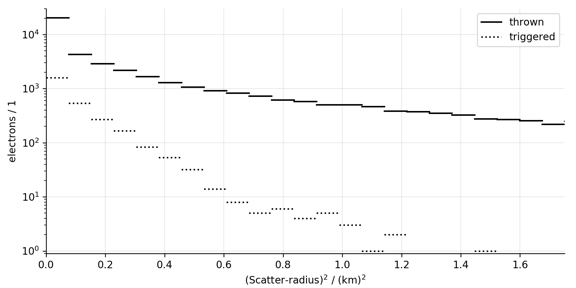}
    \includegraphics[width=1\textwidth]{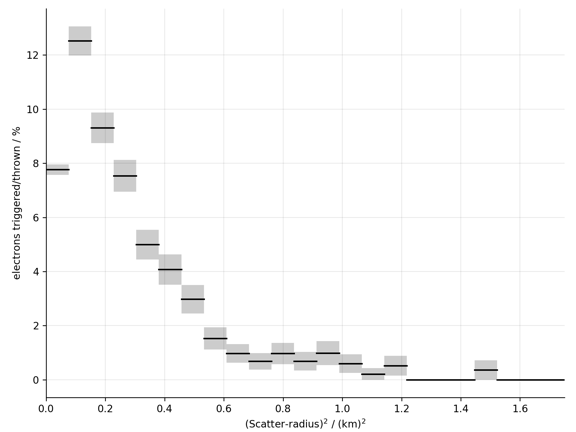}
    \caption[Scatter-radii of diffuse electrons and positrons]{Absolute (upper), and relative (lower) distributions of squared
        scatter-radii of diffuse electrons, and positrons.
        Integrated over all energies, and over all scatter-angles.
    }
    \label{FigScatterRadiusElectrons}
\end{figure}

\begin{figure}
    \centering
    \includegraphics[width=1\textwidth]{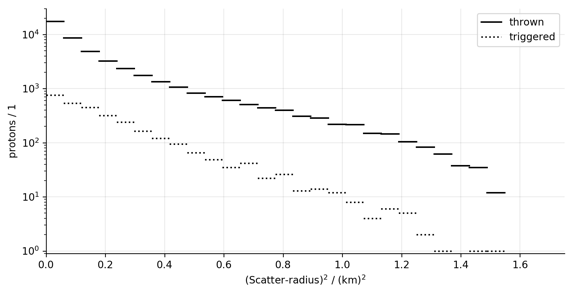}
    \includegraphics[width=1\textwidth]{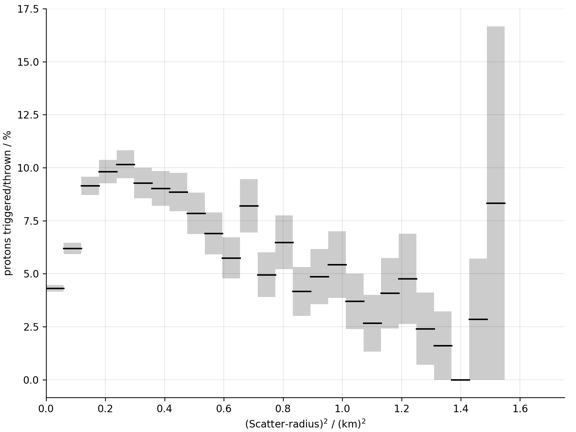}
    \caption[Scatter-radii of diffuse protons]{Absolute (upper), and relative (lower) distributions of squared
        scatter-radii of diffuse protons.
        Integrated over all energies, and over all scatter-angles.
    }
    \label{FigScatterRadiusProtons}
\end{figure}

To reduce the amount of simulations needed for this first estimate of \NameAcp{}'s performance, we apply three compromises:
\begin{itemize}
\item We only simulate gamma-rays, electrons, positrons and protons.
\item For the diffuse electrons, and protons we only draw scatter-angles within a maximum scatter-angle which is twice the radius of \NameAcp{}'s field-of-view.
\item Depending on the particle-type we only draw scatter-radii within a maximum scatter-radius which we make to depend on the energy of the particle, see Figure \ref{FigScatterRadiiVsEnergy}.
\end{itemize}
The Figures \ref{FigScatterAngleElectrons}, and \ref{FigScatterAngleProtons} show the distributions of the squared scatter-angles for diffuse electrons, and diffuse protons respectively.
Since the scatter-angles are drawn uniformly in the solid-angle of the opening-cone defined by the maximum scatter-angle, the distributions of the square of the thrown scatter-angles are flat in the Figures \ref{FigScatterAngleElectrons}, and \ref{FigScatterAngleProtons}.
We find the ratio of triggered over thrown events versus the squared scatter-angles to approach zero for larger scatter-angles.\\
The Figures \ref{FigScatterRadiusGammaRays}, \ref{FigScatterRadiusElectrons}, and \ref{FigScatterAngleProtons} show the distributions of the squared scatter-radii.
We find that unlike in the distributions for the scatter-angle, the ratio of triggered over thrown events does not peak at zero a scatter-radius of zero.
This is probably because of the extended area on the principal-aperture-plane which is illuminated by Cherenkov-photons produced by each air-shower.
We also find that all the ratios of triggered over thrown events approach zero for larger scatter-radii.
In the Figures \ref{FigScatterRadiusElectrons}, and \ref{FigScatterRadiusProtons} we find that for electrons the ratio of triggered over thrown events decreases faster for larger scatter-radii than for protons.
The non uniform absolute distributions of the scatter-radii for thrown events is the result of our energy depended choice of the maximum scatter-radius shown in Figure \ref{FigScatterRadiiVsEnergy}.
%
% Scatter-angle
% -------------
%
%
% Scatter-radii
% -------------
%
\subsection*{Conclusion and improvements for future simulations}
We identify the distributions for the ratios of triggered over thrown events as a benchmark for our choices of maximum scatter-angles, and maximum scatter-radii.
Ideally we want to see these distributions to vanish towards zero for larger scatter-angles, and scatter-radii.
In the current simulations, the vanishing for the ratio of triggered over thrown events is least well estimated for the scatter-radius of protons in Figure \ref{FigScatterRadiusProtons}.
But still, for all distributions the vanishing itself is clearly visible.\\
From this figures alone we do not see a method yet which allows us to conclude that the compromises made on the maximum scatter-angles and maximum scatter-radii do not spoil the instrument-response-functions too much.
In the end, it is also important how often the corresponding events occur in nature, what can not be concluded from the figures here.
The conclusion here is, that a more quantitative procedure based on the instrument-response-functions themselves is needed in the future to estimate this in more detail.\\
\section{Data Availability}%
\label{SecDataAvailability}
The source-code developed and used in this thesis is available on:
\begin{center}
    \begin{Large}
        \url{https://github.com/cherenkov-plenoscope}
    \end{Large}
\end{center}
%
%------------------------------------------------------------------------------
%
%
%
%
%
%
%
%------------------------------------------------------------------------------
%
\chapter{Comparing the plenoscope with other methods}
\label{ChComparingOtherMethods}
We compare how the Cherenkov-plenoscope samples the light-field in contrast to how existing methods, like the Cherenkov-telescopes, sample the light-field.
First, we introduce the light-field-sampling-graph to illustrate how each particular method samples the light-field.
And second, we list existing and future planned methods together with the Cherenkov-plenoscope to point out differences and similarities.
\section*{The light-field-sampling-graph}
\begin{figure}
    \centering
    \includegraphics[width=0.7\textwidth]{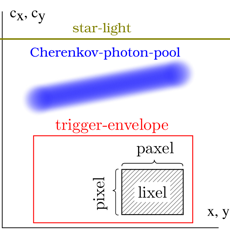}
    \caption[Light-field-sampling, legend]{
The light-field-sampling-graph.
Black boxes correspond to read-out-channels.
The vertical extension of the read-out-channel in $c_x, c_y$ represents the coverage of incident-directions.
This is often called the picture-cell, or \textit{pixel}.
The horizontal extension of the read-out-channel in $x, y$ represents the coverage of support-positions.
We call this the principal-aperture-cell, or \textit{paxel}.
The hatched area in the read-out-channel represents the coverage of the light-field.
We call this area the light-field-cell, or \textit{lixel}.
We put a red curve which illustrates the space enveloped and thus instantaneously accessible to the trigger.
The further a \textit{single} trigger-envelope is spread along the axis of support-positions, the larger is the collection-area for Cherenkov-photons, and thus the lower is the energy-threshold.
In yellow we show parallel star-light.
Star-light shows up as a horizontal line because it has only a single incident-direction in $c_x$, and $c_y$, but it shines on the entire aperture-plane in $x$, and $y$.
In blue we show a typical pool of Cherenkov-photons.
Cherenkov-photons from an air-shower are tilted in this graph to illustrate that their incident-directions correlate with their support-positions.
This is why stereo-observations with multiple telescopes see different images of the same air-shower.
    }
    \label{FigLightFieldSamplingLegend}
\end{figure}

The light-field-sampling-graph is a two-dimensional graph which shows the space of support-positions $x$, $y$ on the horizontal axis, and the space of incident-directions $c_x$, $c_y$ on the vertical axis, see Figure \ref{FigLightFieldSamplingLegend}.
We only show one observation-method at a time in the light-field-sampling-graph.
Our light-field-sampling-graph is not to scale, and is only meant as a guidance.
\newpage
\thispagestyle{plain} % empty
\mbox{}
%
%==============================================================================
\newpage
\section{Areal-sampling (sparse)}
\label{SecSamplingArealSparse}
\begin{figure}[H]
    \centering
    \includegraphics[width=0.6\textwidth]{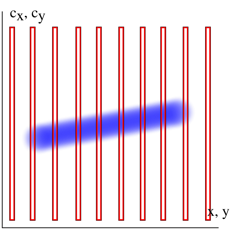}
    \begin{tabular}{lrr}
        % up to 25deg zd-> 50deg fov
        % 1000m^2 / 133 sensors -> 85m sensor distances
        & pixel & paxel \\
        \toprule
        number & 1 & 133\\
        coverage & $\approx 2000$\,deg$^2$ & $\approx 0.4\,$m$^2$ \\
        spacing & -- & 85\,m\\
        \bottomrule
    \end{tabular}
    \caption[Light-field-sampling, areal-sampling (sparse)]{
        Tunka-133 \cite{lubsandorzhiev2008tunka}, compare Figure \ref{FigTunka}.
    }
    \label{FigArealSamplingSparse}
\end{figure}
\begin{figure}[H]
    \centering
    \includegraphics[width=0.6\textwidth]{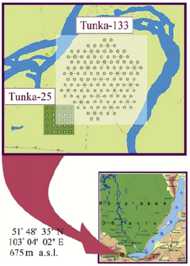}
    \includegraphics[width=0.7\textwidth]{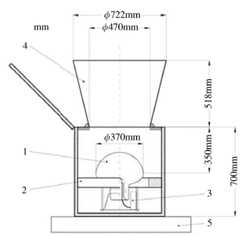}
    \caption[Tunka-133, overview]{
        Figures taken from \cite{lubsandorzhiev2008tunka}.
        The Tunka-133 array for sparse areal-sampling in Tunka-valley.
        A total of 133 large photo-sensors (1) with light-guides (4) (lower figure) are arranged in a hexagonal grid in the Tunka-valley (upper figure).
    }
    \label{FigTunka}
\end{figure}

Sparse arrays of photo-sensors on ground are used to reconstruct the incoming Cherenkov-light-front based on the different arrival-times of the photons at different support-positions.
There is no imaging-optics.
The whole wide field-of-view is just one pixel.
See in Figure \ref{FigArealSamplingSparse} the read-out-channels being wide extended in $c_x$, $c_y$, but being narrow and sparse in $x$, $y$.
The lack of sampling in incident-directions $c_x$, $c_y$ limits the power to tell apart gamma-rays from cosmic-rays.
Because of the wide spacing of support-positions, the trigger has only access the collection-area of single photo-sensors.
This raises the energy-threshold up to $\approx 10^{5}\,$GeV.
% {lubsandorzhiev2008tunka,Introduction}
Yet the sparse design allows for large collection-areas at energies close to the 'knee' at $3 \times 10^{6}\,$GeV.
Implementations are Tunka-133 \cite{lubsandorzhiev2008tunka} in Figure \ref{FigTunka}, and AIROBICC \cite{karle1995design}.
%-----------
%
%==============================================================================
\newpage
\section{Areal-sampling (dense)}
\label{SecSamplingArealDense}
\begin{figure}[H]
    \centering
    \includegraphics[width=0.6\textwidth]{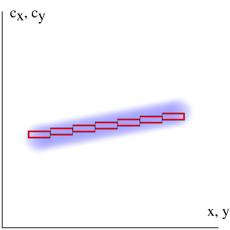}
    \begin{tabular}{lrr}
        & pixel & paxel \\
        \toprule
        number & $\approx 1$ & 48\\
        coverage & $\approx 49\times10^{-3}$deg$^{2}$ & $36\,$m$^2$ \\
        spacing & -- & $\approx 10\,$m \\
        \bottomrule
    \end{tabular}
    \caption[Light-field-sampling, areal-sampling (dense)]{
        STACEE-48 \cite{chantell1998prototype,covault2001status}, compare \ref{FigSamplingStacee}.
        %number $c_x$, $c_y$ bins:\hfill$1$\\
        %number $x$, $y$ bins:\hfill$48$\\
        %solid-angle $c_x$, $c_y$ bin:\hfill$ \approx 49\,$m\,sq\,deg\\
        %area $x$, $y$ bin:\hfill$36\,$m$^2$\\
        %spacing $x$, $y$ bins: \hfill $\approx 10\,$m\\
        %spacing $c_x$, $c_y$ bins: \hfill --
    }
    \label{FigArealSamplingDense}
\end{figure}
\begin{figure}[H]
    \centering
    \includegraphics[width=0.75\textwidth]{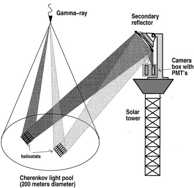}
    \includegraphics[width=1.\textwidth]{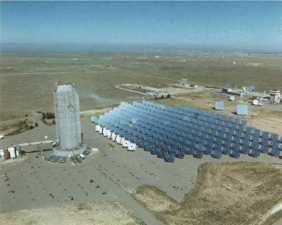}
    \caption[STACEE, overview]{
        Figures taken from \cite{covault2001status, chantell1998prototype}.
        The STACEE solar-concentrator for dense areal-sampling.
        The upper drawing shows the canting of the mirror-facets.
        Cherenkov-light reaches different mirrors (heliostats) on ground but was produced in the same point.
    }
    \label{FigSamplingStacee}
\end{figure}

A dense array of tiltable mirrors on ground reflects the light from a narrow region of incident-directions onto an array of photo-sensors mounted on a nearby tower.
Each mirror reflects its light to a different photo-sensor.
This allows to densely sample the support-positions.
Often \cite{chantell1998prototype} the mirror-facets are slightly canted so that each mirror reflects light from a slightly different incident-direction.
This is done to focus onto a certain object-distance $\approx 10\,$km above ground where Cherenkov-photons are most likely to be produced.
Pointing dependend time-delays between the read-out-channels allow the trigger to access all mirrors which lowers the energy-threshold.
Figure \ref{FigArealSamplingDense} shows the read-out-channels for dense areal-sampling.
Note the canting which makes the read-out-channels follow the shape of the Cherenkov-photon-pool.
Figure \ref{FigSamplingStacee} shows the STACEE solar-concentrator.
Similar implementations are CELESTE \cite{pare2002celeste}, and CACTUS \cite{lizarazo2006data}.
%
%-----------
%
%==============================================================================
\newpage
\section{Telescope}
\label{SecSamplingTelescope}
\begin{figure}[H]
    \centering
    \includegraphics[width=0.6\textwidth]{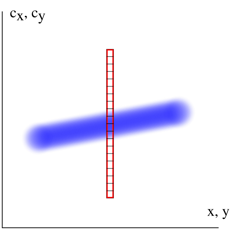}
    \begin{minipage}{0.45\textwidth}
        \begin{tabular}{lrr}
            & pixel & paxel \\
            \toprule
            number & $\approx 250$ & 1\\
            coverage & $\approx 49\times10^{-3}$\,deg$^2$ & $80\,$m$^2$ \\
            spacing & dense, 0.25\,deg & -- \\
            \bottomrule
        \end{tabular}
    \end{minipage}
    \hspace{0.5cm}
    \begin{minipage}{0.45\textwidth}
        \caption[Light-field-sampling, telescope]{
            Whipple-telescope \cite{kildea2007whipple}, Mount-Hopkins 2,600\,m a.s.l, Arizona.
            Here T.\,Weeks, A.\,M.\,Hillas, and colleagues started a new field.
            %number $c_x$, $c_y$ bins:\hfill$\approx 250$\\
            %number $x$, $y$ bins:\hfill$1$\\
            %solid-angle $c_x$, $c_y$ bin:\hfill$49\,$m\,sq\,deg\\
            %area $x$, $y$ bin:\hfill$80\,$m$^2$\\
            %spacing $x$, $y$ bins: \hfill --\\
            %spacing $c_x$, $c_y$ bins: \hfill dense, 250\,m\,deg
        }
        \label{FigSamplingTelescope}
    \end{minipage}
\end{figure}
\begin{figure}[H]
    \centering
    \includegraphics[width=0.7\textwidth]{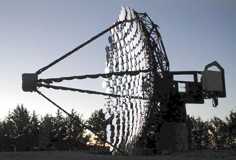}
    \caption[Whipple Cherenkov-telescope]{
        Figure taken from \cite{kildea2007whipple}.
        The pioneer of the imaging-atmospheric-Cherenkov-method.
        The Whipple Cherenkov-telescope on Mount Hopkins had $10\,$m effective aperture-diameter.
    }
    \label{FigWhipplePhotograph}
\end{figure}
\begin{figure}[H]
    \centering
    \includegraphics[width=1\textwidth]{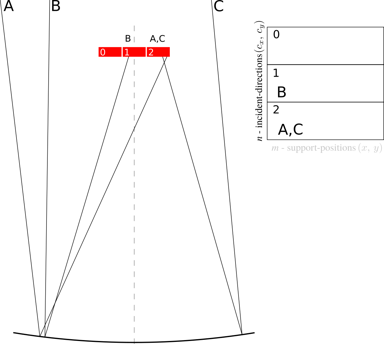}
    \caption[Optics of the telescope]{
        A telescope observing the thee photons A,B, and C.
        Compare the telescope with the plenoscope in the Figures \ref{FigOpticsOverview}, and \ref{FigOpticsOverviewCloseUp}.
    }
    \label{FigOpticsOverviewTelescope}
\end{figure}

The telescope collects light only from a small region of support-positions.
An imaging-system maps the photons based on their incident-directions to different positions on an array of photo-sensors which is called image-sensor.
Telescopes have fields-of-views large enough to comfortably contain the photons produced in one air-shower.
A high resolution in incident-directions allows the telescope to tell apart gamma-rays from cosmic-rays based on the structure of the Cherenkov-photons in the space of incident-directions (in the image) \cite{hillas1985cerenkov}.
Figure \ref{FigSamplingTelescope} shows how the telescope samples the light-field.
The Whipple-telescope in Figure \ref{FigWhipplePhotograph} pioneered the method, and many successors improved it ever since.
Figure \ref{FigOpticsOverviewTelescope} shows the optics of Cherenkov-telescopes using the same three photons A,B, and C which are in the Figures \ref{FigOpticsOverview}, and \ref{FigOpticsOverviewCloseUp}.
Similar implementations are CAT-France \cite{barrau1998cat}, FACT \cite{anderhub2013design}, and all the Cherenkov-telescopes used in telescope-arrays, see Section \ref{SecSamplingTelescopeArraySparse}.
%
%-----------
%
%==============================================================================
\newpage
\section{Telescope-array (sparse)}
\label{SecSamplingTelescopeArraySparse}
\begin{figure}[H]
    \centering
    \includegraphics[width=0.6\textwidth]{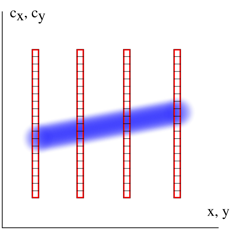}
    \begin{minipage}{0.45\textwidth}
        \begin{tabular}{lrr}
            & pixel & paxel \\
            \toprule
            number & $\approx 250$ & $\approx 4$\\
            coverage & $\approx 49\times10^{-3}$deg$^{2}$ & $10\,$m$^2$ \\
            spacing & dense, 0.25\,deg & 70\,m \\
            \bottomrule
        \end{tabular}
    \end{minipage}
    \hspace{0.5cm}
    \begin{minipage}{0.45\textwidth}
        \caption[Light-field-sampling, telescope-array (sparse)]{
            HEGRA \cite{bulian1998characteristics}, Roque de los Muchachos 2,200\,m a.s.l, Canary island La Palma, Spain.
            %number $c_x$, $c_y$ bins:\hfill$\approx 250$\\
            %number $x$, $y$ bins:\hfill$4$\\
            %solid-angle $c_x$, $c_y$ bin:\hfill$49\,$m\,sq\,deg\\
            %area $x$, $y$ bin:\hfill$10\,$m$^2$\\
            %spacing $x$, $y$ bins: \hfill 70\,m\\
            %spacing $c_x$, $c_y$ bins: \hfill dense, 250\,m\,deg
        }
        \label{FigSamplingCherenkovTelescope}
    \end{minipage}
\end{figure}
About four Cherenkov-telescopes are sparsely positioned on ground such that they can collect photons from the same air-shower.
The array of telescopes can sample the incident-directions like a usual telescope.
But at the same time it can sparsely sample the support-positions.
The array of telescopes is very powerful and cost-effective to observe gamma-rays with energies above 100\,GeV.
Unfortunately, the energy-threshold for the trigger in an array of telescopes does not go below the energy-threshold for the trigger of the individual telescope.
Figure \ref{FigSamplingCherenkovTelescope} shows how a sparse array of Cherenkov-telescopes samples the light-field.
Implementations are H.E.S.S. \cite{funk2004trigger}, MAGIC \cite{garcia2014status}, VERITAS \cite{weinstein2007veritas}, and the upcoming CTA \cite{cta2013introducing} and TAIGA \cite{budnev2017taiga}.
%
%-----------
%
%==============================================================================
\newpage
\section{Meridian-telescope}
\label{SecSamplingMeridianTelescope}
\begin{figure}[H]
   \centering
    \includegraphics[width=0.6\textwidth]{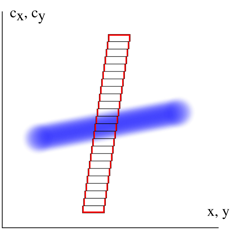}
    \begin{minipage}{0.45\textwidth}
        \begin{tabular}{lrr}
            & pixel & paxel \\
            \toprule
            number & $\approx 15,000$ & $\approx 1$\\
            coverage & $\approx 20\times10^{-3}$deg$^{2}$ & $177\,$m$^2$ \\
            spacing & dense, 0.1\,deg & --\\
            \bottomrule
        \end{tabular}
    \end{minipage}
    \hspace{0.5cm}
    \begin{minipage}{0.45\textwidth}
        \caption[Light-field-sampling, meridian-telescope]{
            MACHETE \cite{cortina2016}, see also Figures \ref{FigMacheteDesign}, and \ref{FigMacheteOptics}.
        }
        \label{FigSamplingMeridianTelescope}
    \end{minipage}
\end{figure}
\begin{figure}[H]
    \centering
    \includegraphics[width=1.\textwidth]{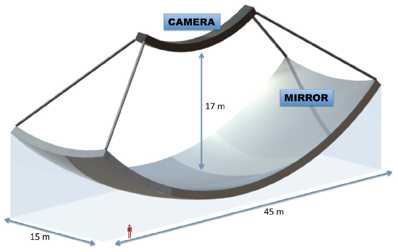}
    \caption[MACHETE, meridian-Cherenkov-telescope]{
        Figure taken from \cite{cortina2016}.
        The meridian-telescope MACHETE is optimized to nightly monitor a large fraction of the gamma-ray-sky at energies below 1\,TeV down to 150\,GeV.
    }
    \label{FigMacheteDesign}
\end{figure}
\begin{figure}[H]
    \centering
    \includegraphics[width=0.6\textwidth]{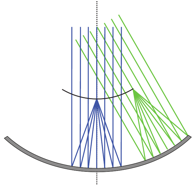}
    \caption[MACHETE, optics]{
        Figure taken from \cite{cortina2016}.
        MACHETE uses a spherical imaging-reflector and an image-sensor which only accepts photons in a limited range of incident-directions relative to its surface-normal.
        This is why different photo-sensors in the image-sensor collect photons from different parts of the imaging-reflector which leads to the slight skew in the light-field-sampling-graph in Figure \ref{FigSamplingMeridianTelescope}.
    }
    \label{FigMacheteOptics}
\end{figure}

The meridian-telescope is intended to observe a large part (60\,deg $\times$ 5\,deg) of the sky.
Facing zenith all the time, the meridian-telescope waits for the sources in the sky to pass by its field-of-view while the earth is rotating.
The meridian-telescope samples the light-field in a similar way to the telescope.
But in the meridian-telescope different incident-direction-bins (pixels) are exposed to different parts of the aperture.
This creates a slight skew in the light-field-sampling-graph in Figure \ref{FigSamplingMeridianTelescope}.
The slight skew in the light-field-sampling-graph does not reduce the performance of the meridian-telescope, but it is a unique quirk which put the meridian-telescope into this list for comparison.
The Figures \ref{FigMacheteDesign}, and \ref{FigMacheteOptics} show the MACHETE meridian-telescope.
MACHETE is not yet build.
%
%-----------
%
%
%==============================================================================
\newpage
\section{Large telescope}
\label{SecSamplingLargeTelescope}
\begin{figure}[H]
   \centering
    \includegraphics[width=0.6\textwidth]{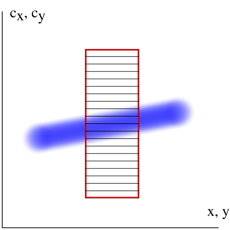}
    \begin{minipage}{0.45\textwidth}
        \begin{tabular}{lrr}
            & pixel & paxel \\
            \toprule
            number & $\approx 2,048$ & $\approx 1$\\
            coverage & $\approx 3.6\times10^{-3}$deg$^{2}$ & $614\,$m$^2$ \\
            spacing & dense, 0.067\,deg & --\\
            \bottomrule
        \end{tabular}
    \end{minipage}
    \hspace{0.5cm}
    \begin{minipage}{0.45\textwidth}
        \caption[Light-field-sampling, large telescope]{
            H.E.S.S. II \cite{cornils2005optical}, see also Figure \ref{FigHessIi}.
        }
        \label{FigSamplingLargeTelescope}
    \end{minipage}
\end{figure}
Large telescopes are like telescopes but with a wider sampling of support-positions.
Since there is only one bin to sample the support-positions in $x$, $y$, the support-position of the individual photon-trajectory is not well known.
In Figure \ref{FigSamplingLargeTelescope} we see that different read-out-channels sample light from different parts of the large telescope's aperture.
The slope of the Cherenkov-pool in Figure \ref{FigSamplingLargeTelescope} indicates that different parts of the wide aperture in $x$, and $y$ create different images.
This is the reason for the narrow depth-of-field on large telescopes.
Large telescopes investigate focusing \cite{trichard2015enhanced} to minimize the blurring caused by their narrow depth-of-field.
The energy-threshold of the trigger can go below $50\,$GeV for large telescopes, what makes them today the best choice for observations of low energetic gamma-rays on ground.
Similar proposals are \mbox{ECO-1000} \cite{baixeras2004}, Large-Size-Telescopes in the Cherenkov-Telescope-Array \cite{cta2013introducing}, and MACE \cite{borwankar2016simulation}.
%
%-----------
%
%
%==============================================================================
\newpage
\section{Telescope-array with topological trigger}
\label{SecSamplingTelescopeArrayTopoTrigger}
\begin{figure}[H]
   \centering
    \includegraphics[width=0.6\textwidth]{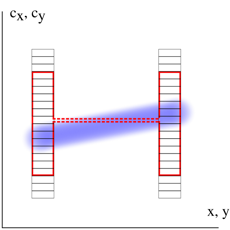}
    \begin{minipage}{0.45\textwidth}
        \begin{tabular}{lrr}
            & pixel & paxel \\
            \toprule
            number & $\approx 1,200$ & $2$\\
            coverage & $\approx 7.9\times10^{-3}$deg$^{2}$ & $226\,$m$^2$ \\
            spacing & dense, 0.1\,deg & 70\,m\\
            \bottomrule
        \end{tabular}
    \end{minipage}
    \hspace{0.5cm}
    \begin{minipage}{0.45\textwidth}
        \caption[Light-field-sampling, telescope-array with topological trigger]{
            MAGIC I and II with topological trigger \cite{lopez2016topo}, see also Figures \ref{FigMagic}, and \ref{FigMagicTopoTrigger}.
        }
        \label{FigSamplingTelescopeArrayTopoTrigger}
    \end{minipage}
\end{figure}
The topological trigger tries to make the the decision to read-out or not to read-out the array of telescopes as early as possible by combining as much information from multiple telescopes as technology allows.
The information about the arrival of photons is reduced by lowering the resolution of incident-directions and time.
Further, time-delays are applied depending of the pointing of the individual telescopes.
Even with these reductions, current topological triggers can still not handle the flood of information to be the first stage of the trigger in a telescope-array.
There is still a first layer of the trigger which only acts on the individual telescope.
In simulations, the topological trigger for telescope-arrays shows to lower the energy-threshold for gamma-rays.
Although current implementations of topological triggers are held back by technology, the red curve of the trigger-envelope in the light-field-sampling-graph indicates the intended goal which is spreading the aperture-area of the trigger onto multiple telescopes.
Similar work is done in VERITAS \cite{schroedter2009topological}.
%
%-----------
%
\begin{figure}[H]
    \centering
    \includegraphics[width=0.9\textwidth]{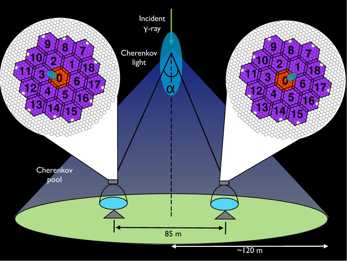}
    \caption[MAGIC topological trigger]{
        Figure taken from \cite{lopez2016topo}.
        Schematics of the topological trigger in the MAGIC telescope-array.
        Based on the pointing of the incident-direction of the gamma-ray and the location of the telescopes, the topological trigger expects certain regions in the images of both telescopes to be correlated to the air-shower.
        Current implementation act on coarse regions composed from small pixels to reduce the demands for processing.
    }
    \label{FigMagicTopoTrigger}
\end{figure}
%==============================================================================
\newpage
\section{Telescope-array (dense)}
\label{SecSamplingTelescopeArrayDense}
\begin{figure}[H]
   \centering
    \includegraphics[width=0.6\textwidth]{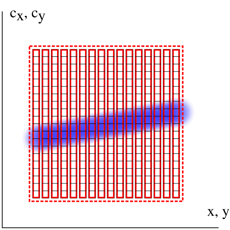}
        \begin{tabular}{lrr}
            & pixel & paxel \\
            \toprule
            number & $\approx 1,280$ & $400$\\
            coverage & $\approx 15.3\times10^{-3}$deg$^{2}$ & $5\,$m$^2$ \\
            spacing & dense, 0.14\,deg & dense 2.5\,m\\
            \bottomrule
        \end{tabular}
        \caption[Light-field-sampling, telescope-array with topological trigger]{
            The STAR \cite{jung2005star} proposal, see also Figure \ref{FigTelescopeArrayDenseStar}.
        }
        \label{FigSamplingTelescopeArrayDense}
\end{figure}
A dense array of telescopes can sample the light-field densely in both incident-directions $c_x$, $c_y$ and support-positions $x$, $y$.
It can sample the light-field in a very similar way as the plenoscope, \cite{wilburn2005high}.
The STAR-proposal \cite{jung2005star} pointed out the need for a trigger which can act instantaneously on all the apertures of the individual telescopes.
In STAT, possible solutions like time-delays which depend on the pointing of the telescopes were investigated.
The trigger-envelope intended by the STAR-proposal is shown by the dashed, red line in the light-field-sampling-graph in Figure \ref{FigSamplingTelescopeArrayDense}.
However, we added the many but much smaller trigger-envelopes of each individual telescope which we think are possible to implement today.
For a low energy-threshold of the trigger, it is best to have the read-out-channels, which belong to similar incident-directions, physically close together.
But in a telescope-array this is not the case which confronts proposals like STAR with huge technological challenges.
Somewhat related are the Durham telescopes Mrk\,III and IV in Narrabri \cite{brazier1990narrabri}.
%
%-----------
%
\begin{figure}[H]
    \centering
    \includegraphics[width=1.\textwidth]{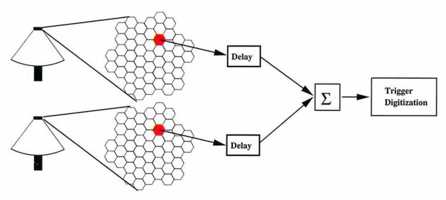}
    \caption[STAR telescope-array, schematics]{
        Figure taken from \cite{jung2005star}.
        A schematics of the STAR-proposal illustrating the important flow of signals coming from near-by pixels in different telescopes.
        On the left are two telescopes of the array.
        Further to the right are hexagonal grids which illustrate the two images recorded by the two telescopes.
        Now the signal-flow of the trigger is shown as it tries to make the trigger-decision based on the sum of pixels which are near to each other in the space of incident-directions, but come from different telescopes.
        The signal-flow here is only shown for one incident-direction, see the red pixels.
        The signals from the different telescopes need to be delayed depending of the pointing of the array before the sum of near-by pixels is created.
        The problem illustrated by this simple but brilliant figure turned out to be a huge technological challenge.
    }
    \label{FigTelescopeArrayDenseStar}
\end{figure}
%==============================================================================
\newpage
\section{Plenoscope}
\label{SecSamplingPlenoscope}
\begin{figure}[H]
   \centering
    \includegraphics[width=0.6\textwidth]{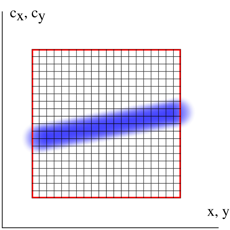}
        \begin{tabular}{lrr}
            & pixel & paxel \\
            \toprule
            number & $ \NumPix{} $ & \NumPax{}\\
            coverage & $3.6\times10^{-3}$deg$^{2}$ & $57\,$m$^2$ \\
            spacing & dense, 0.067\,deg & dense 8\,m\\
            \bottomrule
        \end{tabular}
        \caption[Light-field-sampling, telescope-array with topological trigger]{
            The \NameAcp{} Cherenkov-plenoscope.
        }
        \label{FigSamplingPlenoscope}
\end{figure}
The plenoscope can sample the light-field densely in both incident-directions $c_x$, $c_y$ and support-positions $x$, $y$, see Figure \ref{FigSamplingPlenoscope}.
It can sample the light-field in the same way as the dense telescope-array.
However, in the plenoscope all the photo-sensors which belong to similar incident-directions are physically close together, and do not need variable time-delays to compensate different pointing-directions.
This allows the plenoscope to have a trigger which acts on its full aperture to reach the lowest possible energy-threshold for gamma-rays.
The plenoscope is able to sample the entire (plenum) state of Cherenkov-photons relevant for the reconstruction of air-showers (support-positions, incident-directions, and arrival-times).
The sampling of the light-field provided by the plenoscope shown in Figure \ref{FigSamplingPlenoscope} is the best possible result which can be reached in the light-field-sampling-graph.
\part[Every single photon counts]
     {Every single photon counts\\[\bigskipamount]
      \large Pushing Gamma-Ray-Astronomy on Cherenkov-Telescopes by sensing Light in the Quantum-Regime}

\label{PartPhotonStream}
\chapter{Introduction}
We identify Cherenkov-telescopes to have unused potential.
First, we find that established Cherenkov-telescopes are limited by the way they represent their recorded air-shower-events.
And second, we find that the image-sensors of future Cherenkov-telescopes can be build more cost-effectively by going to a true digital read-out and leaving all analog processing behind.\\
Established Cherenkov-telescopes represent air-shower-events not by describing observables of photons, but by describing electric-pulses which are the responses of specific photo-sensors to these photons.
Established Cherenkov-telescopes have photo-sensors which respond with analog pulses to incoming photons where the shape and the arrival-time of the pulse need to be interpreted.
Such a representation of air-shower-events based on analog pulses limits the classification of Cherenkov-photons and night-sky-background-photons which prevents the detection of air-showers induced by cosmic particles with lower energies.
Detecting air-showers induced by cosmic particles with lower energies is of great interest.
It means either, that the costly aperture-area for Cherenkov-photons of future telescopes can be reduced while recording particles with the same energies, or that future telescopes will have more statistics of gamma-rays when recording the more abundant gamma-rays with lower energies.\\
Here we propose a more natural representation of air-shower-events to overcome present limitations.
We propose to break with describing the largest electric pulses but to describe the recorded air-shower-event as what it naturally is: The arrival-times of individual photons in the pixels of an image-sensor.
%
%We propose to leave all analog processing and cabling of established image-sensors behind and build a true digital image-sensor which replaces the quasi-digital-counter for photons with a true digital parallel-counter and outputs the arrival-times of individual photons in its pixels instead of vague pulses.
%
In this part of the thesis:
\begin{itemize}
\item We discuss the established procedure to detect gamma-rays with Cherenkov-telescopes on the example of the FACT Cherenkov-telescope.
\item We discuss the established representation of air-shower-events which is based on the largest pulses found on analog time-series.
\item We propose a natural representation for air-shower-events which is based on observables and not on the specific responses of specific sensors.
\item We implement our natural representation for air-shower-events on FACT and call it photon-stream.
\item We quantify the gain in performance for the classification of Cherenkov-photons when using the natural representation with the photon-stream over the established representation of largest-pulses.
\item We estimate the performance and limitations of our photon-stream on both simulations and observations of FACT.
\item Finally, we propose a digital read-out for Cherenkov-telescopes that does not need costly processing and routing of analog signals, and that records the natural photon-stream directly.
\end{itemize}
%
%=============================================================================
%
%  WWWWWWWWWW  WWWWWWWWW  WWWWWWWWWW  WWWWWWWWWW
%  WW          W       W  WW             WWW
%  WW          W       W  WW             WWW
%  WW          W       W  WW             WWW
%  WWWWWWWW    W       W  WW             WWW
%  WW          WWWWWWWWW  WW             WWW
%  WW          W       W  WW             WWW
%  WW          W       W  WW             WWW
%  WW          W       W  WW             WWW
%  WW          W       W  WW             WWW
%  WW          W       W  WW             WWW
%  WW          W       W  WWWWWWWWWW     WWW
%
%
\chapter{FACT -- A demonstration}
\label{ChFactDemonstrator}
\begin{figure}
    \begin{center}
        \includegraphics[width=1.0\textwidth]{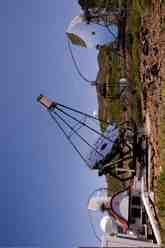}
        \caption[FACT on Rouque\,de\,los\,Muchachos]{FACT during the night on Rouque\,de\,los\,Muchachos 2,200\,m above sea-level.
            The two large Cherenkov-telescopes MAGIC$\,$I (left) and MAGIC$\,$II (right) can be seen in the background.
            Image by Thomas Kraehenbuehl.
        }
        \label{FigFactOverview}
    \end{center}
\end{figure}

The First Geiger-mode-Avalanche-photo-diode Cherenkov-Telescope (FACT) is a small Cherenkov-telescope on Canary island La Palma, Spain, see Figures \ref{FigFactOverview}, and \ref{FigSchematicsOverview}.
FACT demonstrates the usage of novel photo-sensors based on silicon outside the lab, but in the field.
FACT monitors bright sources of cosmic gamma-rays such as Markarian\,421 and Markarian\,501.
FACT also monitors the steady Crab nebula to validate that its own sensitivity does not degrade over time.
Sensing photons with Silicon-Photo-Multipliers (SiPMs)\footnote{%
Originally called Geiger-mode-Avalanche-Photo-Diode (GAPD).
Nowadays often called Silicon-Photo-Multiplier (SiPM), or Single-Photon-Avalanche-Diode (SPAD).
} instead of Photo-Multiplier-Tubes (PMTs) is a novelty \cite{FACT_design} for Cherenkov-telescopes.
FACT has a $9.5$m$^2$ aperture for Cherenkov-photons provided by a segmented imaging-reflector with $4.889\,$m focal-length.
The image-sensor of FACT has $4.5^\circ$ field-of-view, and has 1440 hexagonal pixels with $\approx 0.1^\circ$ field-of-view each, see Figure \ref{FigSchematicsImageSensor}.
As established Cherenkov-telescopes, FACT reconstructs the direction, the energy and the type of individual cosmic particles from air-shower-events to do gamma-ray-astronomy.\\
We take a close look at FACT and the imaging-atmospheric-Cherenkov-technique in general in two approaches.
First, we will tell the story of the observation of one single air-shower-event and we will follow the information as it flows through the telescope.
Second we will describe the general response-behavior of FACT in a quantitative way using not the statistics of one, but of many air-shower-events.
\section{Story of a single event}
\label{SecSingleEventStory}
\begin{figure}
    \begin{center}
        \includegraphics[width=1.1\textwidth]{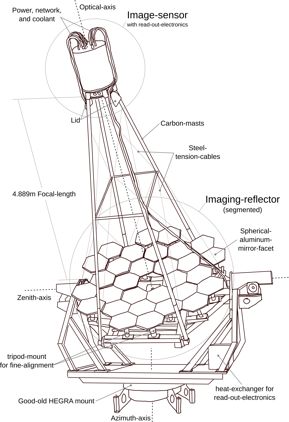}
        \caption[Schematic of FACT, a small telescope]{The FACT Telescope.
            See also Figure \ref{FigFactOverview}.
        }
        \label{FigSchematicsOverview}
    \end{center}
\end{figure}

\begin{figure}
    \begin{center}
        \includegraphics[width=1.1\textwidth]{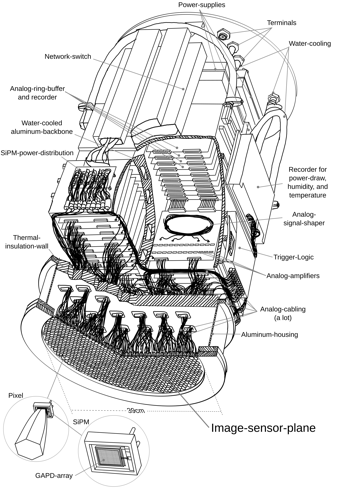}
        \caption[Schematic of the image-sensor of FACT]{The inner components of the image-sensor of FACT.
            Cut open to see the inner workings.
            The actual image-sensor has much more cabling going on.
        }
        \label{FigSchematicsImageSensor}
    \end{center}
\end{figure}

\begin{figure}
    \begin{minipage}{0.5\textwidth}
        \begin{figure}[H]
            \includegraphics[width=1.0\textwidth]{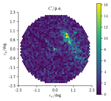}
        \end{figure}
    \end{minipage}
    \hfill
    \begin{minipage}{0.5\textwidth}
        \begin{figure}[H]
            \includegraphics[width=1.0\textwidth]{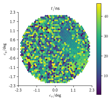}
        \end{figure}
    \end{minipage}
    \caption[Air-shower-event represented with largest-pulses]{%
        An air-shower-event recorded with FACT in the night of the 13th of November in 2015 (Run:\,109, Event:\,1506).
        Note that largest-pulses with small photon-equivalents (p.e.) have random arrival-times $t$.
        The hexagonal arrangement of the pixels here is because of FACT's  image-sensor-plane, see Figure \ref{FigSchematicsImageSensor}.
    }
    \label{FigLargetsPulseExampleEvent}
\end{figure}

It begins with a cosmic particle entering earth's atmosphere.
The cosmic particle has the three properties: Direction, energy and type.
The first interaction of the cosmic particle in the atmosphere at $25\pm15 \,$km above sea-level starts an air-shower.
From now on, the properties of the cosmic particle can not be measured directly anymore.
All following steps have the sole goal to reconstruct the three properties again from the observation of the air-shower.\\
In the atmosphere the air-shower grows towards ground and emits, among others, bluish Cherenkov-photons.
These Cherenkov-photons rush down the atmosphere and move with the local speed-of-light head-to-head with the other highly relativistic interaction-products of the air-shower.
In a narrow front of $\approx 2\,$m height, the interaction-products of the air-shower reach ground, and illuminate an area with the shape of a disc with  about $\approx 200\,$m diameter.
A small fraction of the Cherenkov-photons fall into the $3.5\,$m diameter imaging-reflector of FACT which is looking up into the night-sky, approximately anti-parallel to the trajectory of the cosmic particle.\\
The Cherenkov-photons, together with the night-sky-background-photons fall into the imaging-reflector, see Figure \ref{FigSchematicsOverview}.
The imaging-reflector bins the photons, depending on their incident-directions $c_x$ and $c_y$, into the pixels of the image-sensor with $0.1^\circ$ resolution, see Figure \ref{FigSchematicsImageSensor}.
The Cherenkov-photons reach the image-sensor within a narrow time-window of $\approx 5\,$ns.\\
The SiPM-photo-sensors in the pixels produce electric pulses corresponding to the arriving Cherenkov-photons and night-sky-background-photons.\\
The coincident arrival-times of the pulses related to the air-shower, together with the directional-coincidence of near-by pixels, creates a pattern in the recorded video that sets itself apart from the pattern of the night-sky-background-pulses alone in the video.
When the pattern of an air-shower is found in the video from the ensemble of pixels, the recording of the video is triggered.
Today, the time-series of electric amplitudes of the pixels can not be recorded continuously to permanent-storage\footnote{1440\,pixel $\times$ 12\,bit resolution $\times$ $2 \times 10^9$ samples/s gives $4.32\,$TByte/s. Today $500\,$GByte/night is the most FACT can tolerate for permanent-storage.}.
FACT records the time-series of all pixels simultaneously for a time-window of $150\,$ns with a sampling-rate of $2\,$GHz.
The trigger-threshold is continuously adjusted such that FACT records $\approx 70$ air-shower related videos per second.\\
On each time-series of a pixel, the charge-integral $C$ and the arrival-time $t$ of the largest pulse is extracted.
As the charge-integral $C$ is correlated to the number of photons arriving in the time-window of the integration, it is called photon-equivalent.
The resulting representation of the air-shower-event with large pulses has two values $C$ and $t$ for each pixel and can be visualized in two images, see Figure \ref{FigLargetsPulseExampleEvent}.
The identification of large pulses is already the first of multiple stages to classify Cherenkov-photons in the pool of night-sky-background-photons.
A largest-pulse corresponds to the highest density of photon-arrivals in a pixel.
In further stages, a set of pixels with a high density of photon-equivalents is searched in the image and associated with the air-shower\footnote{This is often called image-cleaning.}.
About $\approx 10$ geometric features are generated from the image of photon-equivalents.
\begin{figure}
    \begin{center}
        \includegraphics[width=0.7\textwidth]{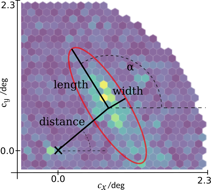}
        \caption[Hillas air-shower model based on ellipse]{
            A simple ellipse-model for the intensity of the photon-equivalents in the image.
            The most common features are the length, and the width of the ellipse.
            Those are often computed using the covariance-matrix of the intensity-distribution in the image.
            Different Cherenkov-telescopes use different additional features to describe both the photon-intensity and the arrival-time-distribution of the largest-pulses in more detail.
            This is the same air-shower-event as shown in Figure \ref{FigLargetsPulseExampleEvent}.
        }
        \label{FigHillasFeatures}
    \end{center}
\end{figure}
Figure \ref{FigHillasFeatures} shows the basics of the popular ellipse-model for air-showers proposed by Micheal Hillas \cite{hillas1985cerenkov} which is also used in FACT.
Such so called \textit{Hillas}-features describe the air-shower in the image using an ellipse model.\\
For each recorded air-shower-event where \textit{Hillas}-features could be generated, these \textit{Hillas}-features%
\footnote{%
    Also called Hillas-parameters. FACT uses a flavor Hillas-features which are independent of the position of a hypothetical source of gamma-rays. Other Cherenkov-telescopes might use different definitions.
}
are compared with the \textit{Hillas}-features of simulated air-shower-events of which the properties of the cosmic particle are known.
From this comparison, the type, the direction, and the energy of the observed cosmic particle is estimated.\\
By comparing the incident-directions of gamma-rays with known directions of cosmic sources in the sky, FACT estimates the flux of gamma-rays from these sources.
Finally, FACT publishes the estimated flux of gamma-rays of cosmic sources on its \textit{Quick-Look-Analysis}-webpage \cite{fact2017icrc} about $10\,$min after the cosmic gamma-ray entered the atmosphere.
FACT also emits alerts for increased fluxes of gamma-rays to its subscribers such as other telescopes on ground or in space.
\section{Statistics of responses to cosmic particles}
\begin{figure}
    \begin{center}
        \includegraphics[width=1.0\textwidth]{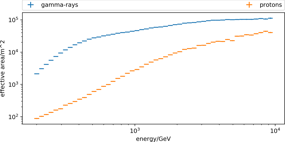}
        \caption[Effective collection-area of the trigger of FACT]{
            Based on simulations.
            The effective collection-area of the trigger of FACT.
        }
        \label{FigTriggerEffectiveArea}
    \end{center}
\end{figure}

\begin{figure}
    \begin{center}
        \includegraphics[width=1.0\textwidth]{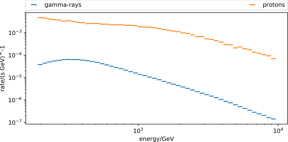}
        \caption[Differential rate of the trigger of FACT]{
            The differential rate of the trigger of FACT.
            Based on the effective collection-area of the trigger of FACT, see Figure \ref{FigTriggerEffectiveArea}.
        }
        \label{FigTriggerDifferentialRate}
    \end{center}
\end{figure}

\begin{figure}
    \begin{center}
        \includegraphics[width=1.0\textwidth]{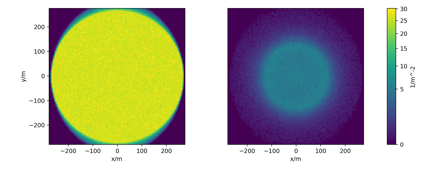}
        \caption[Distribution of the effective area of the trigger of FACT on ground]{
            Areal-density of the support-position of the trajectory of gamma-rays on ground.
            Left is the density of all thrown gamma-rays.
            Right is the density of the gamma-rays which triggered the readout of FACT.
            Based on the same simulations as Figure \ref{FigTriggerEffectiveArea}.
            The radial truncation for the thrown events is only to reduce the need for computation.
        }
        \label{FigGammaImpact}
    \end{center}
\end{figure}

The statistics of the responses of FACT to different cosmic particles are a quantitative description of the performance of FACT.
For Cherenkov-telescopes, such statistics of responses\footnote{Also called response-function, instrument-response-function, or depending on the context also called effective area} are estimated in simulations.
Calibrations in the lab with artificial sources of gamma-rays and charged particles are not feasible today at energies in the $1\,$TeV regime.
For FACT, we simulate the observation of individual air-shower-events while specific properties of the cosmic particle are drawn randomly for each simulated event\footnote{We use CORSIKA \cite{heck1998corsika} to simulate air-showers and the emission of Cherenkov-photons \cite{bernlohr2008AtmoIact}.
}
The incident-direction and position of the trajectory
\footnote{%
The trajectory is a ray with a support-vector (position) and a direction-vector (direction) in the three-dimensional frame of the telescope.
}
of the cosmic particle relative to the telescope is drawn randomly as well as the energy of the cosmic particle is drawn randomly.
We estimate the statistics of the responses of FACT by counting the number of thrown events and counting the number of events that lead to a response in the telescope.
In this case, the response we are looking for is the activation of the trigger for the read-out in the image-sensor of FACT.
Figure \ref{FigTriggerEffectiveArea} shows the effective collection-areas for both cosmic gamma-rays and cosmic protons which triggered the readout of FACT.
These effective collection-areas of the trigger give an upper-limit on the largest possible area for the collection of gamma-rays that can be obtained from the recorded air-shower-events.
For the detection of gamma-rays, hadronic cosmic-rays need to be rejected which reduces the effective collection-area for gamma-rays by about one order-of-magnitude \cite{nothe2017fact}.
When we convolve the effective collection-area of the trigger of FACT with the fluxes for cosmic gamma-rays \cite{hillas1998spectrum} and cosmic protons \cite{olive2014Review}, we obtain the expected rate of the trigger of FACT, see Figure \ref{FigTriggerDifferentialRate}.
The integrated expected rates of the trigger over the range of simulated energies of cosmic particles are $5.7\,$s$^{-1}$ for cosmic protons and only $0.045\,$s$^{-1}$ for cosmic gamma-rays.
The flux of gamma-rays here is the flux of the Crab nebula, which is one of the brightest source in the gamma-ray-sky.
In our simulations we use the flux of night-sky-background-photons shown in Figure \ref{FigPhotonSpectra}, which is the dark night on Canary island La Palma.
This flux for night-sky-background-photons gives $\approx 30\times10^6\,$s$^{-1}$ photons in each pixel of FACT's image-sensor.
FACT's trigger is discussed in \cite{vogler2011trigger}.
Figure \ref{FigGammaImpact} illustrates a mayor advantage of Cherenkov-telescopes over direct detectors for gamma-rays in space.
Although the aperture for Cherenkov-photons of FACT is only $\approx 3.5\,$m in diameter, FACT is well able to record air-showers where the trajectory of the cosmic gamma-ray intersects the ground $\approx 100$\,m off the telescope.
This is because Cherenkov-telescopes rely on the air-showers illuminating large areas on ground with $\approx 100$\,m radius.
The trigger of FACT has an effective collection-area of $>10^4$\,m$^2$ for gamma-rays which exceeds its aperture-area for Cherenkov-photons by $\approx 3$ orders-of-magnitude.
This makes Cherenkov-telescopes cost efficient to investigate cosmic-rays and gamma-rays.
\chapter{The image-sensor of FACT}
The photo-sensors, trigger, and read-out-electronics in the image-sensors of Cherenkov-telescopes are the most complex component of a Cherenkov-telescope.
The image-sensor of a Cherenkov-telescope records videos with $\approx 10^9\,$ images per second at a resolution of $\approx 10^3\,$pixel.
The photo-sensors in the image-sensor respond already to single-photons within nano-seconds\footnote{On some Cherenkov-telescopes the number of photons needed to resolve the response of a photo-sensor might be higher, but is usually below 10.}.
The trigger of the image-sensor identifies the pattern of air-showers in the video within nano-seconds to allow the read-out of the video to permanent storage.
Image-sensors for Cherenkov-telescopes have to work in the field where weather and light-conditions might change unexpectedly.
Image-sensors are expensive and very demanding to develop.
They use custom-build, high-bandwidth electronics for the processing of analog signals.
It is difficult to find the best compromise of power-consumption, heat-removal, signal-to-noise-ratio, event-throughput, weight, part-availability, and cost.
The design of the image-sensor also effects the rest of the telescope which for example must be built more rigid when the weight of the image-sensor increases.
From the experience with FACT, we learned that routing of analog signals with high-bandwidth in cables or on circuit-boards is possible, but very demanding and costly.
Because FACT is small, it can tolerate a relatively heavy but compact image-sensor with all its read-out within one housing.
But the large MAGIC Cherenkov-telescopes for example are forced to put their read-out-electronics off the moving telescope what creates additional challenges \cite{bartko2005tests} and costs for cabling and the transmission of analog signals.
\section{SiPMs in FACT}
\begin{figure}
    \begin{center}
        \includegraphics[width=1.0\textwidth]{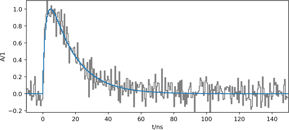}
        \caption[Response of an SiPM in FACT to a single-photon]{%
            A single-photon arriving in an SiPM-pixel of FACT.
            The average pulse observed in the image-sensor of FACT is blue.
            The example time-series of an SiPM-pixel of FACT with its electronic-noise is gray.
        }
        \label{FigSipmPulseNoise}
    \end{center}
\end{figure}

\begin{figure}
    \begin{center}
        \includegraphics[width=1\textwidth]{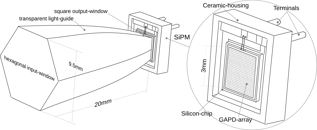}
        \caption[SiPM-pixel and SiPM schematics]{
            A pixel of the image-sensor of FACT, compare Figure \ref{FigSchematicsImageSensor}.
            A transparent light-guide concentrates the hexagonal aperture of the pixel down to a square output-window glued to the silicon-chip of the SiPM.
            The light-guide is made out of poly-methyl-methacrylate (PMMA), and was mass produced using injection-molding.
            The SiPM part designation is Hamamatsu\,MPPC\,S10362-33-50C.
            The GAPD-array has 60 $\times$ 60 = 3600 GAPDs.
            All GAPDs are connected in parallel to the terminals of the SiPM.
        }
        \label{FigSchematicsSipmPixel}
    \end{center}
\end{figure}

FACT senses photons using Silicon-Photo-Multipliers (SiPMs) of the type Hamamatsu\,MPPC\,S10362-33-50C.
SiPMs are two-dimensional arrays of Geiger-mode-avalanche-photo-diodes
(GAPDs) which are typically \cite{golovin1999avalanche, sadygov1998avalanche}  all connected in parallel and operated with reverse bias.
SiPMs respond to incoming photons with an electric pulse on their analog signal-output.
Figure \ref{FigSchematicsSipmPixel} shows a pixel of the image-sensor of FACT with its SiPM, and Figure \ref{FigSipmDrs} shows the circuit diagram of the SiPMs used in FACT.
In reverse bias, all GAPDs act like capacitors, each storing the same amount of charge.
The SiPMs used in FACT operate at $\approx 70\,$V of bias-potential.
When a GAPD is hit by a photon it might discharge thus creating a well defined, and easy to sense electric-pulse on the time-series of the terminal of the SiPM.
When multiple GAPDs are hit in short time, their pulses add up on the time-series.
Figure \ref{FigSipmPulseNoise} shows an electric-pulse on the time-series of a SiPM-pixel used in FACT when a single-photon arrives.
%
%Figure \ref{FigSipmCurrentsVsNighSkyBackground} shows how much current is flowing through the GAPDs in an SiPM-pixel for different fluxes of photons.
%
%From the Figures \ref{FigSipmCurrentsVsNighSkyBackground} and \ref{FigExtractedRateVsNighSkyBackground} we learn that
The gain%
\footnote{This is comparing the power falling into the sensor in form of visible photons ($100\times10^6$\,photons\,s$^{-1}$ $\times$ $3\,$eV\,photon$^{-1}$) to the power from the bias supply which is released by the GAPDs in form of current ($7.5\,$uA $\times$ $70\,$V). The values are taken from the upcoming Figures \ref{FigSipmCurrentsVsNighSkyBackground} and \ref{FigExtractedRateVsNighSkyBackground}.}%
of the SiPMs used in FACT is $\approx 1.1\times10^7$.
\section{PMTs in established Cherenkov-telescopes}
Established Cherenkov-telescopes use Photo-Multiplier-Tubes\footnote{Also called photo-multiplier without the tube.} (PMTs) to sense photons.
PMTs are photo-sensors which respond to photons with electric pulses on their analog signal-output.
PMTs are vacuum-tubes with a transparent entrance window.
The inner surface of the entrance window is coated with a material where incoming photons can release electrons into the vacuum of the tube.
A static electric-field inside the vacuum-tube accelerates the electron which falls into a plate of metal where additional electrons are released from the impact on the metal-plate.
This way the number of released electrons is multiplied.
In a cascade of metal-plates and electric-fields, the process of acceleration and release is repeated until the number of released electrons is large enough to be processed and read out.
As we will show in the following, dedicated PMTs are optimized to offer a well suited, and high photon-detection-efficiency for blueish photons what makes them a great choice for Cherenkov-telescopes.
However, we will also see that PMTs lack single-photon-resolution which is due to the statistical process of the multiplication of the released electrons.
\section{Alternative photo-sensors}
Two alternative photo-sensors relevant for Cherenkov-Telescopes must be acknowledged as well.
First, there are Hybrid-Photo-Detectors (HPDs) which are a hybrid of the entrance-window from a photo-multiplier-tube and a single avalanche-photo-diode within the same vacuum-tube \cite{tridon2010magic}.
HPDs have great potential for Cherenkov-telescopes as they have similar, or even higher, photon-detection-efficiencies as PMTs, but also have a high single-photon-resolution like SiPMs.
According to private communication with Adrian Biland (MAGIC-collaboration), HPDs were not deployed on MAGIC due to extensive aging of at least the first production-batch during typical observation-conditions.
Second, there are photo-sensors using electric discharges in gas filled multi-wire-proportional-chambers.
The CLUE-experiment \cite{alexandreas1995status} successfully recorded images of air-showers using such chambers.
The gas-chambers are most sensitive to photons with wavelength in the range between $150\,$nm and $350\,$nm which naturally results in strong rejection of reddish night-sky-background-photons.
We are not sure why, but apparently no other Cherenkov-telescope used gas filled multi-wire-proportional-chambers after CLUE.
\section{Photon-detection-efficiency}
\begin{figure}
    \begin{center}
        \includegraphics[width=1\textwidth]{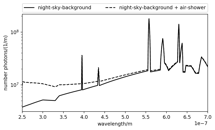}
        \caption[Photon-spectra during air-shower-observations]{
            Initial amount of photons before reaching the telescope.
            Reconstructed from a simulated air-shower-event of FACT with $50\,$ns exposure-time.
            Night-sky-brightness is $\approx 21\,$mag\,arcsec$^{-2}$ and $\approx 12\times10^3$\,photons total in the shown interval of wavelengths.
            The spectrum of the night-sky-brightness is taken from \cite{Benn1998503}.
            About $1.1\times10^3$\,Cherenkov-photons produced by a $3\,$TeV gamma-ray from direction zenith.
        }
        \label{FigPhotonSpectra}
    \end{center}
\end{figure}

\begin{figure}
    \begin{center}
        \includegraphics[width=1\textwidth]{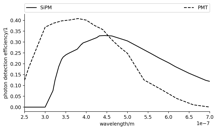}
        \caption[Photon-detection-efficiencies of SiPMs and PMTs]{
            Photon-detection-efficiency for a state-of-the-art PMT \cite{toyama2013novel} and the SiPM used in FACT.
            Photon-detection-efficiency of the SiPM is based on claims of the manufacturer on the spectral-distribution \cite{hamamatsu2009mppc} and our own experiences \cite{FACT_design} with the peak-efficiency .
        }
        \label{FigSensorPde}
    \end{center}
\end{figure}

Exploiting the different spectra of the wavelengths of blueish Cherenkov-photons and of reddish night-sky-background-photons can be a first step to achieve a classification of Cherenkov-photons, see Figure \ref{FigPhotonSpectra}.
Although Cherenkov-telescopes do not have photo-sensors which can measure the wavelength of the incoming photons, Cherenkov-telescopes use photo-sensors which are optimized to sense blue photons and to reject red photons.
Here, the photon-detection-efficiency of the first generation of mass-produced SiPMs used in FACT falls behind the photon-detection-efficiency of state-of-the-art PMTs, see Figure \ref{FigSensorPde}.
Table \ref{TabSesnorResponseNsbCherenkov} estimates how much the ratio of Cherenkov-photons and night-sky-back-ground-photons changes if FACT had sensors with the photon-detection-efficiency of state-of-the-art PMTs.
\begin{table}
    \begin{center}
    \begin{tabular}{ l  r  r  r}
                & Night-sky-background/1 & Air-shower/1 & ratio/$\%$\\
        \midrule
        before image-sensor & 11856 & 1132 & $9.5$\\
        SiPM & 2500 & 200 & $8.0$\\
        PMT & 1414 & 313 & $22.1$\\
    \end{tabular}
    \caption[Ratio of air-shower- and nigh-sky-background-photons]{
        Ratio of Cherenkov-photons to night-sky-background-photons and absolute amounts during $50\,$ns of exposure-time.
        Observing an air-shower induced by a $3\,$TeV gamma-ray from direction zenith.
        Based on Figures \ref{FigSensorPde}, and \ref{FigPhotonSpectra}.
        Note: This estimate is limited to the range of wavelengths of current simulation-tools \cite{heck1998corsika,bernlohr2008AtmoIact}.
    }
    \label{TabSesnorResponseNsbCherenkov}
    \end{center}
\end{table}
\section{Single-photon-resolution}
\begin{figure}
\includegraphics[width=1\textwidth]{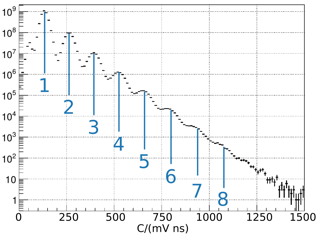}
\caption[Single-photon-resolution, SiPM in FACT, only dark-noise]{Based on \cite{fact-performance}.
    Single-photon-resolution of the SiPM used in FACT.
    Internally triggered on incoming electric pulses, hence no '0' peak, compare Figure \ref{FigPmtSpe}.
    Across all 1440\,SiPMs of the image-sensor of FACT, across all temperatures during operation, across several month of operation.
}
\label{FigSipmSpe}
\end{figure}

\begin{figure}
\includegraphics[width=1\textwidth]{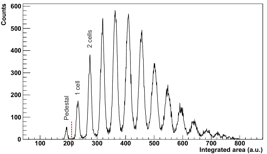}
\caption[Single-photon-resolution, SiPM in FACT, light-source]{Figure taken from \cite{kraehenbuehl2013diss}.
    In contrast to Figure \ref{FigSipmSpe} where only the rare accidental discharges of the SiPMs are used, here for each event a light-source deposited about $5\,$photons in the SiPM.
    Here 'cell' is the Geiger-mode-avalanche-diode.
    The left most pedestal-bump indicates that this Figure was created with a very different procedure than Figure \ref{FigSipmSpe}.
    Here a method was used similar to the one shown in Figure \ref{FigPmtSpe} where one does not wait for sensor-responses to show up randomly, but reads out the sensor unconditionally after photons have been emitted by a controlled light-source which in rare cases might not emit any photon.
    Anyhow, for our purposes here it is only relevant that many distinct peaks can be found what indicates a high single-photon-resolution.
}
\label{FigSipmSpeWithLightSource}
\end{figure}

\begin{figure}%
\includegraphics[width=1\textwidth]{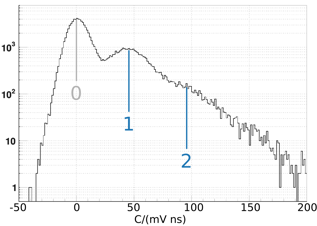}
\caption[Single-photon-resolution of a state-of-the-art PMT]{Based on \cite{masuda2015development}.
    Single-photon-resolution of a state-of-the-art PMT.
    Externally triggered by a source of photons, hence the '0' peak when no electric pulse was released in the PMT.
    One specific PMT only.
}
\label{FigPmtSpe}
\end{figure}

\begin{figure}
    \begin{center}
        \includegraphics[width=1.0\textwidth]{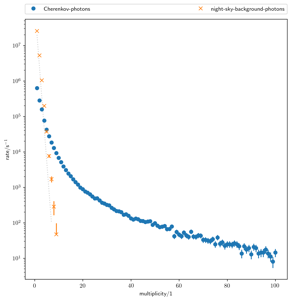}
        \caption[Rate of photon-multiplicities on FACT]{
            The rate of different photon-multiplicities for a time-window of $1\,$ns.
            Recorded in the night of 11th November 2013 in run 160.
            Multiplicities of Cherenkov-photons in blue.
            Multiplicities of night-sky-background photons in orange.
            The exposure-time-equivalent for a single pixel is 1,440\,pixels $\times$ 17,302\,events $\times$ 50\,ns/event = 1.2\,s for the Cherenkov-photons and 1,440\,pixels $\times$ 294\,events $\times$ 50\,ns/event = 21\,ms for night-sky-background-photons.
            Total integrated photon-rate is $43.6\times10^6$\,s$^{-1}$\,pixel$^{-1}$.
            Compare with Table \ref{TabPhotonMultiplicity}.
            The gray, dotted line indicates the decay of the peaks in Figure \ref{FigSipmSpe} which are dominated by an artifact of the SiPM called optical-cross-talk which we describe later in Section \ref{SecArtifacts}.
        }
        \label{FigPhotonMultiplicity}
    \end{center}
\end{figure}

SiPMs have a good single-photon-resolution.
For photo-sensors which output analog signals, a good single-photon-resolution means that a certain number of simultaneously arriving photons, will create a charge-integral $C$ of a certain amplitude which can be told apart from the amplitudes of other charge-integrals created by different numbers of photons.
We call the number of photons arriving within a small, simultaneous time-window the multiplicity of the arriving photons.
The Figures \ref{FigSipmSpe} and \ref{FigPmtSpe} show the histogram of the charge-integrals of the SiPMs used in FACT and a state-of-the-art PMT.
Because of the individual, but identical GAPDs in the SiPM, we find several distinct peaks of distinct photon-multiplicities in the histogram of charge-integrals in Figure \ref{FigSipmSpe}.\\
The Figure \ref{FigPhotonMultiplicity} and Table \ref{TabPhotonMultiplicity} show how often different photon-multiplicities occur%
\footnote{The Figure \ref{FigPhotonMultiplicity} and Table \ref{TabPhotonMultiplicity} show the rates of photon-multiplicities separately for both night-sky-background-photons and Cherenkov-photons.
The multiplicities of Cherenkov-photons are estimated by subtracting the multiplicities found in events which were randomly triggered from the multiplicities found in events which were triggered because the pattern of an air-shower was found in the video.
Randomly triggered events are very unlikely to contain Cherenkov-photons.
The multiplicities of the night-sky-background-photons are estimated from the randomly triggered events only.
}%
during the observations of FACT.
We find that the SiPMs used in FACT can tell apart the number of the arriving photons.
In the measurement for the histogram in Figure \ref{FigSipmSpe}, the probability for larger multiplicities drops exponentially which might create the impression that the single-photon-resolution of the SiPM vanishes for multiplicities larger than eight.
But the measurement is just running out of statistics for larger multiplicities.
Figure \ref{FigSipmSpeWithLightSource} shows an different measurement \cite{kraehenbuehl2013diss} which shows that the single-photon-resolution does not vanish beyond a multiplicity of eight.\\
We conclude that FACT does not need to represent photons vaguely by describing the continuous charge-integral $C$ of its sensor.
We conclude that FACT can describe the photons directly in a quantized way.
\begin{table}
    \begin{center}
    \begin{tabular}{p{1.8cm}  r r  r r}
        &
        \multicolumn{2}{c}{night-sky-background-photons} &
        \multicolumn{2}{c}{Cherenkov-photons} \\
        %night-sky-background-photons & & & Cherenkov-photons & \\
        multiplicity & rate/$10^6$\,s$^{-1}$ & ratio/$\%$ & rate/$10^6$\,s$^{-1}$ & ratio/$\%$\\
        \midrule
        1 & 25.599 & 63.513 & 0.634 & 18.520\\
        2 & 5.263 & 26.117 & 0.281 & 16.437\\
        3 & 1.045 & 7.780 & 0.158 & 13.877\\
        4 & 0.199 & 1.976 & 0.077 & 8.945\\
        5 & 0.037 & 0.465 & 0.042 & 6.195\\
        6 & 0.008 & 0.114 & 0.027 & 4.821\\
        7 & 0.002 & 0.030 & 0.018 & 3.743\\
        8 & 0.000 & 0.006 & 0.013 & 3.017\\
        9 & 0.000 & 0.001 & 0.009 & 2.427\\
        10 & -- & -- & 0.007 & 1.975\\
    \end{tabular}
    \caption[Rate of photon-multiplicities on FACT]{
    The rate of different photon-multiplicities for a time-window of $1\,$ns.
    The ratio indicates that $18.520\%$ of all Cherenkov-photons arrive alone as single-photons, $16.437\%$ of all Cherenkov-photons arrive in a tuple of two simultaneous photons, and so on.
    The majority of photons detected by FACT arrive alone as single-photons.
    Only $20\%$ of all Cherenkov-photons arrive in tuples with multiplicities $>10$.
    Compare and see uncertainties in Figure \ref{FigPhotonMultiplicity}.
    The decay of the rate of the multiplicity of night-sky-background-photons is dominated by an artifact called optical-cross-talk which we describe later in Section \ref{SecArtifacts}.
    }
    \label{TabPhotonMultiplicity}
    \end{center}
\end{table}
\section{Wear and aging}
While Cherenkov-telescopes with PMTs have to use additional filters \cite{archambault2017gamma} for observations during bright moonlight in order to not degrade the sensitivity of their PMTs over several years \cite{gazda2016photon} of observations, the regular FACT Cherenkov-telescope with its SiPMs can observe Cherenkov-photons while the full-moon is in same field-of-view \cite{knoetig2013fact} without degrading the sensitivity of its SiPMs.
After more than five years of observations with the SiPMs used in FACT, a charge of over 370\,C was released in each SiPM-pixel \cite{neise2017fact} and we do not see any hint for a change in the performance to detect photons yet.
%
%% switch off pmts on moon light \cite{kildea2007whipple}
% PMT inconsistency, \cite{sitarek2013analysis}
% modern pmts are stable, old one died rather fast \cite{kildea2007whipple}
%
\section{Artifacts}
\label{SecArtifacts}
The SiPMs used in FACT have two artifacts which are relevant for Cherenkov-telescopes.
First, the SiPMs used in FACT discharge their GAPDs accidentally even when it is dark and no photons arrive.
This dark-noise increases with temperature, and is $\approx 4\times10^6$\,s$^{-1}$\,pixel$^{-1}$ during room-temperature.
On FACT, this accidental rate of discharges of GAPDs is only a minor issue because the rate of night-sky-background-photons in a pixel is about one order-of-magnitude larger at $\approx 30\times10^6$\,s$^{-1}$\,pixel$^{-1}$.
For example, the sensor-responses used in Figure \ref{FigSipmSpe} are only caused by accidental discharges of GAPDs at a rate of $\approx 4\times10^6$\,s$^{-1}$\,pixel$^{-1}$.\\
Second, the discharge of a GAPD can cause the discharge of other GAPDs on the same SiPM \cite{otte2009efficiency}.
This so called optical-cross-talk of GAPDs causes false responses by the SiPMs which appear to be incoming photons but actually are not.
For large rates of arriving photons, the optical-cross-talk results simply in an increase in gain of the sensor-response, but for the low rates of photons on FACT, where we are approaching the quantized regime of single-photons, optical-cross-talk becomes a noise \cite{buss2015fact} which is worked on to be minimized \cite{buzhan2009cross}.
For FACT, the probability for a GAPD to optically cross-talk into another GAPD is $9.5\%$ \cite{fact-performance}.
\section{Read-out}
\begin{figure}
    \begin{center}
        \includegraphics[width=1\textwidth]{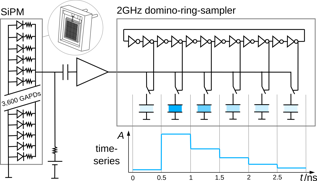}
        \caption[Readout-electronics of FACT]{
            A photon falling into the SiPM-pixel can discharge a GAPD.
            The electric-pulse of the discharge is stored in the array of capacitors in the domino-ring-sampler.
            When the trigger detects the pattern of an air-shower in the pixels, the charge in the capacitors is converted to digital values which are stored permanently.
            This configuration is called quasi-digital-counter.
            Compare SiPM with Figures \ref{FigSchematicsSipmPixel} and \ref{FigSchematicsImageSensor} and compare the time-series with Figure \ref{FigSipmPulseNoise}.
            Inspired by schematics by Stefan Ritt of the domino-ring-sampler.
        }
        \label{FigSipmDrs}
    \end{center}
\end{figure}

In the image-sensor of FACT, the analog signals of all pixels are continuously written into a ring-buffer with a depth of $512\,$ns.
When the trigger of the image-sensor detects the pattern of an air-shower in the video, the trigger stops the continuous writing into the ring-buffer and copies the recorded time-series from the ring-buffers to the slow permanent-storage.
The ring-buffer of a pixel in FACT is an array of 1024 capacitors which are connected to the output of the pixel one after another for a time of $500\,$ps each.
During these $500\,$ps when the capacitor is connected to the output of a pixel, the potential of the capacitor floats towards the average potential of the analog signal of the pixel.
When the next capacitor in the array is connected to the pixel, the first capacitor holds its potential until it is overwritten again $512\,$ns later when all 1024 capacitors have been connected to the pixel once.
When the trigger decides to read out the ring-buffer, each capacitor is connected to an analog-to-digital-converter for a time of $\approx 1\,\mu$s.
The output of the analog-to-digital-converter is then written to a permanent storage.
Figure \ref{FigSipmDrs} shows the simplified schematics of the read-out-electronics used in FACT.
The array of sample-and-hold-capacitors is called 'switched-capacitor-array', or 'domino-ring-sampler' \cite{ritt2008design}.
%
%%%%%%%%%%%%%%%%%%%%%%%%%%%%%%%%%%%%%%%%%%%%%%%%%%%%%%%%%%%%%%%%%%%%%%%%%%%%%%%
%  ______                                    _   _
%  | ___ \                                  | | (_)
%  | |_/ /___ _ __  _ __ ___  ___  ___ _ __ | |_ _ _ __   __ _
%  |    // _ \ '_ \| '__/ _ \/ __|/ _ \ '_ \| __| | '_ \ / _` |
%  | |\ \  __/ |_) | | |  __/\__ \  __/ | | | |_| | | | | (_| |
%  \_| \_\___| .__/|_|  \___||___/\___|_| |_|\__|_|_| |_|\__, |
%            | |                                          __/ |
%            |_|                                         |___/
%
%%%%%%%%%%%%%%%%%%%%%%%%%%%%%%%%%%%%%%%%%%%%%%%%%%%%%%%%%%%%%%%%%%%%%%%%%%%%%%%
%
\chapter{Representing air-shower-events}
Cherenkov-telescopes record a pool of night-sky-background-photons with a few Cherenkov-photons in it.
Cherenkov-telescopes record three observables for each of these photons:
\begin{itemize}
\item The incident-direction $c_x$,
\item the incident-direction $c_y$,
\item and the arrival-time $t$.
\end{itemize}
The different pixels in the image-sensor measure the incident-directions $c_x$, and $c_y$, and the time-slices on the time-series of a pixel measure the arrival-time $t$.
Current Cherenkov-telescopes do not record additional observables.
As a consequence, the only way to classify Cherenkov-photons and night-sky-background-photons is to investigate the spatial structure of the photons in the three-dimensional space of observables.
A limited description of these observables limits the performance to classify Cherenkov-photons and night-sky-background-photons, and thus prevents us from observing low energetic gamma-rays.
\section{Largest-pulses}
We name the group of established representations for air-shower-events the representation of largest-pulses.
The representation of largest-pulses describes not the observables of the incoming photons, but the electric-pulses created by these photons in photo-sensors such as PMTs or SiPMs.
Usually Cherenkov-telescopes describe the largest-pulse with two attributes.
First a charge-integral, called photon-equivalent $C$, and second an arrival-time $t$.
The representation of largest-pulses can be visualized in two images as shown in Figures \ref{FigLargetsPulseExampleEvent}.
As the largest-pulse on a time-series of a pixel in a Cherenkov-telescope is associated with the Cherenkov-photons, the representation of largest-pulses is already the result of a first classification of Cherenkov-photons along the arrival-times of the photons.
%
% __Past__
% puhlhofer2003technical, 3.1. Pulse shape analysis (pixel level)
% kildea2007whipple, 2.3.3. Signal recording
% barrau1998cat, 5.3. Charge measurement
%   noise-amplitude: 0.4
%
% __Current__
% holder2006first, 2.5. Data acquisition
%   noise-amplitude: 0.5
% sitarek2013analysis, 3.3. Signal extraction
% holler2013status, 1 Introduction
%
All past \cite{kildea2007whipple,barrau1998cat,puhlhofer2003technical}, current \cite{holder2006first,sitarek2013analysis,holler2013status,albert2008fadc}, and future \cite{catalano2013astri,yashin2015imaging} Cherenkov-telescopes represent air-shower-events using the charge-integrals of analog photo-sensors in their pixels.
Some Cherenkov-telescopes \cite{kildea2007whipple,barrau1998cat,puhlhofer2003technical,yashin2015imaging} do not have additional arrival-time informations, but only the charge-integral.
All the listed references focus on describing the air-shower-event using the response of their individual sensors and read-out.
None of the listed references interprets their sensor-responses in order to restore all the observables of the individual photons.
All references listed here stuff single-photons into one large pulse.
\section{Photon-stream}
\begin{figure}
    \begin{center}
        \includegraphics[width=1\textwidth]{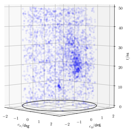}
        \caption[Air-shower-event represented with the photon-stream]{
            An air-shower-event recorded by FACT and represented as a stream of photons in the three-dimensional space of observables.
            Each blue dot represents one photon.
            All blue photon-dots have the same semi-transparent color, so that dense regions of photons look darker.
            The black ring on the bottom represents the field-of-view.
            This is the same air-shower-event as shown in Figures \ref{FigLargetsPulseExampleEvent}, \ref{FigDistanceMetrik}, and \ref{FigDbscanExampleEvent}.
        }
        \label{FigPhotonStreamExampleEvent}
    \end{center}
\end{figure}

When we describe all the observables of the incoming photons as they stream through the image-sensor-plane of a Cherenkov-telescope, we end up with a list of photon-arrival-times for each pixel.
We call this representation the photon-stream.
The photon-stream is the most natural representation for air-shower-events possible for a Cherenkov-telescope.
With the photon-stream it is not longer the question whether the representation of the air-shower-event itself is limited but whether we are able to create a Cherenkov-telescope which can record the natural representation of the photon-stream with sufficient accuracy to take advantage over the established representation of largest-pulses.
As we will show later on, we believe that FACT with its SiPM-pixels, its low electronics-noise, and fast read-out is about to provide this accuracy.
Figure \ref{FigPhotonStreamExampleEvent} shows an air-shower-event recorded on FACT and represented in the photon-stream.\\
The photon-stream is not dedicated to a specific photo-sensor or read-out-electronic.
The photon-stream is decoupled from the hardware and only describes the observables of the incoming photons.
We introduce here the photon-stream on the example of FACT with its SiPM photo-sensors, but the photon-stream is not limited to SiPM sensors.
Although not practical at the moment, the quantized photon-stream could also be recorded with PMTs\cite{catalano2008single} when the number of pixels is increased to lower the rate of photons in the individual pixels.
%
%%%%%%%%%%%%%%%%%%%%%%%%%%%%%%%%%%%%%%%%%%%%%%%%%%%%%%%%%%%%%%%%%%%%%%%%%%%%%%%
%   _____     _                  _   _
%  |  ___|   | |                | | (_)
%  | |____  _| |_ _ __ __ _  ___| |_ _  ___  _ __
%  |  __\ \/ / __| '__/ _` |/ __| __| |/ _ \| '_ \
%  | |___>  <| |_| | | (_| | (__| |_| | (_) | | | |
%  \____/_/\_\\__|_|  \__,_|\___|\__|_|\___/|_| |_|
%
%%%%%%%%%%%%%%%%%%%%%%%%%%%%%%%%%%%%%%%%%%%%%%%%%%%%%%%%%%%%%%%%%%%%%%%%%%%%%%%
%
\chapter{Single-Photon-Extractor}
\label{ChSinglePhotonExtractor}
To create the photon-stream for FACT, we have to extract the arrival-time of each single-photon from the recorded time-series of a pixel.
Our first implementation of a single-photon-extractor for the SiPMs in FACT is based on two assumptions:
\begin{itemize}
    \item First, we assume that the amplitude on the time-series of a pixel is only the sum of discharge-pulses of individual GAPDs.
    \item And second, we assume that discharge-pulses of all GAPDs are the same for all pixels, and time-epochs.
\end{itemize}
\section{Algorithm}
We run an iterative detection and subtraction of single-pulses on the time-series of a pixel.
In each iteration we append the extracted arrival-time of a single-pulse to the resulting list of arrival-times and subtract the pulse we just found from the time-series.
We stop the iteration when the time-series of the pixel is flat and does not contain anymore single-pulses.
\subsection*{Preparation}
\begin{figure}[H]
    \begin{center}
        \includegraphics[width=1.0\textwidth]{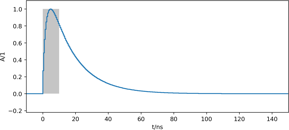}
        \caption[Template pulse for the extraction]{The full-template-pulse $A_\text{full}[t]$.
            The gray box from 0\,ns to 10\,ns marks the range of the rising-edge-template-pulse $A_\text{edge}[t]$.
            Compare Figure \ref{FigSipmPulseNoise} to see the electronic-noise again.
        }
        \label{FigTemplateFull}
    \end{center}
\end{figure}

Before the extraction, we prepare two templates of discharge-pulses of our GAPDs.
The first, called full-template-pulse $A_\text{full}[t]$
\footnote{We use brackets in $A[t]$ instead of parentheses to stress the discrete array-character with 2\,GHz sampling-rate and $500\,$ps sample-duration respectively of our implementation.}
, contains the full, and uncut single-pulse.
The second, called rising-edge-template-pulse $A_\text{edge}[t]$ contains a truncated single-pulse which only contains the rising-edge and the apex of the pulse.
Find both pulses in Figure \ref{FigTemplateFull}.
The rising-edge-template-pulse will be used to detect candidates of single-pulses.
The full-template-pulse on the other hand will be used to subtract a detected single-pulse from the time-series of the pixel.
%
% 1.626* (1.0 - \exp(-0.3803 * t)) \exp(-0.0649 * t)
%
%\begin{eqnarray}
%a[t] &=& 1.626 (1 - \exp(-0.3803\,t)) \exp(-0.0649\,t)\\
%A[t] &=&
%    \begin{cases}
%        a[t] ,& \text{if } a[t] > 0\\
%        0,              & \text{otherwise}
%    \end{cases}
%\end{eqnarray}
%%
%where the time $t$ is in nano-seconds.
%
Further, we create an empty list where the arrival-times of the single-pulses will be appended during the extraction.
\subsection*{Input}
The input is the time-series of an individual pixel $A_\text{pixel}^{0}[t]$.
We expect the calibration and the removal of artifacts of the read-out to be done already.
However, we know and accept that rare artifacts, which we do not address yet, can still be present on the time-series.
The most common of the not addressed artifacts is a sinusoidal resonance, see 'ringing' in \cite{vogler2015diss}.
\subsection*{Detecting a single-pulse}
We detect the $i$-th single-pulse by correlating the current $(i-1)$-th input-time-series of a pixel $A_\text{pixel}^{i-1}[t]$ with the rising-edge-template-pulse
\begin{eqnarray}
R^{i}[t] &=& A_\text{pixel}^{i-1}[t] \ast A_\text{edge}[t].
\end{eqnarray}
In the correlation with the rising-edge-template-pulse, frequencies on the time-series contained in the rising-edge-template-pulse are likely to pass on into $R^{i}[t]$.
But other frequencies as e.g. the frequencies of the electronic-white-noise of our read-out are likely to be suppressed in $R^{i}[t]$.
On the correlation-response, we identify the time-slice of the maximum
\begin{eqnarray}
t_i &=& \text{argmax}(R^{i}[t])
\end{eqnarray}
and define its associated time $t_i$ to be the arrival-time of the $i$-th single-pulse.
We append this arrival-time $t_i$ to the list of arrival-times for this pixel.
%
% puhlhofer2003technical, 3.1. Pulse shape analysis (pixel level)
%
Correlating the input-time-series with a template-pulse\cite{puhlhofer2003technical} or simply with a rectangular-box\cite{sitarek2013analysis} is already done on Cherenkov-telescopes to identify and restore the largest-pulse on a time-series.
\subsection*{Subtracting a single-pulse}
After the detection of the $i$-th single-pulse, we subtract the full-template-pulse $A_\text{full}[t]$ from the current $i$-th iteration of the pixel-time-series
\begin{eqnarray}
A_\text{pixel}^{i}[t] &=& A_\text{pixel}^{i-1}[t] - A_\text{full}[t - t_i]
\end{eqnarray}
at the corresponding arrival-time $t_i$.
\subsection*{Stopping-criteria}
On the remaining time-series of the pixel $A_\text{pixel}^{i}[t]$
we further detect and subtract single-pulses until a stopping-criteria is fulfilled.
The stopping-criteria is fulfilled when the maximum of the response $R^{i}[t]$ drops below $1/2$ of the maximum of the response to a single-pulse.
When the stopping-criteria was reached, the final time-series of the pixel is flat.
\section{Example}
\label{SecSinglePulseExtractorExample}
\begin{figure}
    \begin{center}
        \includegraphics[width=1.0\textwidth]{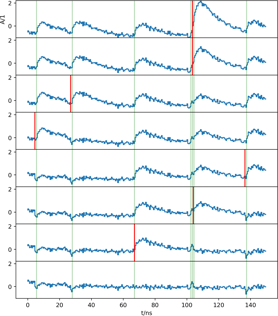}
        \caption[Extraction of single pulses from a time-series]{The time-series after each iteration of the single-photon-extractor.
            Starting on the top, going down to the bottom.
            The extracted arrival-times are in red, and the true arrival-times are in green.
            Extracted and true arrival-times are expected to be close to each other.
            In the end of the extraction, the time-series is flat.
        }
        \label{FigSinglePulseSubtraction1}
    \end{center}
\end{figure}

We run our extractor on simulated time-series where the arrival-times of the photons are known.
Figure \ref{FigSinglePulseSubtraction1} demonstrates the iterative procedure of our single-photon-extractor.
The simulated amplitude of electronic-white-noise is typical for FACT.
The simulated rate of photons is $\approx 50\times10^6$\,s$^{-1}$\,pixel$^{-1}$, which is $\approx 30\%$ higher than the rate of photons during a very dark night at $\approx 21\,$Mag\,arcsec$^{-2}$.
\section{Benefits of iterative subtraction}
\label{SecBenefitsOfIterativeSubtraction}
During our investigations, the iterative subtraction of single-pulses was the only method to identify the arrival of single-photons right after the arrival of a bunch of multiple, simultaneous photons.
In technical terms: Only the iterative subtraction is able to identify single-pulses on the steeply falling edge of large, piled-up pulses.
Investigations with non iterative methods, in which only a single convolution-response is calculated, are not able to identify such photons, as the convolution-response is dominated by the steeply falling edge of the piled-up pulses.
Tn such scenarios, the rising edge of a single pulse is not enough to raise the convolution-response above the detection-threshold.
However, with an iterative subtraction, the large, piled-up pulse is flattened out one pulse after another until a trailing single-pulse on its falling edge can raise the convolution-response above the detection-threshold.
\chapter{Performance of extraction}
\label{ChPerformanceOfExtraction}
We estimate the performance and limitations of our single-photon-extractor in three different ways.
First, we evaluate the precision of our extracted arrival-times using simulated time-series.
Second, we crosscheck if the photon-stream for FACT created by our single-photon-extractor still contains the information that is relevant for our established reconstruction of cosmic particles from largest-pulses.
Third, we look for a saturation of the rate of extracted photons under different brightnesses of the night-sky.
%
%==============================================================================
%      _                 _       _   _
%     (_)               | |     | | (_)
%  ___ _ _ __ ___  _   _| | __ _| |_ _  ___  _ __  ___
% / __| | '_ ` _ \| | | | |/ _` | __| |/ _ \| '_ \/ __|
% \__ \ | | | | | | |_| | | (_| | |_| | (_) | | | \__ \
% |___/_|_| |_| |_|\__,_|_|\__,_|\__|_|\___/|_| |_|___/
%
%==============================================================================
\section{Validation on Simulations}
\label{SecValidationOnSimulations}
We perform two tests on simulated time-series.
First, we investigate the reconstruction of the arrival-time for one isolated single-photon with different amplitudes of electronic-white-noise.
Second, we investigate the reconstruction of the arrival-times of multiple photons during typical fluxes of night-sky-background-photons on FACT.
\subsection*{Arrival-time of a single, isolated photon}
\begin{figure}
    \begin{center}
        \includegraphics[width=1\textwidth]{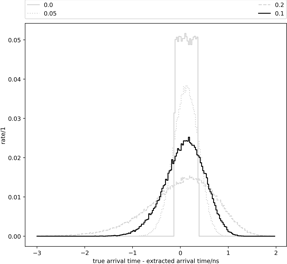}
        \caption[Arrival-time-resolution for a single-photon]{
            Performance of the reconstruction of the arrival-time of single-photons for four different amplitudes of electronic-white-noise.
            The corresponding amplitude of electronic-white-noise for FACT is shown in black.
            Table \ref{TabSinglePulseArrivalTimePerformance} shows the standard-deviation of the distribution of the residual arrival-times.
        }
        \label{FigSinglePulseArrivalTimePerformance}
    \end{center}
\end{figure}

We put one isolated pulse $A_\text{full}[t]$ on a time-series and add electronic-white-noise with different amplitudes.
The true arrival-time of the pulse is drawn uniformly.
Figure \ref{FigSinglePulseArrivalTimePerformance} shows the distribution of the
{\small
\begin{eqnarray}
\text{residual-arrival-time} &=& \text{true-arrival-time} - \text{extracted-arrival-time}
\end{eqnarray}}
for four different amplitudes of electronic-white-noise.
For each amplitude of electronic-white-noise we simulate $10^5$ trials.
Table \ref{TabSinglePulseArrivalTimePerformance} shows the standard-deviation of the distribution of the residual arrival-times depicted in Figure \ref{FigSinglePulseArrivalTimePerformance}.
We express the amplitudes of electronic-white-noise using the standard-deviation of the electronic-white-noise in units of the maximum amplitude $A$ of our single-pulse-template $A_\text{full}[t]$.
FACT has a measured amplitude of electronic-white-noise of $\approx 0.1\,A$.\\
We find that our single-photon-extractor reaches $416\,$ps resolution for arrival-times of isolated single-photons on time-series with amplitudes of electronic-white-noise that are typical for FACT.
We also find that our single-photon-extractor's intrinsic limitation of a fix sample-time-duration of $500\,$ps becomes the dominant limitation for the reconstruction of arrival-times when there is zero electronic-white-noise.
Finally, we find that the performance of the reconstruction of arrival-times of our single-photon-extractor scales inverse to the amplitude of the electronic-white-noise as can be seen in the gray colored findings in Figure \ref{FigSinglePulseArrivalTimePerformance} and Table \ref{TabSinglePulseArrivalTimePerformance}.
For an amplitude of electronic-white-noise $\leq 0.2\,A$, our single-photon-extractor detects exactly one photon in each trial.
Only if the electronic-white-noise increases beyond $0.4\,A$, our single-photon-extractor starts to detect false photons, as can be seen in the lower part of Table \ref{TabSinglePulseArrivalTimePerformance}.
We notice that the residual arrival-time of our single-photon-extractor has a small systematic offset of about $125\,$ps, see Figure \ref{FigSinglePulseArrivalTimePerformance}.
For the reconstruction of the air-shower, such absolute resolution of the arrival-times is not relevant for FACT.
We suspect this systematic to show up because of the asymmetric shape of the pulse and the limited sample-duration of $500\,$ps.
\definecolor{light-gray}{gray}{0.5}
\begin{table}
    \begin{center}
    \begin{tabular}{c c c c}
        Electronic-white-noise & Arrival-time & number of & number of\\
        amlitude/$A$ & resolution/ps & trials & detections\\
        \midrule
        \color{light-gray}{0.0} &
        \color{light-gray}{144} &
        \color{light-gray}{100,000} &
        \color{light-gray}{100,000}\\

        \color{light-gray}{0.05} &
        \color{light-gray}{260} &
        \color{light-gray}{100,000} &
        \color{light-gray}{100,000}\\
        0.1 & 416 & 100,000 & 100,000\\

        \color{light-gray}{0.2} &
        \color{light-gray}{718} &
        \color{light-gray}{100,000} &
        \color{light-gray}{100,000}\\
        \midrule

        \color{light-gray}{0.4} &
         &
        \color{light-gray}{1,000} &
        \color{light-gray}{1,021}\\

        \color{light-gray}{0.6} &
         &
        \color{light-gray}{1,000} &
        \color{light-gray}{1,483}\\

        \color{light-gray}{0.8} &
         &
        \color{light-gray}{1,000} &
        \color{light-gray}{5,477}\\
        \midrule
    \end{tabular}
    \caption[Performance of the reconstruction of the arrival-time for single-photons]{Resolution of the arrival-time of an isolated photon for four different amplitudes of electronic-white-noise.
        The resolution of the reconstruction of the arrival-times is the standard-deviation of the distribution of the residual arrival-times presented in Figure \ref{FigSinglePulseArrivalTimePerformance}.
        The corresponding amplitude of electronic-white-noise for FACT is shown in black.
        For zero noise, we find the standard-deviation of a uniform distribution within the $500\,$ps sampling-duration which is $500\,$ps$/\sqrt{12} \approx 144\,$ps.
    }
    \label{TabSinglePulseArrivalTimePerformance}
    \end{center}
\end{table}
\subsection*{Arrival-times of multiple photons}
\begin{figure}
    \begin{center}
        \includegraphics[width=1.0\textwidth]{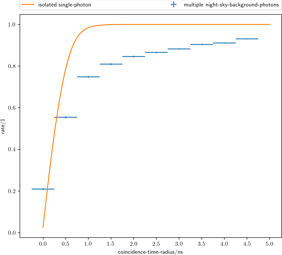}
        \caption[Arrival-time-resolution for multiple photons]{In blue is the true-positive-rate of our single-photon-extractor during typical observations on FACT at a rate of nigh-sky-background-photons of $50\times10^6$\,s$^{-1}$\,pixel$^{-1}$.
            In orange is the estimated true-positive-rate of our single-photon-extractor for isolated single-photons based on Figure \ref{FigSinglePulseArrivalTimePerformance}.
            The dark night at $21\,$mag\,argsec$^{-2}$ has only $37\times10^6$\,s$^{-1}$\,pixel$^{-1}$, see Figures \ref{FigFluxOfNightSkyBackgroundPhotons}, and \ref{FigExtractedRateVsNighSkyBackground}.
            For each coincidence-time-radius we simulate $216\mu$s exposure-time.
        }
        \label{FigBenchmark}
    \end{center}
\end{figure}

We expect the resolution of the arrival-times for single-photons to drop when the rate of photons increases.
For example, when the rate of photons approaches the sampling-frequency of the time-series, the time-series will be full of photons, but also it will be rather flat.
In such a case, large photon-multiplicities will still be recognized, but the arrival-times of the photons which form the base of the time-series will not be recognizable anymore with our single-pulse-extractor.\\
Here we simulate time-series with photons which represent the flux of night-sky-background-photons in FACT, such as shown and described in Section \ref{SecSinglePulseExtractorExample}, and Figure \ref{FigSinglePulseSubtraction1}.
On simulated time-series, we know the true arrival-times of the photons.
To quantify the performance, we define \textit{true-positive}-, \textit{false-negative}-, and \textit{false-positive}-matches between a true and an extracted arrival-time.
When counting the occurrences of these matches, we quantify the true-positive-rate \footnote{Also known as sensitivity.} of our single-photon-extractor.
After we ran our single-photon-extractor on a simulated time-series, we have both the list of true arrival-times and the list of extracted arrival-times.
We loop over the list of true arrival-times and will remove matching extracted arrival-times from the list of extracted arrival-times in the process.
In case the list of extracted arrival-times is already empty, we say it is a \textit{false-negative}-match.
Else, we calculate the residual arrival-time between the true arrival-time and its nearest match among the extracted arrival-times.
When this residual arrival-time is smaller than, or equal to a certain coincidence-time-radius, we call this match a \textit{true-positive}.
On the other hand, when the residual arrival-time is bigger than our coincidence-time-radius, we call this match a \textit{false-negative}.
Finally, when we have looped over the list of true arrival-times, we call each remaining extracted arrival-time a \textit{false-positive}-match.\\
Figure \ref{FigBenchmark} shows the true-positive-rate of our single-photon-extractor.
We find $75\%$ probability to extract the arrival-time of a single-photon within $1\,$ns of its true arrival-time.
We can not stress enough that this is about individual photons expected while staring into the night-sky-brightness on Canary island La Palma.
This is not about intense bunches of coincident photons from a calibration-source found in a lab.
%
%==============================================================================
%  _                _                                    _           _
% | |              | |                                  | |         | |
% | |__   __ _  ___| | __   ___ ___  _ ____   _____ _ __| |_ ___  __| |
% | '_ \ / _` |/ __| |/ /  / __/ _ \| '_ \ \ / / _ \ '__| __/ _ \/ _` |
% | |_) | (_| | (__|   <  | (_| (_) | | | \ V /  __/ |  | ||  __/ (_| |
% |_.__/ \__,_|\___|_|\_\  \___\___/|_| |_|\_/ \___|_|   \__\___|\__,_|
%
%==============================================================================
\section{Crosscheck with classic analysis}
\label{SecCrosscheckOnClassicAnalysisChain}
The list of arrival-times of single-photons can be converted back to an emulation of the original time-series.
We add the full-template-pulse $A_\text{full}[t]$ to a time-series at each arrival-time where a single-photon was extracted.
This back-converted time-series can be analyzed for air-shower-features by our classic extractor which uses largest-pulses on time-series.
When feeding back-converted time-series into our classic extractor for air-shower-features, we can compare the features from the back-converted time-series to the features extracted from the original time-series.
This way we can estimate if the photon-stream still contains the information that is relevant for our classic extractor for air-shower-features.
This crosscheck of course does not use any of the additional information in the photon-stream which is not accessible with largest-pulses.\\
So we perform a crosscheck with a sample of observations of FACT with $18\,$h exposure-time and 4,233,265 recorded air-shower-events.
This sample was simultaneously \cite{ahnen2016diss} recorded with the MAGIC telescopes, see Figure \ref{FigFactOverview}.
This means, that both MAGIC telescopes and FACT observed the same air-showers induced by the same cosmic-particles.
The records of this sample are sparsely scattered between November 2013 and December 2014.
MAGIC observes air-showers with two telescopes for a stereo-view.
Each MAGIC telescope has $230\,$m$^2$ aperture-area for Cherenkov-photons.
MAGIC's lower energy-threshold for gamma-rays is below $100\,$GeV which is about one-order-of-magnitude below the lower energy-threshold of FACT.
Because of this, the properties of the cosmic particle (direction, energy, and type) reconstructed by the two MAGIC telescopes are of excellent quality for cosmic particles with energies close to the energy-threshold of FACT at $\approx 1\,$TeV.
Therefore, we assume that the properties of the cosmic particle reconstructed by MAGIC are the \textit{true} properties and thus can serve as a benchmark for FACT.
Further, we will not only look at the high-level benchmark of the properties of the cosmic particle, but we will also take a brief look at the air-shower-features extracted and used in between.
\subsection*{Air-shower features}
\begin{figure}
    \centering
    \begin{subfigure}[b]{0.42\textwidth}
        \includegraphics[width=\textwidth]{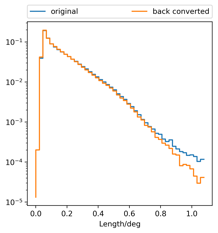}
    \end{subfigure}%
    \begin{subfigure}[b]{0.57\textwidth}
        \includegraphics[width=\textwidth]{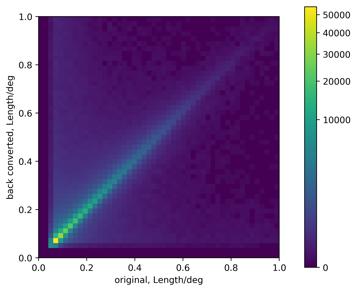}
    \end{subfigure}
    \caption[Distribution of Hillas-Length]{\textit{Hillas}-Length, the long radius of the ellipse of the air-shower.
    The deviation for \textit{Hillas}-Length$ > 0.8\,$deg is not understood yet.}
    \label{FigHillasLength}
\end{figure}

\begin{figure}
    \centering
    \begin{subfigure}[b]{0.42\textwidth}
        \includegraphics[width=\textwidth]{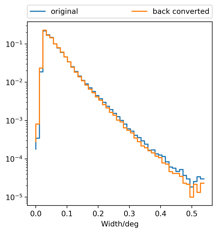}
    \end{subfigure}%
    \begin{subfigure}[b]{0.57\textwidth}
        \includegraphics[width=\textwidth]{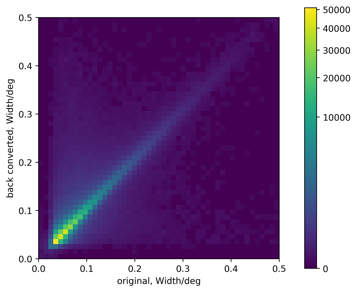}
    \end{subfigure}
    \caption[Distribution of Hillas-Width]{\textit{Hillas}-Width, the short radius of the ellipse of the air-shower.}
    \label{FigHillasWidth}
\end{figure}

\begin{figure}
    \centering
    \begin{subfigure}[b]{0.42\textwidth}
        \includegraphics[width=\textwidth]{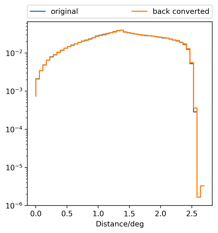}
    \end{subfigure}%
    \begin{subfigure}[b]{0.57\textwidth}
        \includegraphics[width=\textwidth]{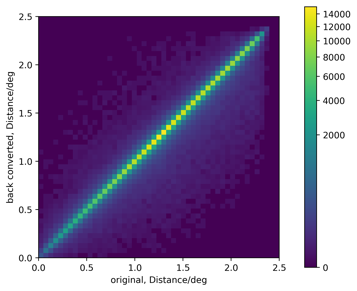}
    \end{subfigure}
    \caption[Distribution of Hillas-Distance]{\textit{Hillas}-Distance between the center of the air-shower-ellipse and the center of the image.}
    \label{FigHillasDistance}
\end{figure}

\begin{figure}
    \centering
    \begin{subfigure}[b]{0.42\textwidth}
        \includegraphics[width=\textwidth]{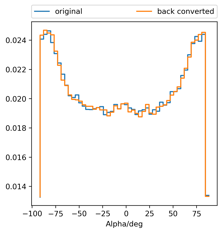}
    \end{subfigure}%
    \begin{subfigure}[b]{0.57\textwidth}
        \includegraphics[width=\textwidth]{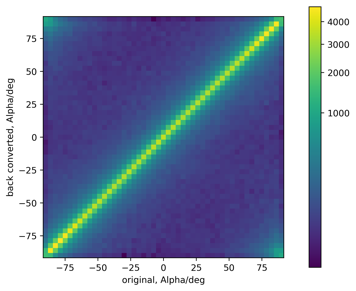}
    \end{subfigure}
    \caption[Distribution of Hillas-Alpha]{\textit{Hillas}-Alpha, orientation of air-shower-ellipse in the image.}
    \label{FigHillasAlpha}
\end{figure}

\begin{figure}
    \centering
    \begin{subfigure}[b]{0.42\textwidth}
        \includegraphics[width=\textwidth]{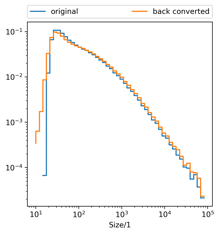}
    \end{subfigure}%
    \begin{subfigure}[b]{0.57\textwidth}
        \includegraphics[width=\textwidth]{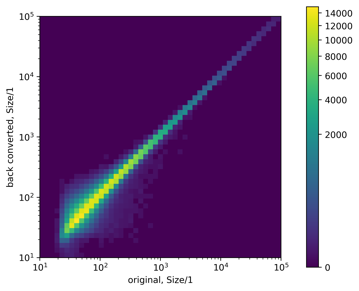}
    \end{subfigure}
    \caption[Distribution of Hillas-Size]{\textit{Hillas}-Size, number of photons reconstructed from the air-shower.}
    \label{FigHillasSize}
\end{figure}

\begin{figure}
    \centering
    \begin{subfigure}[b]{0.42\textwidth}
        \includegraphics[width=\textwidth]{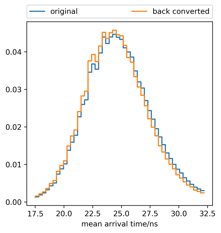}
    \end{subfigure}%
    \begin{subfigure}[b]{0.57\textwidth}
        \includegraphics[width=\textwidth]{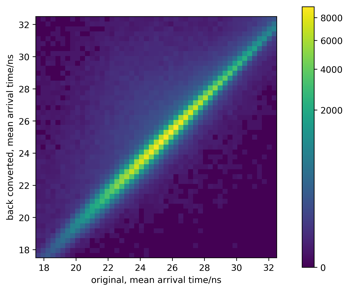}
    \end{subfigure}
    \caption[Distribution of arrival-time-mean]{Arrival-time-mean of the largest-pulses associated with the air-shower.}
    \label{FigHillasArrivalTime}
\end{figure}

\begin{figure}
    \centering
    \begin{subfigure}[b]{0.42\textwidth}
        \includegraphics[width=\textwidth]{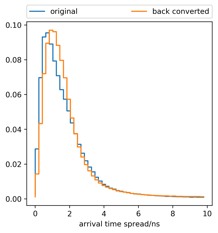}
    \end{subfigure}%
    \begin{subfigure}[b]{0.57\textwidth}
        \includegraphics[width=\textwidth]{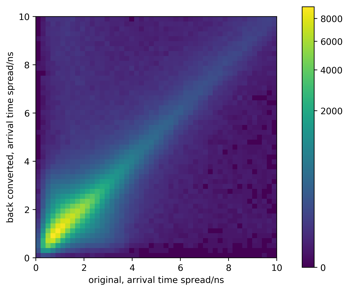}
    \end{subfigure}
    \caption[Distribution of arrival-time-spread]{Arrival-time-spread of the largest-pulses associated with the air-shower.}
    \label{FigHillasTimespread}
\end{figure}

The intermediate representation of the air-shower is done with \textit{Hillas}-features, see Figure \ref{FigHillasFeatures}.
The Figures \ref{FigHillasLength}, \ref{FigHillasWidth}, \ref{FigHillasDistance}, \ref{FigHillasAlpha} and \ref{FigHillasSize} show the \textit{Hillas}-features which model the photon-distribution in the image using a simple ellipse.
The one-dimensional histogram on the left side of the figures shows the distributions of the features, and the two-dimensional histogram on the right side of the figures shows the confusion between the two time-series.
The figures show that the distributions of the \textit{Hillas}-features are similar for the original and the back-converted time-series.
Also the figures show that the confusion is low between the features reconstructed from the original and from the back-converted time-series.
The spill-over in the confusion of \textit{Hillas}-Alpha in Figure \ref{FigHillasAlpha} at $\pm 90^\circ$ is because of the $180^\circ$ rotation-invariance of the ellipse-model.
The small deviation for long air-showers in \textit{Hillas}-Length, see Figures \ref{FigHillasLength}, is not understood.
Since there is not such a large deviation for air-showers with many photons in \textit{Hillas}-Size, see Figures \ref{FigHillasSize}, the deviations in \textit{Hillas}-Length do not seem to be a saturation-effect of the single-pulse-extractor.
In \textit{Hillas}-Size we find more air-showers with low numbers of photons reconstructed from the back-converted time-series.
This might be because the generation of the air-shower-ellipse is sensitive to electronic-white-noise, which is not present on the back-converted time-series.
The Figures \ref{FigHillasArrivalTime}, \ref{FigHillasTimespread} show the air-shower-features related to time.
The mean arrival-time of the largest-pulses associated with the air-shower have similar distributions with even similar substructures for both the original and back-converted time-series.
The spread of the arrival-times of the largest-pulses associated with the air-shower, see Figure \ref{FigHillasTimespread}, is a bit wider for the back-converted time-series.
This might be either a hint for a loss in precision of the arrival-time from the back-converted time-series, see Section \ref{SecValidationOnSimulations}, or a hint for different, probably more pixels in the air-shower-ellipses which have a wider time-spread.
Overall the similarities between the air-shower-features extracted from both time-series are very high, and we conclude that our classic reconstruction of the cosmic particle should be able to use the back-converted time-series.
\subsection*{Reconstructed properties of the cosmic particle}
\begin{figure}
    \centering
    \begin{subfigure}[b]{0.42\textwidth}
        \includegraphics[width=\textwidth]{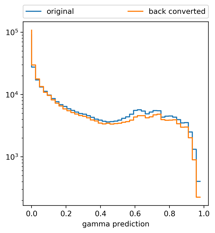}
    \end{subfigure}%
    \begin{subfigure}[b]{0.57\textwidth}
        \includegraphics[width=\textwidth]{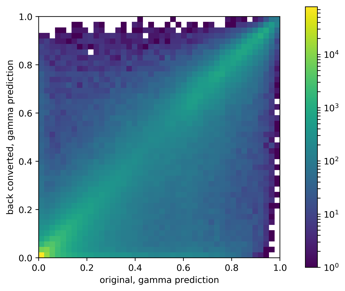}
    \end{subfigure}
    \caption[Distribution of gamma-ray-prediction]{Gamma-prediction ranging from not a gamma-ray $0$ to gamma-ray $1$.}
    \label{FigGammaPrediction}
\end{figure}

\begin{figure}
    \includegraphics[width=1.0\textwidth]{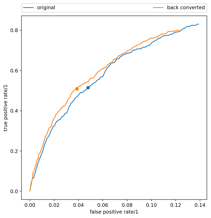}
    \caption[Receiver-Operating-Characteristic on gamma-ray-prediction]{Receiver-Operating-Characteristic (ROC) curves for original and back-converted time-series.
        Here the gamma-prediction cut, see Figure \ref{FigGammaPrediction} is varied from 0.7 to 1.0.
        The dots mark the standard gamma-prediction-cut of $0.85$.
    }
    \label{FigRoc}
\end{figure}

\begin{figure}
    \includegraphics[width=1.0\textwidth]{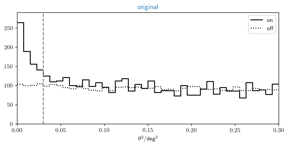}
    \includegraphics[width=1.0\textwidth]{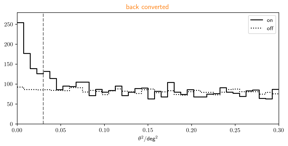}
    \caption[Gamma-ray-source-direction-reconstruction]{Great-circle-distance $\theta$ between the reconstructed position of the cosmic source and the true position of the source 'on'.
        As well as distance $\theta$ between the reconstructed and a false position of the source 'off'.
        When counting the events in both 'on' and 'off' regions, the significance of the detection can be estimated\cite{nothe2017fact}.
        Upper graph shows $\theta$ for the original time-series, and the lower graph shows $\theta$ for the back-converted time-series.
        Vertical and dashed line at $\theta^2 = 0.03\,$deg$^2$ marks the border of on- and off-regions.
    }
    \label{FigTheta2}
\end{figure}

The gamma-prediction is the outcome of a machine-learner which uses the ensemble of extracted air-shower-features to estimate the properties of the cosmic particle, in this case the type of the particle.
Figure \ref{FigGammaPrediction} shows the gamma-prediction to have a similar distribution and confusion for the original and back-converted time-series.
The machine-learner here was only trained on features extracted from the original-time-series but is found to be able to assign features from the back-converted time-series appropriately.
Since this sample consists vastly out of protons and not out of gamma-rays, the distribution and confusion in Figure \ref{FigGammaPrediction} show a great excess for low gamma-predictions.
Based on the gamma-prediction estimated from the extracted air-shower-features, FACT classifies the cosmic particle to be of type gamma-ray or not.
The standard\cite{nothe2017fact} cut for the gamma-prediction is $\geq 0.85$ to be classified as a gamma-ray.
Based on the claims on the detection of gamma-rays in both MAGIC and FACT, we define matching-scenarios, see Table \ref{TabMatchingScenarios}.
\begin{table}
    \begin{center}
    \begin{adjustbox}{angle=-90}
    \begin{tabular}{ l  r  r  r  r  r  r  r}
        scenario  &    &    &    & \rotatebox{90}{\textit{true-negative}} & \rotatebox{90}{\textit{false-positive}}  & \rotatebox{90}{\textit{true-positive}}  & \rotatebox{90}{\textit{false-negative}} \\
        \midrule
        trigger                & $\bullet$ & $\bullet$ & $\bullet$  & $\bullet$  & $\bullet$  & $\bullet$ & $\bullet$ \\
        image-cleaning          &    & $\bullet$ & $\bullet$ & $\bullet$  & $\bullet$  & $\bullet$ & $\bullet$ \\
        ellipse-requirements             &    &  & $\bullet$ & $\bullet$  & $\bullet$  & $\bullet$ & $\bullet$ \\
        gamma-prediction $\geq 0.85$ &    &    &  &  & $\bullet$  & $\bullet$ &   \\
        gamma-prediction $< 0.85$ &    &  &  &  $\bullet$  &   &  &  $\bullet$ \\
        is gamma &    &    &  &  &    & $\bullet$ & $\bullet$ \\
        is not gamma &  &  &  & $\bullet$ & $\bullet$ &  &  \\
        \midrule
        Number of events\\
        original       &
        4,233,265 &
        700,045 &
        373,109 &
        354,938 &
        17,812 &
        185 &
        174 \\
        back-converted &
        4,233,265 &
        589,802 &
        397,034 &
        381,273 &
        15,372 &
        198 &
        191 \\
        \color{light-gray}{overlap} &
        \color{light-gray}{4,233,265} &
        \color{light-gray}{584,560} &
        \color{light-gray}{346,100} &
        \color{light-gray}{324,531} &
        \color{light-gray}{11,166} &
        \color{light-gray}{164} &
        \color{light-gray}{134} \\
    \end{tabular}
    \end{adjustbox}
    \caption[Air-showers simultaneously observed by MAGIC and FACT]{The four possible matching-scenarios for each simultaneously observed air-shower-event of MAGIC and FACT.
        Here the number of events is shown for a gamma-prediction $\geq 0.85$, see dot in Figure \ref{FigRoc}.
        'trigger' means that these events have activated the trigger in the image-sensor of FACT.
        'image-cleaning' means that after the night-sky-background was removed, these events still contained the minimal structures necessary to apply the \textit{Hillas}-ellipse-model.
        'ellipse-requirements' means that these events have air-shower-ellipses which do not leak too far out of the image and have not too many sub-structures beside the ellipse.
    }
    \label{TabMatchingScenarios}
    \end{center}
\end{table}
When counting the matching-scenarios, we can estimate the true-positive-rate and the false-positive-rate for both the events in the original time-series  and the events in the back-converted time-series.
Figure \ref{FigRoc} shows the ratio of the true-positive-rate and the false-positive-rate for both the original and the back-converted time-series.
The gamma-prediction cut parameter is varied from 0.7 to 1.
For a gamma-prediction threshold of $0.85$, the true-positive-rate of the original time-series is $0.515$ and $0.509$ for the back-converted time-series.
However, the back-converted time-series reaches $0.0388$ false-positive-rate while the original time-series only reaches $0.0478$.
This means that when we use the back-converted time-series from the photon-stream, the signal-rate of gamma-rays drops by $1\%$, but at the same time the background-rate of cosmic-rays drops by $23\%$ which will be beneficial for gamma-ray-astronomy.
\\Finally, we take a look at the reconstruction of the incident-direction of the cosmic particle.
Together with the gamma-ray-prediction, the incident-direction of a cosmic particle are important to detect cosmic sources of gamma-rays.
Here we follow the standard\cite{nothe2017fact} procedure to detect a cosmic source of gamma-rays with FACT.
Table \ref{TabSignificance} and Figure \ref{FigTheta2} show the significance of a detection of a cosmic source of gamma-rays for both the events processed from the original time-series and the events processed from the back-converted time-series.
\begin{table}
    \begin{center}
    \begin{tabular}{ l r r r r}
        & Significance Li\&Ma\cite{li1983analysis} & $N_\text{on}$ & $N_\text{off}$ & $t_\text{on}/t_\text{off}$ \\
        \midrule
        original       & 14.41 & 750 & 2046 & 0.2 \\
        back-converted & 15.58 & 696 & 1755 & 0.2 \\
    \end{tabular}
    \caption[Significance of a detection of a source of gamma-rays]{
        Significance of a detection of a cosmic source of gamma-rays for both original and back-converted time-series.
        Higher significance is better.
        Intermediate results $N_\text{on}$, $N_\text{off}$ are treated according to the standard\cite{nothe2017fact} procedure for detection of cosmic sources of gamma-rays with FACT.
        Figure \ref{FigTheta2} shows the reconstructed directions of the cosmic particle which are the basis for the significance of the detection of a source.
    }
    \label{TabSignificance}
    \end{center}
\end{table}
\begin{figure}
    \begin{minipage}{0.475\textwidth}
        \begin{figure}[H]
            \includegraphics[width=1.0\textwidth]{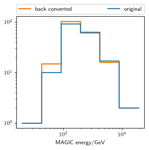}
            \caption[Energy reconstruction for true-positive matches]{Gamma-ray-energy reconstructed by MAGIC for \textit{true-positive} events in Table \ref{TabMatchingScenarios}.}
            \label{FigEnergyTruePositives}
        \end{figure}
    \end{minipage}
    \hfill
    \begin{minipage}{0.475\textwidth}
        \begin{figure}[H]
            \includegraphics[width=1.0\textwidth]{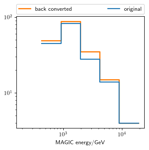}
            \caption[Energy reconstruction for false-negative matches]{Gamma-ray-energy reconstructed by MAGIC for \textit{false-negative} events in Table \ref{TabMatchingScenarios}.}
            \label{FigEnergyFalseNegatives}
        \end{figure}
    \end{minipage}
\end{figure}
\subsection*{Summary on back-conversion}
We conclude that the back-converted time-series from the photon-stream contain the information for gamma-ray-astronomy that is relevant to our established extraction of air-shower-features from largest-pulses.
The drop in false-positive-rate with the back-converted time-series in contrast to the original time-series makes the back-converted time-series from the photon-stream even preferable as the observations are dominated by cosmic-rays by $\approx 3$ orders-of-magnitude.
Figure \ref{FigRoc} shows a gain in 'area under the ROC curve' for the back-converted time-series across the whole relevant space of gamma-prediction cuts.
As a result we find that the significance to detect a cosmic source of gamma-rays increases when using the back-converted time-series from the photon-stream, see Table \ref{TabSignificance}.
The Figures \ref{FigEnergyTruePositives} and \ref{FigEnergyFalseNegatives} show that the reconstruction-performance from both the original and the back-converted time-series follow a similar dependency of the energy.
For \textit{true-positive} gamma-rays in Figure \ref{FigEnergyTruePositives}, the back-converted time-series offer slightly better performance for energies below $1\,$TeV.
Such a good reproduction of air-shower-features, yet even an indication for a gain in the power to detect gamma-rays are not anticipated when keeping in mind that the back-converted time-series from the photon-stream is just an emulation of the original time-series.\\
However, the single-photon-extractor is first of all a filter.
We suspect that the filtering of the single-photon-extractor, which only passes on signals which can be represented by adding up single-pulses, is beneficial for the established classification of Cherenkov-photons on largest-pulses.
During the development we found a strong rejection-power of our single-photon-extractor to artifacts as e.g a sinusoidal resonances which occurs occasionally on the original time-series of FACT (See 'ringing', \cite{vogler2015diss}).
Further, our established classification of Cherenkov-photons on largest-pulses uses statistical estimators which are biased like a search for the maximum.
These biased estimators profit from the lack of electronic-noise on the back-converted time-series in contrast to the original time-series which has electronic-noise.\\
Finally, we want to stress again that this crosscheck does not show the full potential of the photon-stream.
This crosscheck only shows that the photon-stream contains at least the information for gamma-ray-astronomy which is relevant to our established methods based on the representation of air-shower-events using largest-pulses.
\section{Saturation}
\begin{figure}
    \includegraphics[width=1.0\textwidth]{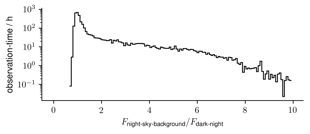}
    \caption[Flux of night-sky-background-photons]{The flux $F_\text{night-sky-background}$ of night-sky-background-photons observed with the sky-quality-meter\cite{birriel2010simple} of FACT.
        Even though FACT observes during strong moon-light, the majority of the observations is done during dark nights.
    }
    \label{FigFluxOfNightSkyBackgroundPhotons}
\end{figure}

\begin{figure}
    \includegraphics[width=1.0\textwidth]{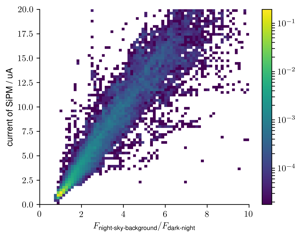}
    \caption[Current in a SiPM vs. flux of photons]{The current released in the GAPDs of the SiPMs used in FACT is proportional to the flux of night-sky-background-photons observed by the sky-quality-meter of FACT.
    }
    \label{FigSipmCurrentsVsNighSkyBackground}
\end{figure}

As mentioned in Section \ref{SecValidationOnSimulations}, we expect our single-photon-extractor to miss photons when the rate of photons becomes too high.
For the first implementation of the single-photon-extractor we focused on a good arrival-time resolution for single-photons during low fluxes of night-sky-background-photons.
Figure \ref{FigFluxOfNightSkyBackgroundPhotons} shows how often FACT observes during different fluxes of night-sky-background-photons.
The typical flux of night-sky-background-photons is $F_\text{dark-night}$ and corresponds to a surface-brightness of $\approx 21$\,mag/arcsec$^2$.
Figure \ref{FigSipmCurrentsVsNighSkyBackground} shows how much current is flowing through the GAPDs in an SiPM-pixel for different fluxes of photons.
Figure \ref{FigSipmCurrentsVsNighSkyBackground} shows that the current through the SiPMs does not saturate in the relevant range of fluxes of night-sky-background-photons.
\begin{figure}
    \includegraphics[width=1.0\textwidth]{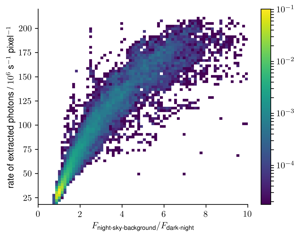}
    \caption[Rate of extracted photons vs. flux of photons]{The rate of extracted photons from a time-series of a pixel in FACT correlates with the flux of night-sky-background-photons, but is not exactly proportional.
        The higher the flux of photons, the more photons are missed by our single-photon-extractor.
    }
    \label{FigExtractedRateVsNighSkyBackground}
\end{figure}
The current flowing through the SiPMs in FACT is rather proportional to the flux of night-sky-background-photons.
Figure \ref{FigExtractedRateVsNighSkyBackground} shows that the rate of extracted photons correlates with the flux of night-sky-background-photons and is rather proportional in the range of fluxes of night-sky-background-photons which are most common on FACT.
All Figures \ref{FigFluxOfNightSkyBackgroundPhotons}, \ref{FigSipmCurrentsVsNighSkyBackground}, and \ref{FigExtractedRateVsNighSkyBackground} are based on the same sample which spans the time from July 2014 to September 2016.
Each entry for the rate of extracted photons is estimated for a time-window of 5\,min and is without the Cherenkov-photons.\\
We find that our single-photon-extractor misses more and more photons when the flux of photons increases.
We find the saturation takes place in our single-photon-extractor since the current released by the SiPMs does not show such a saturation.
However, our first implementation of the single-photon-extractor extracts photons sufficiently proportional to the flux of photons.
Especially in the range of fluxes of night-sky-background-photons from $1$ to $3 \times F_\text{dark-night}$, where FACT observes most, our single-photon-extractor responses proportional to the flux of photons.
Up to fluxes of $8 \times F_\text{dark-night}$, we do not see a hard limit of saturation of our single-photon-extractor yet.
%
%%%%%%%%%%%%%%%%%%%%%%%%%%%%%%%%%%%%%%%%%%%%%%%%%%%%%%%%%%%%%%%%%%%%%%%%%%%%%%%
%
%  ______               _ _
%  |  _  \             (_) |
%  | | | |___ _ __  ___ _| |_ _   _
%  | | | / _ \ '_ \/ __| | __| | | |
%  | |/ /  __/ | | \__ \ | |_| |_| |
%  |___/ \___|_| |_|___/_|\__|\__, |
%                              __/ |
%                             |___/
%
%%%%%%%%%%%%%%%%%%%%%%%%%%%%%%%%%%%%%%%%%%%%%%%%%%%%%%%%%%%%%%%%%%%%%%%%%%%%%%%
%
\chapter{Classifying Cherenkov-photons}
\label{ChaDensityBasedClustering}
Cherenkov-photons from one specific air-shower are likely to have similar incident-directions and arrival-times as they are produced in a limited volume in the atmosphere and as they reach the ground almost at the same time.
Night-sky-background-photons on the other hand are present all the time.
And, with the exception of starlight, night-sky-background-photons have random incident-directions.
Here we try to take advantage of the photon-stream-representation to classify Cherenkov-photons.
In this first attempt, we use solely the density of photons in the three-dimensional space of observables in the photon-stream.\\
We will first propose how to express the density of photons in the photon-stream, and second we will discuss an established algorithm which we find to be well suited for finding dense clusters of Cherenkov-photons within the pool of night-sky-background-photons.
\section{Density in the photon-stream}
\begin{figure}
    \begin{minipage}{0.5\textwidth}
        \begin{figure}[H]
            \includegraphics[width=1.0\textwidth]{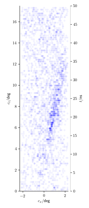}
        \end{figure}
    \end{minipage}
    \hfill
    \begin{minipage}{0.5\textwidth}
        \begin{figure}[H]
            \includegraphics[width=1.0\textwidth]{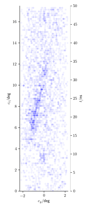}
        \end{figure}
    \end{minipage}
    \caption[The time-axis of the photon-stream to scale]{Two-dimensional projection of the photon-stream onto $c_x$ and $c_t$ on the left, and $c_y$ and $c_t$ on the right.
        The $c_t$-axis is scaled accordingly to Equation \ref{EqTimeScaling}.
        Here, the distances in $c_x$, and $c_y$ have the same importance as the distances in $c_t$.
        This is the same air-shower-event as shown in Figures \ref{FigLargetsPulseExampleEvent}, \ref{FigPhotonStreamExampleEvent} and \ref{FigDbscanExampleEvent}.}
    \label{FigDistanceMetrik}
\end{figure}

The three-dimensional space of the photon-stream mixes incident-directions $c_x$, $c_y$, and arrival-time $t$.
Therefore it is not obvious how to express a distance $d_{i,j}$ in the photon-stream between two photons $i$ and $j$.
To express the density of photons in the photon-stream, we have to introduce a metric which e.g. scales the time-axis to mediate between incident-directions and arrival-time.
In our implementation for FACT, we define
\begin{eqnarray}
c_t &=& \alpha t
\label{EqTimeScaling}
\end{eqnarray}
and choose the scaling of the time-axis
\begin{eqnarray}
\alpha &=& 0.35 \times 10^9 \frac{\text{Deg}}{\text{s}}
\label{EqAlphaForFact}
\end{eqnarray}
such that the one-dimensional density of photons in the scaled space of $c_x, c_y,$ and $c_t$ is the same along all three axes for a typical rate of photons  arriving during a dark night.
Depending whether one wants to prefer or penalize the coincidence among incident-directions over the coincidence among arrival-times, one can alter the scaling $\alpha$ accordingly.
With the scaling we define
\begin{eqnarray}
d_{i,j} &=& \sqrt{(c_x^i - c_x^j)^2 + (c_y^i - c_y^j)^2 + (c_t^i - c_t^j)^2}
\label{EqDistanceInPhotonStream}
\end{eqnarray}
to be the distance between two photons $i$, and $j$.
Figure \ref{FigDistanceMetrik} shows an example air-shower-event with the axes $c_x$, $c_y$, and $c_t$ to scale.
\section{DBSCAN Algorithm}
\begin{figure}
    \begin{center}
        \includegraphics[width=1\textwidth]{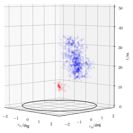}
        \caption[An event in photon-stream-representation]{
            The air-shower-classification based on density identified two clusters in the photon-stream.
            The larger cluster is blue, and the smaller cluster is red.
            In the established classification of Cherenkov-photons using the representation of largest-pulses and image-cleaning, such separate clusters were often called 'islands'.
            This is the same air-shower-event as shown in Figures \ref{FigLargetsPulseExampleEvent}, and \ref{FigPhotonStreamExampleEvent}.
        }
        \label{FigDbscanExampleEvent}
    \end{center}
\end{figure}

We find that the 'density based algorithm for discovering clusters with noise' (DBSCAN \cite{ester1996density}) is well suited to identify dense clusters of photons in the three-dimensional space of observables in the photon-stream
\footnote{In FACT we use the DBSCAN implementation of the scientific-python-kit\cite{scikit-learn}.}
.
Here we briefly describe DBSCAN from the photon-stream and Cherenkov-telescope point-of-view.\\
DBSCAN assigns each photon in the stream either to a cluster of dense Cherenkov-photons, or to the night-sky-background-photons.
DBSCAN can identify multiple dense clusters of Cherenkov-photons at once, and makes no assumptions on the shape of the clusters.
DBSCAN has only two parameters.
First, there is the minimal number of photons $m$ a dense cluster of Cherenkov-photons must have.
On FACT we found $m=20$ to be a decent first choice.
Second, there is the maximum distance $\epsilon$ between each two photons which must not be exceeded for each two photons to be considered to be dense.
On FACT, where we decided to express the time $t$ using the scaled $c_t$, we found $\epsilon=0.45^\circ$ to be a decent choice.
The algorithm is split into two repeating processes.
First, there is cluster-discovery which loops over the photons until it finds a photon which has at least $m$ photons in its neighborhood of radius $\epsilon$.
Then second, there is cluster-expansion which adds photons to the newly discovered cluster when these photons are either directly density-reachable with respect to $m$ and $\epsilon$, or when these photons can be reached via a chain of photons in the cluster which are directly density-reachable.
When all directly reachable, and iteratively reachable photons have been added to the cluster, the cluster-discovery starts again on the photons which are not yet assigned to a cluster.
When all photons where looped, the photons which do not belong to a dense cluster are defined to be night-sky-background-photons.
See Figure \ref{FigDbscanExampleEvent} to see two dense clusters of photons discovered by DBSCAN in an example event in the representation of the photon-stream.
\section{Summary}
We propose to classify Cherenkov-photons in the photon-stream with the DBSCAN algorithm.
DBSCAN has a concept of background, and can discover multiple dense clusters.
The classification of Cherenkov-photons in the photon-stream with DBSCAN depends only on three parameters.
First, the scaling $\alpha$ between the axes in the photon-stream.
Second, the minimum number $m$ of photons in a cluster.
And third, the maximum distance $\epsilon$ between two photons to be considered to be dense.
%
%%%%%%%%%%%%%%%%%%%%%%%%%%%%%%%%%%%%%%%%%%%%%%%%%%%%%%%%%%%%%%%%%%%%%%%%%%%%%%%
%
%  ______          __                                           ______
%  | ___ \        / _|                                          |  _  \
%  | |_/ /__ _ __| |_ ___  _ __ _ __ ___   __ _ _ __   ___ ___  | | | |
%  |  __/ _ \ '__|  _/ _ \| '__| '_ ` _ \ / _` | '_ \ / __/ _ \ | | | |
%  | | |  __/ |  | || (_) | |  | | | | | | (_| | | | | (_|  __/ | |/ /
%  \_|  \___|_|  |_| \___/|_|  |_| |_| |_|\__,_|_| |_|\___\___| |___/
%
%%%%%%%%%%%%%%%%%%%%%%%%%%%%%%%%%%%%%%%%%%%%%%%%%%%%%%%%%%%%%%%%%%%%%%%%%%%%%%%
%
\chapter{Performance of classification}
\label{ChAirShowerClassification}
We evaluate simulated air-shower-events of FACT to compare the performance of the classification of Cherenkov-photons.
We compare the classification of Cherenkov-photons in the photon-stream based on density with the established classification of Cherenkov-photons based on image-cleaning on largest-pulses.
Both methods use the density of photons in the space of observables $c_x$, $c_y$, and $t$ to classify Cherenkov-photons.
However, the established classification on largest-pulses with image-cleaning has multiple and consecutive stages where in each stage only a sub-set of the observables is taken into account.
Typically first a stage in time $t$, second a stage in incident directions $c_x$, and $c_y$, and third again a stage in time $t$.
Our novel classification in the photon-stream on the other hand has only one single stage where all observables are taken into account at once.\\
Here we will first introduce a measure of performance to discuss the performance of the two methods in a quantitative way.
Second, we will outline the procedure of the simulation of the air-shower-events.
Third, we will briefly describe our implementations of the two methods.
And fourth, we will show the resulting classification of Cherenkov-photons in example events, and discuss the results based on a sample of simulated air-shower-events which is representative for the observations of FACT.
\section{Measure of performance on images}
We estimate the performance of the classification of Cherenkov-photons by comparing the set of photons which was classified to be Cherenkov-photons with the set of true Cherenkov-photons.
For simplicity, we compare images of photon-intensities.
An image of photon-intensities is a vector $I$ of photon-intensities for all pixels.
An image of photon-intensities has as many dimensions as there are pixels, which is dim($I_\text{FACT}$) = 1440 in the case of FACT.
Images of photon-intensities are easy to visualize, see Figures \ref{FigLargetsPulseExampleEvent}, and are currently the basis of the generation of air-shower-features like the Hillas-features.
However, images of photon-intensities alone neglect the time-structure of the air-shower.
For each event we create three images.
First, we create an image of the true Cherenkov-photons.
Second, we create an image of the Cherenkov-photons classified in the photon-stream using clustering based on density.
And third, we create an image of the Cherenkov-photons classified on largest-pulses using image-cleaning.
We quantify the performance of the reconstruction of the Cherenkov-photons using two performance-measures.
First we use the angle
\newcommand{\norm}[1]{\left\lVert#1\right\rVert}
\begin{eqnarray}
\delta_{1,2} &=& \arccos \left( \frac{I_1 \cdot I_2}{\norm{I_1} \norm{I_2}} \right)
\end{eqnarray}
between the images $I_1$ and $I_2$ to compare the structure in the images independently of the intensity.
Second, we use the euclidean distance
\begin{eqnarray}
D_{1,2} &=& \norm{I_1 - I_2}
\end{eqnarray}
between the images $I_1$ and $I_2$ to compare the overall similarity.
The smaller the angle $\delta_{\text{reconstructed},\,\text{true}}$ and the smaller the distance $D_{\text{reconstructed},\,\text{true}}$ are between a true image of the air-shower $I_\text{true}$ and a reconstructed image of the air-shower $I_\text{reconstructed}$, the higher is the performance of the reconstruction.
\section{Procedure of the simulation}
\begin{figure}
    \begin{center}
        \includegraphics[width=1\textwidth]{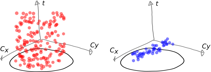}
        \caption[Representing origins of Cherenkov-photons]{Simulated telescope response.
        The true arrival-times $t$ of both night-sky-background-photons (red), and Cherenkov-photons (blue) across the image-plane in $c_x$ and $c_y$.
        Black ring illustrates the telescope's field-of-view.}
        \label{FigAclTrueResponse}{}
    \end{center}
\end{figure}

We simulate the observation of air-showers on the FACT telescope.
For air-showers which are bright enough to trigger the telescope, we take the true arrival-times of photons detected in the image-sensor.
In the simulation, we know whether a photon originated from the air-shower or from the night-sky-background.
The arrival-times of the these photons can be represented in two three-dimensional point-clouds, see Figure \ref{FigAclTrueResponse}.
From the arrival times of the photons, we emulate the electric response of the image-sensor.
For each pixel we emulate the time-series of the response of the photo-sensor, compare section \ref{SecCrosscheckOnClassicAnalysisChain}.
From these emulated representation of the air-shower-event using time-series, we proceed in two branches.
First, we run the established\cite{nothe2017fact} reconstruction for air-showers of FACT which describes the photons using largest-pulses and classifies Cherenkov-photons in multiple, separate stages of density clustering along different dimensions of observables.
And second we run the novel reconstruction for air-showers based on the photon-stream which reconstructs the arrival-times of single-photons with our single-photon-extractor, and classifies Cherenkov-photons in a single stage based on the density in the photon-stream.
\section{Classification on the largest-pulses}
We use the standard\cite{nothe2017fact} classification for Cherenkov-photons in FACT.
In the implementation of FACT, the classification of Cherenkov-photons on largest-pulses is split into three stages.
In the first stage, largest-pulses are identified on the time-series of each pixel individually, which corresponds to a search of high photon-density along the dimension of time $t$.
In the second stage, the photon-equivalents of the largest-pulses are compared with the photon-equivalents of the largest-pulses found in neighboring pixels to identify regions of high photon-density in the dimensions of the incident-directions $c_x$, and $c_y$.
In the implementation for FACT, pixels with photon-equivalents above a threshold of $5.5\,$p.e., and neighboring pixels with photon-equivalents above a threshold of $3\,$p.e. are declared to be candidates for pixels containing the Cherenkov-photons.
Other implementations might set the thresholds for the pixels dynamically based on the variance of the occurrence of night-sky-background-photons \cite{krawczynski2006gamma}.
In the third stage, the arrival-times of the largest-pulses are taken into account to reject candidate-pixels which have largest-pulses arriving more than $10\,$ns off the median arrival-time of the largest-pulses of all the candidate-pixels \cite{aliu2009improving}.
The combination of the second and the third stage are often called image-cleaning.
Figure \ref{FigLargetsPulseExampleEvent} shows the full input of information to the image-cleaning on largest-pulses used in FACT.
Finally, we create an image of photon-intensities only from those largest-pulses which are considered to contain Cherenkov-photons as they passed the cleaning of the image.
\section{Classification in the photon-stream}
From the time-series of each pixel, we extract single-pulses, as shown in Chapter \ref{ChSinglePhotonExtractor}.
On the resulting photon-stream we run the clustering based on density, see Chapter \ref{ChaDensityBasedClustering}, to classify Cherenkov-photons and night-sky-background-photons.
Finally, we project the cluster of Cherenkov-photons into the plane of incident-directions to have an image of the photon-intensities associated with the air-shower.
\section{Results}
\begin{figure}
    \begin{center}
        \includegraphics[width=1\textwidth]{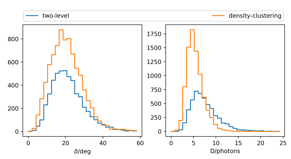}
        \caption[Photon-classification, gamma-rays]{
            Gamma-rays.
            Density-clustering in the photon-stream finds 63\% more air-showers with same or even better distributions of performance-measures $\delta$ and $D$.
            See Table \ref{TabAclPerformance}.
        }
        \label{FigAclIndependentPhsVsMp_gamma}
    \end{center}
\end{figure}

\begin{figure}
    \begin{center}
        \includegraphics[width=1\textwidth]{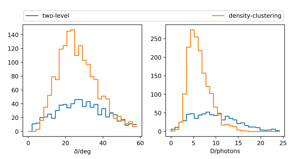}
        \caption[Photon-classification, protons]{
            Protons.
            Density-clustering in the photon-stream finds 126\% more air-showers with same or even better distributions of performance-measures $\delta$ and $D$.
            See Table \ref{TabAclPerformance}.
        }
        \label{FigAclIndependentPhsVsMp_proton}
    \end{center}
\end{figure}

\begin{figure}
    \begin{center}
        \includegraphics[width=1\textwidth]{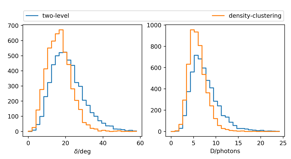}
        \caption[Photon-classification, gamma-rays, overlap]{
            Overlap, air-showers found by both methods
            Gamma-rays.
            See Table \ref{TabAclPerformance}.
        }
        \label{FigAclBothPhsVsMp_gamma}
    \end{center}
\end{figure}

\begin{figure}
    \begin{center}
        \includegraphics[width=1\textwidth]{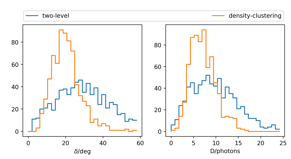}
        \caption[Photon-classification, protons, overlap]{
            Overlap, air-showers found by both methods.
            Protons.
            See Table \ref{TabAclPerformance}.
        }
        \label{FigAclBothPhsVsMp_proton}
    \end{center}
\end{figure}

\begin{figure}
    \begin{center}
        \includegraphics[width=1\textwidth]{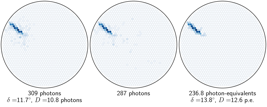}
        \includegraphics[width=1\textwidth]{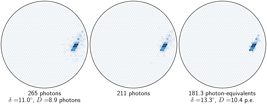}
        \includegraphics[width=1\textwidth]{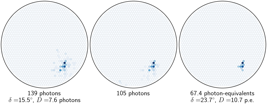}
        \includegraphics[width=1\textwidth]{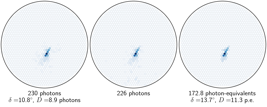}
        \caption[Photon-classification, bright gamma-rays 1 of 2]{Middle is simulation truth, left is density-clustering in photon-stream, right is two-level-time-neighbor-cleaning on largest-pulses. Rare but intense ($>100$ air-shower photons) gamma-rays on FACT.}
        \label{FigAclExampleGammaBright1}
    \end{center}
\end{figure}

\begin{figure}
    \begin{center}
        \includegraphics[width=1\textwidth]{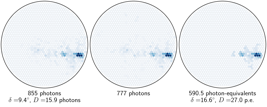}
        \includegraphics[width=1\textwidth]{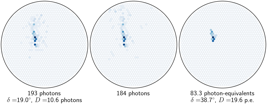}
        \includegraphics[width=1\textwidth]{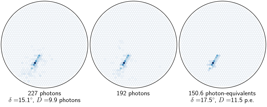}
        \includegraphics[width=1\textwidth]{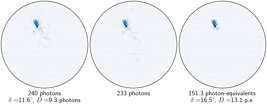}
        \caption[Photon-classification, bright gamma-rays 2 of 2]{Additional intense gamma-rays as in Figure \ref{FigAclExampleGammaBright1}.}
        \label{FigAclExampleGammaBright2}
    \end{center}
\end{figure}

\begin{figure}
    \begin{center}
        \includegraphics[width=1\textwidth]{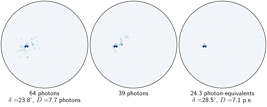}
        \includegraphics[width=1\textwidth]{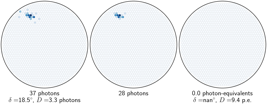}
        \includegraphics[width=1\textwidth]{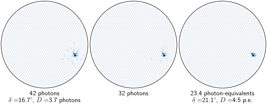}
        \includegraphics[width=1\textwidth]{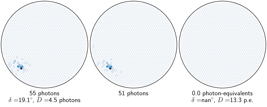}
        \caption[Photon-classification, typical gamma-rays 1 of 2]{Typical gamma-rays close to the trigger threshold.}
        \label{FigAclExampleGammaTypical1}
    \end{center}
\end{figure}

\begin{figure}
    \begin{center}
        \includegraphics[width=1\textwidth]{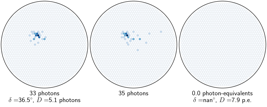}
        \includegraphics[width=1\textwidth]{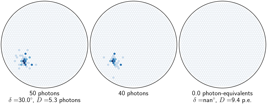}
        \includegraphics[width=1\textwidth]{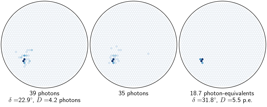}
        \includegraphics[width=1\textwidth]{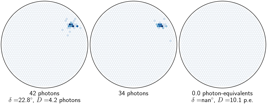}
        \caption[Photon-classification, typical gamma-rays 2 of 2]{Additional typical gamma-rays, same as Figure \ref{FigAclExampleGammaTypical1}.}
        \label{FigAclExampleGammaTypical2}
    \end{center}
\end{figure}

\begin{figure}
    \begin{center}
        \includegraphics[width=1\textwidth]{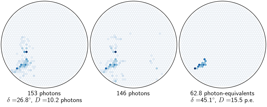}
        \includegraphics[width=1\textwidth]{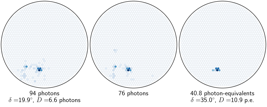}
        \includegraphics[width=1\textwidth]{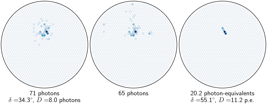}
        \includegraphics[width=1\textwidth]{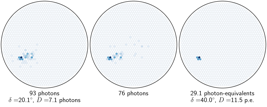}
        \caption[Photon-classification, typical protons 1 of 1]{Typical cosmic protons close to the trigger threshold.}
        \label{FigAclExampleProtonTypical1}
    \end{center}
\end{figure}

For FACT, the performance of the classification of Cherenkov-photons in the photon-stream turns out to be clearly superior to the classification of Cherenkov-photons on largest-pulses.
This can be found in two ways.
First, the clustering based on density in the photon-stream finds much more air-showers for the same measures of performance $\delta$ and $D$, see Table \ref{TabAclPerformance} and the Figures \ref{FigAclIndependentPhsVsMp_gamma}, and \ref{FigAclIndependentPhsVsMp_proton}.
Second, for the air-showers which are both found on largest-pulses and in the photon-stream, the air-showers of the photon-stream have much better measures of performance, as shown in Table \ref{TabAclPerformance} and the Figures \ref{FigAclBothPhsVsMp_gamma}, and \ref{FigAclBothPhsVsMp_proton}.
\begin{table}
    \begin{center}
    \begin{tabular}{l | r r r | r r r}
        & \multicolumn{3}{c}{Gamma-rays} & \multicolumn{3}{c}{Protons} \\
        Method &
        \rotatebox{90}{photon-stream} &
        \rotatebox{90}{largest-pulses} &
        \rotatebox{90}{\color{light-gray}{overlap}} &
        \rotatebox{90}{photon-stream} &
        \rotatebox{90}{largest-pulses} &
        \rotatebox{90}{\color{light-gray}{overlap}}\\
        \midrule
        Number events & & & & & &\\
        simulated &
        12,228 & 12,228 & \color{light-gray}{12,228} & 12,228 & 12,228 & \color{light-gray}{12,228} \\
        passed trigger &
        8,725 & 8,725 & \color{light-gray}{8,725} & 2,134 & 2,134 & \color{light-gray}{2,134}\\
        found air-shower &
        8,300 & 5,103 & \color{light-gray}{5,078} &  1,668 & 739 & \color{light-gray}{726} \\
        \midrule
        Performance & & & & & &\\
        mean $\delta /$deg & 22.6 & 23.3 & & 29.3 & 31.4 & \\
        mean $D /$p.e. & 6.2 & 8.4 & & 7.2 & 10.5 & \\
        \midrule
        Performance & & & & & &\\
        on overlap & & & & & &\\
        mean $\delta /$deg &
        18.6 &
        23.3 & &
        22.8 &
        31.3 & \\
        mean $D /$p.e. &
        6.8 &
        8.4 & &
        8.0 &
        10.6 & \\
    \end{tabular}
    \caption[Performance of the classification of Cherenkov-photons]{
    Performance of the classification of Cherenkov-photons.
    Comparing air-shower-classification on largest-pulses and in the photon-stream.
    Compare Figures \ref{FigAclIndependentPhsVsMp_gamma}, \ref{FigAclIndependentPhsVsMp_proton}, \ref{FigAclBothPhsVsMp_gamma}, and \ref{FigAclBothPhsVsMp_proton}.
    }
    \label{TabAclPerformance}
    \end{center}
\end{table}
The Figures \ref{FigAclExampleGammaBright1}, \ref{FigAclExampleGammaBright2}, \ref{FigAclExampleGammaTypical1}, \ref{FigAclExampleGammaTypical2}, and \ref{FigAclExampleProtonTypical1} show examples of air-shower-events used in the comparison.
Each row in the figures shows the three images of the same air-shower-event simulated on FACT.
The middle image shows the true Cherenkov-photons.
The image on the right shows the reconstructed Cherenkov-photons using largest-pulses, and the image on the left shows the reconstructed Cherenkov-photons using the photon-stream.
Below each image of reconstructed Cherenkov-photons, the measures of performance $\delta$ and $D$ are written.
For air-shower-events with a high intensities of Cherenkov-photons, the performance of the classification on largest-pulses comes close to the performance of the classification in the photon-stream, see Figures \ref{FigAclExampleGammaBright1}, and \ref{FigAclExampleGammaBright2}.
However, when the intensity of the Cherenkov-photons is low, the classification in the photon-stream is superior, see Figures \ref{FigAclExampleGammaTypical1}, \ref{FigAclExampleGammaTypical2}, and \ref{FigAclExampleProtonTypical1}.
Since gamma-rays create higher densities of Cherenkov-photons in the images than protons \cite{badran1997improvement}, the classification in the photon-stream finds 126\% more proton-events, but only 63\% more gamma-ray-events than the classification on largest-pulses with image-cleaning, see Table \ref{TabAclPerformance}.
To take advantage of the improvement in classifying Cherenkov-photons, the reconstruction of the properties of the cosmic particle (direction, energy, and type) will most likely have to be adjusted.
Therefore, although the benefits of the photon-stream for the reconstruction of Cherenkov-photons is undeniable, the gain in performance for gamma-ray-astronomy can not be estimated from this investigation alone.
%
%
%%%%%%%%%%%%%%%%%%%%%%%%%%%%%%%%%%%%%%%%%%%%%%%%%%%%%%%%%%%%%%%%%%%%%%%%%%%%%%%
%
%              ##     ## ##     ##  #######  ##    ##  ######
%              ###   ### ##     ## ##     ## ###   ## ##    ##
%              #### #### ##     ## ##     ## ####  ## ##
%              ## ### ## ##     ## ##     ## ## ## ##  ######
%              ##     ## ##     ## ##     ## ##  ####       ##
%              ##     ## ##     ## ##     ## ##   ### ##    ##
%              ##     ##  #######   #######  ##    ##  ######
%
%%%%%%%%%%%%%%%%%%%%%%%%%%%%%%%%%%%%%%%%%%%%%%%%%%%%%%%%%%%%%%%%%%%%%%%%%%%%%%%
%
\chapter{A first test on muons}
\label{ChaMuons}
Relativistic muons which come close to the telescope can produce a ring like distribution of photons in the images.
We find about $10\,$k events/night in the observations of FACT which show structures with the shape of a ring or the segment of a ring.
Events induced by muons on a small telescope like FACT are a tough test for any classification of Cherenkov-photons.
The directional density of photons from the muon is only a low $\approx 2.3$\,photons/pixel, and thus makes muons difficult to find in the presence of night-sky-background-photons.
Muons allow us to test the limits of the photon-stream on observations of FACT.
We will first discuss the origin of the rings in the air-shower-events and why we call such events 'muon-events'.
Second, we will present our algorithm to find muon-events in the photon-stream of FACT.
Third, we will show and discuss examples of muon-events, and we will show side-by-side the Cherenkov-photons found in the photon-stream with clustering, and the photons found on largest-pulses with image-cleaning.
\section{Origin of muon-events}
When faster than the local speed of light, any charged particle in the air-shower emits Cherenkov-photons.
These Cherenkov-photons are emitted in a cone with a radial symmetry relative to the trajectory of the charged particle.
The opening-angle of this cone depends on the speed of the charged particle and the refractive-index of the air the particle is traversing.
Opening-angles observed on FACT range from $\approx 0.45^\circ$ up to $1.6^\circ$.
To see a large part of the ring, the charged particle must be close to the telescope.
At the typical opening-angle of $1.1^\circ$, the charged particle must be closer than $\approx 105\,$m to FACT to create full ring structures in the image.
When the imaging-reflector maps the directions of the photons onto the image-sensor, a sharp ring structure emerges.
Since muons are the most likely charged particles to come close to the telescope, we assume that all events with a ring-structure are caused by muons.
Therefore we call such events 'muon-events' although every isolated, charged particle will induce the same ring-structures.
\section{Algorithm}
We start with the observed air-shower-event represented in the photon-stream.
First, we classify Cherenkov-photons and night-sky-background-photons based on the density in the photon-stream as shown in Chapter \ref{ChaDensityBasedClustering}.
We drop all night-sky-background-photons and continue only with the dense Cherenkov-photons.
When multiple dense clusters of Cherenkov-photons were found, we combine these clusters into one cluster for simplicity\footnote{This is not optimal, and we loose muon-events where the air-shower itself is within the photon-stream as a dense, central and late cluster.}.
Second, we fit the model of a ring to the distribution of Cherenkov-photons in the three-dimensional space of the photon-stream.
For resistance against outliers, we use the Random-Sample-Consensus (RANSAC) \cite{scikit-learn} algorithm to fit the model of the ring.
Third, we assert that the model of the ring fulfills the criteria listed in Table \ref{TabMounCriteria}.
\begin{table}
    \begin{center}
    \begin{tabular}{ l  r}
        \midrule
        Max. trails of RANSAC & 15\\
        Max. std.-dev. of arrival-times of the photons & $5\,$ns\\
        Min. radius of ring & $0.45^\circ$\\
        Max. radius of ring & $1.6^\circ$\\
        Min. part of photons to agree with the ring & $60\%$\\
        Min. areal overlap of ring and the field-of-view & $20\%$\\
        Min. circumference of ring in the field-of-view & $1.5^\circ$\\
        \midrule
    \end{tabular}
    \caption[Criteria for muon-events]{Criteria for muon-events.%
    }
    \label{TabMounCriteria}
    \end{center}
\end{table}
\section{Results}
\begin{figure}
    \vspace{-1cm}
    \begin{minipage}{0.5\textwidth}
        \begin{figure}[H]
            \includegraphics[width=1.0\textwidth]{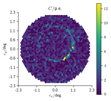}
        \end{figure}
    \end{minipage}
    \hfill
    \begin{minipage}{0.5\textwidth}
        \begin{figure}[H]
            \includegraphics[width=1.0\textwidth]{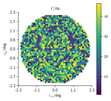}
        \end{figure}
    \end{minipage}
    \begin{center}
        \includegraphics[width=1\textwidth]{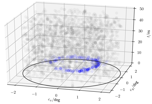}
    \end{center}
    \vspace{-1cm}
    \begin{minipage}{0.475\textwidth}
        \begin{figure}[H]
            \includegraphics[width=1.0\textwidth]{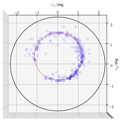}
        \end{figure}
    \end{minipage}
    \hfill
    \begin{minipage}{0.52\textwidth}
        \begin{figure}[H]
            \includegraphics[width=1.0\textwidth]{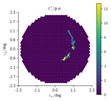}
        \end{figure}
    \end{minipage}
    \caption[Muon example event 1]{
        Observed in the night of 23\,July\,2014, run\,192, event\,1180.
        1\,cluster of 320\,photons in 135\,pixels. 67\,pixels have only one photon.
    }
    \label{FigMuonExample1180}
\end{figure}

\begin{figure}
    \vspace{-1cm}
    \begin{minipage}{0.5\textwidth}
        \begin{figure}[H]
            \includegraphics[width=1.0\textwidth]{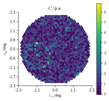}
        \end{figure}
    \end{minipage}
    \hfill
    \begin{minipage}{0.5\textwidth}
        \begin{figure}[H]
            \includegraphics[width=1.0\textwidth]{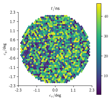}
        \end{figure}
    \end{minipage}
    \begin{center}
        \includegraphics[width=1\textwidth]{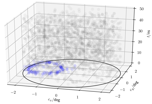}
    \end{center}
    \vspace{-1cm}
    \begin{minipage}{0.475\textwidth}
        \begin{figure}[H]
            \includegraphics[width=1.0\textwidth]{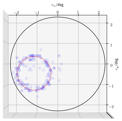}
        \end{figure}
    \end{minipage}
    \hfill
    \begin{minipage}{0.52\textwidth}
        \begin{figure}[H]
            \includegraphics[width=1.0\textwidth]{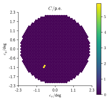}
        \end{figure}
    \end{minipage}
    \caption[Muon example event 2]{
        Observed in the night of 23\,July\,2014, run\,192, event\,1444.
        1\,cluster of 199\,photons in 103\,pixels. 54\,pixels have only one photon.
    }
    \label{FigMuonExample1444}
\end{figure}

\begin{figure}
    \vspace{-1cm}
    \begin{minipage}{0.5\textwidth}
        \begin{figure}[H]
            \includegraphics[width=1.0\textwidth]{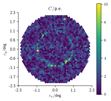}
        \end{figure}
    \end{minipage}
    \hfill
    \begin{minipage}{0.5\textwidth}
        \begin{figure}[H]
            \includegraphics[width=1.0\textwidth]{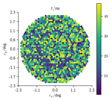}
        \end{figure}
    \end{minipage}
    \begin{center}
        \includegraphics[width=1\textwidth]{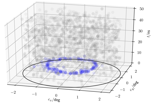}
    \end{center}
    \vspace{-1cm}
    \begin{minipage}{0.475\textwidth}
        \begin{figure}[H]
            \includegraphics[width=1.0\textwidth]{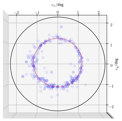}
        \end{figure}
    \end{minipage}
    \hfill
    \begin{minipage}{0.52\textwidth}
        \begin{figure}[H]
            \includegraphics[width=1.0\textwidth]{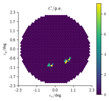}
        \end{figure}
    \end{minipage}
    \caption[Muon example event 3]{
        Observed in the night of 23\,July\,2014, run\,192, event\,9146.
        1\,cluster of 347\,photons in 148\,pixels. 63\,pixels have only one photon.
    }
    \label{FigMuonExample9146}
\end{figure}

\begin{figure}
    \vspace{-1cm}
    \begin{minipage}{0.5\textwidth}
        \begin{figure}[H]
            \includegraphics[width=1.0\textwidth]{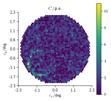}
        \end{figure}
    \end{minipage}
    \hfill
    \begin{minipage}{0.5\textwidth}
        \begin{figure}[H]
            \includegraphics[width=1.0\textwidth]{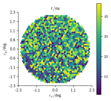}
        \end{figure}
    \end{minipage}
    \begin{center}
        \includegraphics[width=1\textwidth]{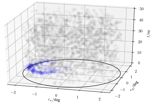}
    \end{center}
    \vspace{-1cm}
    \begin{minipage}{0.475\textwidth}
        \begin{figure}[H]
            \includegraphics[width=1.0\textwidth]{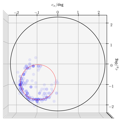}
        \end{figure}
    \end{minipage}
    \hfill
    \begin{minipage}{0.52\textwidth}
        \begin{figure}[H]
            \includegraphics[width=1.0\textwidth]{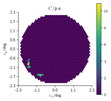}
        \end{figure}
    \end{minipage}
    \caption[Muon example event 4]{
        Observed in the night of 23\,July\,2014, run\,192, event\,15065.
        1\,cluster of 219\,photons in 92\,pixels. 45\,pixels have only one photon.
    }
    \label{FigMuonExample15065}
\end{figure}

\begin{figure}
    \vspace{-1cm}
    \begin{minipage}{0.5\textwidth}
        \begin{figure}[H]
            \includegraphics[width=1.0\textwidth]{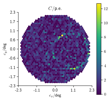}
        \end{figure}
    \end{minipage}
    \hfill
    \begin{minipage}{0.5\textwidth}
        \begin{figure}[H]
            \includegraphics[width=1.0\textwidth]{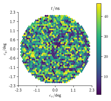}
        \end{figure}
    \end{minipage}
    \begin{center}
        \includegraphics[width=1\textwidth]{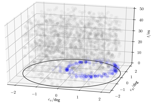}
    \end{center}
    \vspace{-1cm}
    \begin{minipage}{0.475\textwidth}
        \begin{figure}[H]
            \includegraphics[width=1.0\textwidth]{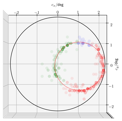}
        \end{figure}
    \end{minipage}
    \hfill
    \begin{minipage}{0.52\textwidth}
        \begin{figure}[H]
            \includegraphics[width=1.0\textwidth]{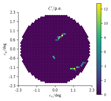}
        \end{figure}
    \end{minipage}
    \caption[Muon example event 4]{
        Observed in the night of 23\,July\,2014, run\,192, event\,17408.
        3\,clusters of 334\,photons in 132\,pixels. 62\,pixels have only one photon.
    }
    \label{FigMuonExample17408}
\end{figure}

The Figures \ref{FigMuonExample1180}, \ref{FigMuonExample1444}, \ref{FigMuonExample9146}, \ref{FigMuonExample15065}, and \ref{FigMuonExample17408} show the examples of muon-events which were found in the photon-stream of FACT.
Each figure has five panels.
The two panels on the top show the event represented in largest-pulses.
The central panel shows the event represented in the photon-stream, and the photons are colored based on the density-clustering.
Gray photons are classified to the night-sky-background, and blue photons are classified to the muon.
The two panels at the bottom show the photons which were classified to the muon.
First, in the left panel at the bottom is the projection of the photons in the $c_x$-$c_y$-plane of the photon-stream.
The red ring was found with our RANSAC ring-model.
Second, in the right panel at the bottom is the result of our standard\cite{nothe2017fact} image-cleaning algorithm on the largest-pulses.\\
We find that muons can be detected very well in the photon-stream of FACT with a classification of Cherenkov-photons based on density.
We find that the ring of the muons is reconstructed better in the photon-stream, and conclude that this is a hint for the photon-stream to perform better in the classification of Cherenkov-photons.\\
For FACT, the search for muons is important.
On a Cherenkov-telescope which observes mono without other Cherenkov-telescopes which could rise a veto, muon-events have to be identified and rejected in the search for gamma-rays.
Also muon-events can serve as a powerful calibration-device.
With the photon-stream we have produced the largest sample of muon-events which was ever found on FACT.
The muon-sample in the photon-stream has over $18\times10^6$\,muons.
It is the first sample of muon-events which spans all epochs of FACT.
%
%-------------------------------------------------------------------------------
% 1180
%-------------------------------------------------------------------------------
% 1444
%-------------------------------------------------------------------------------
% 9146
%-------------------------------------------------------------------------------
% 15065
%-------------------------------------------------------------------------------
% 17408
%-------------------------------------------------------------------------------
\chapter{Outlook}
\label{ChapOutlook}
\begin{figure}
    \includegraphics[width=1.0\textwidth]{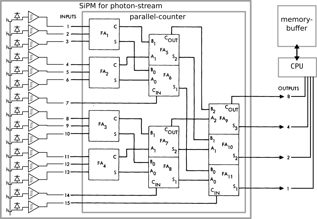}
    \caption[Our proposal for an SiPM with parallel-counter]{%
        Draft of our proposed SiPM design to output the photon-stream directly.
        On the left is the GAPD-array, compare Figure \ref{FigSchematicsSipmPixel}.
        All GAPDs have individual bias-supplies as implemented here \cite{frach2009digital}.
        Right of the GAPDs are the Schmitt-triggers which serve as amplifiers to reach bias-levels for digital processing.
        The logic for clocking the individual GAPDs into the parallel-counter is omitted here.
        When individual GAPDs are faulty and have high rates of accidental discharges, adjustments of the Schmitt-trigger thresholds can exclude these SiPMs, similar as demonstrated here \cite{frach2009digital}.
        Right of the array of GAPDs is the parallel-counter.
        The parallel-counter here is constructed from a tree of 11 full-adders (FA).
        Figure of parallel-counter is taken from \cite{swartzlander1973parallel}.
        This particular SiPM has only $N=15$ GAPDs and $M=4$ bit output-width for simplicity.
        Actual SiPMs for Cherenkov-telescopes will have $N > 1\,$k GAPDs.
        Also, actual parallel-counters might need buffers after each stage in order to be clocked at speed of $\approx 10^9$s$^{-1}$.
        Such buffers are omitted in this figure.
    }
    \label{FigPhotonStreamSipm}
\end{figure}

Future Cherenkov-telescopes might be build more cost effective and may output the photon-stream directly.
For FACT, we process the photon-stream on a multi purpose computer after the analog time-series of the pixels have been recorded.
But for future Cherenkov-telescopes we see the potential to lower the cost of read-out-electronics dramatically.
Instead of recording analog time-series of SiPMs, we propose to not record analog time-series at all.
We propose to drop all analog amplifiers, analog ring-buffers, and analog-to-digital-converters which are currently used in the image-sensors of Cherenkov-telescopes.
We propose to start over directly at the GAPDs where the initial electric responses to the photons are produced.
We propose to use a surface-array of $N$ GAPDs, as classic SiPMs do, to receive photons.
But this time we do not connect the GAPDs in parallel.
Instead we connect each GAPD to a Schmitt-trigger.
We adjust the Schmitt-trigger so that it creates digital pulses of a defined amplitude when its GAPD discharges.
Then we clock all $N$ outputs of the Schmitt-triggers into the (N,M) parallel-counter \cite{swartzlander2004review}.
Each GAPD-discharge causes input of the parallel-counter to be only high for a single clock.
The $M$ outputs of the parallel-counter represent the $M$ digits of a binary number counting the number of GAPDs which have released a pulse.
We copy the binary number at the output of the parallel-counter at a fixed, and clocked rate of e.g. $10^9\,$s$^{-1}$ into a digital memory-buffer such as a shift-register.
We do this for each pixel in the image-sensor.
Now we have the photon-stream in a digital memory-buffer in our image-sensor.
There will be no need for costly high bandwidth electronics to process analog signals.
Figure \ref{FigPhotonStreamSipm} shows a draft of our proposed SiPM design.
\begin{figure}
    \includegraphics[width=1.0\textwidth]{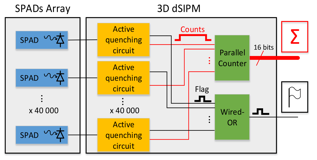}
    \caption[SiPM with parallel-counter, by J. F. Pratte and F. Retiere]{%
        The prototype schematics for a digital SiPM with integrated parallel-counter, by Jean-Francois Pratte and Fabrice Retiere \cite{retiere2017digital}.
        Here Single-Photon-Avalanche-Diode (SPAD) is the same what we use to call GAPD.
        It is reported, that prototype SiPMs were already build and are under evaluation.
        An arrival-time resolution $< 1\,$ns is reported for the prototype SiPMs.
        This SiPM has 1\,cm$^2$ photon-aperture, and was developed for the nEXO experiment which will cover a surface of 4\,m$^2$ with these SiPMs.
        See the great similarity with our proposal in Figure \ref{FigPhotonStreamSipm}.
    }
    \label{Fig3dSipm}
\end{figure}
Previous studies \cite{catalano2008single} already pointed out the benefits of a digital representation of the air-shower-events with single-photons and proposed to shrink the size of the pixels while increasing their number to reach the necessary single-photon-resolution.
Our proposal takes this previous study to the extremes by reading out each GAPD in the SiPM-pixel individually, but in contrast to this previous study we do not propose to increase the number of pixels and read-out-electronics but to condense the arrivals of single-photons in the parallel-counter of each SiPM-pixel to reduce the costs.
In computer science, our current implementation of recording time-series and later using analog to digital converters is known to be a possible implementation of a parallel-counter called the 'quasi-digital-counter' (See section 2.5, \cite{swartzlander2004review}).
Other users of SiPMs have also realized that the true potential of SiPMs goes way beyond PMTs when they are combined with dedicated electronics on the chip of the SiPM directly \cite{frach2009digital}.
Our colleagues in particle-physics have a similar need for digital SiPM with $\approx 10^9\,$Hz bandwidth and started developing first prototypes \cite{retiere2017digital} which are very similar to our proposal in Figure \ref{FigPhotonStreamSipm}.
Their advanced prototype of a digital SiPM is reported to cost $\approx 2\times10^6$\,USD\,m$^{-2}$ which results in $\approx 26\times10^3$\,USD for the sensitive area used in the image-sensor of FACT.
The cost of the image-sensor including the read-out for the FACT prototype was $\approx 0.5\times10^6$\,USD.
Given the omission of the costly routing and processing of analog signals, there is valid hope to reduce the cost of the FACT's image-sensor by one order-of-magnitude when using digital SiPMs with integrated parallel-counters.\\
We propose to team up with the SiPM-experts \cite{retiere2017digital} and create a dedicated SiPM for Cherenkov-telescopes to build inexpensive and powerful Cherenkov-telescopes which output the photon-stream directly.
\section{Outlook for FACT}
We provide access to all simulated events of FACT in photon-stream-representation to the general public.
We provide access to a sample of observations of FACT of the Crab-Nebula from November 2013 in photon-stream-representation to the general public.
To the members of the FACT-collaboration we provide access to all\footnote{About $80\%$ of the time spent on this thesis was spent here. The computation of the fourth and final production-pass itself took 8\,month.} observed events of FACT in photon-stream-representation.
We automatically add the latest observed events in photon-stream-representation.
We allow the members of FACT to access our events without the need for accounts on specific computing-sites, which is a novelty.
Go and visit
\begin{center}
\url{http://fact-project.org/photon-stream}\\
\end{center}
to access the photon-stream of FACT.
We implemented a format for storage and exchange of the photon-stream and implemented readers and writers in different programming languages.
See the appendix in Chapter \ref{SecPhotonStreamReadMe} for details on the implementation and a brief discussion on 'Known Limitations and Outlook'.\\
Because of the accessibility, the photon-stream already contributes to projects within the FACT-collaboration beside the activities of the author of this thesis.
For the very immediate outlook on the photon-stream in FACT, here are a few projects which investigate the photon-stream and its possible benefits for gamma-ray-astronomy:
\begin{itemize}
    \item Kevin Sedlaczek investigates the advantages of the additional timing information in the photon-stream for gamma-ray-astronomy.
    In his master-thesis, Kevin uses both simulations and observations of FACT in the photon-stream-representation to improve the sensitivity to gamma-rays of the Crab-Nebula.
    \item Jan Moritz Behnken submitted a bachelor-thesis on the prospects of using machine-learning with deep neural-networks to reconstruct the particle-type from air-shower-events.
    He used the observations and simulations of FACT represented with the photon-stream.
    \item Dorothee Hildebrand, and Adrian Biland investigate muon-events to crosscheck novel methods for monitoring the production-efficiency of Cherenkov-photons in the atmosphere\cite{hildebrand2017using} and a possible aging of the SiPMs in FACT .
    The muon-events were found in the photon-stream for FACT using the methods presented in Chapter \ref{ChaMuons}.
    \item Kai Bruegge investigates the performance of observing gamma-rays with the novel photon-stream by implementing the generation of novel features from the photon-stream in his doctoral-thesis.
    \item Jacob Bieker investigates to improve the reconstruction of the direction of the cosmic particles using machine-learning and the photon-stream for FACT in his bachelor-thesis.
    \item Amandeep Singh uses the photon-stream to investigate the rates of different photon-multiplicities to search for correlations with properties of the atmosphere, compare Figure \ref{FigPhotonMultiplicity}.
    \item Laurits Tani uses the muon-sample in the photon-stream representation.
    In his master-thesis, he investigates how FACT's optical point-spread-function changes over time based on the sharpness of muon-rings.
\end{itemize}
\chapter{Conclusion}
\label{ChaConclusion}
We have proposed to lower the energy-threshold of Cherenkov-telescopes for gamma-rays by improving the classification of Cherenkov-photons due to the use of a natural representation of air-shower-events which describes general observables of photons instead of describing special responses of special sensors.
We have implemented such a natural representation for air-shower-events for FACT and called it the photon-stream.
Based on simulations of FACT, we have shown that a density-based classification of Cherenkov-photons in the photon-stream is superior to the established classification of Cherenkov-photons in multiple stages on the largest-pulses.
We have shown that no information relevant for gamma-ray-astronomy is lost when we use the photon-stream instead of the largest-pulses to represent the air-shower-events of FACT.
We have proposed a design of an SiPM which allows to implement image-sensors for Cherenkov-telescopes without costly analog electronics, but with wide spread digital components which output the photon-stream directly.\\
We conclude that the photon-stream representation for air-shower-events has only benefits over the established representation of largest-pulses.
The photon-stream contains more information and is easier to explain%
\footnote{Largest-pulses need two concepts to be explained. First arrival-time, and second photon-equivalent. The photon-stream has only one concept, which is arrival-time. Also largest-pulses can not be explained without explaining the specific photo-sensor and read-out-electronics which is used.}
.
The photon-stream is the most natural representation to store, exchange, and process air-shower-events within and among Cherenkov-telescopes.
\section{Conclusion for FACT}
Before the photon-stream was implemented, air-shower-events of FACT were only accessible in a representation of the raw time-series of all pixels.
The raw representation using the time-series of all pixels is great to investigate the novel SiPMs in combination with the read-out of the image-sensor.
Possible aging of the image-sensor, and artifacts can be investigated in great detail this way.\\
However, the raw representation of air-shower-events with time-series is very inefficient in storage-size and thus inefficient to do gamma-ray-astronomy.
Since easy gamma-ray-astronomy was not the goal of FACT, no format was implemented to store the representation of largest-pulses permanently for easy and compact access.
Some private studies in FACT do gamma-ray-astronomy by permanently storing the air-shower-events represented by their Hillas-features, but these private storages are difficult to access.\\
The photon-stream for FACT is two orders-of-magnitude more compact in storage-size than the raw representation of air-shower-events with time-series even when we apply our own dedicated compression\cite{ahnen2015data} to the time-series.
Instead of $\approx 800\,$TBytes of raw events, which are scattered over several physical storage units on several remote computing-sites, the photon-stream is only $\approx 8\,$TBytes in storage size for all simulations and observations ever done with FACT.
The entire photon-stream for FACT can be stored on a single consumer storage unit in a single computer while it contains at least all the information for gamma-ray-astronomy that is relevant to our established analysis-chain based on largest-pulses and Hillas-features, see Section \ref{SecCrosscheckOnClassicAnalysisChain}.
The photon-stream for FACT is accessible on a web-server to all members of FACT and partly the photon-stream is even accessible to the general public.
Now that the initial phase of demonstration is over for FACT, and the novel SiPMs are well understood, the photon-stream pushes the capability of FACT and its members to focus on gamma-ray-astronomy.
\chapter{Implementing the photon-stream}
\label{SecPhotonStreamReadMe}
This is the read-me of our photon-stream repository.
FACT specific aspects and key implementation aspects are outlined here.
\begin{center}
    \begin{Large}
        \url{https://github.com/fact-project/photon_stream}
    \end{Large}
\end{center}
\newpage
\includegraphics[width=1.25\textwidth]{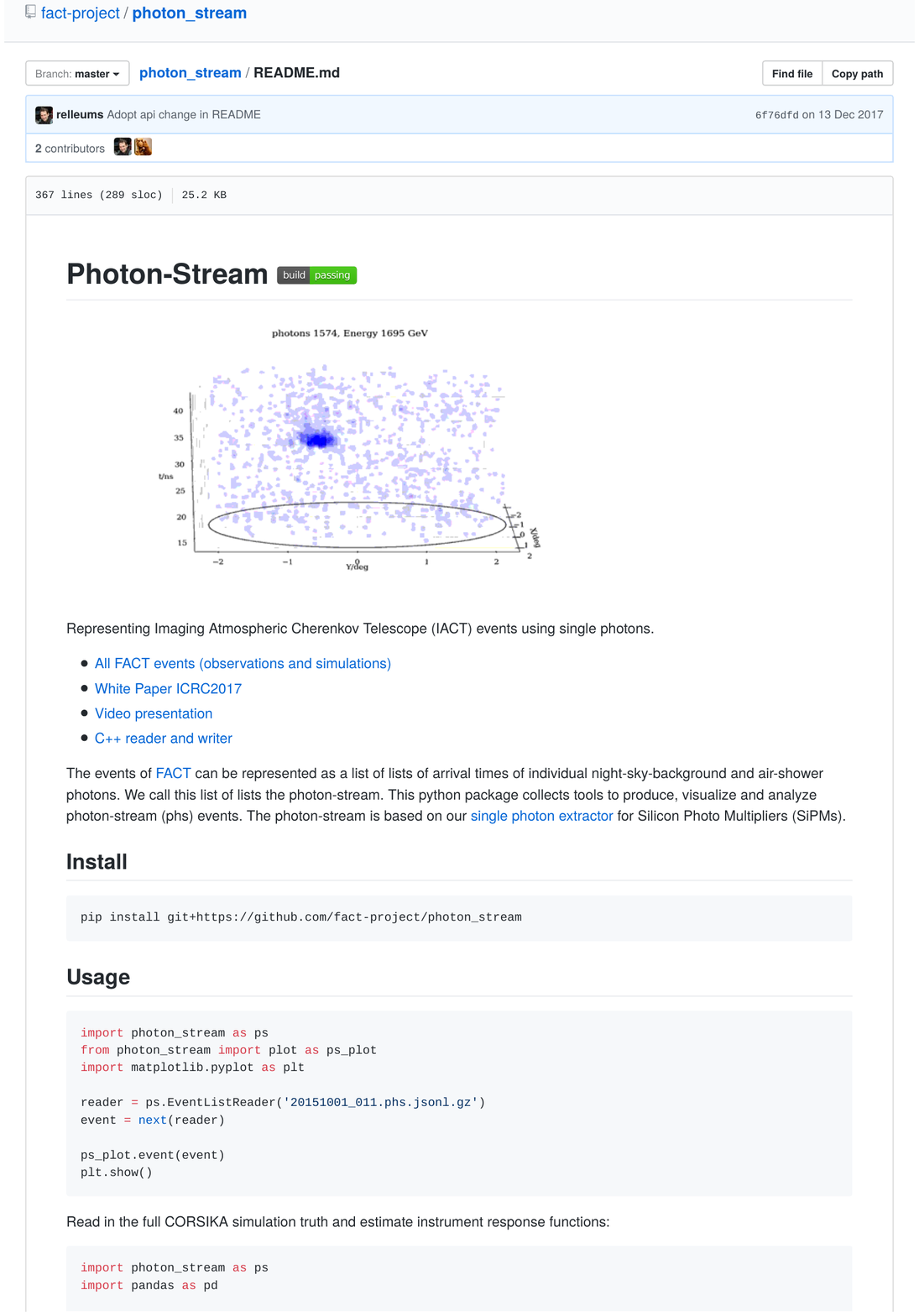}
\includegraphics[width=1.25\textwidth]{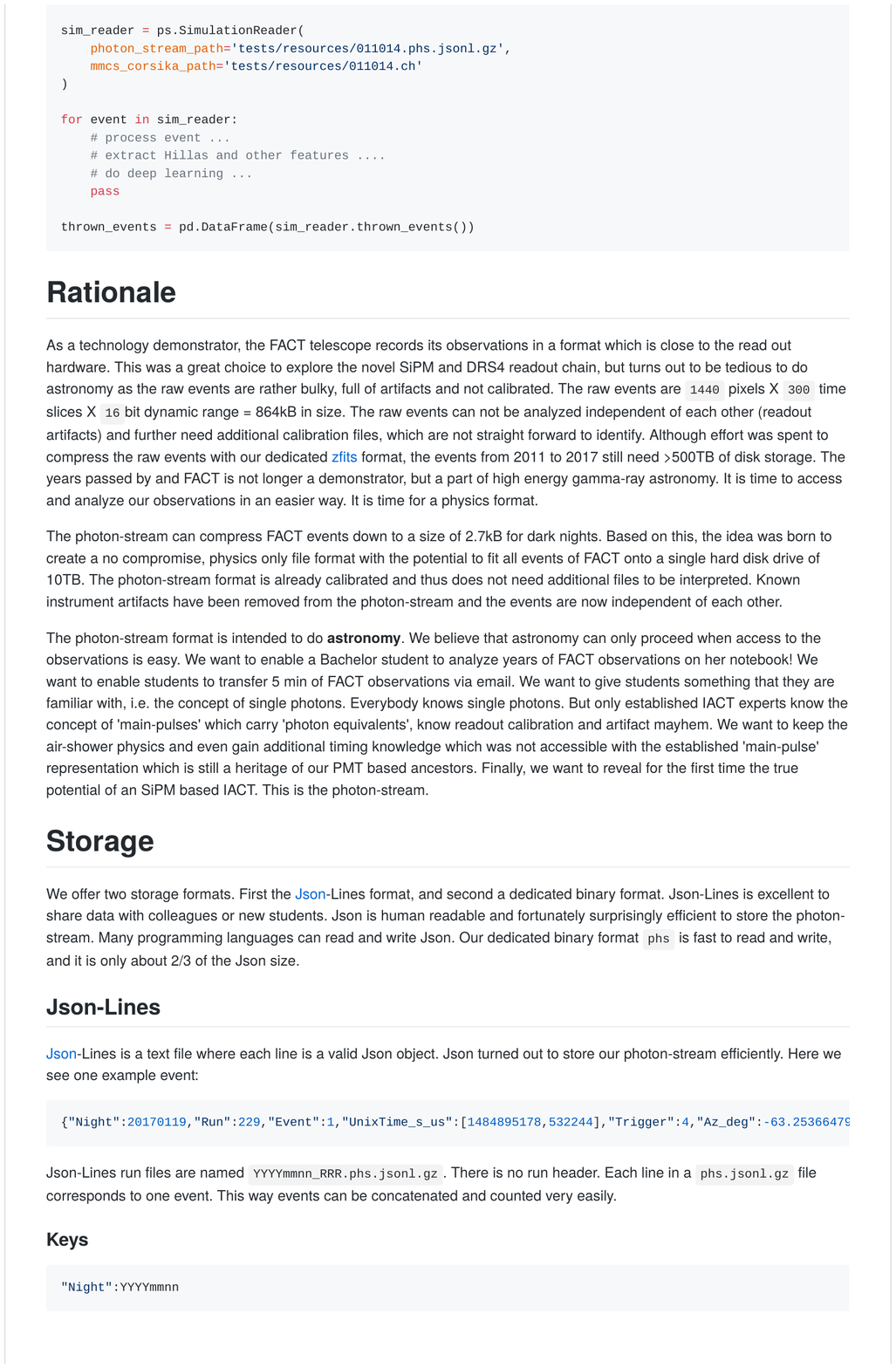}
\includegraphics[width=1.25\textwidth]{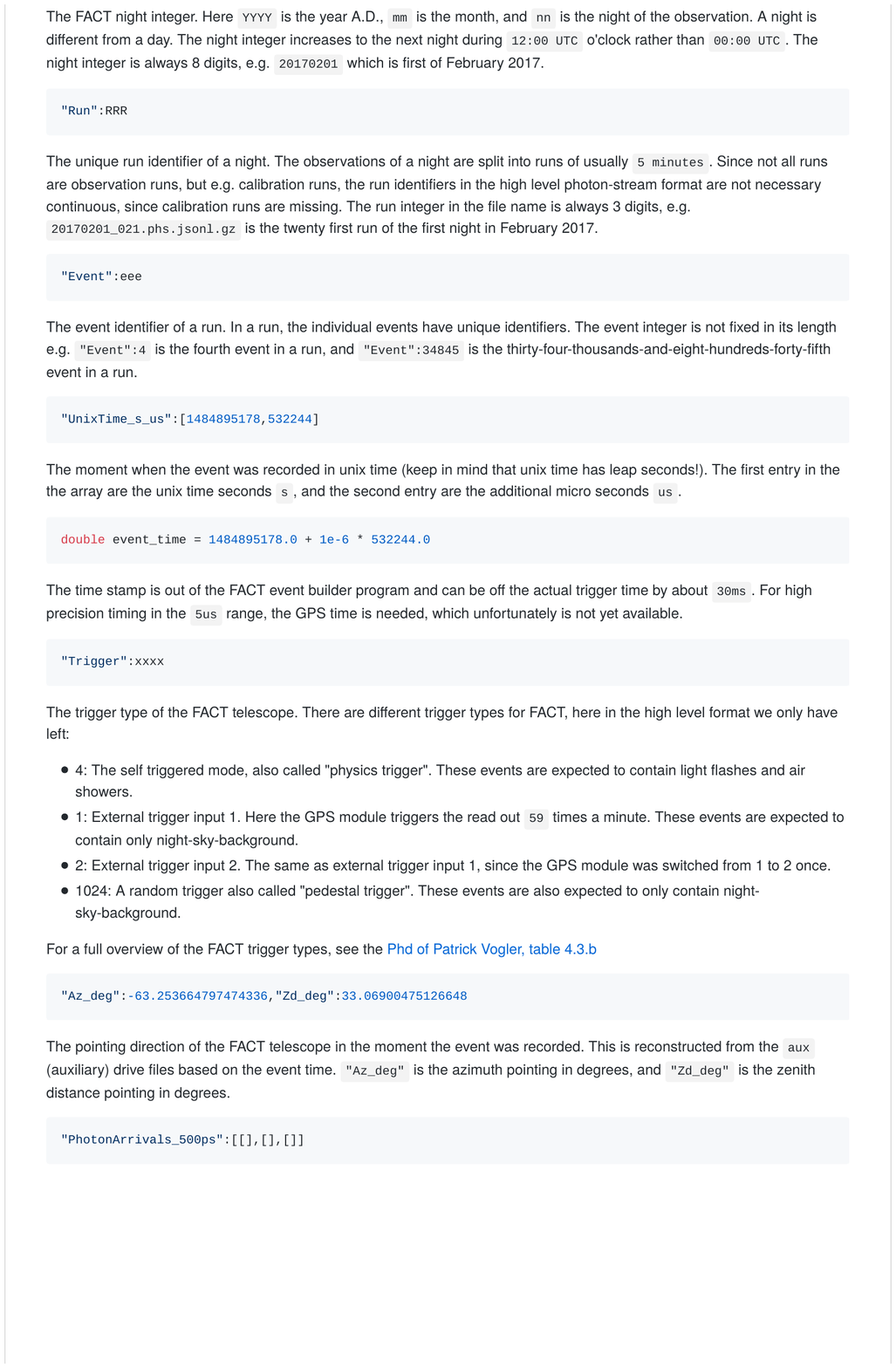}
\includegraphics[width=1.25\textwidth]{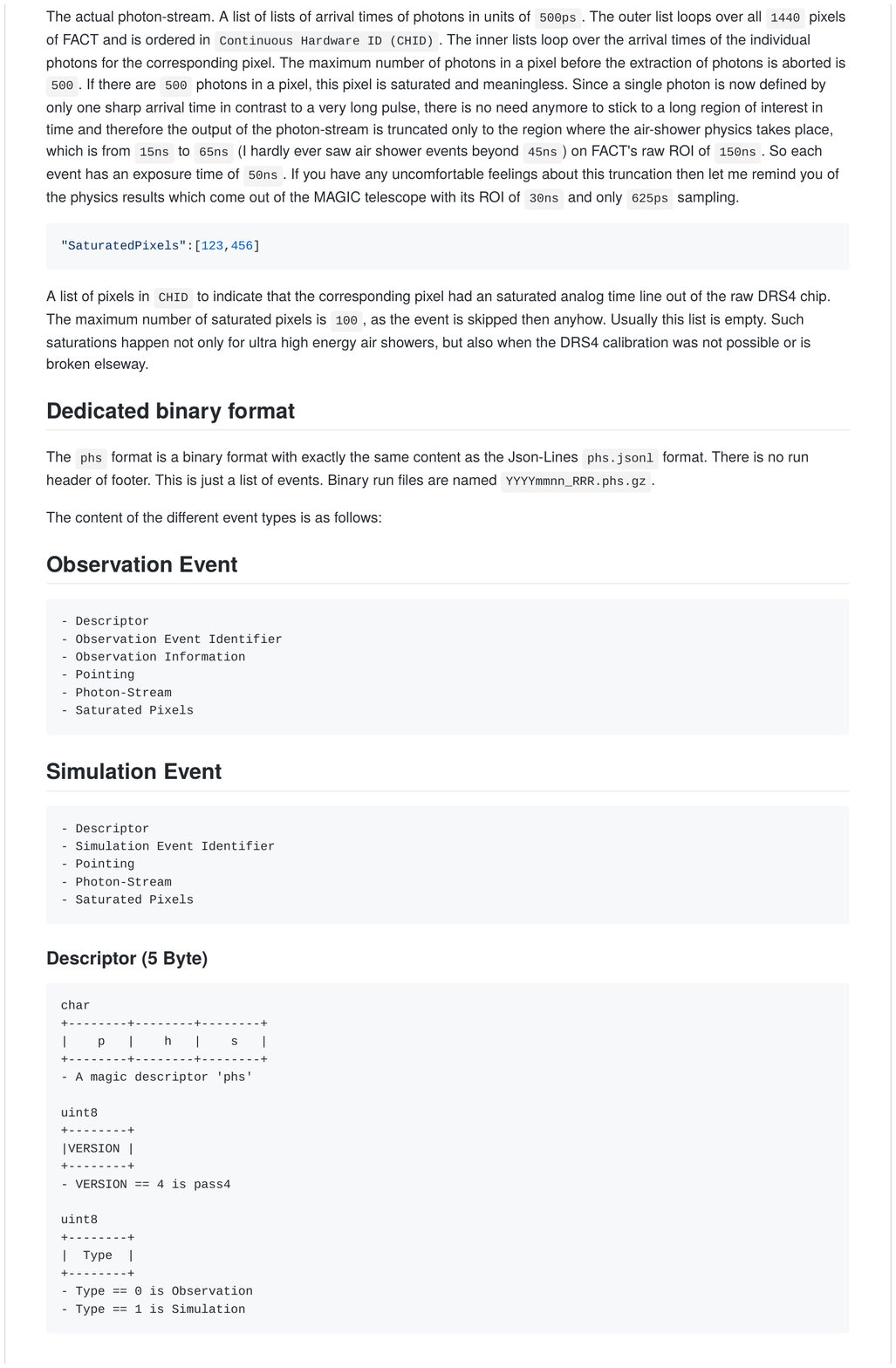}
\includegraphics[width=1.25\textwidth]{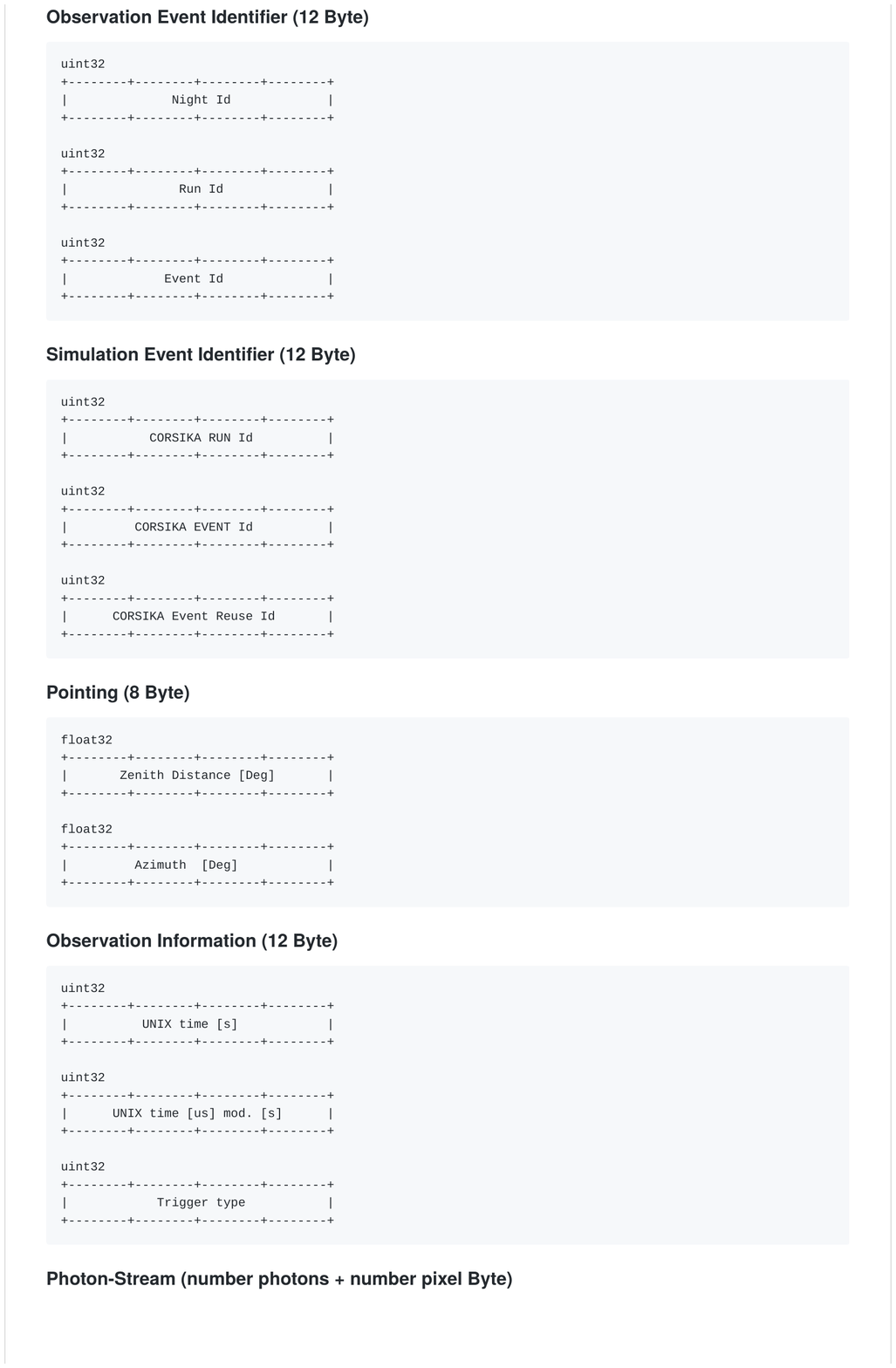}
\includegraphics[width=1.25\textwidth]{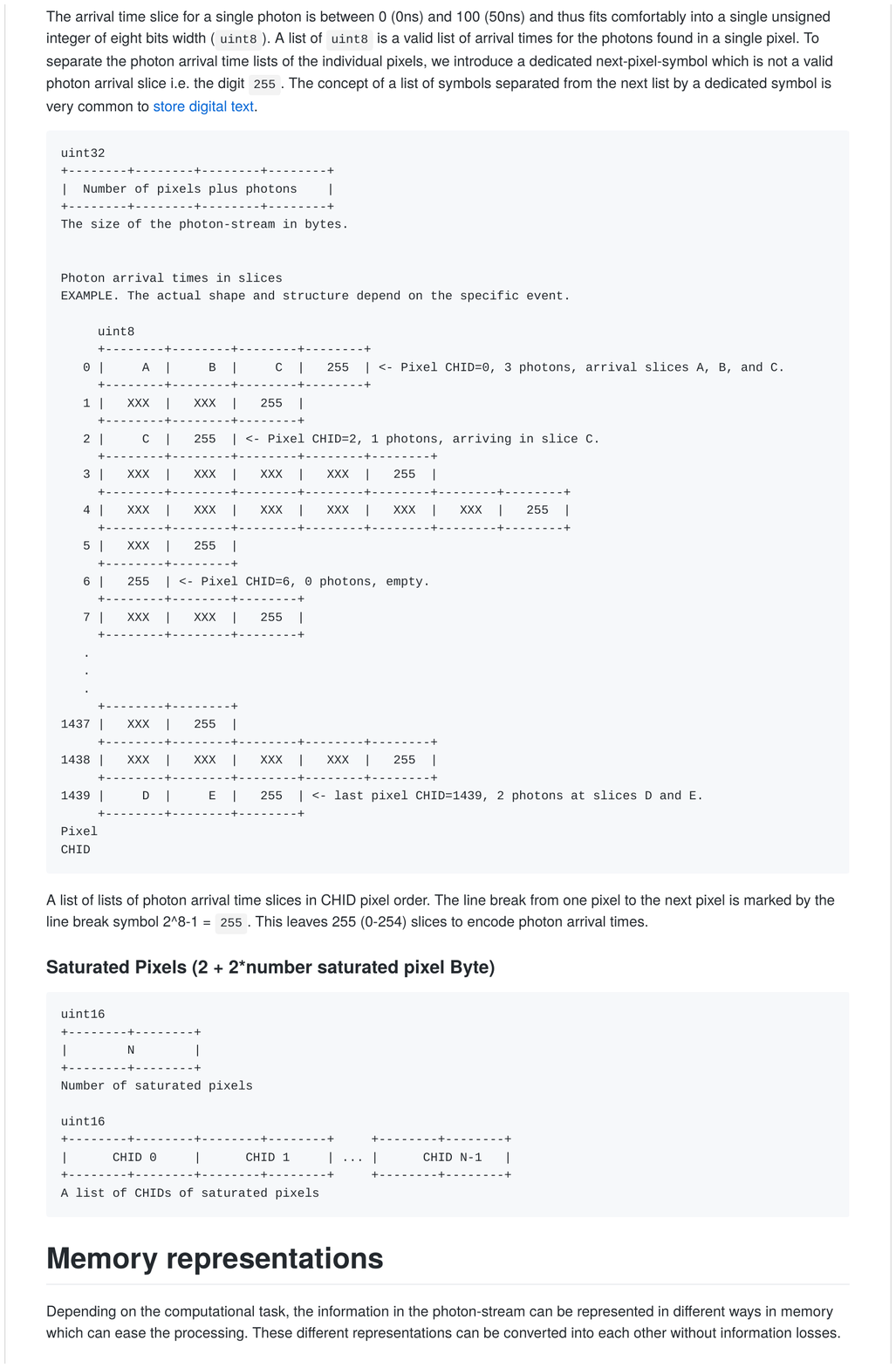}
\includegraphics[width=1.25\textwidth]{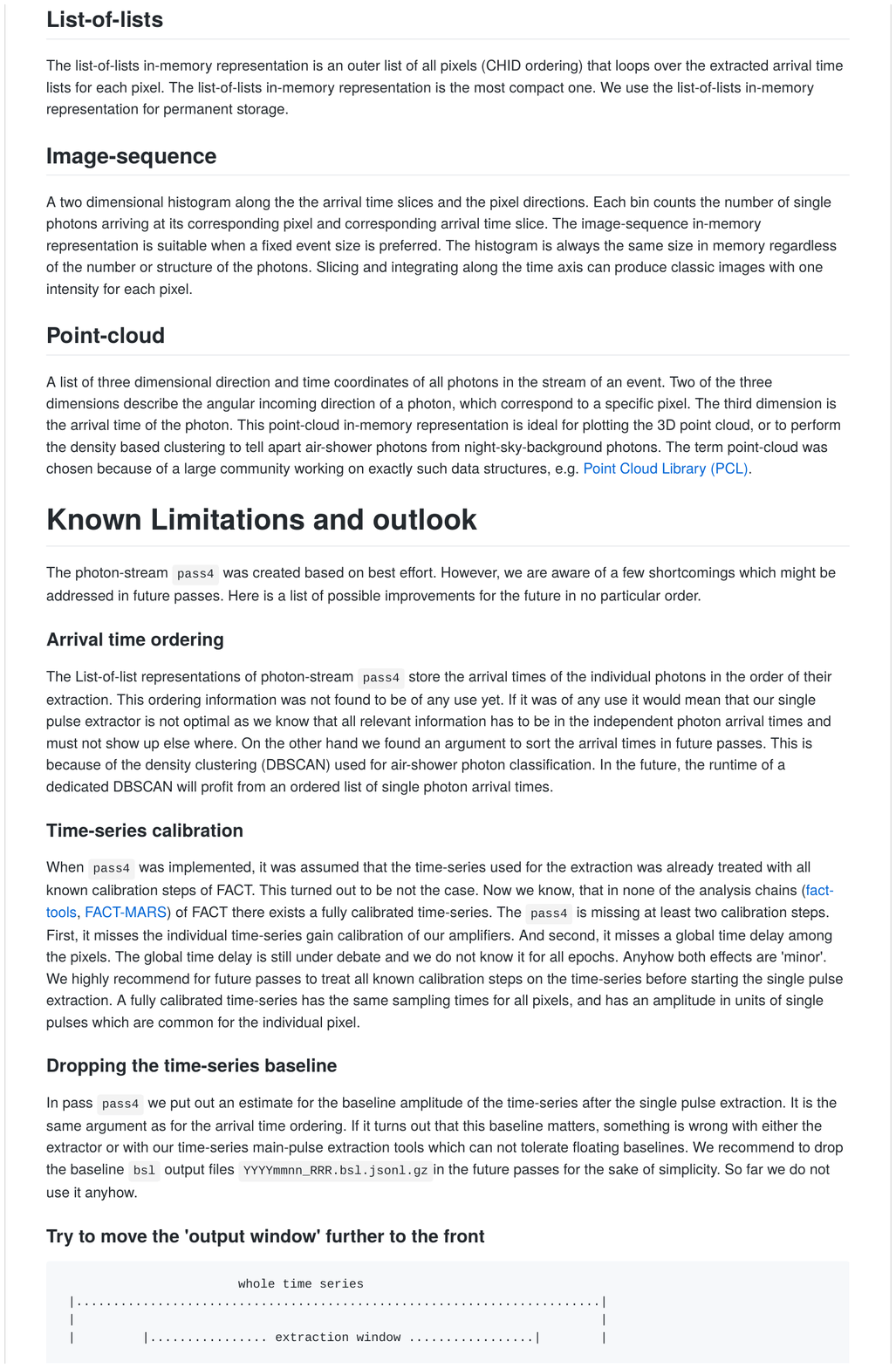}
\includegraphics[width=1.25\textwidth]{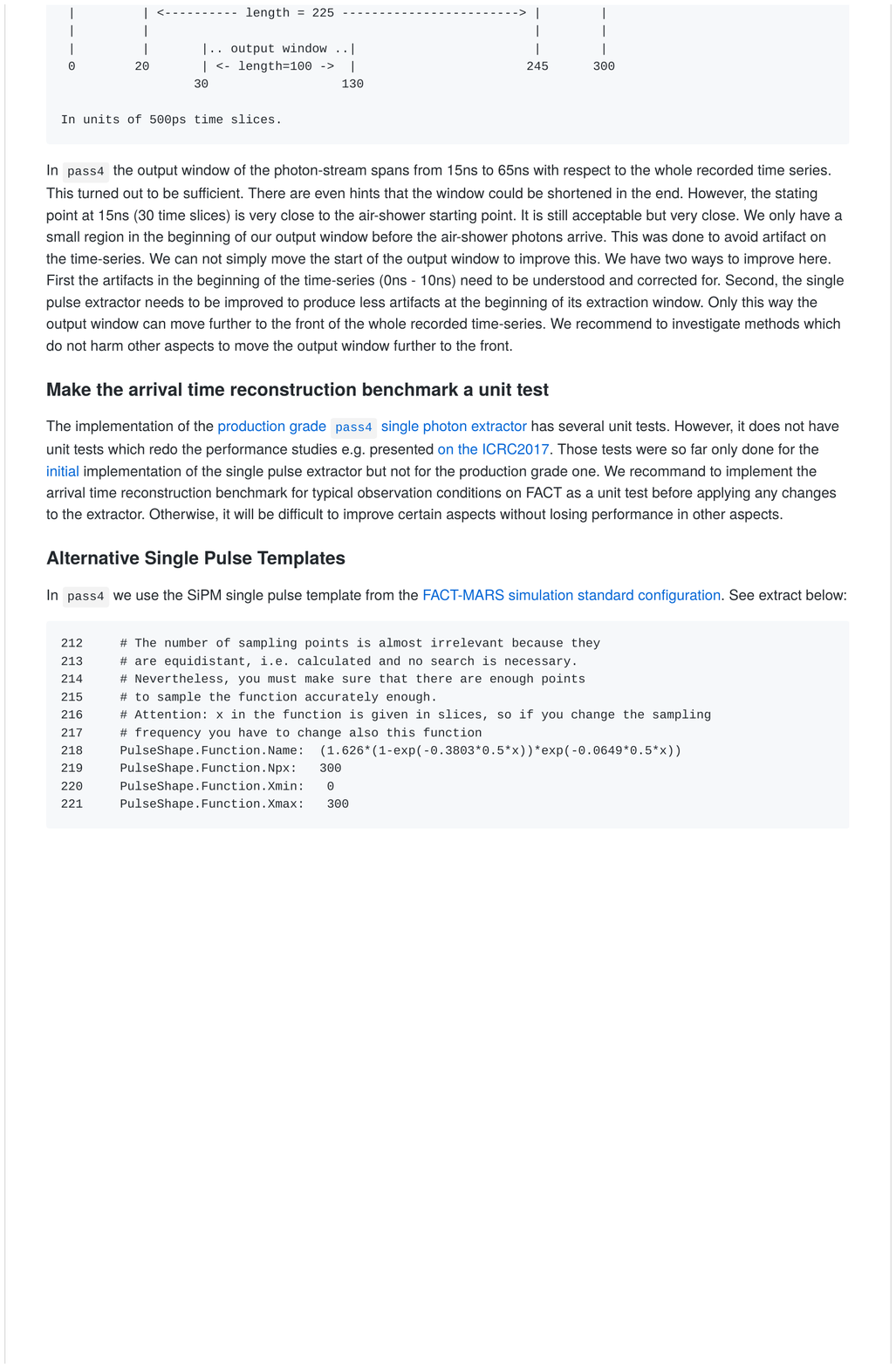}
\includegraphics[width=1.25\textwidth]{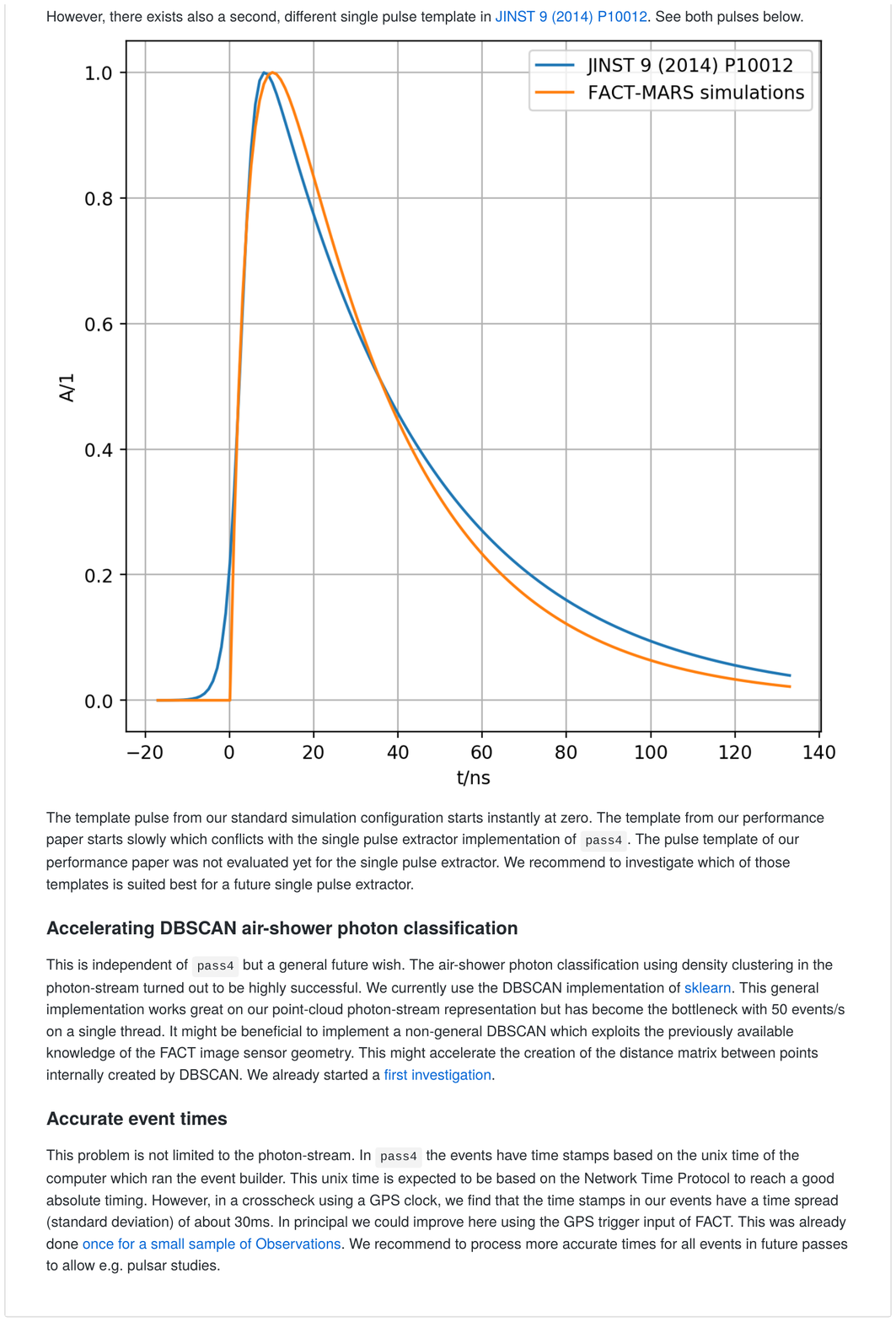}

\bibliographystyle{apalike}
\bibliography{references}

\begin{thebibliography}{}

\bibitem[Aasi et~al., 2014]{aasi2014first}
Aasi, J., Abadie, J., Abbott, B., Abbott, R., Abbott, T., Abernathy, M.,
  Accadia, T., Acernese, F., Adams, C., Adams, T., et~al. (2014).
\newblock First searches for optical counterparts to gravitational-wave
  candidate events.
\newblock {\em The Astrophysical Journal Supplement Series}, 211(1):7.

\bibitem[Aasi et~al., 2015]{aasi2015advanced}
Aasi, J., Abbott, B., Abbott, R., Abbott, T., Abernathy, M., Ackley, K., Adams,
  C., Adams, T., Addesso, P., Adhikari, R., et~al. (2015).
\newblock Advanced ligo.
\newblock {\em Classical and quantum gravity}, 32(7):074001.

\bibitem[Abbott et~al., 2017a]{abbott2017exploring}
Abbott, B.~P., Abbott, R., Abbott, T., Abernathy, M., Ackley, K., Adams, C.,
  Addesso, P., Adhikari, R., Adya, V., Affeldt, C., et~al. (2017a).
\newblock Exploring the sensitivity of next generation gravitational wave
  detectors.
\newblock {\em Classical and Quantum Gravity}, 34(4):044001.

\bibitem[Abbott et~al., 2017b]{abbott2017gravitational}
Abbott, B.~P., Abbott, R., Abbott, T., Acernese, F., Ackley, K., Adams, C.,
  Adams, T., Addesso, P., Adhikari, R., Adya, V., et~al. (2017b).
\newblock Gravitational waves and gamma-rays from a binary neutron star merger:
  Gw170817 and grb 170817a.
\newblock {\em The Astrophysical Journal Letters}, 848(2):L13.

\bibitem[Abdo et~al., 2009a]{abdo2009limit}
Abdo, A., Ackermann, M., Ajello, M., Asano, K., Atwood, W., Axelsson, M.,
  Baldini, L., Ballet, J., Barbiellini, G., Baring, M., et~al. (2009a).
\newblock A limit on the variation of the speed of light arising from quantum
  gravity effects.
\newblock {\em Nature}, 462(7271):331--334.

\bibitem[Abdo et~al., 2009b]{abdo2009population}
Abdo, A., Ackermann, M., Ajello, M., Atwood, W., Axelsson, M., Baldini, L.,
  Ballet, J., Barbiellini, G., Baring, M., Bastieri, D., et~al. (2009b).
\newblock A population of gamma-ray millisecond pulsars seen with the fermi
  large area telescope.
\newblock {\em Science}.

\bibitem[Abdo et~al., 2010]{fermi2010fermi}
Abdo, A., A., Ackermann, M., Ajello, M., Atwood, W., B., Baldini, L., et~al.
  (2010).
\newblock Fermi gamma-ray imaging of a radio galaxy.
\newblock {\em Science}, 328(5979):725--729.

\bibitem[Abdo et~al., 2011]{abdo2011CrabFlare}
Abdo, A.~A., Ackermann, M., Ajello, M., Allafort, A., Baldini, L., Ballet, J.,
  Barbiellini, G., Bastieri, D., Bechtol, K., Bellazzini, R., et~al. (2011).
\newblock {Gamma-ray flares from the Crab Nebula}.
\newblock {\em Science}, 331(6018):739--742.

\bibitem[Abraham et~al., 2010]{abraham2010fluorescence}
Abraham, J., Abreu, P., Aglietta, M., Aguirre, C., Ahn, E.~J., Allard, D.,
  Allekotte, I., Allen, J., Allison, P., Alvarez-Muniz, J., et~al. (2010).
\newblock {The fluorescence detector of the Pierre Auger Observatory}.
\newblock {\em Nuclear Instruments and Methods in Physics Research Section A:
  Accelerators, Spectrometers, Detectors and Associated Equipment},
  620(2):227--251.

\bibitem[Acero et~al., 2015]{acero2015fermi3fgl}
Acero, F., Ackermann, M., Ajello, M., Albert, A.and~Atwood, W.~B., Axelsson,
  M., Baldini, L., Ballet, J., Barbiellini, G., Bastieri, D., et~al. (2015).
\newblock {Fermi large area telescope third source catalog}.
\newblock {\em The Astrophysical Journal Supplement Series}, 218(2):23.

\bibitem[Acharya, 2005]{acharya2005ground}
Acharya, B. (2005).
\newblock Ground based $\gamma$-ray astronomy in india.
\newblock {\em Past, Present and Future. Proc. 29th ICRC, Pune}.

\bibitem[Acharya et~al., 2013]{cta2013introducing}
Acharya, B.~S., Actis, M., Aghajani, T., Agnetta, G., Aguilar, J., Aharonian,
  F., Ajello, M., Akhperjanian, A., Alcubierre, M., Aleksic, J., et~al. (2013).
\newblock Introducing the cta concept.
\newblock {\em Astroparticle Physics}, 43:3--18.

\bibitem[Ackermann et~al., 2014]{ackermann2014fermi}
Ackermann, M., Ajello, M., Asano, K., Atwood, W., Axelsson, M., Baldini, L.,
  Ballet, J., Barbiellini, G., Baring, M., Bastieri, D., et~al. (2014).
\newblock Fermi-lat observations of the gamma-ray burst grb 130427a.
\newblock {\em Science}, 343(6166):42--47.

\bibitem[Adelson and Wang, 1992]{adelson1992single}
Adelson, E.~H. and Wang, J. Y.~A. (1992).
\newblock Single lens stereo with a plenoptic camera.
\newblock {\em IEEE Transactions on Pattern Analysis \& Machine Intelligence},
  2:99--106.

\bibitem[Adriani et~al., 2009]{adriani2009anomalous}
Adriani, O., Barbarino, G., Bazilevskaya, G., Bellotti, R., Boezio, M.,
  Bogomolov, E., Bonechi, L., Bongi, M., Bonvicini, V., Bottai, S., et~al.
  (2009).
\newblock An anomalous positron abundance in cosmic rays with energies 1.5--100
  gev.
\newblock {\em Nature}, 458(7238):607.

\bibitem[Aglae et~al., 2004]{aglea2004owl}
Aglae, K., Andre, E., Andrei, T., Armando, R., Babak, S., Bernhard, B., Bill,
  D., et~al. (2004).
\newblock {\em OWL BLUE BOOK -- Owl concept design report, phase A design
  review, OWL-TRE-ESO-0000-0001 Issue 2}.
\newblock www.eso.org.

\bibitem[Aguilar et~al., 2014]{aguilar2014precision}
Aguilar, M., Aisa, D., Alpat, B., Alvino, A., Ambrosi, G., Andeen, K., Arruda,
  L., Attig, N., Azzarello, P., Bachlechner, A., et~al. (2014).
\newblock Precision measurement of the (e++ e-) flux in primary cosmic rays
  from 0.5 gev to 1 tev with the alpha magnetic spectrometer on the
  international space station.
\newblock {\em Physical review letters}, 113(22):221102.

\bibitem[Aguilar et~al., 2015]{aguilar2015precision}
Aguilar, M., Aisa, D., Alpat, B., Alvino, A., Ambrosi, G., Andeen, K., Arruda,
  L., Attig, N., Azzarello, P., Bachlechner, A., et~al. (2015).
\newblock Precision measurement of the proton flux in primary cosmic rays from
  rigidity 1 gv to 1.8 tv with the alpha magnetic spectrometer on the
  international space station.
\newblock {\em Physical Review Letters}, 114(17):171103.

\bibitem[Aguilar et~al., 2010]{aguilar2010relative}
Aguilar, M., Alcaraz, J., Allaby, J., Alpat, B., Ambrosi, G., Anderhub, H., Ao,
  L., Arefiev, A., Arruda, L., Azzarello, P., et~al. (2010).
\newblock Relative composition and energy spectra of light nuclei in cosmic
  rays: results from ams-01.
\newblock {\em The Astrophysical Journal}, 724(1):329.

\bibitem[Aharonian, 2005]{aharonian2005next}
Aharonian, F. (2005).
\newblock {Next generation of IACT arrays: scientific objectives versus energy
  domains}.
\newblock {\em arXiv preprint astro-ph/0511139}.

\bibitem[Aharonian et~al., 2006a]{aharonian2006hess}
Aharonian, F., Akhperjanian, A., Bazer-Bachi, A., Beilicke, M., Benbow, W.,
  Berge, D., Bernloehr, K., Boisson, C., Bolz, O., Borrel, V., et~al. (2006a).
\newblock Hess observations of the galactic center region and their possible
  dark matter interpretation.
\newblock {\em Physical Review Letters}, 97(22):221102.

\bibitem[Aharonian et~al., 2006b]{aharonian2006low}
Aharonian, F., Akhperjanian, A., Bazer-Bachi, A., Beilicke, M., Benbow, W.,
  Berge, D., Bernloehr, K., Boisson, C., Bolz, O., Borrel, V., et~al. (2006b).
\newblock A low level of extragalactic background light as revealed by
  $\gamma$-rays from blazars.
\newblock {\em Nature}, 440(7087):1018--1021.

\bibitem[Aharonian et~al., 2007]{aharonian2007first}
Aharonian, F., Akhperjanian, A., Bazer-Bachi, A., Beilicke, M., Benbow, W.,
  Berge, D., Bernloehr, K., Boisson, C., Bolz, O., Borrel, V., et~al. (2007).
\newblock {First ground-based measurement of atmospheric Cherenkov light from
  cosmic rays}.
\newblock {\em Physical Review D}, 75(4):042004.

\bibitem[Aharonian et~al., 2008]{aharonian2008limits}
Aharonian, F., Akhperjanian, A.~G., De~Almeida, U.~B., Bazer-Bachi, A.~R.,
  Becherini, Y., Behera, B., Beilicke, M., Benbow, W., Bernloehr, K., Boisson,
  C., et~al. (2008).
\newblock Limits on an energy dependence of the speed of light from a flare of
  the active galaxy pks 2155-304.
\newblock {\em Physical Review Letters}, 101(17):170402.

\bibitem[Aharonian et~al., 2012]{aharonian2012abrupt}
Aharonian, F.~A., Bogovalov, S.~V., and Khangulyan, D. (2012).
\newblock Abrupt acceleration of a/cold/'ultrarelativistic wind from the crab
  pulsar.
\newblock {\em Nature}, 482(7386):507--509.

\bibitem[Aharonian et~al., 2001]{aharonian2001}
Aharonian, F.~A., Konopelko, A.~K., Voelk, H.~J., and Quintana, H. (2001).
\newblock 5@ 5--a 5 gev energy threshold array of imaging atmospheric cherenkov
  telescopes at 5 km altitude.
\newblock {\em Astroparticle Physics}, 15(4):335--356.

\bibitem[Ahnen, 2017a]{ahnen2016diss}
Ahnen, M.~L. (2017a).
\newblock {{Observations of the Crab Pulsar in Gamma Rays Above 1
  Teraelectronvolt}}.
\newblock {\em ETH Zurich, Research Collection}, Department of Physics.

\bibitem[Ahnen, 2017b]{ahnen2017integral}
Ahnen, M.~L. (2017b).
\newblock On integral upper limits assuming power-law spectra and the
  sensitivity in high-energy astronomy.
\newblock {\em The Astrophysical Journal}, 836(2):196.

\bibitem[Ahnen et~al., 2016a]{ahnen2016bokeh}
Ahnen, M.~L., Baack, D., Balbo, M., Bergmann, M., Biland, A., Blank, M., Bretz,
  T., Bruegge, K., Buss, J., Domke, M., et~al. (2016a).
\newblock Bokeh mirror alignment for cherenkov telescopes.
\newblock {\em Astroparticle Physics}, 82:1--9.

\bibitem[Ahnen et~al., 2016b]{ahnen2016normalized}
Ahnen, M.~L., Baack, D., Balbo, M., Bergmann, M., Biland, A., Blank, M., Bretz,
  T., Bruegge, K., Buss, J., Domke, M., et~al. (2016b).
\newblock Normalized and asynchronous mirror alignment for cherenkov
  telescopes.
\newblock {\em Astroparticle Physics}, 82:56--65.

\bibitem[Ahnen et~al., 2015]{ahnen2015data}
Ahnen, M.~L., Balbo, M., Bergmann, M., Biland, A., Bretz, T., Buss, J., Dorner,
  D., Einecke, S., Freiwald, J., Hempfling, C., et~al. (2015).
\newblock {Data compression for the first G-APD Cherenkov Telescope}.
\newblock {\em Astronomy and Computing}, 12:191--199.

\bibitem[Albert et~al., 2008a]{albert2008probing}
Albert, J., Aliu, E., Anderhub, H., Antonelli, L., Antoranz, P., Backes, M.,
  Baixeras, C., Barrio, J., Bartko, H., Bastieri, D., et~al. (2008a).
\newblock Probing quantum gravity using photons from a flare of the active
  galactic nucleus markarian 501 observed by the magic telescope.
\newblock {\em Physics Letters B}, 668(4):253--257.

\bibitem[Albert et~al., 2008b]{albert2008fadc}
Albert, J., Aliu, E., Anderhub, H., Antoranz, P., Armada, A., Asensio, M.,
  Baixeras, C., Barrio, J.~A., Bartko, H., Bastieri, D., et~al. (2008b).
\newblock {FADC signal reconstruction for the MAGIC telescope}.
\newblock {\em Nuclear Instruments and Methods in Physics Research Section A:
  Accelerators, Spectrometers, Detectors and Associated Equipment},
  594(3):407--419.

\bibitem[Albus et~al., 1993]{albus1993nist}
Albus, J., Bostelman, R., and Dagalakis, N. (1993).
\newblock The nist robocrane.
\newblock {\em Journal of Robotic Systems}, 10(5):709--724.

\bibitem[Aleksi{\'c} et~al., 2015]{aleksic2015measurement}
Aleksi{\'c}, J., Ansoldi, S., Antonelli, L., Antoranz, P., Babic, A., Bangale,
  P., Barrio, J., Gonz{\'a}lez, J.~B., Bednarek, W., Bernardini, E., et~al.
  (2015).
\newblock Measurement of the crab nebula spectrum over three decades in energy
  with the magic telescopes.
\newblock {\em Journal of High Energy Astrophysics}, 5:30--38.

\bibitem[Aleksic et~al., 2014]{aleksic2014blackHoleLightning}
Aleksic, J., Ansoldi, S., Antonelli, L.~A., Antoranz, P., Babic, A., Bangale,
  P., Barrio, J.~A., Gonzalez, J.~B., Bednarek, W., Bernardini, E., et~al.
  (2014).
\newblock Black hole lightning due to particle acceleration at subhorizon
  scales.
\newblock {\em Science}, 346(6213):1080--1084.

\bibitem[Alexandreas et~al., 1995]{alexandreas1995status}
Alexandreas, D., Bartoli, B., Bastieri, D., Bedeschi, F., Bertolucci, E.,
  Bigongiari, C., Biral, R., Busetto, G., Centro, S., Chiarelli, G., et~al.
  (1995).
\newblock {Status report on CLUE}.
\newblock {\em Nuclear Instruments and Methods in Physics Research Section A:
  Accelerators, Spectrometers, Detectors and Associated Equipment},
  360(1-2):385--389.

\bibitem[Aliu et~al., 2009]{aliu2009improving}
Aliu, E., Anderhub, H., Antonelli, L.~A., Antoranz, P., Backes, M., Baixeras,
  C., Barrio, J.~A., Bartko, H., Bastieri, D., Becker, J.~K., et~al. (2009).
\newblock {Improving the performance of the single-dish Cherenkov telescope
  MAGIC through the use of signal timing}.
\newblock {\em Astroparticle Physics}, 30(6):293--305.

\bibitem[Allekotte et~al., 2008]{allekotte2008surface}
Allekotte, I., Barbosa, A.~F., Bauleo, P., Bonifazi, C., Civit, B., Escobar,
  C.~O., Garcia, B., Guedes, G., Berisso, M.~G., Harton, J.~L., et~al. (2008).
\newblock {The surface detector system of the Pierre Auger Observatory}.
\newblock {\em Nuclear Instruments and Methods in Physics Research Section A:
  Accelerators, Spectrometers, Detectors and Associated Equipment},
  586(3):409--420.

\bibitem[Altschuler and Nieves, 2002]{altschuler2002national}
Altschuler, D.~R. and Nieves, J.~F. (2002).
\newblock The national astronomy and ionosphere center's (naic) arecibo
  observatory in puerto rico.
\newblock In {\em Single-Dish Radio Astronomy: Techniques and Applications},
  volume 278, pages 1--24.

\bibitem[Anderhub et~al., 2013a]{anderhub2013design}
Anderhub, H., Backes, M., Biland, A., Boccone, V., Braun, I., Bretz, T.,
  Bu{\ss}, J., Cadoux, F., Commichau, V., Djambazov, L., et~al. (2013a).
\newblock Design and operation of fact--the first g-apd cherenkov telescope.
\newblock {\em Journal of Instrumentation}, 8(06):P06008.

\bibitem[Anderhub et~al., 2013b]{FACT_design}
Anderhub, H., Backes, M., Biland, A., Boccone, V., Braun, I., et~al. (2013b).
\newblock {Design and operation of FACT--the First G-APD Cherenkov Telescope}.
\newblock {\em Journal of Instrumentation}, 8(06):P06008.

\bibitem[Andersen and Kak, 1984]{andersen1984simultaneous}
Andersen, A.~H. and Kak, A.~C. (1984).
\newblock Simultaneous algebraic reconstruction technique (sart): a superior
  implementation of the art algorithm.
\newblock {\em Ultrasonic imaging}, 6(1):81--94.

\bibitem[Ansoldi et~al., 2016]{ansoldi2016teraelectronvolt}
Ansoldi, S., Antonelli, L.~A., Antoranz, P., Babic, A., Bangale, P.,
  de~Almeida, U.~B., Barrio, J.~A., Gonzalez, J.~B., Bednarek, W., Bernardini,
  E., et~al. (2016).
\newblock Teraelectronvolt pulsed emission from the crab pulsar detected by
  magic.
\newblock {\em Astronomy \& Astrophysics}, 585:A133.

\bibitem[Archambault et~al., 2017]{archambault2017gamma}
Archambault, S., Archer, A., Benbow, W., Bird, R., Bourbeau, E., Bouvier, A.,
  Buchovecky, M., Bugaev, V., Cardenzana, J.~V., Cerruti, M., et~al. (2017).
\newblock {Gamma-ray observations under bright moonlight with VERITAS}.
\newblock {\em Astroparticle Physics}, 91:34--43.

\bibitem[Arqueros et~al., 2003]{arqueros2003novel}
Arqueros, F., Jim{\'e}nez, A., and Valverde, A. (2003).
\newblock A novel procedure for the optical characterization of solar
  concentrators.
\newblock {\em Solar Energy}, 75(2):135--142.

\bibitem[Badran and Weekes, 1997]{badran1997improvement}
Badran, H.~M. and Weekes, T.~C. (1997).
\newblock {Improvement of gamma-hadron discrimination at tev energies using a
  new parameter, image surface brightness}.
\newblock {\em Astroparticle Physics}, 7(4):307--314.

\bibitem[Baixeras et~al., 2004]{baixeras2004}
Baixeras, C. et~al. (2004).
\newblock Design studies for a european gamma-ray observatory.
\newblock {\em arXiv preprint astro-ph/0403180}.

\bibitem[Barrau et~al., 1998]{barrau1998cat}
Barrau, A., Bazer-Bachi, R., Beyer, E., Cabot, H., Cerutti, M., Chounet, L.~M.,
  Debiais, G., Degrange, B., Delchini, H., Denance, J.~P., et~al. (1998).
\newblock {The CAT imaging telescope for very-high-energy gamma-ray astronomy}.
\newblock {\em Nuclear Instruments and Methods in Physics Research Section A:
  Accelerators, Spectrometers, Detectors and Associated Equipment},
  416(2):278--292.

\bibitem[Bartko et~al., 2005]{bartko2005tests}
Bartko, H., Goebel, F., Mirzoyan, R., Pimpl, W., and Teshima, M. (2005).
\newblock {Tests of a prototype multiplexed fiber-optic ultra-fast FADC data
  acquisition system for the MAGIC telescope}.
\newblock {\em Nuclear Instruments and Methods in Physics Research Section A:
  Accelerators, Spectrometers, Detectors and Associated Equipment},
  548(3):464--486.

\bibitem[Benn and Ellison, 1998]{Benn1998503}
Benn, C.~R. and Ellison, S.~L. (1998).
\newblock Brightness of the night sky over la palma.
\newblock {\em New Astronomy Reviews}, 42(6–8):503 -- 507.

\bibitem[Bergholm et~al., 2002]{bergholm2002plenoscope}
Bergholm, F., Renlund, E., and Karl, F. (2002).
\newblock The plenoscope concept and image formation.
\newblock In {\em Swedish Society for Automated Image Analysis Symposium
  (SSAB)}, volume~1, pages 75--78. Uppsala Universitet, Centre for Mathematical
  Sciences, Lund University.

\bibitem[Bergstr{\"o}m, 2013]{bergstrom2013dark}
Bergstr{\"o}m, L. (2013).
\newblock Dark matter and imaging air cherenkov arrays.
\newblock {\em Astroparticle Physics}, 43:44--49.

\bibitem[Bergstr{\"o}m et~al., 2011]{bergstrom2011}
Bergstr{\"o}m, L., Bringmann, T., and Edsj{\"o}, J. (2011).
\newblock Complementarity of direct dark matter detection and indirect
  detection through gamma rays.
\newblock {\em Physical Review D}, 83(4):045024.

\bibitem[Bernl{\"o}hr, 2008]{bernlohr2008AtmoIact}
Bernl{\"o}hr, K. (2008).
\newblock Simulation of imaging atmospheric cherenkov telescopes with corsika
  and sim\_telarray.
\newblock {\em Astroparticle Physics}, 30(3):149--158.

\bibitem[Bernl{\"o}hr et~al., 2013]{bernlohr2013monte}
Bernl{\"o}hr, K., Barnacka, A., Becherini, Y., Bigas, O.~B., Carmona, E.,
  Colin, P., Decerprit, G., Di~Pierro, F., Dubois, F., Farnier, C., et~al.
  (2013).
\newblock Monte carlo design studies for the cherenkov telescope array.
\newblock {\em Astroparticle Physics}, 43:171--188.

\bibitem[Bernl{\"o}hr et~al., 2003]{bernlohr2003optical}
Bernl{\"o}hr, K., Carrol, O., Cornils, R., Elfahem, S., Espigat, P., Gillessen,
  S., Heinzelmann, G., Hermann, G., Hofmann, W., Horns, D., et~al. (2003).
\newblock The optical system of the hess imaging atmospheric cherenkov
  telescopes. part i: layout and components of the system.
\newblock {\em Astroparticle Physics}, 20(2):111--128.

\bibitem[Bertone, 2010]{bertone2010moment}
Bertone, G. (2010).
\newblock {The moment of truth for WIMP dark matter}.
\newblock {\em Nature}, 468(7322):389--393.

\bibitem[Biland et~al., 2014]{fact-performance}
Biland, A., Bretz, T., Buß, J., Commichau, V., Djambazov, L., Dorner, D.,
  et~al. (2014).
\newblock {Calibration and performance of the photon sensor response of FACT
  – the first G-APD Cherenkov telescope}.
\newblock {\em JINST}, 9(10):P10012.

\bibitem[Biland et~al., 2007]{biland2007active}
Biland, A., Garczarczyk, M., Anderhub, H., Danielyan, V., Hakobyan, D., Lorenz,
  E., Mirzoyan, R., and Collaboration, M. (2007).
\newblock The active mirror control of the magic telescope.
\newblock In {\em Proceedings of the 30th ICRC}, volume 533, pages 1353--1356.

\bibitem[Birriel and Adkins, 2010]{birriel2010simple}
Birriel, J. and Adkins, J.~K. (2010).
\newblock A simple, portable apparatus to measure night sky brightness at
  various zenith angles.
\newblock {\em Journal of the American Association of Variable Star Observers},
  38:221--229.

\bibitem[Borkowski, 1987]{borkowski1987near}
Borkowski, K. (1987).
\newblock Near zenith tracking limits for altitude-azimuth telescopes.
\newblock {\em Acta astronomica}, 37:79.

\bibitem[Borwankar et~al., 2016]{borwankar2016simulation}
Borwankar, C., Bhatt, N., Bhattacharyya, S., Rannot, R., Tickoo, A., Koul, R.,
  and Thoudam, S. (2016).
\newblock Simulation studies of mace-i: Trigger rates and energy thresholds.
\newblock {\em Astroparticle Physics}, 84:97--106.

\bibitem[Brazier et~al., 1990]{brazier1990narrabri}
Brazier, S., Carraminana, A., Chadwick, M., Dipper, A., Lincoln, W., McComb,
  L., Orford, J., Rayner, M., Turver, E., and Williams, G. (1990).
\newblock The narrabri and la palma cerenkov gamma ray telescope.
\newblock In {\em International Cosmic Ray Conference}, volume~4, page 270.

\bibitem[Brunetto et~al., 2004]{brunetto2004progress}
Brunetto, E.~T., Dierickx, P., Gilmozzi, R., Le~Louarn, M., Koch, F., Noethe,
  L., Verinaud, C., and Yaitskova, N. (2004).
\newblock Progress of eso's 100-m owl optical telescope design.
\newblock In {\em Second Backaskog Workshop on Extremely Large Telescopes},
  volume 5382, pages 159--169. International Society for Optics and Photonics.

\bibitem[Budnev et~al., 2017]{budnev2017taiga}
Budnev, N., Astapov, I., Barbashina, N., Barnyakov, A., Bezyazeekov, P.,
  Bogdanov, A., Boreyko, V., Br{\"u}ckner, M., Chiavassa, A., Chvalaev, O.,
  et~al. (2017).
\newblock The taiga experiment: From cosmic-ray to gamma-ray astronomy in the
  tunka valley.
\newblock {\em Nuclear Instruments and Methods in Physics Research Section A:
  Accelerators, Spectrometers, Detectors and Associated Equipment},
  845:330--333.

\bibitem[Bulian et~al., 1998]{bulian1998characteristics}
Bulian, N., Daum, A., Hermann, G., Hess, M., Hofmann, W., Lampeitl, H.,
  Puehlhofer, G., Koehler, C., Panter, M., Stein, M., et~al. (1998).
\newblock Characteristics of the multi-telescope coincidence trigger of the
  hegra iact system.
\newblock {\em Astroparticle Physics}, 8(4):223--233.

\bibitem[Buss et~al., 2015]{buss2015fact}
Buss, J., Noethe, M., Walter, R., Morik, K., Bockermann, C., Wilbert, A.,
  Dorner, D., Balbo, M., Mannheim, K., Meier, K., et~al. (2015).
\newblock {Influence of SiPM Crosstalk on the Performance of an Operating
  Cherenkov Telescope}.
\newblock {\em PoS}, page 863.

\bibitem[Buzhan et~al., 2009]{buzhan2009cross}
Buzhan, P., Dolgoshein, B., Ilyin, A., Kaplin, V., Klemin, S., Mirzoyan, R.,
  Popova, E., and Teshima, M. (2009).
\newblock {The cross-talk problem in SiPMs and their use as light sensors for
  imaging atmospheric Cherenkov telescopes}.
\newblock {\em Nuclear Instruments and Methods in Physics Research Section A:
  Accelerators, Spectrometers, Detectors and Associated Equipment},
  610(1):131--134.

\bibitem[Cannon et~al., 2012]{cannon2012toward}
Cannon, K., Cariou, R., Chapman, A., Crispin-Ortuzar, M., Fotopoulos, N., Frei,
  M., Hanna, C., Kara, E., Keppel, D., Liao, L., et~al. (2012).
\newblock Toward early-warning detection of gravitational waves from compact
  binary coalescence.
\newblock {\em The Astrophysical Journal}, 748(2):136.

\bibitem[Catalano et~al., 2013]{catalano2013astri}
Catalano, O., Giarrusso, S., La~Rosa, G., Maccarone, M.~C., Mineo, T., Russo,
  F., Sottile, G., Impiombato, D., Bonanno, G., Belluso, M., et~al. (2013).
\newblock {The ASTRI SST-2M prototype: Camera and electronics}.
\newblock {\em arXiv preprint arXiv:1307.5142}.

\bibitem[Catalano et~al., 2008]{catalano2008single}
Catalano, O., Maccarone, M.~C., and Sacco, B. (2008).
\newblock {Single photon counting approach for imaging atmospheric Cherenkov
  telescopes}.
\newblock {\em Astroparticle Physics}, 29(2):104--116.

\bibitem[Chantell et~al., 1998]{chantell1998prototype}
Chantell, M., Bhattacharya, D., Covault, C., Dragovan, M., Fernholz, R.,
  Gregorich, D., Hanna, D., Marion, G., Ong, R., Oser, S., et~al. (1998).
\newblock Prototype test results of the solar tower atmospheric cherenkov
  effect experiment (stacee).
\newblock {\em Nuclear Instruments and Methods in Physics Research Section A:
  Accelerators, Spectrometers, Detectors and Associated Equipment},
  408(2-3):468--485.

\bibitem[Clowe et~al., 2006]{clowe2006direct}
Clowe, D., Brada{\v{c}}, M., Gonzalez, A.~H., Markevitch, M., Randall, S.~W.,
  Jones, C., and Zaritsky, D. (2006).
\newblock A direct empirical proof of the existence of dark matter.
\newblock {\em The Astrophysical Journal Letters}, 648(2):L109.

\bibitem[{Cmglee}, 2014]{cmglee2014comparison}
{Cmglee} (2014).
\newblock Comparison of nominal sizes of primary mirrors of notable optical
  telescopes --- {W}ikipedia{,} the free encyclopedia.
\newblock [Online; accessed 4 June 2018].

\bibitem[Conti, 2009]{conti2009state}
Conti, M. (2009).
\newblock State of the art and challenges of time-of-flight pet.
\newblock {\em Physica Medica}, 25(1):1--11.

\bibitem[Cornils et~al., 2005]{cornils2005optical}
Cornils, R., Bernl{\"o}hr, K., Heinzelmann, G., Hofmann, W., and Panter, M.
  (2005).
\newblock The optical system of the hess ii telescope.
\newblock In {\em International Cosmic Ray Conference}, volume~5, page 171.

\bibitem[Cortina et~al., 2016]{cortina2016}
Cortina, J., L{\'o}pez-Coto, R., and Moralejo, A. (2016).
\newblock Machete: A transit imaging atmospheric cherenkov telescope to survey
  half of the very high energy $\gamma$-ray sky.
\newblock {\em Astroparticle Physics}, 72:46--54.

\bibitem[Covault et~al., 2001]{covault2001status}
Covault, C., Boone, L., Bramel, D., Chae, E., Fortin, P., Gingrich, D., Hinton,
  J., Hanna, D., Mukherjee, R., Ong, C., et~al. (2001).
\newblock The status of the stacee observatory.
\newblock {\em arXiv preprint astro-ph/0107427}.

\bibitem[{CTAO gGmbH}, 2018]{cta2018baseline}
{CTAO gGmbH} (2018).
\newblock {CTA's expected baseline performance}.
\newblock \url{https://www.cta-observatory.org/science/cta-performance/},
  accessed August 2018.

\bibitem[Daglas, 2017]{daglas2015master}
Daglas, S. (2017).
\newblock {{Structural Optimization of a Next Generation Gamma-Ray Telescope}}.
\newblock {\em ETH Zurich, Research Collection}, Department of Civil
  Engineering.

\bibitem[Davies and Cotton, 1957]{Davies_Cotton_1957}
Davies, J.~M. and Cotton, E.~S. (1957).
\newblock Design of the quartermaster solar furnace.
\newblock {\em Solar Energy}, 1(2):16--22.

\bibitem[de~Naurois and Mazin, 2015]{naurois2015}
de~Naurois, M. and Mazin, D. (2015).
\newblock Ground-based detectors in very-high-energy gamma-ray astronomy.
\newblock {\em Comptes Rendus Physique}, 16(6):610--627.

\bibitem[Dean et~al., 2008]{dean2008polarized}
Dean, A.~J., Clark, D.~J., Stephen, J.~B., McBride, V.~A., Bassani, L.,
  Bazzano, A., Bird, A.~J., Hill, A.~B., Shaw, S.~E., and Ubertini, P. (2008).
\newblock {Polarized gamma-ray emission from the Crab}.
\newblock {\em Science}, 321(5893):1183--1185.

\bibitem[Delagnes et~al., 2006]{delagnes2006sam}
Delagnes, E., Degerli, Y., Goret, P., Nayman, P., Toussenel, F., and Vincent,
  P. (2006).
\newblock Sam: A new ghz sampling asic for the hess-ii front-end electronics.
\newblock {\em Nuclear Instruments and Methods in Physics Research Section A:
  Accelerators, Spectrometers, Detectors and Associated Equipment},
  567(1):21--26.

\bibitem[DeYoung et~al., 2012]{deyoung2012hawc}
DeYoung, T., collaboration, H., et~al. (2012).
\newblock {The HAWC observatory}.
\newblock {\em Nuclear Instruments and Methods in Physics Research Section A:
  Accelerators, Spectrometers, Detectors and Associated Equipment}, 692:72--76.

\bibitem[Dole et~al., 2006]{dole2006cosmic}
Dole, H., Lagache, G., Puget, J.-L., Caputi, K.~I., Fernandez-Conde, N.,
  Le~Floc'h, E., Papovich, C., P{\'e}rez-Gonz{\'a}lez, P.~G., Rieke, G.~H., and
  Blaylock, M. (2006).
\newblock The cosmic infrared background resolved by spitzer-contributions of
  mid-infrared galaxies to the far-infrared background.
\newblock {\em Astronomy \& Astrophysics}, 451(2):417--429.

\bibitem[Dorner et~al., 2017]{fact2017icrc}
Dorner, D. et~al. (2017).
\newblock Fact-highlights from more than five years of unbiased monitoring at
  tev energies.
\newblock {\em PoS (ICRC2017)}, 609.

\bibitem[Dravins et~al., 2012]{dravins2012stellar}
Dravins, D., LeBohec, S., Jensen, H., and Nu{\~n}ez, P.~D. (2012).
\newblock Stellar intensity interferometry: Prospects for sub-milliarcsecond
  optical imaging.
\newblock {\em New Astronomy Reviews}, 56(5):143--167.

\bibitem[Dravins et~al., 2013]{dravins2013optical}
Dravins, D., LeBohec, S., Jensen, H., Nu{\~n}ez, P.~D., Consortium, C., et~al.
  (2013).
\newblock Optical intensity interferometry with the cherenkov telescope array.
\newblock {\em Astroparticle Physics}, 43:331--347.

\bibitem[Ebisuzaki et~al., 2014]{ebisuzaki2014jem}
Ebisuzaki, T., Medina-Tanco, G., Santangelo, A., Collaboration, J.-E., et~al.
  (2014).
\newblock {The JEM-EUSO mission}.
\newblock {\em Advances in Space Research}, 53(10):1499--1505.

\bibitem[Engels, 2017]{engels2017master}
Engels, A.~A. (2017).
\newblock Cosmic ray composition measurements via direct {C}herenkov light with
  the atmospheric {C}herenkov plenoscope.
\newblock {\em ETH Zurich, Research Collection}, Institute for Particlephysics
  and Astrophysics.

\bibitem[Ester et~al., 1996]{ester1996density}
Ester, M., Kriegel, H.-P., Sander, J., Xu, X., et~al. (1996).
\newblock {A density-based algorithm for discovering clusters in large spatial
  databases with noise}.
\newblock In {\em Kdd}, volume~96, pages 226--231.

\bibitem[{Fermi-LAT Collaboration} et~al., 2018]{fermi2018gamma}
{Fermi-LAT Collaboration} et~al. (2018).
\newblock A gamma-ray determination of the universe’s star formation history.
\newblock {\em Science}, 362(6418):1031--1034.

\bibitem[Finnegan et~al., 2011]{finnegan2011orbit}
Finnegan, G. et~al. (2011).
\newblock Orbit mode observation technique developed for veritas.
\newblock {\em arXiv preprint arXiv:1111.0121}.

\bibitem[Fioretti et~al., 2016]{fioretti2016cherenkov}
Fioretti, V., Bulgarelli, A., and Sch{\"u}ssler, F. (2016).
\newblock The cherenkov telescope array on-site integral sensitivity: observing
  the crab.
\newblock In {\em Ground-based and Airborne Telescopes VI}, volume 9906, page
  99063O. International Society for Optics and Photonics.

\bibitem[Fontaine et~al., 1990]{fontaine1990aims}
Fontaine, G., Baillon, P., Behr, L., Dudelzak, B., Eschstruth, P., Espigat, P.,
  Fabre, J., Fontaine, G., George, R., Ghesqui{\`e}re, C., et~al. (1990).
\newblock Aims and status of the themistocle physics experiment.
\newblock {\em Nuclear Physics B-Proceedings Supplements}, 14(2):79--94.

\bibitem[Fors et~al., 2013]{fors2013telescope}
Fors, O., N{\'u}{\~n}ez, J., Muinos, J.~L., Montojo, F.~J., Baena-Gall{\'e},
  R., Boloix, J., Morcillo, R., Merino, M.~T., Downey, E.~C., and Mazur, M.~J.
  (2013).
\newblock {Telescope Fabra ROA Montsec: A New Robotic Wide Field Baker--Nunn
  Facility}.
\newblock {\em Publications of the Astronomical Society of the Pacific},
  125(927):522.

\bibitem[Forsyth and Ponce, 2003]{forsyth2003modern}
Forsyth, D.~A. and Ponce, J. (2003).
\newblock A modern approach.
\newblock {\em Computer vision: a modern approach}, pages 88--101.

\bibitem[Frach et~al., 2009]{frach2009digital}
Frach, T., Prescher, G., Degenhardt, C., de~Gruyter, R., Schmitz, A., and
  Ballizany, R. (2009).
\newblock {The digital silicon photomultiplier—Principle of operation and
  intrinsic detector performance}.
\newblock In {\em Nuclear Science Symposium Conference Record (NSS/MIC), 2009
  IEEE}, pages 1959--1965. IEEE.

\bibitem[Funk et~al., 2004]{funk2004trigger}
Funk, S., Hermann, G., Hinton, J., Berge, D., Bernl{\"o}hr, K., Hofmann, W.,
  Nayman, P., Toussenel, F., and Vincent, P. (2004).
\newblock The trigger system of the hess telescope array.
\newblock {\em Astroparticle Physics}, 22(3):285--296.

\bibitem[Gaidos et~al., 1996]{gaidos1996extremely}
Gaidos, J.~A., Akerlof, C.~W., Biller, S., Boyles, P.~J., et~al. (1996).
\newblock {Extremely rapid bursts of TeV photons from the active galaxy
  Markarian 421}.
\newblock {\em Nature}, 383(6598):319.

\bibitem[Galilei, 1638]{square_qube_law}
Galilei, G. (1638).
\newblock {\em Discorsi e dimostrazioni matematiche, intorno {\`a} due nuove
  scienze attenenti alla mecanica \& i movimenti locali}.
\newblock Appresso gli Elsevirii, Leida.

\bibitem[Garcia et~al., 2014]{garcia2014status}
Garcia, J.~R., Dazzi, F., H{\"a}fner, D., Herranz, D., L{\'o}pez, M., Mariotti,
  M., Mirzoyan, R., Nakajima, D., Schweizer, T., and Teshima, M. (2014).
\newblock Status of the new sum-trigger system for the magic telescopes.
\newblock {\em arXiv preprint arXiv:1404.4219}.

\bibitem[Gaug et~al., 2013]{gaug2013night}
Gaug, M. et~al. (2013).
\newblock Night sky background analysis for the cherenkov telescope array using
  the atmoscope instrument.
\newblock In {\em In Proceedings of the 33th International Cosmic Ray
  Conference (ICRC2013)}.

\bibitem[Gazda et~al., 2016]{gazda2016photon}
Gazda, E., Nguyen, T., Otte, N., and Richards, G. (2016).
\newblock {Photon detection efficiency measurements of the VERITAS Cherenkov
  telescope photomultipliers after four years of operation}.
\newblock {\em Journal of Instrumentation}, 11(11):P11015.

\bibitem[Golovin, 1998]{golovin1999avalanche}
Golovin, V. (1998).
\newblock {Avalanche Photodetector}.
\newblock {\em Russian Agency for Patents and Trademarks}, 2142175.

\bibitem[Hamamatsu, 2009]{hamamatsu2009mppc}
Hamamatsu (2009).
\newblock {\em MPPC (multi-pixel photon Counter), S10362-33 series}.
\newblock HAMAMATSU PHOTONICS K.K., Solid State Division.

\bibitem[Hanrahan and Ng, 2006]{hanrahan2006digital}
Hanrahan, P. and Ng, R. (2006).
\newblock Digital correction of lens aberrations in light field photography.
\newblock In {\em International Optical Design Conference}, page WB2. Optical
  Society of America.

\bibitem[Heck et~al., 1998]{heck1998corsika}
Heck, D., Capdevielle, J., Schatz, G., Thouw, T., et~al. (1998).
\newblock {\em CORSIKA: A Monte Carlo Code to Simulate Extensive Air Showers,''
  Report FZKA 6019, Forschungszentrum Karlsruhe}.

\bibitem[{Heraeus}, 2018]{heraeus2018quarz}
{Heraeus} (2018).
\newblock {Quarzglas fuer die Optik, Daten und Eigenschaften, Suprasil 311,
  312, 313}.
\newblock {\em {Quarzglas GmbH und Co. KG, Optics, Quarzstr. 8, 63450 Hanau,
  Germany}}, HQS-MO 02.6 D04.2018.

\bibitem[Hess, 1912]{hess1912uber}
Hess, V.~F. (1912).
\newblock Uber beobachtungen der durchdringenden strahlung bei sieben
  freiballonfahrten.
\newblock {\em Phys. Zeits.}, 13:1084--1091.

\bibitem[H.E.S.S.~collaboration, 2018]{hess2018images}
H.E.S.S.~collaboration, M. (2018).
\newblock {Photographs of the H.E.S.S. and H.E.S.S. II telescopes}.
\newblock
  \url{https://www.mpi-hd.mpg.de/hfm/HESS/pages/about/pictures/HESS_IMG/},
  accessed 2018 June 29.

\bibitem[Hildebrand et~al., 2017]{hildebrand2017using}
Hildebrand, D. et~al. (2017).
\newblock {Using Charged Cosmic Ray Particles to Monitor the Data Quality of
  FACT}.
\newblock In {\em Proceedings of the 35th International Cosmic Ray Conference.
  PoS (ICRC2017)}, volume 301.

\bibitem[Hillas, 1985]{hillas1985cerenkov}
Hillas, A.~M. (1985).
\newblock Cerenkov light images of eas produced by primary gamma.
\newblock In {\em International Cosmic Ray Conference}, volume~3.

\bibitem[Hillas et~al., 1998]{hillas1998spectrum}
Hillas, A.~M., Akerlof, C.~W., Biller, S.~D., Buckley, J.~H., Carter-Lewis,
  D.~A., Catanese, M., Cawley, M.~F., Fegan, D.~J., Finley, J.~P., Gaidos,
  J.~A., et~al. (1998).
\newblock {The Spectrum of TeV gamma-rays from the Crab Nebula}.
\newblock {\em The Astrophysical Journal}, 503(2):744.

\bibitem[Hofmann, 2001]{hofmann2001focus}
Hofmann, W. (2001).
\newblock How to focus a cherenkov telescope.
\newblock {\em Journal of Physics G: Nuclear and Particle Physics},
  27(4):933--939.

\bibitem[Hofmann, 2006]{hofmann2006performance}
Hofmann, W. (2006).
\newblock Performance limits for cherenkov instruments.
\newblock {\em arXiv preprint astro-ph/0603076}.

\bibitem[Holder et~al., 2006]{holder2006first}
Holder, J., Atkins, R.~W., Badran, H.~M., Blaylock, G., Bradbury, S.~M.,
  Buckley, J.~H., Byrum, K.~L., Carter-Lewis, D.~A., Celik, O., Chow, Y. C.~K.,
  et~al. (2006).
\newblock {The first VERITAS telescope}.
\newblock {\em Astroparticle Physics}, 25(6):391--401.

\bibitem[Holler et~al., 2013]{holler2013status}
Holler, M., Balzer, A., Becherini, Y., Klepser, S., Murach, T., de~Naurois, M.,
  Parsons, R., et~al. (2013).
\newblock {Status of the Monoscopic Analysis Chains for HESS II}.
\newblock {\em arXiv preprint arXiv:1307.6003}.

\bibitem[Jean-Francois and Retiere, 2017]{retiere2017digital}
Jean-Francois, P. and Retiere, F. (2017).
\newblock {3D integrated digital SiPM}.
\newblock {\em 10th International Meeting on Front-End Electronics}.

\bibitem[Jung et~al., 2005]{jung2005star}
Jung, I., Krawczynski, H., Buckley, J., and Falcone, A. (2005).
\newblock {STAR a next generation Cherenkov telescope}.
\newblock In {\em Towards a Network of Atmospheric Cherenkov Detectors VII},
  pages 463--6.

\bibitem[Karle et~al., 1995]{karle1995design}
Karle, A., Merck, M., Plaga, R. {\aa}. a.~a., Arqueros, F., Haustein, V.,
  Heinzelmann, G., Holl, I., Fonseca, V., Lorenz, E., Martinez, S., et~al.
  (1995).
\newblock Design and performance of the angle integrating {\v{c}}erenkov array
  airobicc.
\newblock {\em Astroparticle Physics}, 3(4):321--347.

\bibitem[Kieda et~al., 2001]{kieda2001high}
Kieda, D.~B., Swordy, S., and Wakely, S. (2001).
\newblock {A high resolution method for measuring cosmic ray composition beyond
  10 TeV}.
\newblock {\em Astroparticle Physics}, 15(3):287--303.

\bibitem[Kildea et~al., 2007]{kildea2007whipple}
Kildea, J., Atkins, R.~W., Badran, H.~M., Blaylock, G., Bond, I.~H., Bradbury,
  S.~M., Buckley, J.~H., Carter-Lewis, D.~A., Celik, O., Chow, Y. C.~K., et~al.
  (2007).
\newblock {The Whipple Observatory 10m gamma-ray telescope, 1997--2006}.
\newblock {\em Astroparticle Physics}, 28(2):182--195.

\bibitem[Knoetig et~al., 2013]{knoetig2013fact}
Knoetig, M.~L., Biland, A., Bretz, T., Buss, J., Dorner, D., Einecke, S.,
  Eisenacher, D., Hildebrand, D., Kraehenbuehl, T., Lustermann, W., et~al.
  (2013).
\newblock Fact-long-term stability and observations during strong moon light.
\newblock {\em arXiv preprint arXiv:1307.6116}.

\bibitem[Kraehenbuehl, 2017]{kraehenbuehl2013diss}
Kraehenbuehl, T. (2017).
\newblock {{The first semiconductor-based camera for imaging atmospheric
  Cherenkov telescopes}}.
\newblock {\em ETH Zurich, Research Collection}, Department of Physics.

\bibitem[Krawczynski et~al., 2006]{krawczynski2006gamma}
Krawczynski, H., Carter-Lewis, D.~A., Duke, C., Holder, J., Maier, G.,
  Le~Bohec, S., and Sembroski, G. (2006).
\newblock {Gamma--hadron separation methods for the VERITAS array of four
  imaging atmospheric Cherenkov telescopes}.
\newblock {\em Astroparticle Physics}, 25(6):380--390.

\bibitem[Kubo et~al., 2004]{kubo2004status}
Kubo, H., Asahara, A., Bicknell, G., Clay, R.~W., Doi, Y., Edwards, P.,
  Enomoto, R., Gunji, S., Hara, S., Hara, T., et~al. (2004).
\newblock Status of the cangaroo-iii project.
\newblock {\em New Astronomy Reviews}, 48(5):323--329.

\bibitem[Kuldeep, 2013]{yadav2013mace}
Kuldeep, Y. (2013).
\newblock Status of the mace telescope.

\bibitem[Kuo, 2017]{kuo2017assessments}
Kuo, C.-L. (2017).
\newblock Assessments of ali, dome a, and summit camp for mm-wave observations
  using merra-2 reanalysis.
\newblock {\em The Astrophysical Journal}, 848(1):64.

\bibitem[Le~Bohec et~al., 1998]{le1998new}
Le~Bohec, S., Degrange, B., Punch, M., Barrau, A., Bazer-Bachi, R., Cabot, H.,
  Chounet, L., Debiais, G., Dezalay, J., Djannati-Atai, A., et~al. (1998).
\newblock A new analysis method for very high definition imaging atmospheric
  cherenkov telescopes as applied to the cat telescope.
\newblock {\em Nuclear Instruments and Methods in Physics Research Section A:
  Accelerators, Spectrometers, Detectors and Associated Equipment},
  416(2-3):425--437.

\bibitem[Levoy et~al., 2006]{ng2006lightfieldmicroscopy}
Levoy, M., Ng, R., Adams, A., Footer, M., and Horowitz, M. (2006).
\newblock {Light Field Microscopy}.
\newblock {\em ACM Transactions on Graphics (TOG), SIGGRAPH}, 25(3):924--934.

\bibitem[Lewis, 1990]{lewis1990optical}
Lewis, D. (1990).
\newblock Optical characteristics of the whipple observatory tev gamma-ray
  imaging telescope.
\newblock {\em Experimental Astronomy}, 1(4):213--226.

\bibitem[Li and Ma, 1983]{li1983analysis}
Li, T.-P. and Ma, Y.-Q. (1983).
\newblock {Analysis methods for results in gamma-ray astronomy}.
\newblock {\em The Astrophysical Journal}, 272:317--324.

\bibitem[Linden and Profumo, 2013]{linden2013probing}
Linden, T. and Profumo, S. (2013).
\newblock {Probing the pulsar origin of the anomalous positron fraction with
  AMS-02 and atmospheric Cherenkov telescopes}.
\newblock {\em The Astrophysical Journal}, 772(1):18.

\bibitem[Lipari, 2002]{lipari2002fluxes}
Lipari, P. (2002).
\newblock The fluxes of sub-cutoff particles detected by ams, the cosmic ray
  albedo and atmospheric neutrinos.
\newblock {\em Astroparticle physics}, 16(3):295--323.

\bibitem[Lippmann, 1908]{lippmann1908}
Lippmann, G. (1908).
\newblock {\'E}preuves r{\'e}versibles donnant la sensation du relief.
\newblock {\em J. Phys. Theor. Appl. 7 821-825}.

\bibitem[Lizarazo et~al., 2006]{lizarazo2006data}
Lizarazo, J., Afonso, P., Chertok, M., Marleau, P., Maruyama, S., Stilley, J.,
  and Tripathi, S. (2006).
\newblock Data acquisition system and trigger electronics for cactus.
\newblock In {\em Astroparticle, Particle and Space Physics, Detectors and
  Medical Physics Applications: Proceedings of the 9th Conference: Villa Olmo,
  Como, Italy, 17-21 October 2005}, page 262. World Scientific.

\bibitem[L{\'o}pez-Coto et~al., 2016]{lopez2016topo}
L{\'o}pez-Coto, R., Mazin, D., Paoletti, R., Bigas, O.~B., and Cortina, J.
  (2016).
\newblock The topo-trigger: a new concept of stereo trigger system for imaging
  atmospheric cherenkov telescopes.
\newblock {\em Journal of Instrumentation}, 11(04):P04005.

\bibitem[Lubsandorzhiev et~al., 2008]{lubsandorzhiev2008tunka}
Lubsandorzhiev, B., Collaboration, T., et~al. (2008).
\newblock Tunka-eas cherenkov experiment in the tunka valley.
\newblock {\em Nuclear Instruments and Methods in Physics Research Section A:
  Accelerators, Spectrometers, Detectors and Associated Equipment},
  595(1):73--76.

\bibitem[Masuda et~al., 2015]{masuda2015development}
Masuda, S., Konno, Y., Barrio, J.~A., Bigas, O.~B., Delgado, C., Coromina,
  L.~F., Gunji, S., Hadasch, D., Hatanaka, K., Ikeno, M., et~al. (2015).
\newblock Development of the photomultiplier tube readout system for the first
  large-sized telescope of the cherenkov telescope array.
\newblock {\em arXiv preprint arXiv:1509.00548}.

\bibitem[McCann et~al., 2010]{mccann2010new}
McCann, A., Hanna, D., Kildea, J., and McCutcheon, M. (2010).
\newblock A new mirror alignment system for the veritas telescopes.
\newblock {\em Astroparticle Physics}, 32(6):325--329.

\bibitem[McNally et~al., 1999]{mcnally1999three}
McNally, J.~G., Karpova, T., Cooper, J., and Conchello, J.~A. (1999).
\newblock Three-dimensional imaging by deconvolution microscopy.
\newblock {\em Methods}, 19(3):373--385.

\bibitem[Merklinger, 1997]{merklinger1997bokeh}
Merklinger, H.~M. (1997).
\newblock A technical view of bokeh.
\newblock {\em Photo Techniques}, 18(3).

\bibitem[Miermeister et~al., 2016]{miermeister2016cablerobot}
Miermeister, P., L{\"a}chele, M., Boss, R., Masone, C., Schenk, C., Tesch, J.,
  Kerger, M., Teufel, H., Pott, A., and B{\"u}lthoff, H. (2016).
\newblock {The CableRobot Simulator: Large Scale Motion Platform Based on Cable
  Robot Technology}.
\newblock In {\em {IEEE/RSJ International Conference on Intelligent Robots and
  Systems}}. IEEE/RSJ.

\bibitem[Mirzoyan et~al., 1996]{mirzoyan1996optical}
Mirzoyan, R., Fomin, V., and Stepanian, A. (1996).
\newblock On the optical design of vhe gamma ray imaging cherenkov telescopes.
\newblock {\em Nuclear Instruments and Methods in Physics Research Section A:
  Accelerators, Spectrometers, Detectors and Associated Equipment},
  373(1):153--158.

\bibitem[Miyata et~al., 2008]{miyata2008site}
Miyata, T., Motohara, K., Sako, S., Tanaka, M., Minezaki, T., Mitani, N., Aoki,
  T., Soyano, T., Tanabe, T., Kawara, K., et~al. (2008).
\newblock Site evaluations of the summit of co. chajnantor for infrared
  observations.
\newblock In {\em Ground-based and Airborne Telescopes II}, volume 7012, page
  701243. International Society for Optics and Photonics.

\bibitem[Neise et~al., 2017]{neise2017fact}
Neise, D., Adam, J., Ahnen, M.~L., Baack, D., Balbo, M., Bergmann, M., Biland,
  A., Blank, M., Bretz, T., Bruegge, K.~A., et~al. (2017).
\newblock {FACT--Status and experience from five years of operation of the
  first G-APD Cherenkov Telescope}.
\newblock {\em Nuclear Instruments and Methods in Physics Research Section A:
  Accelerators, Spectrometers, Detectors and Associated Equipment}, 876:17--20.

\bibitem[Neronov and Vovk, 2010]{neronov2010evidence}
Neronov, A. and Vovk, I. (2010).
\newblock {Evidence for strong extragalactic magnetic fields from Fermi
  observations of TeV blazars}.
\newblock {\em Science}, 328(5974):73--75.

\bibitem[Ng et~al., 2005]{ng2005}
Ng, R., Levoy, M., Br{\'e}dif, M., Duval, G., Horowitz, M., and Hanrahan, P.
  (2005).
\newblock Light field photography with a hand-held plenoptic camera.
\newblock {\em Computer Science Technical Report CSTR}, 2(11).

\bibitem[Noethe et~al., 2017]{nothe2017fact}
Noethe, M. et~al. (2017).
\newblock {FACT Performance of the First Cherenkov Telescope Observing with
  SiPMs}.
\newblock In {\em Proceedings of the 35th International Cosmic Ray Conference.
  PoS (ICRC2017)}, volume 791.

\bibitem[Olive et~al., 2014]{olive2014Review}
Olive, K.~A. et~al. (2014).
\newblock {{Review of Particle Physics}}.
\newblock {\em Chin. Phys.}, C38:090001.

\bibitem[Otte, 2009]{otte2009efficiency}
Otte, A.~N. (2009).
\newblock {On the efficiency of photon emission during electrical breakdown in
  silicon}.
\newblock {\em Nuclear Instruments and Methods in Physics Research Section A:
  Accelerators, Spectrometers, Detectors and Associated Equipment},
  610(1):105--109.

\bibitem[Par{\'e} et~al., 2002]{pare2002celeste}
Par{\'e}, E., Balauge, B., Bazer-Bachi, R., Bergeret, H., Berny, F., Briand,
  N., Bruel, P., Cerutti, M., Collon, J., Cordier, A., et~al. (2002).
\newblock Celeste: an atmospheric cherenkov telescope for high energy gamma
  astrophysics.
\newblock {\em Nuclear Instruments and Methods in Physics Research Section A:
  Accelerators, Spectrometers, Detectors and Associated Equipment},
  490(1-2):71--89.

\bibitem[Pareschi et~al., 2013a]{pareschi2013status}
Pareschi, G., Armstrong, T., Baba, H., B{\"a}hr, J., Bonardi, A., Bonnoli, G.,
  Brun, P., Canestrari, R., Chadwick, P., Chikawa, M., et~al. (2013a).
\newblock Status of the technologies for the production of the cherenkov
  telescope array (cta) mirrors.
\newblock In {\em Optics for EUV, X-Ray, and Gamma-Ray Astronomy VI}, volume
  8861, page 886103. International Society for Optics and Photonics.

\bibitem[Pareschi et~al., 2013b]{pareschi2013statusArxiv}
Pareschi, G., Armstrong, T., Baba, H., B{\"a}hr, J., Bonardi, A., Bonnoli, G.,
  Brun, P., Canestrari, R., Chadwick, P., Chikawa, M., et~al. (2013b).
\newblock Status of the technologies for the production of the cherenkov
  telescope array (cta) mirrors.
\newblock {\em arXiv preprint arXiv:1310.1713}.

\bibitem[Pedregosa et~al., 2011]{scikit-learn}
Pedregosa, F., Varoquaux, G., Gramfort, A., Michel, V., Thirion, B., Grisel,
  O., Blondel, M., Prettenhofer, P., Weiss, R., Dubourg, V., Vanderplas, J.,
  Passos, A., Cournapeau, D., Brucher, M., Perrot, M., and Duchesnay, E.
  (2011).
\newblock Scikit-learn: Machine learning in {P}ython.
\newblock {\em Journal of Machine Learning Research}, 12:2825--2830.

\bibitem[Perkins et~al., 2007]{perkins2007mirror}
Perkins, J., Finley, J., Falcone, A., Weekes, T., Harris, K., and Irvin, R.
  (2007).
\newblock Mirror facets for the veritas telescopes.
\newblock In {\em 30th International Cosmic Ray Conference, Mexico}, volume~3,
  page 1397–1400.

\bibitem[Poisson, 1837]{poisson1837recherches}
Poisson, S.~D. (1837).
\newblock {\em {Recherches sur la probabilit{\'e} des jugements en mati{\`e}re
  criminelle et en mati{\`e}re civile prec{\'e}d{\'e}es des r{\`e}gles
  g{\'e}n{\'e}rales du calcul des probabilit{\'e}s par SD Poisson}}.
\newblock Bachelier.

\bibitem[Preuss et~al., 2002]{preuss2002study}
Preuss, S., Hermann, G., Hofmann, W., and Kohnle, A. (2002).
\newblock Study of the photon flux from the night sky at la palma and namibia,
  in the wavelength region relevant for imaging atmospheric cherenkov
  telescopes.
\newblock {\em Nuclear Instruments and Methods in Physics Research Section A:
  Accelerators, Spectrometers, Detectors and Associated Equipment},
  481(1):229--240.

\bibitem[Puehlhofer et~al., 2003]{puhlhofer2003technical}
Puehlhofer, G., Bolz, O., Goetting, N., Heusler, A., Horns, D., Kohnle, A.,
  Lampeitl, H., Panter, M., Tluczykont, M., Aharonian, F., et~al. (2003).
\newblock {The technical performance of the HEGRA system of imaging air
  Cherenkov telescopes}.
\newblock {\em Astroparticle Physics}, 20(3):267--291.

\bibitem[Punch et~al., 2001]{punch2001hess}
Punch, M., Aharonian, F., Bulian, N., Gillessen, S., Hermann, G., Heusler, A.,
  Hofmann, W., Jung, I., Kankanyan, R., Kettler, J., et~al. (2001).
\newblock The hess project camera, tests, and status.
\newblock In {\em International Cosmic Ray Conference 27 ICRC 2001}, pages
  2814--2817. International Union of Pure and Applied Physics.

\bibitem[Punch et~al., 1992]{punch1992detection}
Punch, M., Akerlof, C.~W., Cawley, M.~F., Chantell, M., Fegan, D.~J., Fennell,
  S., Gaidos, J.~A., Hagan, J., Hillas, A.~M., Jiang, Y., et~al. (1992).
\newblock Detection of tev photons from the active galaxy markarian 421.
\newblock {\em nature}, 358(6386):477--478.

\bibitem[Radford et~al., 2008]{radford2008submillimeter}
Radford, S.~J., Giovanelli, R., Gull, G.~E., and Henderson, C.~P. (2008).
\newblock Submillimeter observing conditions on cerro chajnantor.
\newblock In {\em Ground-based and Airborne Telescopes II}, volume 7012, page
  70121Z. International Society for Optics and Photonics.

\bibitem[Ritt, 2008]{ritt2008design}
Ritt, S. (2008).
\newblock {Design and performance of the 6 GHz waveform digitizing chip DRS4}.
\newblock In {\em Nuclear Science Symposium Conference Record, 2008. NSS'08.
  IEEE}, pages 1512--1515. IEEE.

\bibitem[Ritt, 2011]{ritt2011design}
Ritt, S. (2011).
\newblock Design and performance of the 6 gs/s waveform digitizing chip drs4.

\bibitem[Roberts et~al., 1998]{roberts1998tev}
Roberts, M., Dazeley, S., Edwards, P., Hara, T., Hayami, Y., Holder, J.,
  Kakimoto, F., Kamei, S., Kawachi, A., Kifune, T., et~al. (1998).
\newblock Tev gamma-ray observations of southern bl lacs with the cangaroo 3.8
  m imaging telescope.
\newblock {\em arXiv preprint astro-ph/9811260}.

\bibitem[Roscoe et~al., 1988]{roscoe1988stereolithography}
Roscoe, L. et~al. (1988).
\newblock Stereolithography interface specification.
\newblock {\em America-3D Systems Inc}, 27.

\bibitem[Sadygov, 1998]{sadygov1998avalanche}
Sadygov, Z. (1998).
\newblock {Avalanche Detector}.
\newblock {\em Russian Agency for Patents and Trademarks}, 2102820.

\bibitem[Savchenko et~al., 2016]{savchenko2016integral}
Savchenko, V., Ferrigno, C., Mereghetti, S., Natalucci, L., Bazzano, A., Bozzo,
  E., Brandt, S., Courvoisier, T.-L., Diehl, R., Hanlon, L., et~al. (2016).
\newblock Integral upper limits on gamma-ray emission associated with the
  gravitational wave event gw150914.
\newblock {\em The Astrophysical Journal Letters}, 820(2):L36.

\bibitem[Schroedter et~al., 2009]{schroedter2009topological}
Schroedter, M., Anderson, J., Byrum, K., Drake, G., Duke, C., Holder, J.,
  Imran, A., Madhavan, A., Krennrich, F., Kreps, A., et~al. (2009).
\newblock A topological trigger system for imaging atmospheric-cherenkov
  telescopes.
\newblock {\em arXiv preprint arXiv:0908.0179}.

\bibitem[Sinisyna, 2005]{sinisyna2005shalon}
Sinisyna, V. (2005).
\newblock {SHALON: status report}.
\newblock In {\em Towards a Network of Atmospheric Cherenkov Detectors VII},
  pages 57--65.

\bibitem[Sitarek et~al., 2013]{sitarek2013analysis}
Sitarek, J., Gaug, M., Mazin, D., Paoletti, R., and Tescaro, D. (2013).
\newblock {Analysis techniques and performance of the Domino Ring Sampler
  version 4 based readout for the MAGIC telescopes}.
\newblock {\em Nuclear Instruments and Methods in Physics Research Section A:
  Accelerators, Spectrometers, Detectors and Associated Equipment},
  723:109--120.

\bibitem[Supanitsky and Rovero, 2012]{supanitsky2012earth}
Supanitsky, A. and Rovero, A. (2012).
\newblock {Earth magnetic field effects on the cosmic electron flux as
  background for Cherenkov Telescopes at low energies}.
\newblock {\em Astroparticle Physics}, 36(1):123--130.

\bibitem[Swartzlander, 1973]{swartzlander1973parallel}
Swartzlander, E.~E. (1973).
\newblock {Parallel counters}.
\newblock {\em IEEE Transactions on computers}, 100(11):1021--1024.

\bibitem[Swartzlander, 2004]{swartzlander2004review}
Swartzlander, E.~E. (2004).
\newblock {A review of large parallel counter designs}.
\newblock In {\em VLSI, 2004. Proceedings. IEEE Computer society Annual
  Symposium on}, pages 89--98. IEEE.

\bibitem[Tavani et~al., 2011]{tavani2011discovery}
Tavani, M., Bulgarelli, A., Vittorini, V., Pellizzoni, A., Striani, E.,
  Caraveo, P., Weisskopf, M.~C., Tennant, A., Pucella, G., Trois, A., et~al.
  (2011).
\newblock {Discovery of powerful gamma-ray flares from the Crab Nebula}.
\newblock {\em Science}, 331(6018):736--739.

\bibitem[Taylor, 2017]{taylor2017active}
Taylor, A.~M. (2017).
\newblock Active galactic nuclei horizons from the gamma-ray perspective.
\newblock {\em New Astronomy Reviews}, 78:16--25.

\bibitem[Temme et~al., 2015]{ICRC2014_fact_crab_spectrum}
Temme, F. et~al. (2015).
\newblock Fact--first energy spectrum from a sipm cherenkov telescope.
\newblock In {\em Proceedings of the 34th ICRC}, volume 707.

\bibitem[Teshima et~al., 2011]{teshima2011design}
Teshima, M., Schweizer, T., and Blanch, O. (2011).
\newblock {Design study of a CTA large size telescope (LST)}.
\newblock In {\em 32nd International Cosmic Ray Conference}, volume~9, pages
  151--154.

\bibitem[{The MAGIC Collaboration}, 2008a]{magic2008pulsar}
{The MAGIC Collaboration} (2008a).
\newblock Observation of pulsed $\gamma$-rays above 25 gev from the crab pulsar
  with magic.
\newblock {\em Science}, 322(5905):1221--1224.

\bibitem[{The MAGIC Collaboration}, 2008b]{magic2008distantQuasar}
{The MAGIC Collaboration} (2008b).
\newblock Very-high-energy gamma rays from a distant quasar: how transparent is
  the universe?
\newblock {\em Science}, 320(5884):1752--1754.

\bibitem[Toyama et~al., 2013]{toyama2013novel}
Toyama, T., Mirzoyan, R., Dickinson, H., Fruck, C., Hose, J., Kellermann, H.,
  Kn{\"o}tig, M., Lorenz, E., Menzel, U., Nakajima, D., et~al. (2013).
\newblock Novel photo multiplier tubes for the cherenkov telescope array
  project.
\newblock {\em arXiv preprint arXiv:1307.5463}.

\bibitem[Trichard et~al., 2015]{trichard2015enhanced}
Trichard, C., Fiasson, A., Maurin, G., Lamanna, G., and for~the
  H.E.S.S.~Collaboration (2015).
\newblock Enhanced h.e.s.s. ii low energies performance thanks to the focus
  system.
\newblock In {\em In Proceedings of the 34th International Cosmic Ray
  Conference (ICRC2015), The Hague, The Netherlands}.

\bibitem[Tridon et~al., 2010]{tridon2010magic}
Tridon, D.~B., Schweizer, T., Goebel, F., Mirzoyan, R., Teshima, M.,
  collaboration, M., et~al. (2010).
\newblock {The MAGIC-II gamma-ray stereoscopic telescope system}.
\newblock {\em Nuclear Instruments and Methods in Physics Research Section A:
  Accelerators, Spectrometers, Detectors and Associated Equipment},
  623(1):437--439.

\bibitem[Vogler et~al., 2011]{vogler2011trigger}
Vogler, P., Backes, M., and Anderhub, H. (2011).
\newblock Trigger and data acquisition electronics for the geiger-mode
  avalanche photodiode cherenkov telescope camera of fact mount for the high
  energy section of the cherenkov telescope array.
\newblock In {\em 32nd International Cosmic Ray Conference, Beijing, P.R.
  China}, volume~9, pages 11--18.

\bibitem[Vogler, 2017]{vogler2015diss}
Vogler, P.~E. (2017).
\newblock {{Design and commissioning of the trigger electronics for a novel
  Geiger-mode avalanche photodiode based camera for Imaging Atmospheric
  Cherenkov Telescopes}}.
\newblock {\em ETH Zurich, Research Collection}, Department of Physics.

\bibitem[{Wakely} and {Horan}, 2008]{wakely2008tevcat}
{Wakely}, S.~P. and {Horan}, D. (2008).
\newblock {TeVCat: An online catalog for Very High Energy Gamma-Ray Astronomy}.
\newblock {\em International Cosmic Ray Conference}, 3:1341--1344.

\bibitem[Weinstein et~al., 2007]{weinstein2007veritas}
Weinstein, A. et~al. (2007).
\newblock The veritas trigger system.
\newblock {\em arXiv preprint arXiv:0709.4438}.

\bibitem[Wilburn et~al., 2005]{wilburn2005high}
Wilburn, B., Joshi, N., Vaish, V., Talvala, E.-V., Antunez, E., Barth, A.,
  Adams, A., Horowitz, M., and Levoy, M. (2005).
\newblock High performance imaging using large camera arrays.
\newblock In {\em ACM Transactions on Graphics (TOG)}, volume~24, pages
  765--776. ACM.

\bibitem[Williams et~al., 2015]{williams2015cherenkov}
Williams, David, A. et~al. (2015).
\newblock The cherenkov telescope array: The future of gamma-ray astrophysics.

\bibitem[W{\"o}hler, 2012]{wohler20123d}
W{\"o}hler, C. (2012).
\newblock {\em 3D computer vision: efficient methods and applications}.
\newblock Springer Science \& Business Media.

\bibitem[Wood et~al., 2016]{FermiPass8BroadbandSensitivity2016}
Wood, M., Caputo, R., Rando, R., Charles, E., Digel, S., Baldini, L., and
  others for~the Fermi~collaboration (2016).
\newblock Fermi {LAT} performance, {P8R2 SOURCE V6} broadband sensitivity to
  power-law sources.
\newblock
  \url{https://www.slac.stanford.edu/exp/glast/groups/canda/lat_Performance_files/broadband_flux_sensitivity_p8r2_source_v6_all_10yr_zmax100_n03.0_e1.50_ts25.png},
  accessed 2017 Mai 11.

\bibitem[{Wootten}, 2003]{wootten2003alma}
{Wootten}, A. (2003).
\newblock {Atacama Large Millimeter Array (ALMA)}.
\newblock In {Oschmann}, J.~M. and {Stepp}, L.~M., editors, {\em Large
  Ground-based Telescopes}, volume 4837, pages 110--118.

\bibitem[Yashin, 2015]{yashin2015imaging}
Yashin, I. (2015).
\newblock {Imaging Camera and Hardware of TAIGA-IACT Project}.
\newblock {\em PoS}, page 986.

\bibitem[Ye et~al., 2015]{ye2015tibet}
Ye, Q.-Z., Su, M., Li, H., and Zhang, X. (2015).
\newblock Tibet's ali: Asia's atacama?
\newblock {\em Monthly Notices of the Royal Astronomical Society: Letters},
  457(1):L1--L4.

\bibitem[Zuccon et~al., 2003]{zuccon2003atmospheric}
Zuccon, P., Bertucci, B., Alpat, B., Ambrosi, G., Battiston, R., Battistoni,
  G., Burger, W., Caraffini, D., Cecchi, C., Di~Masso, L., et~al. (2003).
\newblock {Atmospheric production of energetic protons, electrons and positrons
  observed in near Earth orbit}.
\newblock {\em Astroparticle Physics}, 20(2):221--234.

\end{thebibliography}
%------------------------------------------------------------------------------
\listoffigures
\listoftables
%------------------------------------------------------------------------------
\newpage{}
\section*{Acknowledgement}
The exciting, yet unforeseen investigations on the Cherenkov-plenoscope would not have been possible without the support of Prof. Adrian Biland, and Prof. Felicitas Pauss.
I thank both Adrian, and Felicitas for not only providing critics, but also resources and patience.
In particular, Adrian's expert-critics on optics, cosmic-rays, and the geomagnetic-cutoff were thankfully received.
I thank Max Ludwig Ahnen for enthusiastic discussions on possible observation-goals, and his inspiring thoughts on estimating Portal's sensitivity without an energy-reconstruction.
For crucial comments on project-management and programming, I want to thank Dominik Neise.
Dominik's advice on social aspects of programming, data-accessibility, and literature, became key-catalysts for Portal's simulations.
I thank Axel Arbet-Engels for his motivating thoughts and careful reasoning while approaching the Cherenkov-plenoscope.
Axel's decision to dedicate his master-thesis to the measurement of the cosmic-ray's composition using the plenoptic methods presented here was a big push for my thesis, and I am very thankful for his choice.
I want to thank Gareth Hughes for his introduction into instrument-response-functions on Cherenkov-telescopes.
I thank Andreas H. Trabesinger for his comments on a possible manuscript-submission, and I thank Werner Lustermann for discussions on possible implementations of read-out-electronics.
The comments by Maximilian Noethe, and Kai Bruegge on programming in general were gladly received.\\
I thank the department for civil, environmental and geomatic engineering at ETH-Zurich.
In our search for expert-opinions, Prof. Eleni Chatzi, Prof. Mario Fontana, Adrian Egger, Georgios Zinas, and Spyridon Daglas got interested in the Cherenkov-Plenoscope.
With ideas, time, and resources, they all contributed substantially to Portal's cable-robot-mount.
The close collaboration with Spyridon Daglas during his master-thesis on Portal's cable-robot-mount was an inspiring introduction into civil-engineering which I am very thankful for.
Rarely I met a person who could learn completely new concepts such as Cherenkov-astronomy, and new skills such as programming with the speed Spyridon Daglas did it.\\
I want to thank Prof. Werner Hofmann, Prof. James A. Hinton, and Prof. Felix Aharonian for sharing their excitement with me on the prospects of a plenoptic approach in Cherenkov-astronomy.
This excitement is what persuaded me to continue my quest in gamma-ray-astronomy, and science in general for a bit longer.
I want to thank Prof. Werner Hofmann in particular for:
First, becoming the co-examiner of this thesis although I initially confronted him with nothing but 'strong claims'.
And second, for encouraging me to present my findings on the Hillas Symposium 2018 in Heidelberg, a symposium in memory of the architect behind the Cherenkov-telescope.
Of course, I am equally thankful for the organizers of the Hillas Symposium for having me as a plenary-speaker.
Also, the critics by Prof. Wolfgang Rhode, and Dominik M. Elsaesser on my contribution for the Hillas Symposium were highly welcome.\\
In general, I want to thank ETH-Zurich for an outstanding phd-experience.
Thanks to the kind supervision of Prof. Adrian Biland, I was free to dedicate my phd-thesis to the Cherenkov-plenoscope.
Good supervisors sometimes have to protect their students from going on too risky, and too ambitious adventures.
And I thank you Adrian for helping me to find the right balance for my adventure.\\
Further, I want to thank Bettina Lareida, Rosa Baechli, Caroline Keufer, and Gabriela Amstutz for administrative support inside and outside of ETH.
I want to thank Bruno Zehr, Patrick Gomez, Ulf H. Roessler, Robert Becker, Diogo di Calafiori, Jean-Pierre Stucki, and Jan Soerensen for advice and discussions.\\
Although its not a part of this thesis, I want to thank Christian A. Monstein for fruitful discussions on a possible detection of cosmic gamma-rays using radio-emission.\\
Finally, I want to thank my family and my wife Katrin Stratmann for sharing my excitement whenever my Cherenkov-plenoscope escalated to the next higher level in the scientific community.
%
%------------------------------------------------------------------------------
\end{document}